\documentclass{emulateapj}
\slugcomment{{\sc Accepted to ApJ 10/2013} }

\newcommand{\Msun}[0]{\mbox{ M}_\odot}

\newcommand{\Zsun}[0]{\mbox{ Z}_\odot}

\newcommand{\hinv}[0]{\mbox{ }h_{\rm 100}^{-1}}
\newcommand{\etal}[0]{et al.}

\newcommand{\eqn}[1]{Equation~(\ref{#1})}	
\newcommand{\fig}[1]{Figure~\ref{#1}}		
\newcommand{\sect}[1]{\S\ref{#1}}			
\newcommand{\tabl}[1]{Table~\ref{#1}}		

\begin{document}

\title{Formation of Compact Clusters from High Resolution Hybrid Cosmological Simulations}
\author{Mark L. A. Richardson\altaffilmark{1}, Evan Scannapieco\altaffilmark{1}, and William J. Gray\altaffilmark{2}}
\altaffiltext{1}{School of Earth and Space Exploration, Arizona State University, Tempe, AZ 85287}
\altaffiltext{2}{Lawrence Livermore National Laboratory, P.O. Box 808, L-038, Livermore, CA 94550, USA}

\setcounter{footnote}{2}

\begin{abstract}

The early Universe hosted a large population of small dark matter `minihalos' that were too small to cool and form stars on their own.
These existed as static objects around larger galaxies until acted upon by some outside influence. Outflows, which have been 
observed around a variety of galaxies, can provide this influence in such a way as to collapse, rather than disperse the
minihalo gas. Gray \& Scannapieco performed an investigation in which idealized spherically-symmetric minihalos were struck
by enriched outflows. Here we perform high-resolution cosmological simulations that form realistic minihalos, which we then 
extract to perform a large suite of simulations of outflow-minihalo interactions including non-equilibrium chemical reactions.
In all models, the shocked minihalo forms molecules through non-equilibrium reactions, and then cools to form
dense chemically homogenous clumps of star-forming gas. The formation of these high-redshift clusters may be observable with the next
generation of telescopes, and the largest of them should survive to the present day,  having properties similar 
to halo globular clusters.

\end{abstract}
\maketitle

\section{Introduction}

The observed history of large-scale structure formation is well explained 
by the cold dark matter model with a cosmological constant  term (e.g., Spergel et al. 2007; Larson et al. 2011). 
This theory posits that  small-scale perturbations in the dark 
matter density merged hierarchically over time, leading to  larger perturbations that continued to coalesce into 
even larger structures, while underdensities became large voids (e.g., White \& Rees 1978; White \& Frenk 1991;
Kauffmann et al. 1993; Cole et al. 2000; Bower et al. 2006). The gas dynamics 
were almost completely dictated by these dark matter potentials. Thus, massive 
baryonic objects formed at later times, while a large population of smaller objects formed early.

At the poorly constrained redshifts before reionization, it is then expected that there existed a large population of 
small gravitationally bound clumps of dark matter and gas, whose masses were much smaller 
than galaxies today. At temperatures below $\approx 10^4$ K, transitions in atomic hydrogen and helium
are not excited, leaving gas to cool radiatively through molecular excitations and dust 
emission. Although some molecules may have survived from recombination and cooled  the 
earliest structures (e.g., Abel et al. 2002; Bromm \& Clarke 2002; Turk et al. 2009; Stacy et al. 2010), 
they would have produced stars that disassociated  these molecules, suppressing cooling in neighboring perturbations
\citep{Galli98}. Furthermore, it is unlikely that even with 
some molecules surviving further, they 
could have effectively cooled such small structures (e.g. Whalen et al. 2008; Ahn et al. 2009). Thus, the subset of these low-mass 
objects with virial temperatures below $10^4$ K, so-called ``minihalos'', persisted as sterile objects, unable to 
cool and form stars without an external influence.

Some of these minihalos may have been located near starbursting galaxies, which can drive 
outflows powered by supernovae, as have been observed around a variety
of galaxies at a range of redshifts (e.g., Lehnert \& Heckman 1996; Franx et al. 1997; Pettini et al. 1998; 
Martin 1999; Heckman et al. 2000; Veilleux et al. 2005; Rupke et al. 2005; Chung et al. 2011). 
It is expected that these observed starbursts are 
a small example of a much larger, earlier population that predated, and likely drove, reionization 
(Scannapieco et al. 2002; Thacker et al. 2002; Ferrara \& Loeb 2013). 
Although these galaxies also produced ionization fronts that disassociated  their environments, and
\citet{Shapiro04} and \citet{Illiev05} have demonstrated that
minihalos that are first struck by the ionization fronts are evaporated on a timescale of 10-100 Myr,
\citet{Fujita03} showed that the outflows of such galaxies can trap this ionizing radiation, shadowing 
regions around the starbursting galaxies. Thus, some neutral minihalos could have been sheltered from 
ionization fronts, and interacted with a kinematic outflow first. 
Furthermore such interactions, through non-equilibrium processes,
could have induced  the formation of molecular gas  that could cool the gas sufficiently to  induce star-formation. 
 
Gray \& Scannapieco (2010; 2011A; 2011B) (hereafter GS10, GS11A, GS11B) performed idealized 
simulations of this interaction, which featured a spherical isothermal gas cloud embedded
in a static analytic dark matter potential. Together the dark matter and gas represented a minihalo, which 
followed an NFW radial density profile \citep{Navarro97}, and it was embedded in a uniform background. 
The starburst outflow, on the other hand, was modeled as a plane-parallel shock of
material inflowing from the $x$ boundary. GS10 also enacted a 14-species 
primordial non-equilibrium chemical network with associated cooling terms. As the shock struck the
minihalo, it catalyzed the creation of H$_{\rm 2}$, while mixing in some of its enriched material. As much
as $100\%$ of the baryonic material of the minihalo collapsed into a small ribbon of gas that 
extended well out of the dark matter potential, that then cooled via H$_{\rm 2}$. For one simulation they included
a UV background, assuming an optically thin medium, and found a reduced abundance of H$_2$, but still sufficient
to cool the minihalo gas. Regardless of a UV background, they found that the ribbon
eventually collapsed into several distinct clumps, whose properties were remarkably similar to present-day
halo globular clusters. GS11A enacted a $K$-$L$ two-equation sub-grid turbulence module and metal-line cooling, which allowed 
for more efficient mixing of enriched shock material into the collapsed minihalo gas, and allowed this enriched
material to add to the net cooling. GS11A found very similar
results as GS10, producing a population of clusters very much like halo globular clusters. Finally, GS11B
performed a parameter suite that looked into the effect of minihalo mass, clumping factor, and angular momentum, 
outflow energy and enrichment, minihalo-starburst separation, redshift, and UV background on the characteristics
of the interaction and its resulting clusters. 

Cosmological simulations show that virialized structures are typically found at the nodes of a fractal-like 
cosmic web (e.g.\ Springel  \etal\ 2005A). Thus minihalos were found at the intersections of cosmic filaments, likely with 
higher-mass objects nearby. Their dark matter was a dynamic background, that responded to gas dynamics. Although
the average radial profile of minihalos likely did follow an NFW profile, they were not perfectly isotropic. These
characteristics make minihalos quite different than the idealized gas clouds simulated in GS10, GS11A, and GS11B. 

In fact, the isotropy in their work may have been the cause of the small collapsed ribbon of material along the $x$-axis,
resulting from the shock wave converging at the  antipodal point. 
It is also unclear how reasonable it is to treat the dark matter as a static analytic term.
As densities in the collapsing gas eventually exceeded those in the dark matter, perhaps a more dynamic treatment 
of the dark matter might uncover motions that significantly affect the future evolution of the gas.

In this work, we address these issues by generating a range of minihalos in high-resolution cosmological simulations. Then, having isolated
the desired objects and their immediate environments into a new simulation volume, we simulate their interactions 
with starburst-driven outflows. In this way we are able to understand this interaction in much more detail
 as it  occurred in the early Universe, and contrast it with the idealized minihalo-outflow simulations previously undertaken.

The structure of this paper is as follows. In \sect{codes} we describe the cosmological
simulations, and
the outflow-minihalo interaction simulations, followed by our post-processing techniques. In \sect{Results} we discuss the results from our parameter suite, and in \sect{images} we present simulated high and low-redshift observables derived from these results.  We summarize our work and 
give conclusions in \sect{Conc}. Throughout this paper we use 
($\Omega_{\Lambda}, \Omega_{\rm M}, \Omega_{\rm b}, n, \sigma_{\rm 8}, \hinv$) = (0.734, 0.266, 0.0449, 0.963, 0.801, 0.71) 
\citep{Larson11}. \vspace{15mm}

\section{Numerical Methods}\label{codes}
\subsection{Particle Simulations}\label{part}
As a first step, we performed a low-resolution cosmological simulation using the smoothed-particle 
hydrodynamic code GADGET-2 (Springel et al. 2001; Springel 2005B) in a box that was 2.57 comoving 
Mpc on a side. This  had 2 spherically nested resolution levels, the lowest resolution level spanning the whole volume, and the next level
 spanning a sphere centered in the box with a radius a quarter of the box size. Each level
had effectively 192$^3$ dark matter particles (with 
masses of $7.39 \times 10^{4}\Msun$ and $9.24 \times 10^{3}\Msun$) and 192$^3$ gas particles 
(with masses of $1.51 \times 10^4\Msun$ and $1.87 \times 10^3\Msun$). At 
each level of resolution the modes of the initial spectrum are truncated to ensure there is no aliasing of high $k$ 
modes into the low $k$ values (as in Thacker \& Couchman 2000; Richardson \etal\ 2013). Initial conditions were 
generated using the transfer function from CAMB \citep{Lewis00}, assuming an initial spectral slope of 
$n=0.963$. CAMB uses a line-of-sight implementation of the linearized equations of the covariant approach to 
cosmic microwave background (CMB) anisotropies. This results in different transfer functions for the dark matter and 
baryon components, with the two weighted together to yield the total transfer function.
This simulation began at $z=199$ and was evolved to $z=15.4$. We did not include star formation 
or feedback, but did include atomic and molecular cooling. We calculated the different ionization states of 
hydrogen and helium from the density and temperature following \citet{Katz96}. Although at such a high-redshift 
collisional ionization equilibrium is perhaps not appropriate, this assumption is irrelevant since the material remains neutral.
Additionally, the structures on which we focus are not particularly rare and have virialized only recently. Rarer objects,
with larger peak densities that virialize at earlier times and are then essentially inert require non-equilibrium chemistry to allow for
the build up of molecular hydrogen. Instead, our objects have lower densities and have had less time to accumulate H$_2$, and thus
the non-equilibrium formation of 
H$_2$ should not be a dominant component of their evolution. This is consistent with \citet{Richardson13}, who compared halos formed in
SPH simulations, which used the same equilibrium chemistry, with the same halos formed in AMR simulations, which consider a fully non-equilibrium chemistry
of H, He, and D.
We assumed a primordial number density fraction of molecular hydrogen of  H$_2$/H $= 1.1\times10^{-6}$ 
following \citet{Palla00} and  a primordial deuterium number density fraction of D/H $= 2.7\times 10^{-5}$ 
following \citet{Steigman09}. Using the baryon-to-light ratio of $\eta = 6.0 \times 10^{-10}$ from 
\citet{Steigman09}, we set the deuterated hydrogen number density fraction at HD/H$_2 = 6\times10^{-4}$ from 
\citet{Palla00}. Given these abundances, we employed the molecular cooling rates of GS10 and cooling rates for 
Compton scattering against CMB photons as given in \citet{Barkana01}. To offset runaway cooling since we do 
not include feedback terms, we implement a cooling temperature floor at 500 K, well below the virial temperature 
of any minihalos we consider. Adiabatic cooling can still cool below this floor.

We then used the friends-of-friends algorithm \citep{Davis85} to determine groups that corresponded to an 
overdensity of 180 with a linking length of 1.19 kpc, expected to be virialized from a spherical top-hat collapse model.
We focused on three groups, with total masses 
of 2.0$\times 10^6 \Msun$, 4.0$\times 10^6 \Msun$, and 8.0$\times 10^6 \Msun,$ respectively. For each group, 
we performed a simulation statistically identical to the low-resolution simulation, but with additional resolution 
centered on the group. We added two additional spherically-nested resolution levels resulting in particle masses 
of $144\Msun$ and $29.3\Msun$ for dark matter and baryons, respectively, at the highest resolution.

These high-resolution initial conditions were evolved to $z=14$. 
We then located groups using the HOP group finder \citep{Eisenstein98} implemented in the yt visualization and analysis toolkit \citep{Turk11} with masses, in units of $10^6 \Msun$, of $M_{\rm 6}$ = 0.716, 1.38, 2.33, 2.72, 7.17, and 18.7. 
For each suitable halo, we isolated regions out to 5 virial radii, $r_{\rm v}$, of the group in the GADGET-2 snapshots and mapped these to the adaptive mesh 
refinement code FLASH3.2 \citep{Fryxell00} using the procedure discussed in \citet{Richardson13}. A summary of the simulations is given in \tabl{tab_sims}. 
For the $z \leq14$ simulations, we extended the $z=14$ datasets by scaling position, velocity and energy by the scale factor difference. The majority of these 
simulations were done at $z=8$ as it is the most likely to be directly observable.

\begin{deluxetable*}{lccccccccccc}[t!]
\tabletypesize{\scriptsize}
\tablewidth{0pc}
\tablecaption{Simulations Summary
\label{tab_sims}}
\startdata
\multicolumn{2}{c}{SPH Parameters:} & SPH $z_{\rm init}$ & SPH $z_{\rm f}$ & SPH $m_{\rm DM} $ &  SPH $m_{\rm gas}$ & Box size \\ \hline
                                     &                            &199                           & 14                          & 144 $\Msun$                                           & 29.3 $\Msun$ & 2.57 Mpc \\ \hline \hline
AMR Model        & $M_{\rm 6}$   & $r_{\rm vir}$ (pc)    & $\Delta x$ (pc)     & Orientation     & $v_{\rm s}$ (km s$^{-1}$) & $\mu_{\rm s}$ & $E_{\rm 55} $    & $ \sigma_{\rm 5}$   & $z$    & $Z (\Zsun)$ & $J_{21}$ \\ \hline \hline
FID                               & 2.72            & 505                      & 5.92                       &  Filament                     &   226         &      60.4    & 10.0                        & 2.62                    & 8      & 0.12  &  0.0 \\  \hline
LR                               & 2.72            & 505                      & 11.8                       &  Filament                      &   226         &      60.4    & 10.0                        & 2.62                    & 8      & 0.12  &  0.0 \\  
HR                               & 2.72            & 505                      & 2.96                       &  Filament                      &   226         &      60.4    & 10.0                        & 2.62                    & 8      & 0.12  &  0.0 \\  \hline 
PO1                             & 2.72            & 505                      & 5.92                       &     IGM                           &   226         &      60.4    & 10.0                       & 2.62                    & 8       & 0.12  &  0.0 \\ \hline 
PM07                          & 0.716    	     & 320.    	          & 3.75                       &  Filament                      &   226         &      60.4    & 10.0                        & 2.62                    & 8      & 0.12  &  0.0 \\  
PM1                             & 1.38            & 402	                   & 4.71                       &  Filament                      &   226         &      60.4    & 10.0                        & 2.62                    & 8      & 0.12  &  0.0 \\  
PM2                             & 2.33            & 470.                     & 5.51                       &  Filament                      &   226         &      60.4    & 10.0                        & 2.62                    & 8      & 0.12  &  0.0 \\
PM7                             & 7.17            & 693                      & 8.12                       &  Filament                      &   226         &      60.4    & 10.0                        & 2.62                    & 8      & 0.12  &  0.0 \\  
PM19                           & 18.7            & 967                      & 11.3                       &  Filament                     &   226         &      60.4     & 10.0                        & 2.62                    & 8      & 0.12  &  0.0 \\  \hline 
Pv75                            & 2.72            & 505                      & 5.92                       &  Filament                     &   75.0         &      60.4     & 30.2                        & 7.91                    & 8      & 0.12  &  0.0 \\  
Pv125                         & 2.72            & 505                      & 5.92                       &  Filament                     &   125         &      60.4     & 18.1                        & 4.74                    & 8      & 0.12  &  0.0 \\  
Pv340                         & 2.72            & 505                      & 5.92                       &  Filament                    &   340.         &      60.4      & 6.66                        & 1.74                    & 8      & 0.12  &  0.0 \\  
Pv510                         & 2.72            & 505                      & 5.92                       &  Filament                    &   510.         &      60.4      & 4.44                        & 1.16                    & 8      & 0.12  &  0.0 \\  \hline 
P$\mu$3                    & 2.72            & 505                      & 5.92                       &  Filament                    &   226         &      32.5      & 1.54                        & 1.14                    & 8      & 0.12  &  0.0 \\  
P$\mu$8                    & 2.72            & 505                      & 5.92                       &  Filament                    &   226         &     77.5       & 4.59                        & 3.35                    & 8      & 0.12  &  0.0 \\  
P$\mu$9                    & 2.72            & 505                      & 5.92                       &  Filament                     &   226         &      90.0     & 5.33                        & 3.89                    & 8      & 0.12  &  0.0 \\  \hline 
Pz10                           & 2.72            & 413                      & 4.84                       &  Filament                    &   226         &      90.7      & 10.0                        & 3.92                    & 10      & 0.12  &  0.0 \\  
Pz14                           & 2.72            & 303                      & 3.55                       &  Filament                     &   226         &      169     & 10.0                        & 7.30                    & 14      & 0.12  &  0.0 \\  \hline 
PZ005                          & 2.72            & 505                      & 5.92                       &  Filament                   &   226         &      60.4       & 10.0                        & 2.62                    & 8      & 0.005  &  0.0 \\  
PZ05                            & 2.72            & 505                      & 5.92                       &  Filament                   &   226         &      60.4       & 10.0                        & 2.62                    & 8      & 0.05  &  0.0 \\  
PZ5                              & 2.72            & 505                      & 5.92                       &  Filament                   &   226         &      60.4       & 10.0                        & 2.62                    & 8      & 0.5  &  0.0  \\ \hline
PJ01                               & 2.72            & 505                      & 5.92                       &  Filament                     &   226         &      60.4    & 10.0                        & 2.62                    & 8      & 0.12  &  0.1 
\enddata
\tablenotetext{}{
\textbf{Notes.} $M_{\rm 6}$ is the minihalo mass in units of $10^6 \Msun$. $r_{\rm vir}$ is the virial radius of the minihalo in units of (physical) pc. $\Delta x$ is the
resolution limit at the highest refinement level in units of (physical) pc. The orientation describes whether the shock travels along a filament or from the lower density
intergalactic medium (IGM). $v_{\rm s}$ is the shock velocity in units of km s$^{-1}$. $\mu_{\rm s}$ is the shock momentum per unit area, in units of $\Msun$ pc$^{-1}$ Myr$^{-1}$. $E_{\rm 55}$ is the energy of the shock in units of $10^{55}$ erg. $\sigma_{\rm 5}$ is the surface density of the shock as it enters the box 
in units of (physical) $10^5 \Msun$ kpc$^{-2}$. $z$ is the redshift of the interaction. In the event that $z \ne 14$ we translate it to the lower redshift by expanding
the $z=14$ minihalo and cooling it. $Z$ is the metallicity of the incoming shock in solar units. $J_{21}$ is the background UV flux at the Lyman limit, with $J(\nu_{\alpha}) = J_{21} \times 10^{-21}$ erg s$^{-1}$ cm$^{-2}$ Hz$^{-1}$ Sr$^{-1}$.}
\end{deluxetable*}

\subsection{AMR Simulations}\label{grid}

\subsubsection{FLASH and the Dark Matter Gravitational Potential}

The mapped GADGET-2 snapshots were treated as initial conditions for simulations performed with FLASH version 3.2, a publicly-available multidimensional adaptive mesh refinement hydrodynamic code  \citep{Fryxell00} that solves the Riemann problem on a Cartesian grid using a directionally-split piecewise parabolic method (PPM) (Colella \& Woodward 1984; Colella \& Glaz 1985; Fryxell \etal\ 1989).  
FLASH also includes particles that we used to represent dark matter, which we simply moved directly from GADGET to FLASH.
Dark matter mass is mapped to the grid after each hydro step using the Cloud-In-Cell (CIC) method (e.g., Birdsall \& Fuss 1997), where the 
particles are assumed to exist in a box of the same size as
the grid at the current refinement. The mass is distributed equally over this box
and thus the percent of mass mapped into a particular cell is the percentage of
the box that overlaps that cell.

To accommodate high-resolution regions with very few  particles, we sometimes
moved dark matter particles to a lower resolution level, mapping them over 
a larger box. This was necessary to prevent any low-density gas in the high-refinement 
region from unrealistically collapsing onto a single dark matter particle, and similar efforts have been made in earlier
work (e.g., Safranek-Shrader et al. 2012, 2013). 
To determine which particles needed this `derefinement', a particle count was monitored for each cell. 
If a cell had three or more neighboring cells that did not contain particles, or if it had two neighboring cells without particles 
in the same direction,  then the particles in that cell were flagged for derefinement, and the same criteria was checked for 
the parent cell.  This criteria was slightly relaxed for particles in cells adjacent to block boundaries, which were always tagged for
derefinement if the neighboring block was at a lower refinement level. This 
was necessary to be consistent with the criteria used by the cells on the lower resolution block to determine if its
neighbors had particles, and it is 
illustrative of the fact that at a lower refinement level a larger fraction of the box associated with the particles  would overlap the neighboring cell. 

\begin{figure}[b!]
\centering
\includegraphics[scale=0.35]{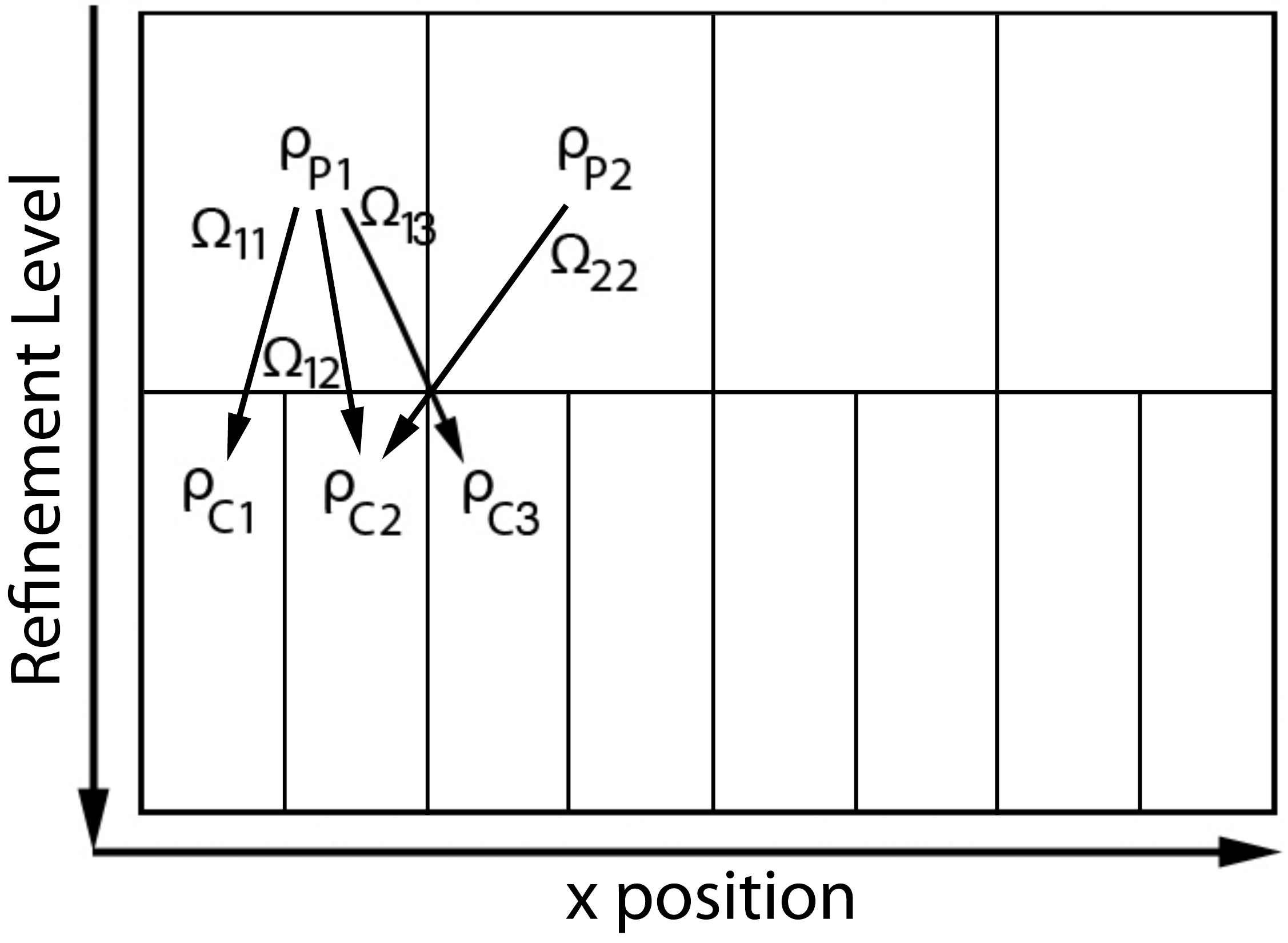}
\caption{\footnotesize{Illustration of our gravity procedure in one dimension. A fraction $\Omega_{\rm ij}$ of the density from parent PI, $\rho_{\rm Pi}$, is prolonged to child CJ and added to its
existing density values, which were mapped from less derefined particles. \vspace{2mm}}}
\label{fig_par_chi}
\end{figure}

\begin{figure*}[t!]
\centering
\includegraphics*[scale=0.55, trim=0 50.5 0 0]{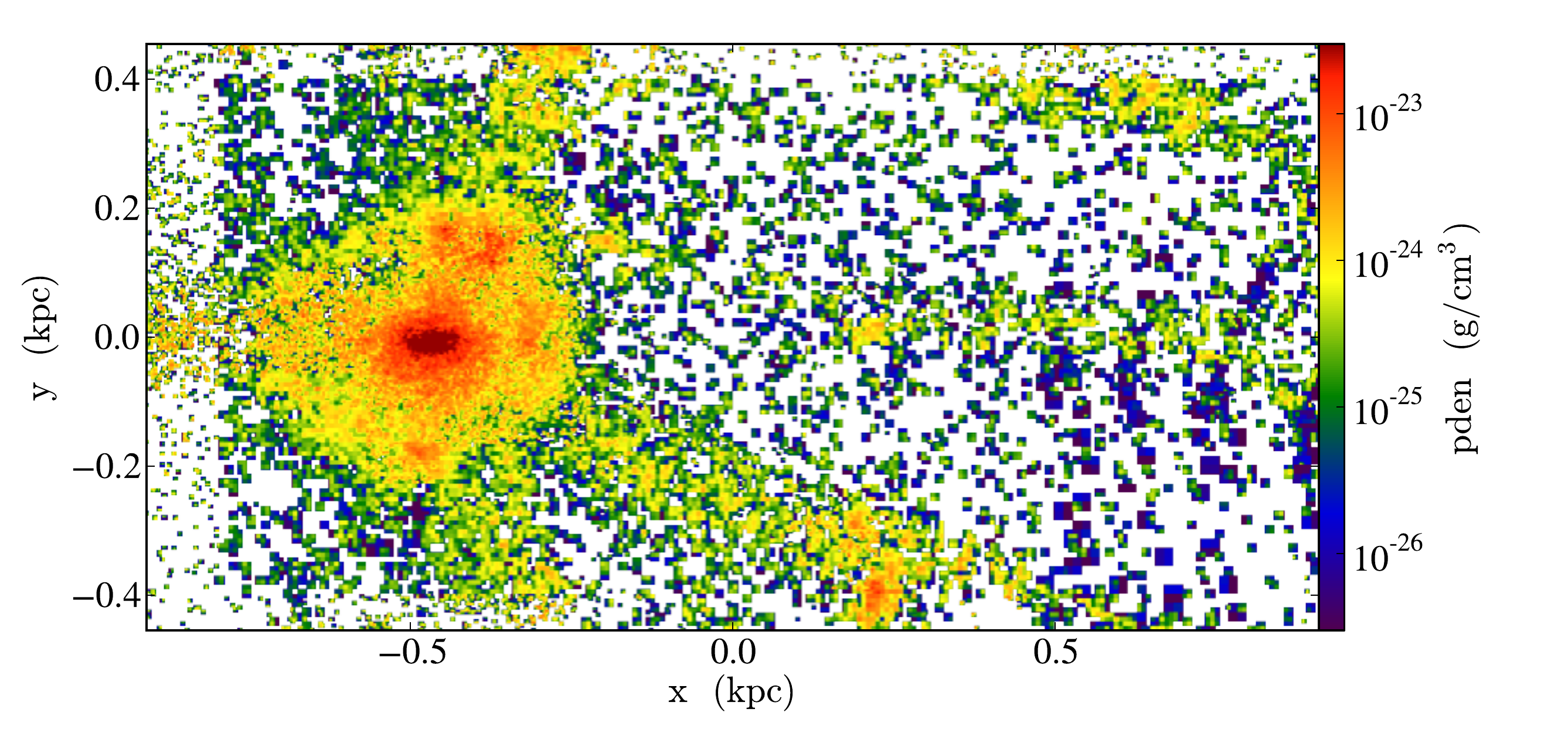}
\includegraphics*[scale=0.55, trim=0 0 0 21.8]{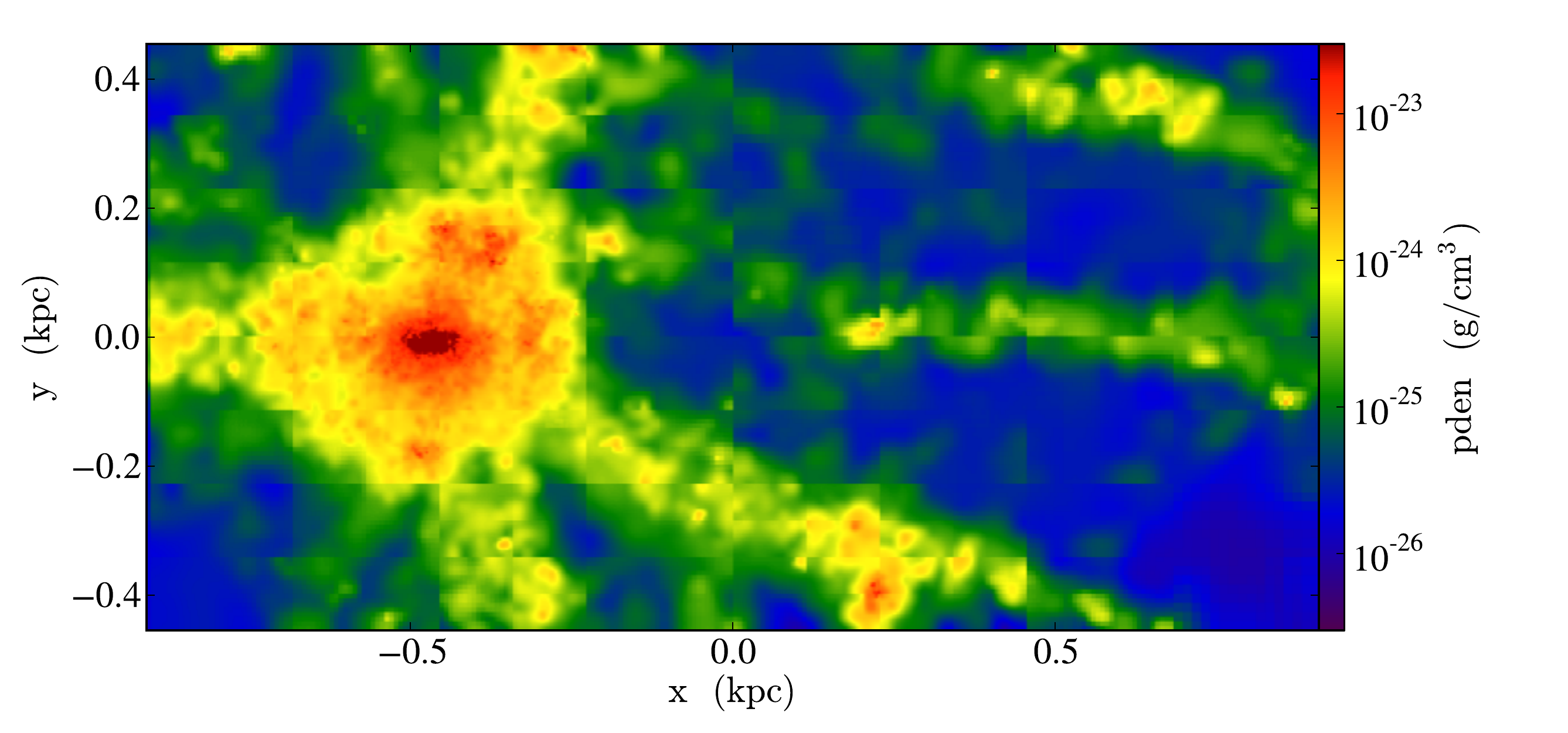}
\caption{\footnotesize{Comparison of the density-weighted projection of the dark matter density field, pden, of the minihalo in the original non-derefined particle gravity routine (top)
and with the new derefinement method that smooths the dark matter field (bottom). Regions where there were few dark matter particles 
moved those particles to lower resolution blocks, increasing the size of their mapped region. Block boundaries are not continuous to
limit the number of inter-processor communications. These projections were made using the yt-toolkit \citep{Turk11} (http://yt-project.org/).}}
\label{fig_dm}
\end{figure*}

In all cases,  once particles were derefined and mapped to that grid level using the standard CIC method, this material must then be `prolonged' 
back onto the high-resolution level, from parent to (higher-resolution) child, and added to any density that was directly mapped to that child.  
The prolongation of the dark matter density into a child cell that was not near the boundary of the block, such that it could see both its parent
cell and its parent's neighbors, was determined by using linear interpolation between this cell's parent density and its closest neighbor's density. A quadratic interpolation scheme could have been used, but in rare circumstances this could lead to negative dark matter mapped in some
cells. \fig{fig_par_chi} shows a plot of the grid layout of a child and parent block with one dimension. If we assume the
density is linearly continuous between parents P1 and P2, then:
\begin{eqnarray} \label{prolong}
\rho_{\rm C2}^{\rm final} &\equiv& \rho_{\rm C2}^{\rm map} + \rho_{\rm C2}^{\rm prolong} \\
\nonumber &\equiv&
\rho_{\rm C2}^{\rm map} + \Omega_{12} \rho_{P1}  + \Omega_{22}\rho_{P2}  \\
\nonumber &=& \rho_{\rm C2}^{\rm map} + \frac{3}{4}\rho_{P1} + \frac{1}{4}\rho_{P2},
\end{eqnarray}
where $\Omega_{\rm ij}$ is the weighting from parent $i$ into child $j$, $\rho_{\rm C2}^{\rm map}$ is the dark matter density mapped into child C2 without requiring derefinement, $\rho_{\rm C2}^{\rm prolong}$ is the amount of dark matter density prolonged into child C2 from derefined particles, and 
$\rho_{\rm C2}^{\rm final}$ is the sum of the mapped and prolonged dark matter density, which is used for the gravity calculation.
At the borders of blocks, where the child cell could not see one of its parent's neighbors, 
we then used a direct mapping from the parents value, 
\begin{eqnarray}\label{prolong2}
\rho_{\rm C1}^{\rm final} &\equiv& \rho_{\rm C1}^{\rm map} + \rho_{\rm C1}^{\rm prolong} \\
\nonumber &\equiv& \rho_{\rm C1}^{\rm map} + \Omega_{11}\rho_{\rm P1} \\
\nonumber&=& \rho_{\rm C1}^{\rm map} + \rho_{\rm P1}.
\end{eqnarray}
This limited the demand of
inter-processor communication, and resulted in a slight error in the dark matter field on the borders of blocks where dark matter was derefined (see Figure \ref{fig_dm}). However, this error was much smaller than the error from leaving the dark matter mass at the highest refinement level, and the method still allowed for a much more efficient runtime. Ideally we would like to have filled the guard cells of parent P1 and 
passed that information to child C1, requiring one interprocessor communication, but in the event that the neighbor
of parent P1 was itself a child block, then we would have been unable to pass the acceleration of child C1 back to the parent's
neighbor when we calculated the particle accelerations. We thus opted for a less complex weighting on the boundaries.
The benefit of using this new technique for handling the dark matter density field is clearly illustrated in Figure \ref{fig_dm},
which shows that individual particles were never mapped to isolated peaks in the dark matter density field.

To cancel out the self-gravitational force from prolonged pieces of the same dark matter particle, the accelerations was 
`restricted' up to the particle's refinement level using identical weightings as the prolongation (see Equations
\ref{prolong}, \ref{prolong2}). Thus:
\begin{eqnarray}
\mathbf{a}_{\rm P1} &\equiv&  \frac{1}{2}(\Omega_{11}\mathbf{a}_{\rm C1} + 
\Omega_{12}\mathbf{a}_{\rm C2} + \Omega_{13}\mathbf{a}_{\rm C3} ) \\
\nonumber &=& \frac{1}{2}(\mathbf{a}_{\rm C1} + 
\frac{3}{4}\mathbf{a}_{\rm C2} + \frac{1}{4}\mathbf{a}_{\rm C3} ),
\end{eqnarray}
 and then the weighting from the Cloud-in-Cell stage was applied to the acceleration from each
cell onto the particle at the same grid level at which it was mapped. This resulted in an exact cancelation
of the self-gravitational force, leaving only the gravitational acceleration from the gas and remaining dark matter particles.

\begin{figure}[h!]
\centering
\includegraphics[scale=0.55]{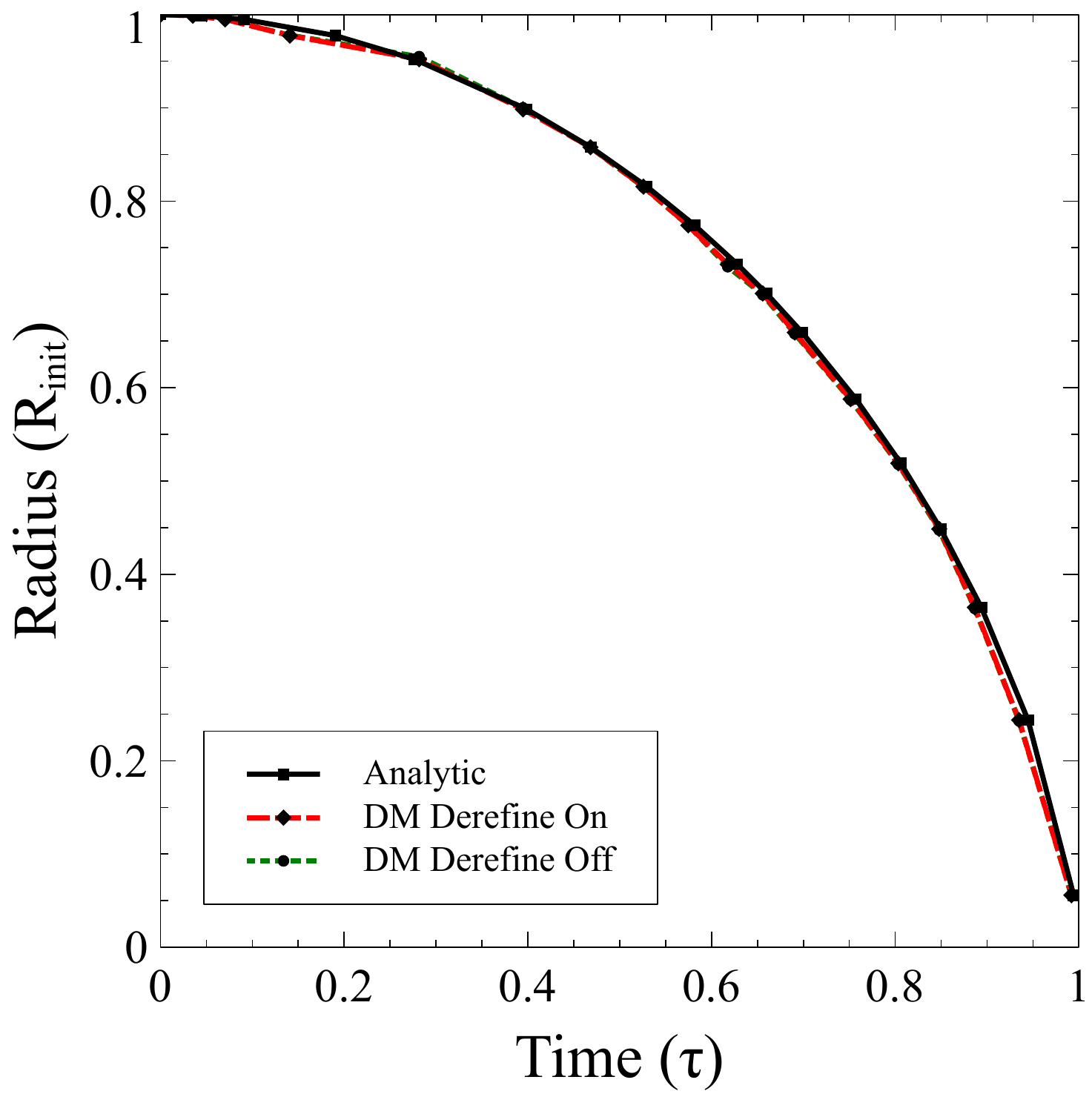}
\caption{\footnotesize{Evolution of the radial extent of the overdensity with time. The solid black line with square points shows the analytic solution while the dotted green line
with circles shows the simulated result from FLASH without derefining the dark matter, and the red dashed line with diamonds is the result with derefining the dark matter.
We see no significant differences between the results of flash regardless of whether the dark matter is derefined or not, and they are consistent with the analytic solution. }}
\label{fig_testSmooth}
\end{figure}

We tested the new particle gravity by tracing the evolution of a pressureless spherical region with a 
density 8 times the background density, the results of which are
shown in \fig{fig_testSmooth}. 
The radius begins to collapse slowly, then accelerates. 
Our new results are indistinguishable from both the results of the earlier particle gravity scheme and  the analytic solution.

\subsubsection{Outflow Simulations}\label{amr_out}

Using FLASH and the selected GADGET groups,
we performed a parameter study on outflows 
interacting with these groups. We mapped the group region into a rectangular box with the group centered at (0,0,0) pc, with $(-1.5r_{\rm v} \le x \le 4.5r_{\rm v})$ and 
 $(-1.5r_{\rm v} \le y,z \le 1.5r_{\rm v})$.
FLASH is an AMR code with blocks of $N_{\rm B,X} \times N_{\rm B,Y} \times N_{\rm B,Z}$ cells divided between processors. Each subsequent level in refinement increases the
spatial resolution by a factor of two in each dimension. Our simulations were run with a maximum of 6 levels of refinement (except for the 
high and low resolution runs), allowing up to 32 blocks in each direction, where the relative difference in resolution between a block and any of its neighbors can be at most a factor of two. For our simulations we used a root grid of $N_{\rm B,X} = 2N_{\rm B,Y} = 2N_{\rm B,Z}=16$, 
accommodating the rectangular box size while maintaining uniform resolution in all directions. This gives a maximum 
resolution of 256 cells per 3$r_{\rm vir}$.

The outflow  with positive \emph{x}-velocity was added to the \emph{x}-boundary, following the  model of GS10. 
A Sedov-Taylor solution was used to estimate the conditions of the galactic outflow.
We decided on appropriate shock velocities, $v_{\rm s}$, and shock surface momentum (per unit area), $\mu_{\rm s}$, and used these to constrain the conditions of the outflow.
We let the initial input energy for the shock be given by $E = \epsilon E_{\rm 55} (10^{55}$ erg $)$, where $E_{\rm 55}$ 
is the energy of the SNe driving the winds in units of $10^{55}$ erg, and the wind efficiency, $\epsilon$, quantifies the coupling between the SNe
and the winds, taken to be $0.3$.  To be consistent with previous work,
we assumed that before reaching our box, the shock had swept up an ambient medium
of material with an overdensity, $\delta_{44}$, in units of 44 times the background density, taken here to be 1, 
over the (physical) separation distance, $R_{\rm s}$, which leads to a surface density, $\sigma_{\rm 5}$, in units of $10^5 \Msun/$kpc$^2$. Thus, for a given shock
velocity and momentum we have: 
\begin{equation}\label{rs}
R_{\rm s} = \left(\frac{13.4 \delta_{44}^{-1}\mu_{\rm s} \mbox{ kpc}}{\Msun \mbox{ pc}^{-1}\mbox{ Myr}^{-1}}\right)\left(\frac{\mbox{km s}^{-1}}{v_{\rm s}}\right)\left(\frac{9}{1+z}\right)^{3},
\end{equation}
\begin{equation}\label{sigma5}
\sigma_5 = \left(\frac{9.77\mu_{\rm s}}{\Msun \mbox{ pc}^{-1}\mbox{ Myr}^{-1}}\right)\left(\frac{\mbox{km s}^{-1}}{v_{\rm s}}\right),
\end{equation}
and
\begin{eqnarray}\label{E55}
E_{55} = 3.04 \times 10^{-3} \epsilon^{-1} \delta_{44}^{-2} \times ... \ \ \ \ \ \ \ \ \ \ \\
\nonumber ... \left(\frac{\mu_{\rm s}}{\Msun \mbox{ pc}^{-1}\mbox{ Myr}^{-1}}\right)^3\left(\frac{\mbox{km s}^{-1}}{v_{\rm s}}\right) \left(\frac{9}{1+z}\right)^{6}.
\end{eqnarray}
The post-shock temperature is $T_{\rm s} = 1.4 \times 10^5 (v_{\rm s}/100 \mbox{ km s}^{-1})$ K, the outflow is fully ionized, and it has an abundance given by $Z$. The fiducial 
values are chosen to match \citet{Scannapieco04}, with $z=8$, $Z=0.12\Zsun$, $v_{\rm s}= 226$ km s$^{-1}$, $\mu_{\rm s} = 60.7 \Msun$ pc$^{-1}$ Myr$^{-1}$, leading to $R_{\rm s} = 3.6$ kpc, 
$E_{\rm 55}=10$, and $\sigma_{\rm 5} = 2.62$, and oriented such that the outflow is propagating along an accretion lane, which is the most likely direction pointing towards a neighboring starbursting galaxy. The shock lifetime is estimated from $\sigma_{\rm s} = \rho_{\rm post}v_{\rm post}t_{\rm s}$, where $\sigma_{\rm s}$
is the surface density of the shock, $\rho_{\rm post}$ is the post-shock density, $v_{\rm post}$ is the post-shock velocity, and $t_{\rm s}$ is the shock lifetime.
After 40\% of the shock time, the shock was tapered off by slowly lowering the density and raising the temperature, keeping pressure constant to inhibit further refinement. The density 
falls off exponentially, with 
\begin{equation}\label{rhos}
\rho(t>0.4t_{\rm s}) = \rho_0(0.01 + 0.99e^{-(t-0.4t_{\rm s})/(0.6t_{\rm s})}),
\end{equation}
while the temperature increases as $1/\rho(t)$ to maintain a constant post-shock pressure.

We allowed for refinement based on the second derivative of density and pressure. Regions that
had their density or pressure profiles vary sufficiently quickly were forced to increase their resolution by a factor of two. After 7 Myr we forced derefinement beyond a cylinder with a radius of 0.8$R_{\rm v}$ aligned with the $x$-axis for material below $3\times10^{-26}$ g cm$^{-3}$. This was to prevent low-density mixing from limiting the time-step.

During the AMR simulations, we used the same cooling source terms as in the SPH runs, although we also accounted for possible non-equillibrium chemistry 
and corresponding cooling following GS10, itself based on \citet{Glover08}. This  is essential as the interaction between the outflow and the minihalo leads to the creation of coolants 
such as H$_{\rm 2}$ and HD. This network follows 84 reactions including the three states of atomic hydrogen (H, H$^+$, H$^-$), atomic deuterium (D, D$^+$, D$^-$), and atomic helium (He, He$^+$, He$^{++}$), two states of molecular hydrogen
(H$_2$, H$_2^+$), and molecular deuterated hydrogen (HD, HD$^+$), and electrons ($e^-$). We do not consider three-body reactions
as these require much denser media than we expect in this work. Reactions that involve free elections are treated in the case B regime. This assumes a medium
optically thick to ionizing photons produced during recombination. GS10 demonstrated that all of their clouds were in the optically thick case. Similarly, our clouds
should be optically thick as well.

For one simulation we include photo-disassociation by a UV background. We model the UV background as originating 
from a $10^5$ K blackbody, and we quantify its intensity at the Lyman limit, with $J(\nu_\alpha) = J_{21}\times10^{-21}$  erg s$^{-1}$ cm$^{-2}$ Hz$^{-1}$ Sr$^{-1}$. We would 
expect dense regions with significant amount of HD and H$_2$ to self-shield from this background, however we assume all regions are optically thin, thus the UV background's 
impact will be viewed as an upper limit to the effect of this background. Including photo-disassociation expands our reaction network to 91 reactions.
 
We also account for metal cooling following GS11A, which can occur once the material is enriched by the outflow. The 
radiative metal cooling assumes an optically thin medium, using the tabulated results from \citet{Weirsma08}, which assumes local thermodynamic equilibrium.

With this package in place, along with the dark matter gravitational potential calculated as described above,
we ran the shock-minihalo simulations until the outflow had collapsed the minihalo material, typically leaving a ribbon of material behind the minihalo. This interaction took between a few million years, and a few tens of million years.

\subsubsection{Ballistic Evolution}\label{ballistic}

The timescale of the evolution of this ribbon of material can be up to ten times as long as the timescale of the shock crossing the simulation volume (GS10). To
avoid modeling this material for such a long period of time, we simplified its evolution by constructing a series of 64 one-dimensional
point-particles,  we call ``cloud particles", whose $x$-positions spanned the space of the ribbon. 
We then added the mass, momentum and metal abundance
of each ribbon segment into a corresponding particle if
it was within half the inter-particle spacing in the $x$-direction from this particle, it  was
denser than 150\% the post-shock density,  and  its radial velocity away from the 
$x$-axis was sufficiently slow as to make it gravitationally bound to this axis on a timescale of 200 Myr. 

\begin{figure*}[t!]
\centering
\includegraphics*[scale=0.23, trim=  0 64.1 163.5   21]{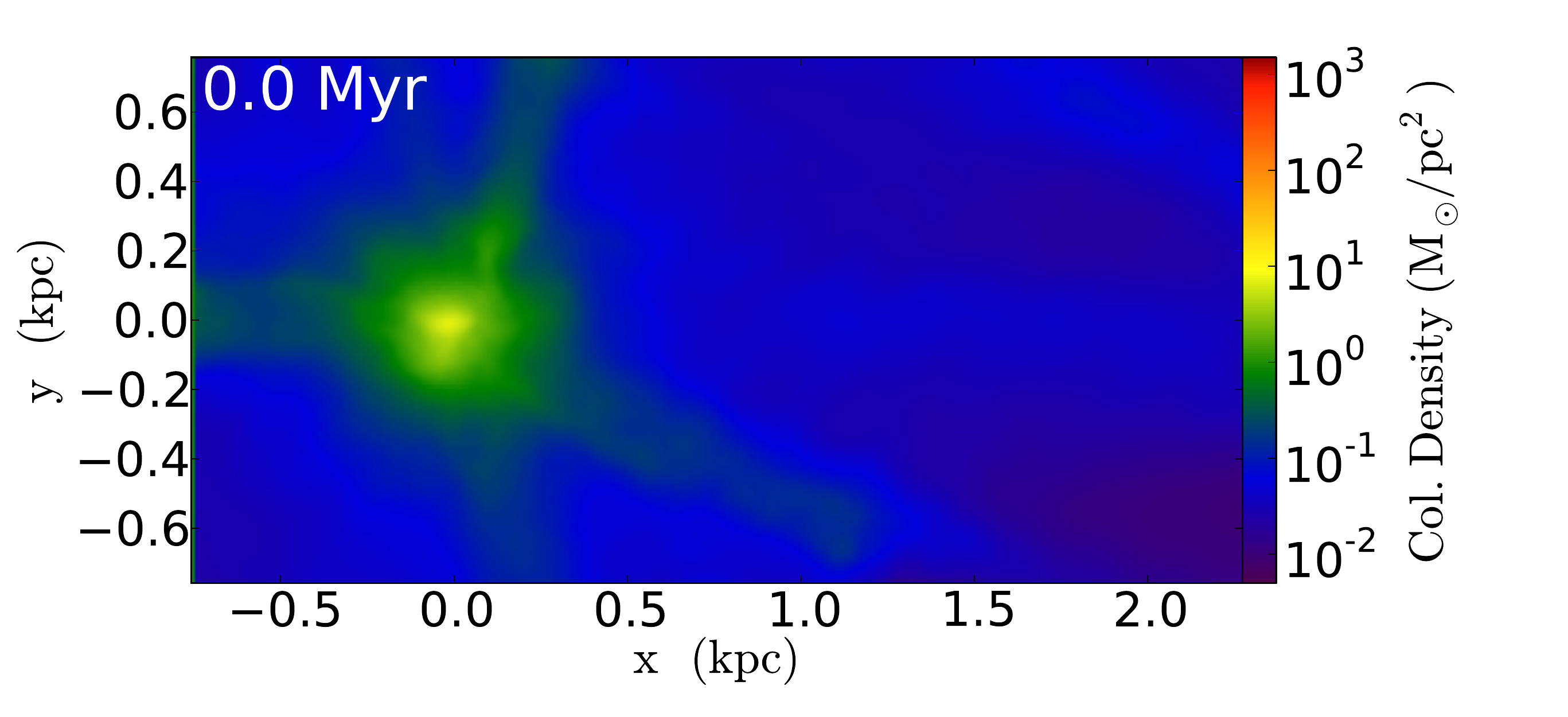}
\includegraphics*[scale=0.23, trim=95 64.1 163.5   21]{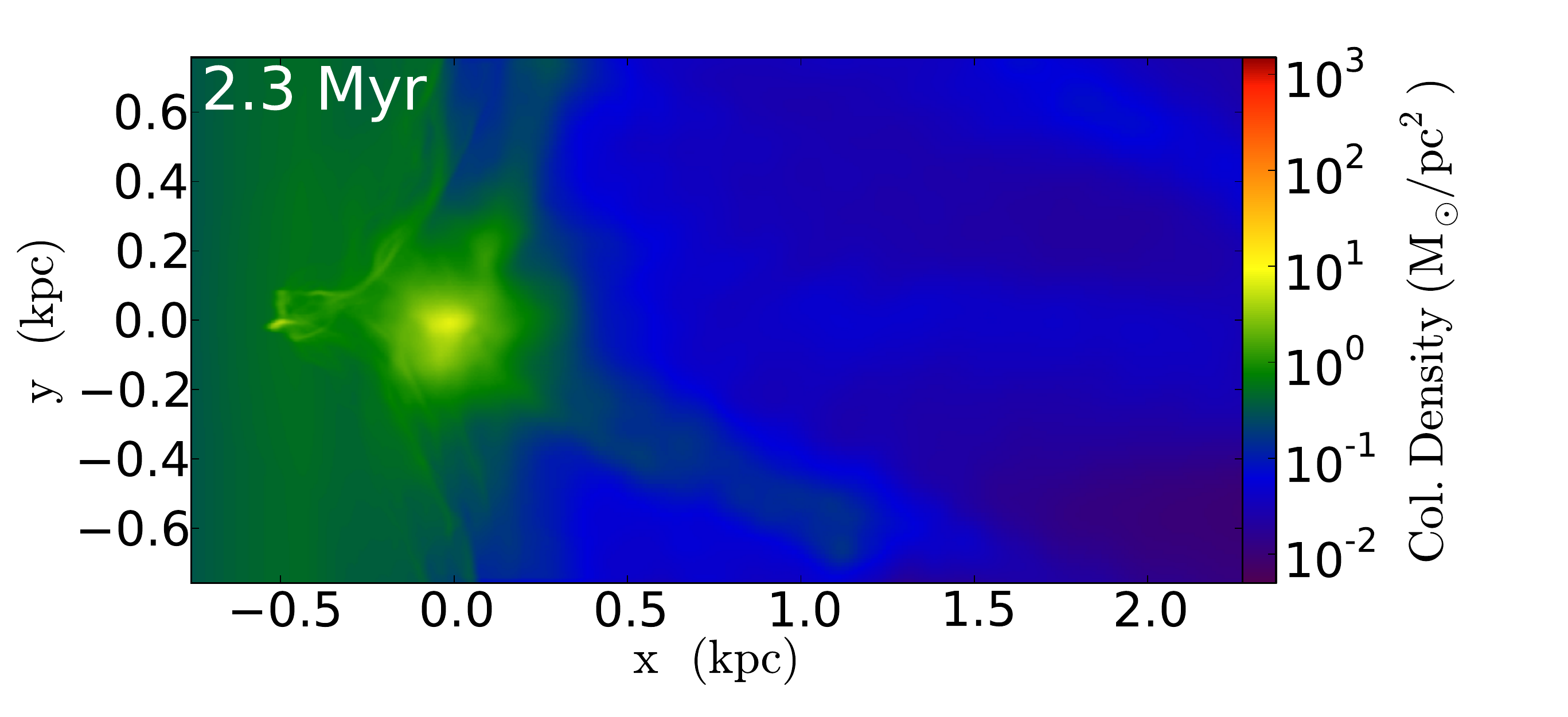}
\includegraphics*[scale=0.23, trim=95 64.1         0   21]{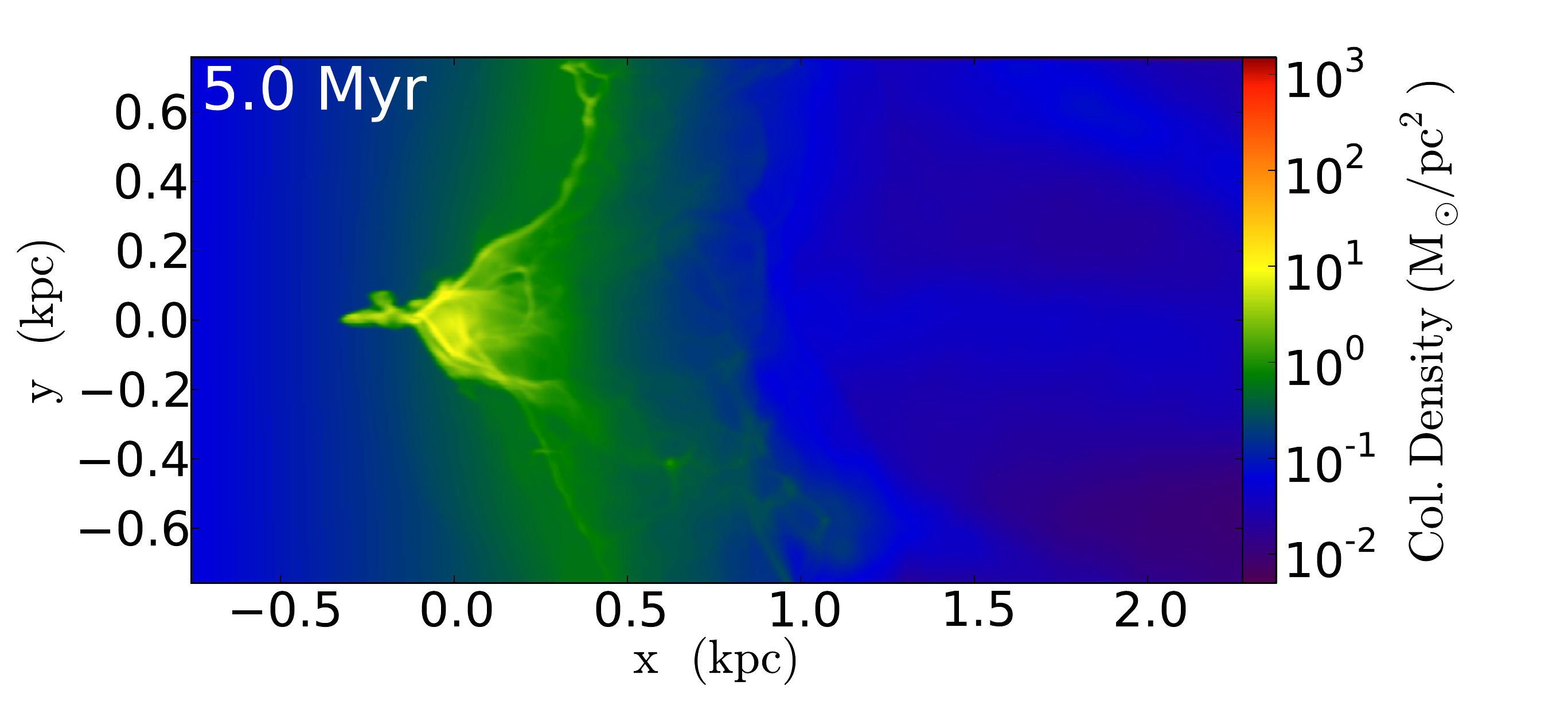}

\includegraphics*[scale=0.23, trim=  0 64.1 163.5 21]{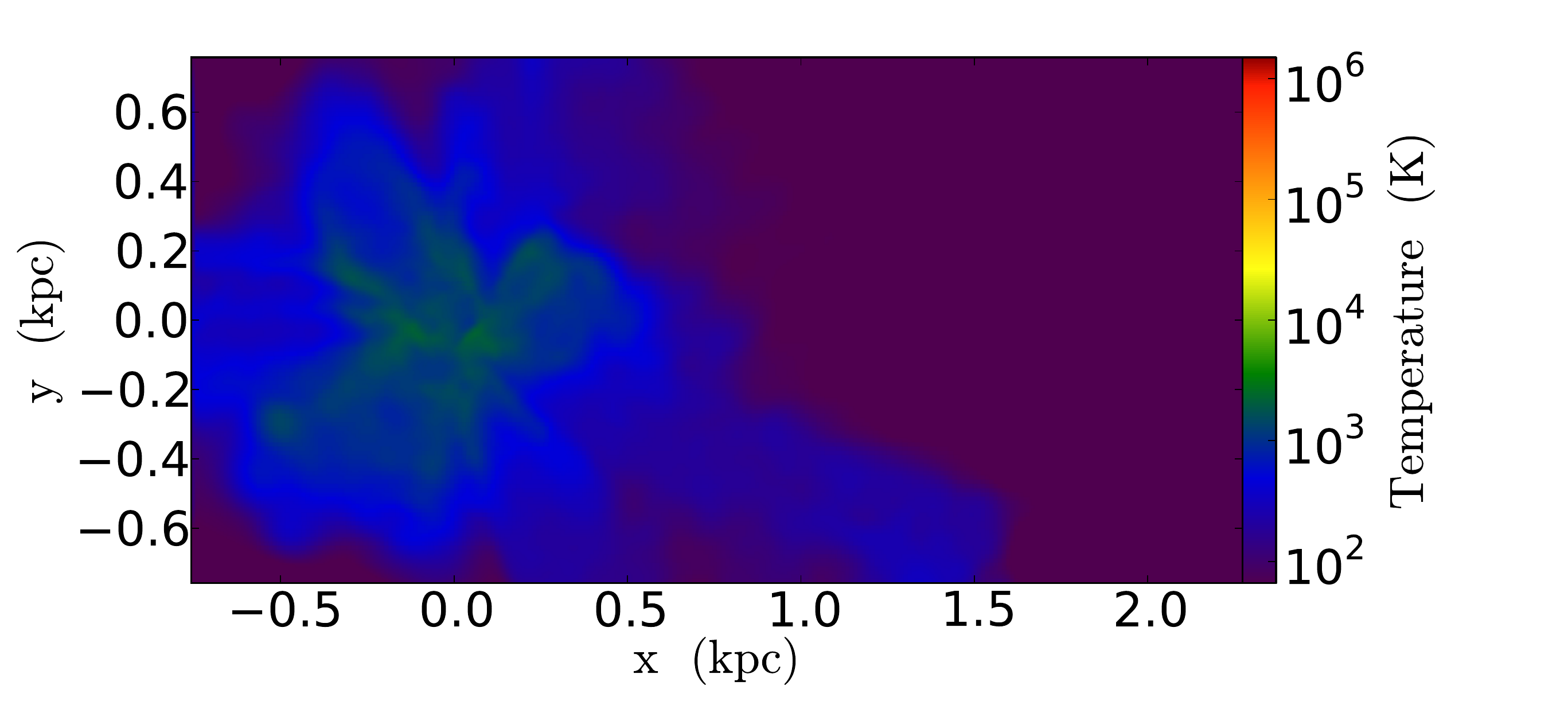}
\includegraphics*[scale=0.23, trim=95 64.1 163.5 21]{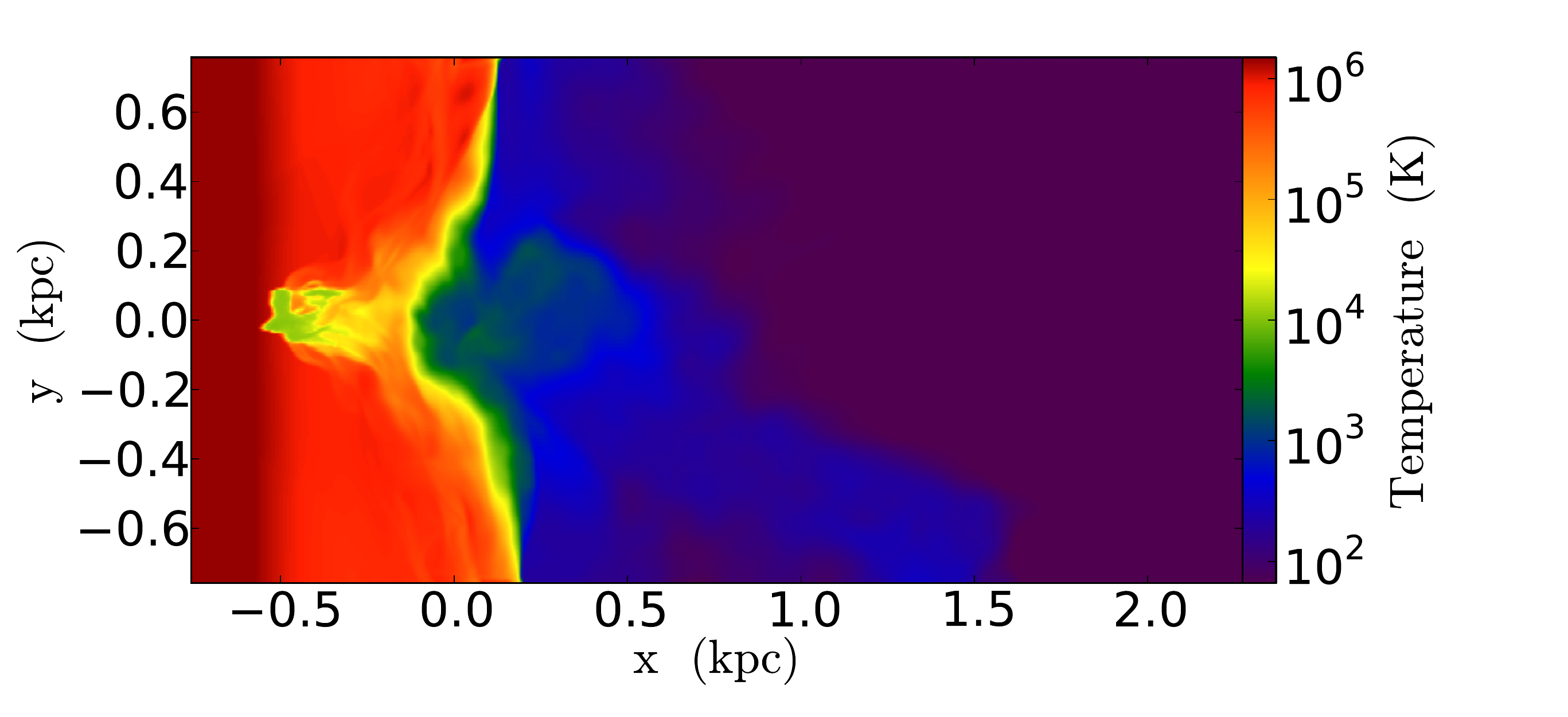}
\includegraphics*[scale=0.23, trim=95 64.1         0 21]{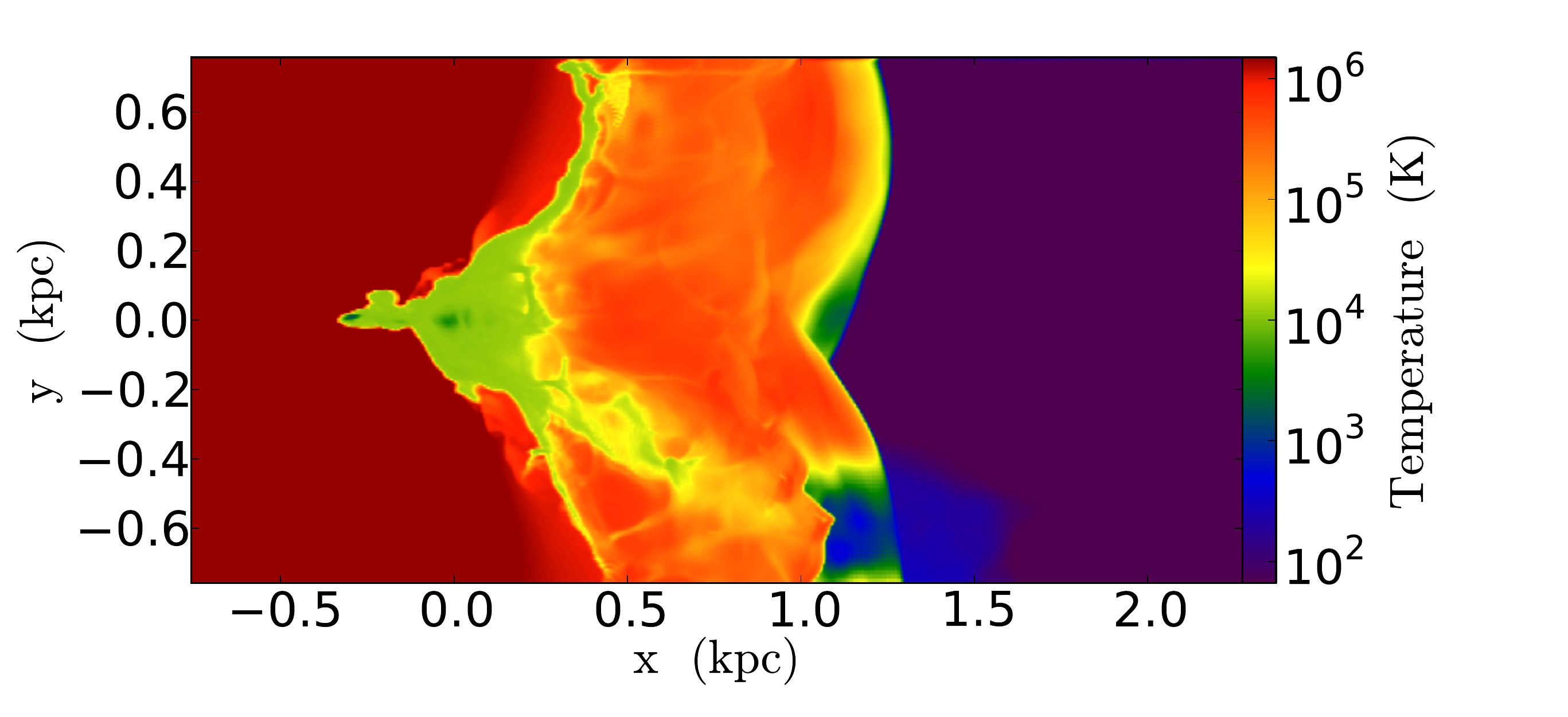}

\includegraphics*[scale=0.23, trim=  0 64.1 163.5 21]{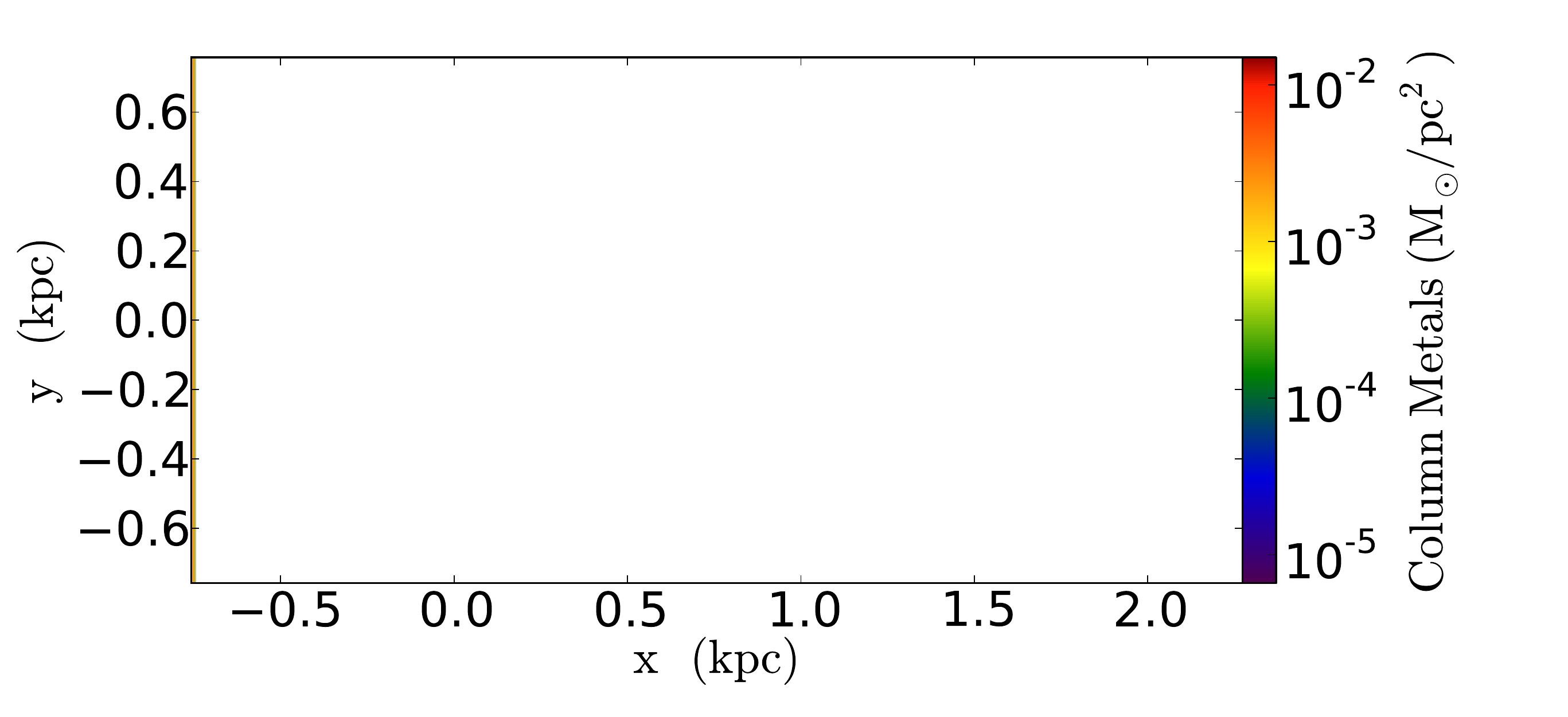}
\includegraphics*[scale=0.23, trim=95 64.1 163.5 21]{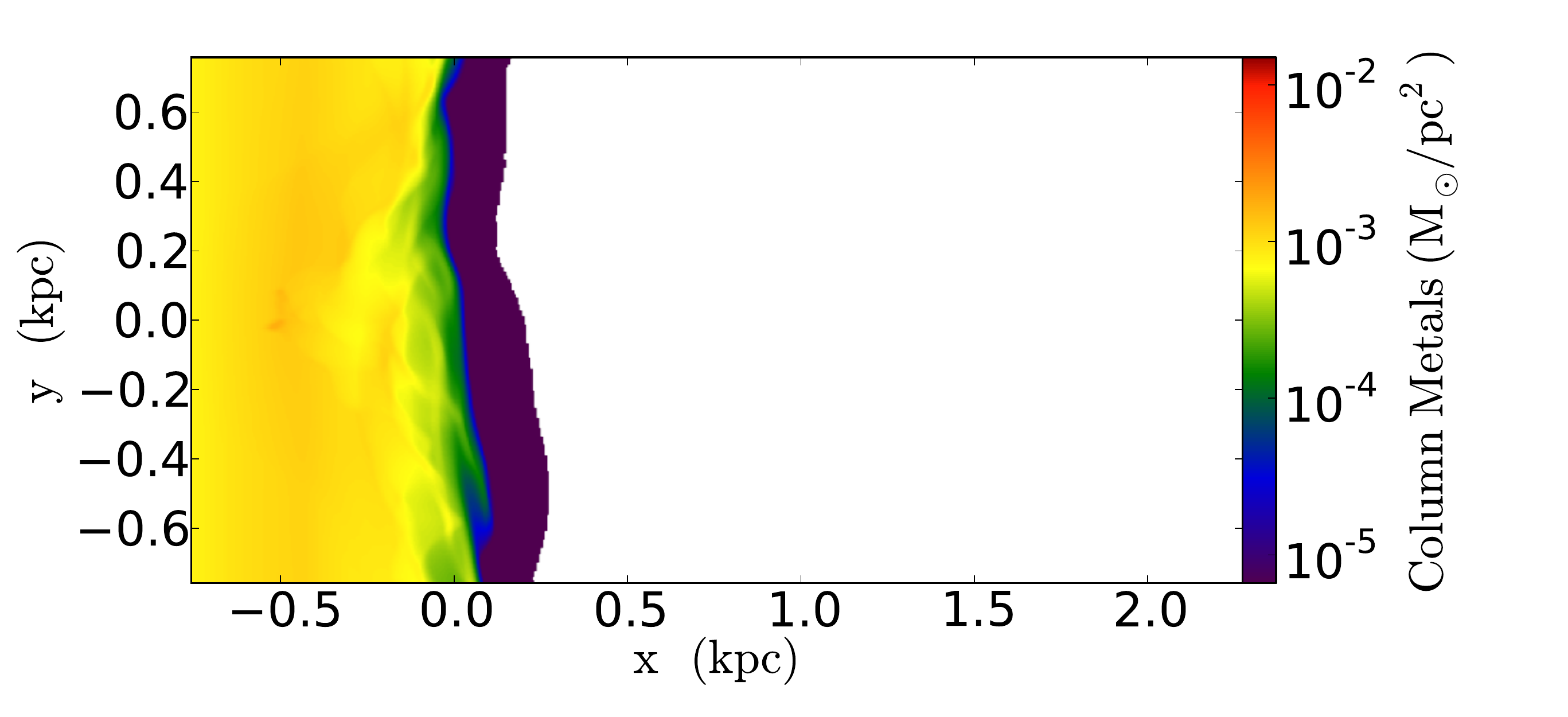}
\includegraphics*[scale=0.23, trim=95 64.1         0 21]{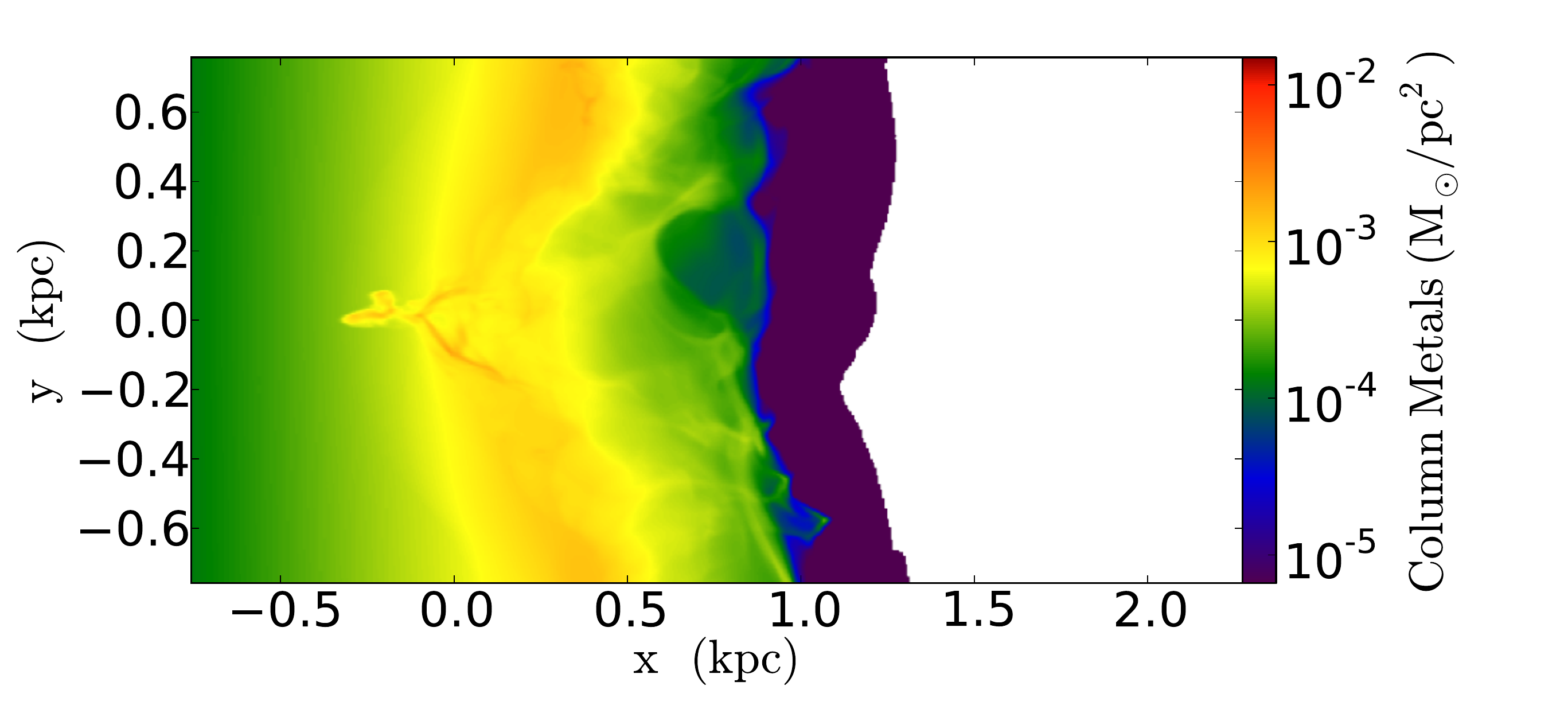}

\includegraphics*[scale=0.23, trim=  0 64.1 163.5 21]{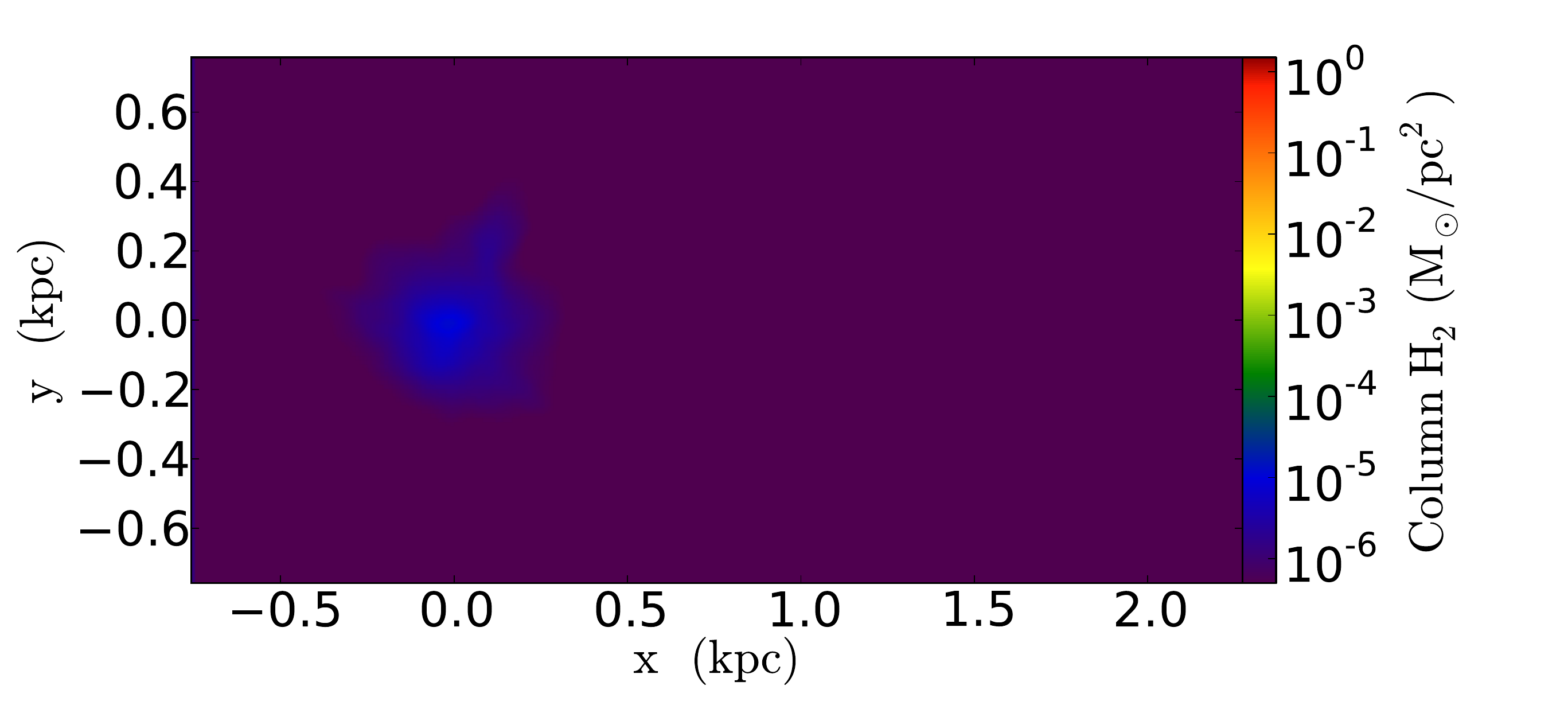}
\includegraphics*[scale=0.23, trim=95 64.1 163.5 21]{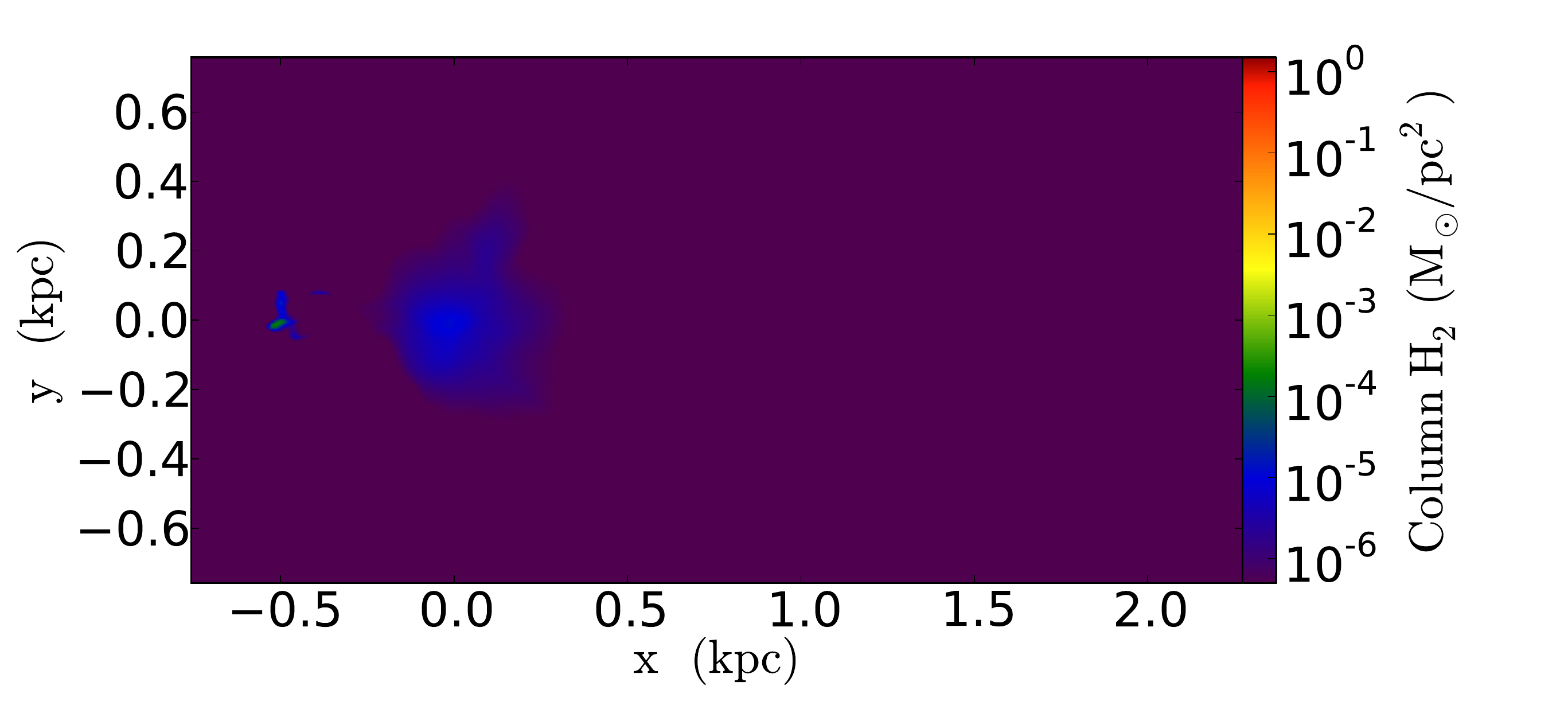}
\includegraphics*[scale=0.23, trim=95 64.1         0 21]{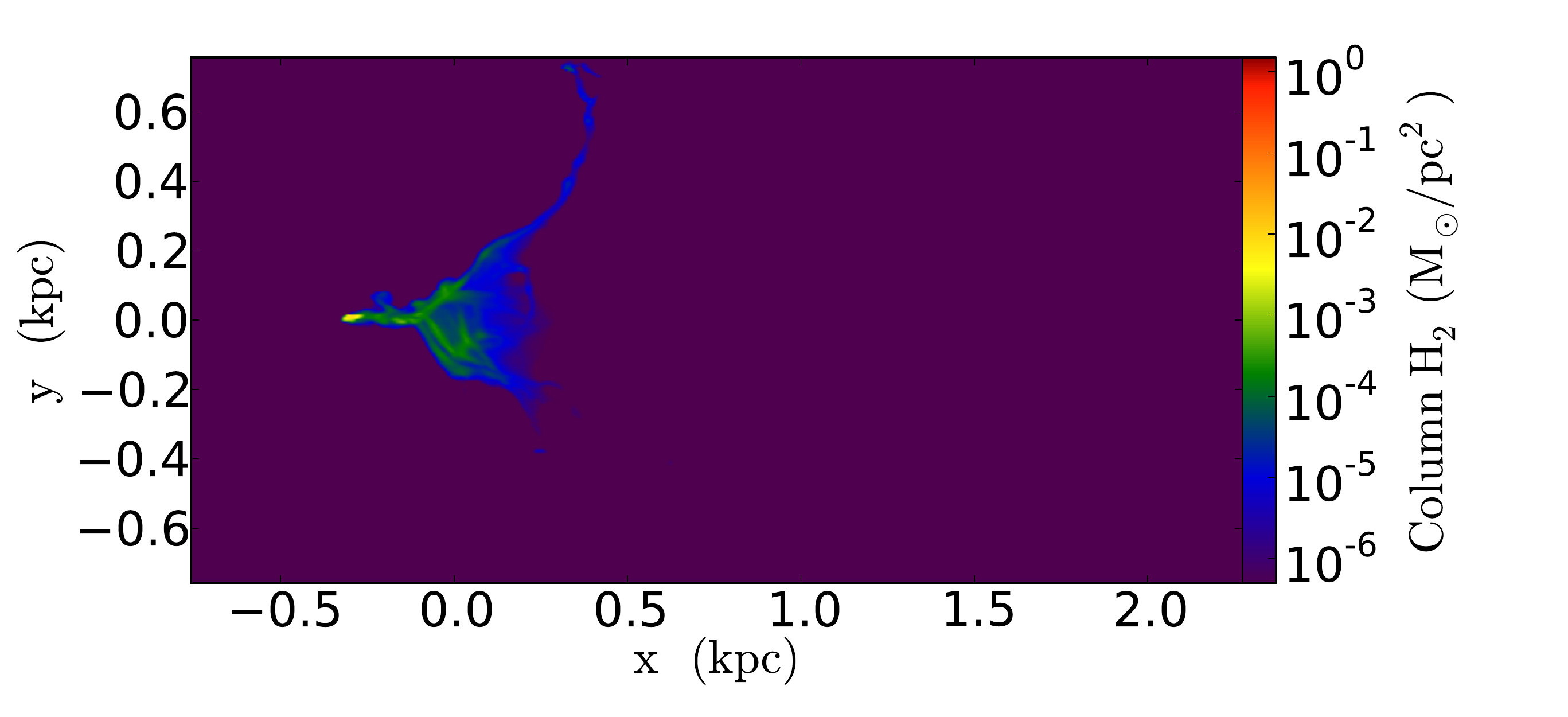}

\includegraphics*[scale=0.23, trim=  0 64.1 162.5  10]{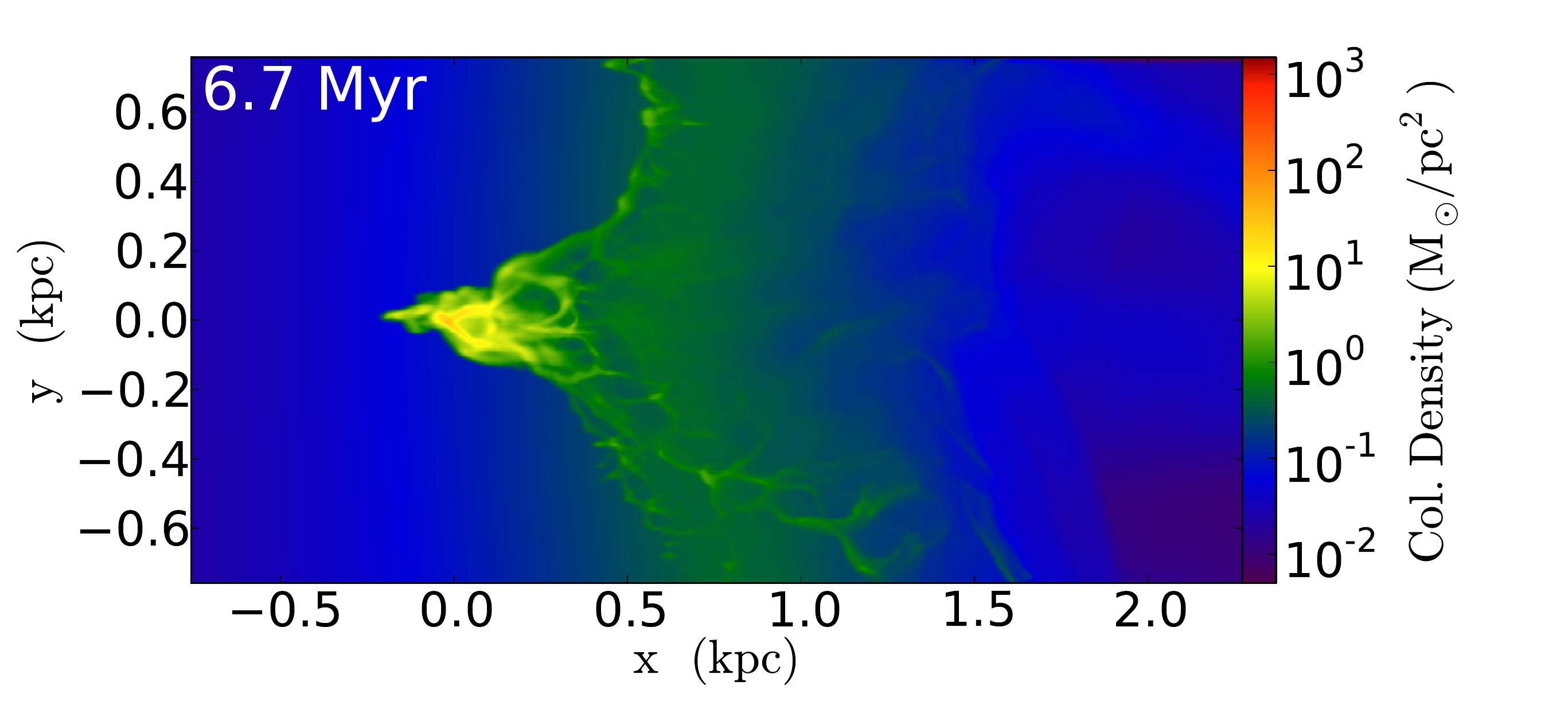}
\includegraphics*[scale=0.23, trim=95 64.1 162.5   10]{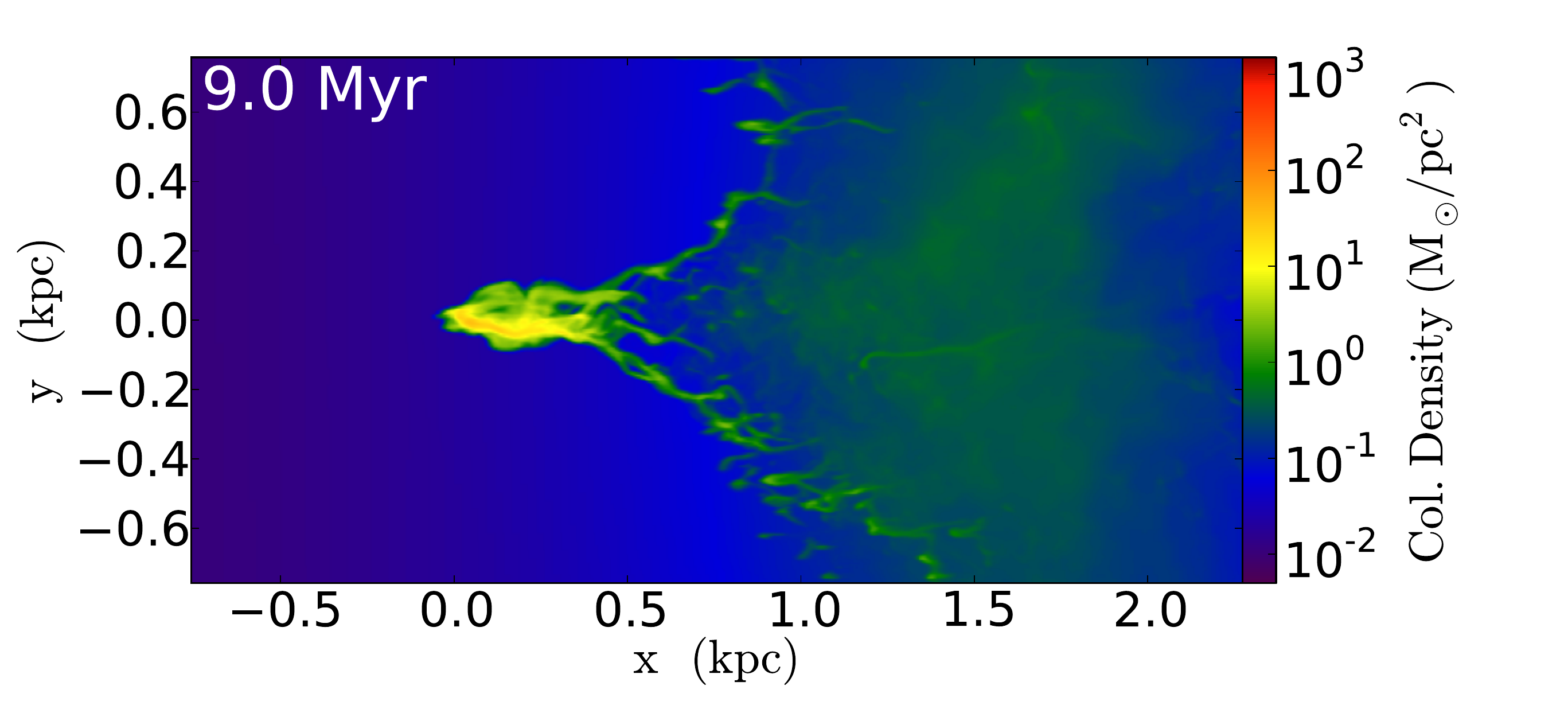}
\includegraphics*[scale=0.23, trim=95 64.1         0   10]{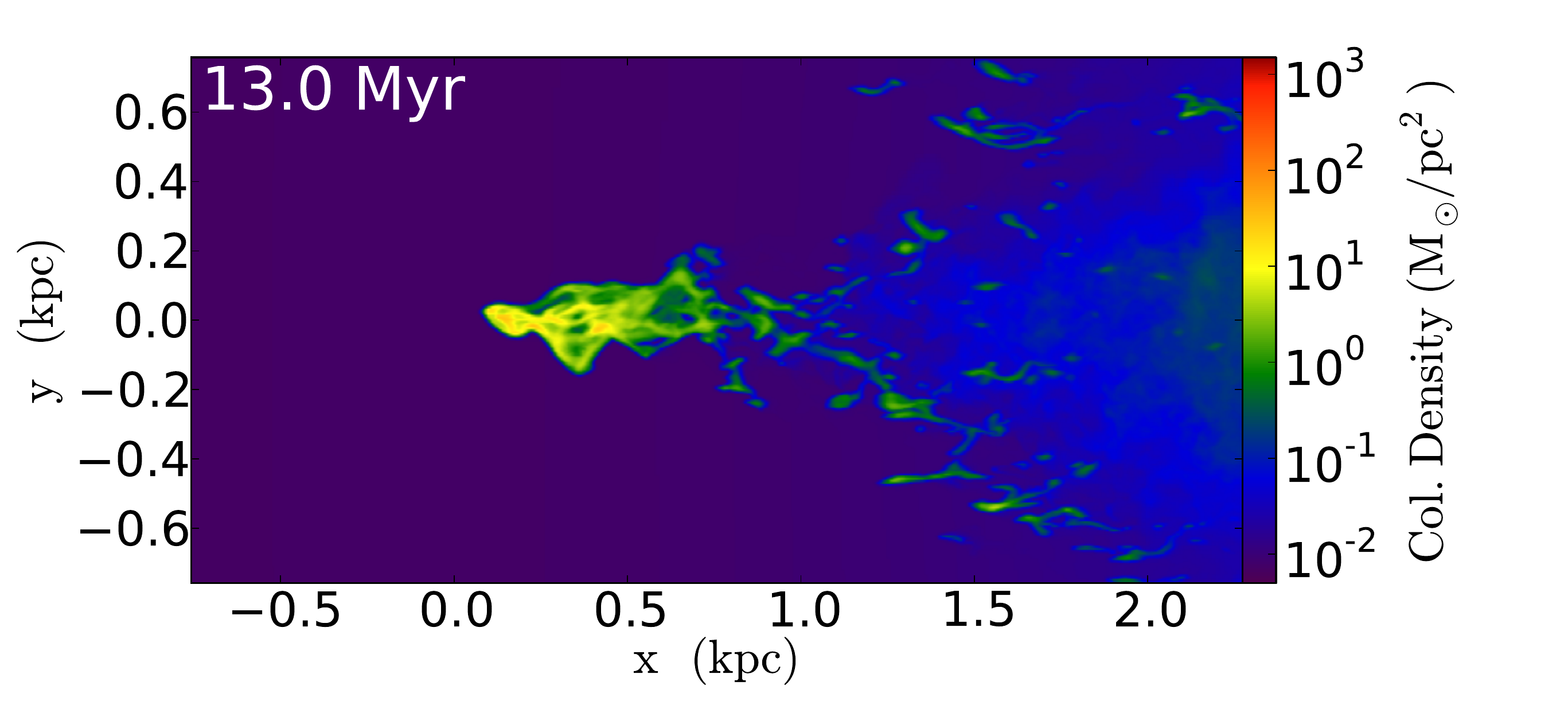}

\includegraphics*[scale=0.23, trim=  0 64.1 162.5 21]{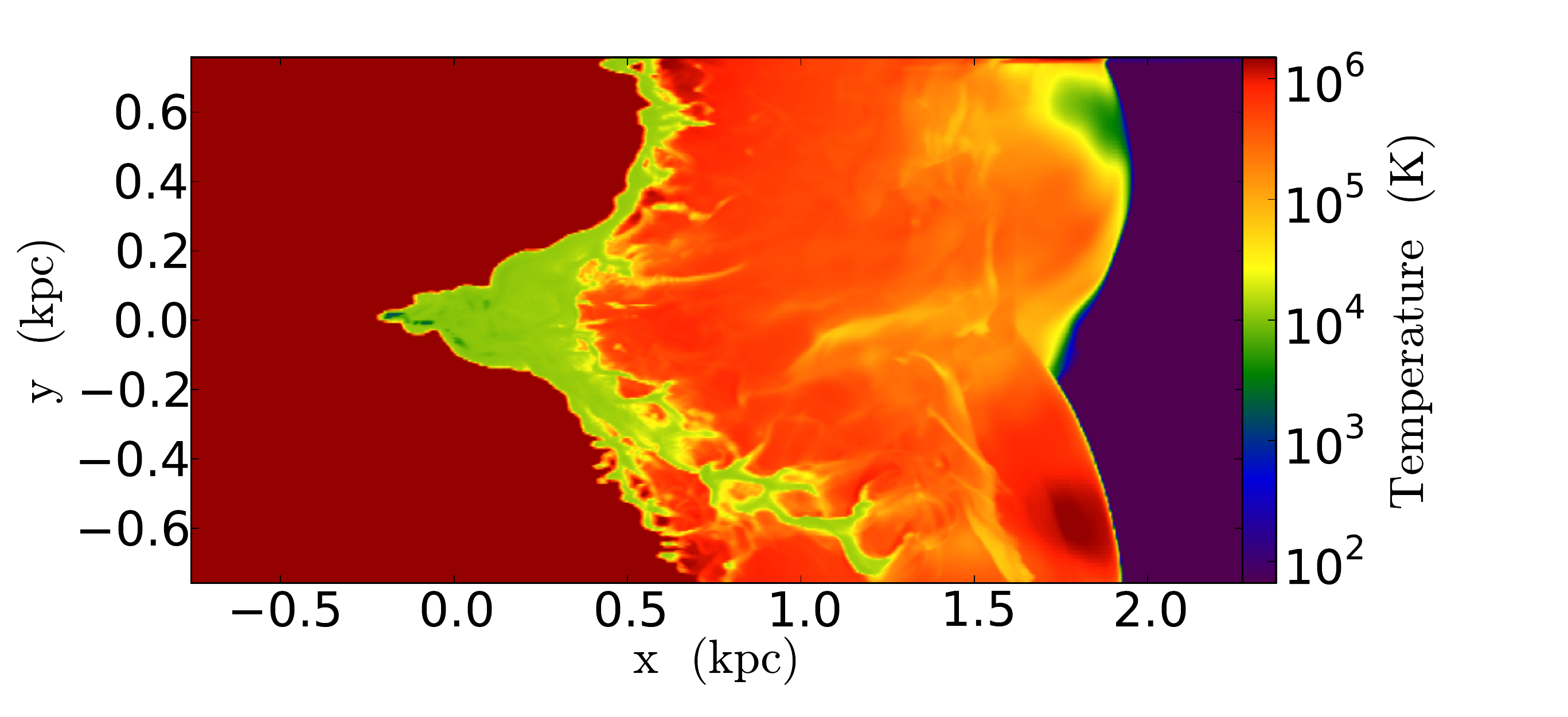}
\includegraphics*[scale=0.23, trim=95 64.1 162.5 21]{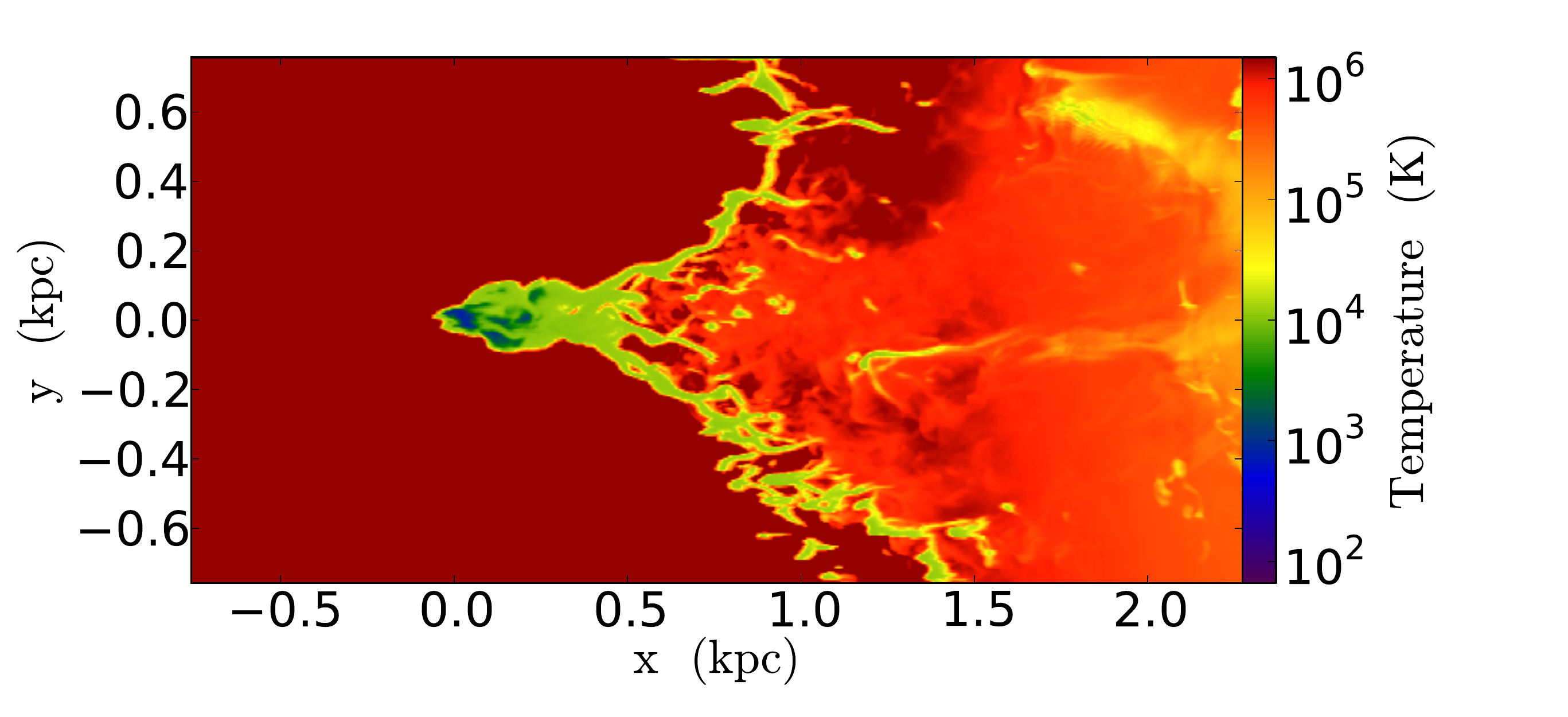}
\includegraphics*[scale=0.23, trim=95 64.1         0 21]{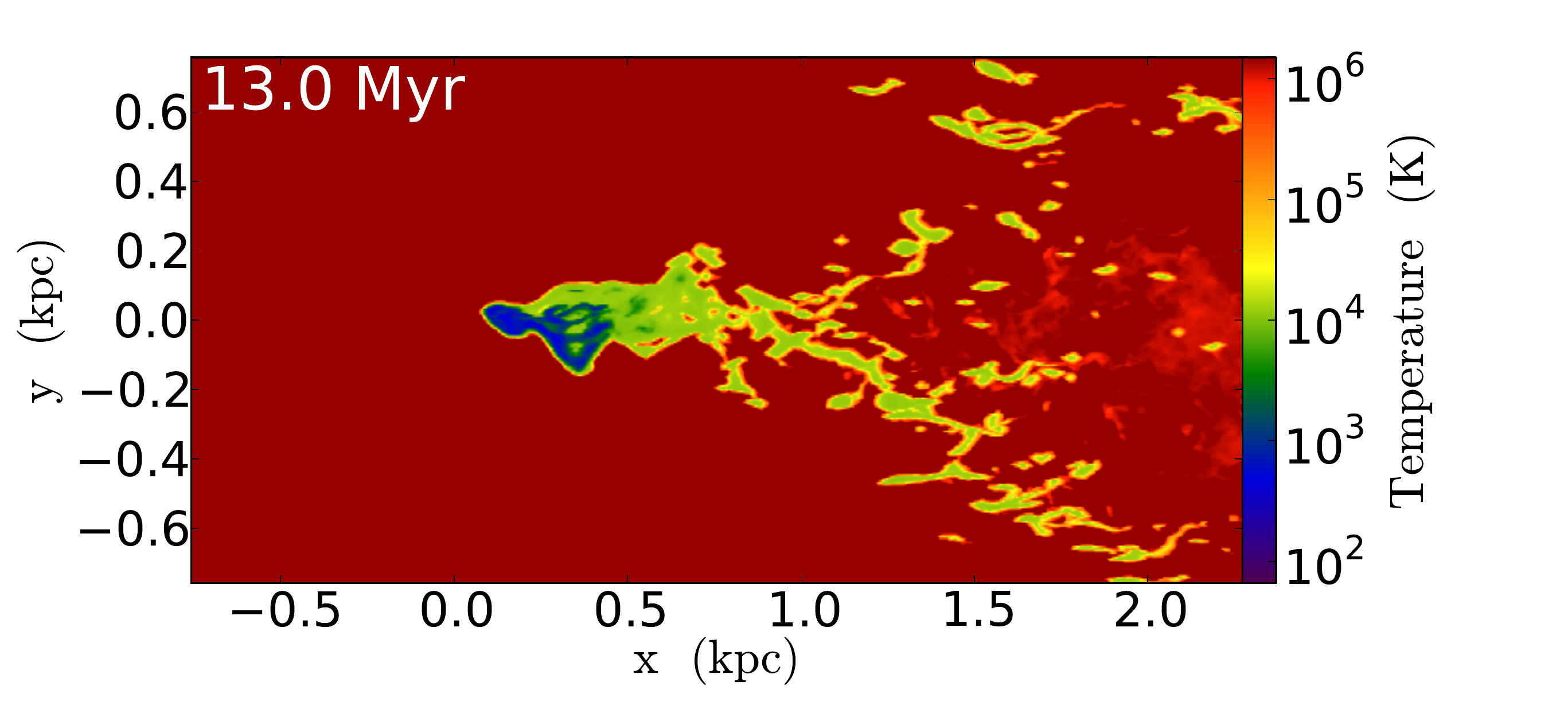}

\includegraphics*[scale=0.23, trim=  0 64.1 162.5 21]{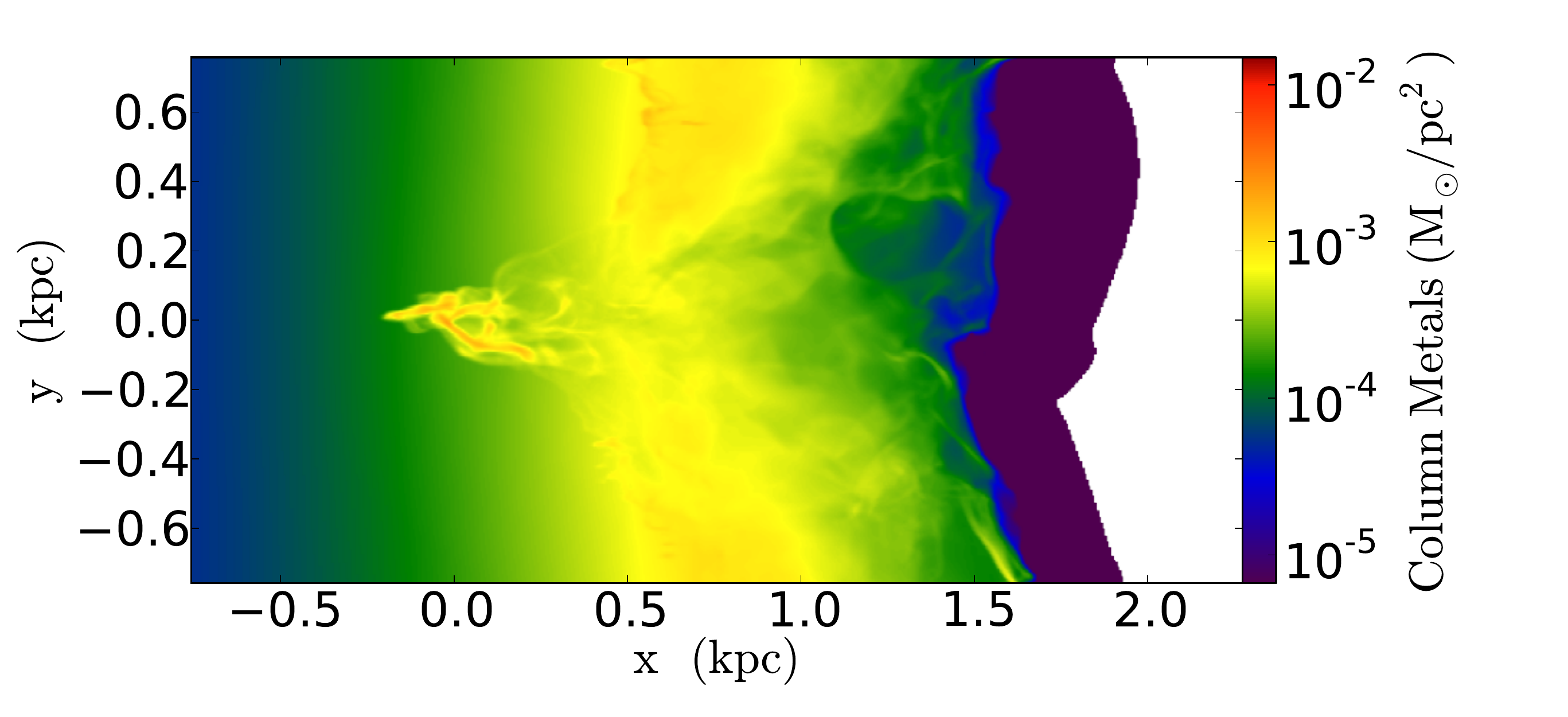}
\includegraphics*[scale=0.23, trim=95 64.1 162.5 21]{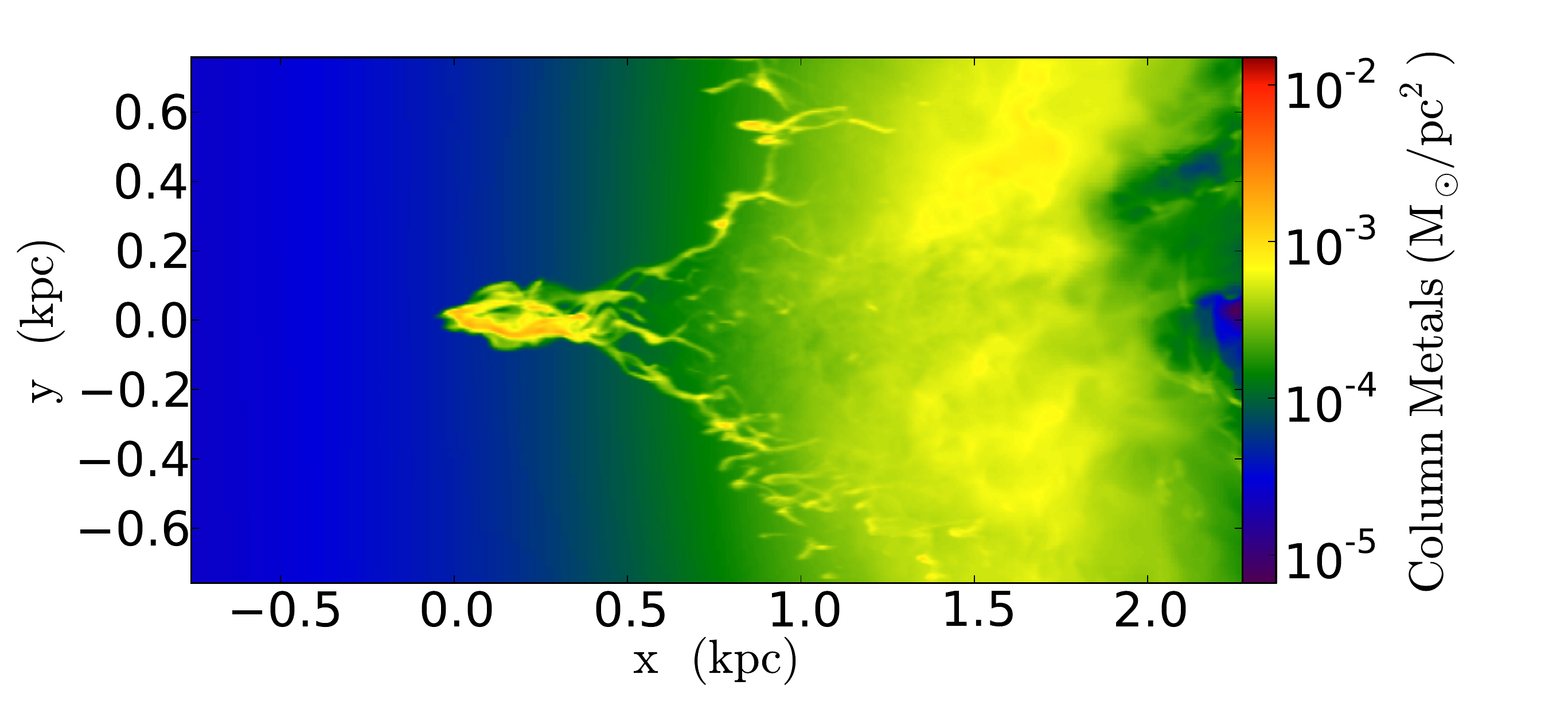}
\includegraphics*[scale=0.23, trim=95 64.1         0 21]{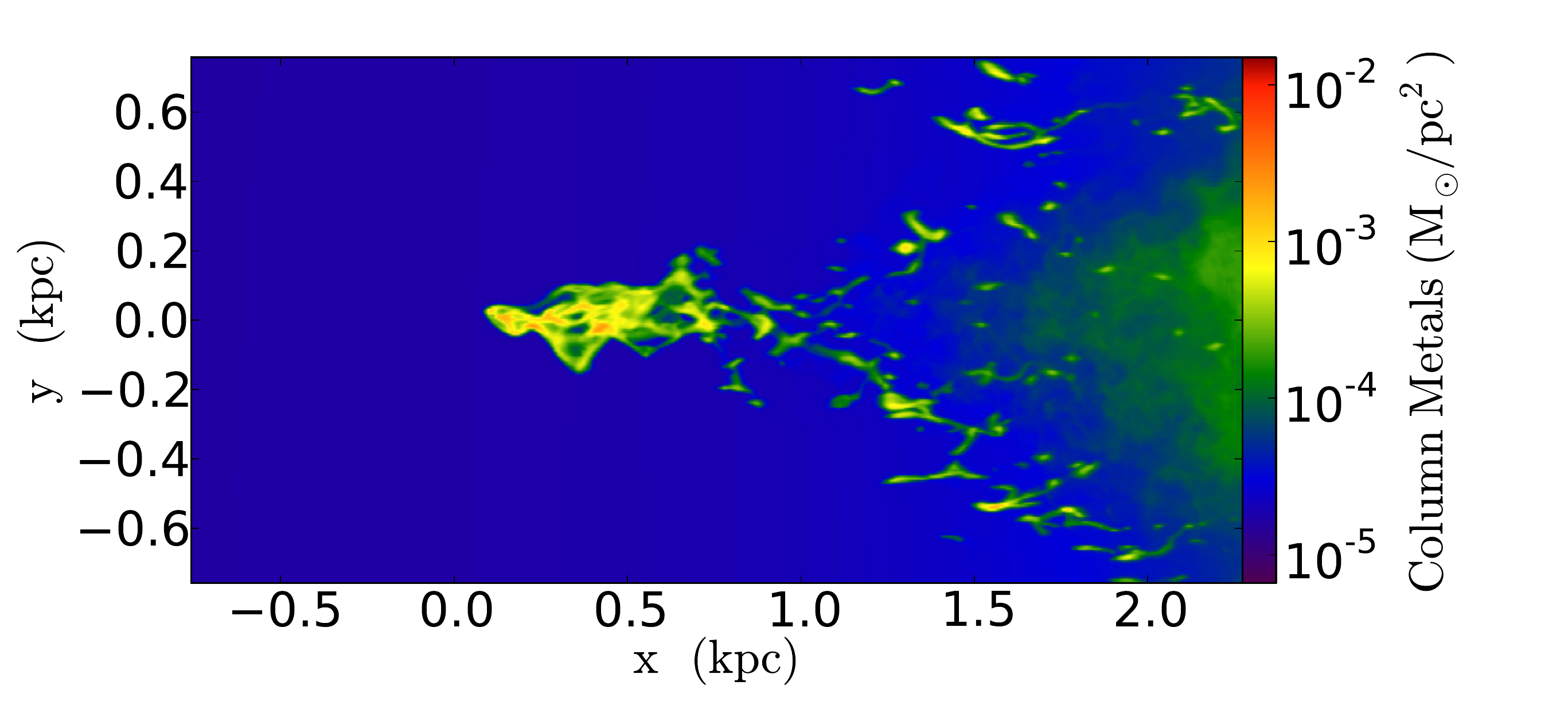}

\includegraphics*[scale=0.23, trim=  0       0 162.5 21]{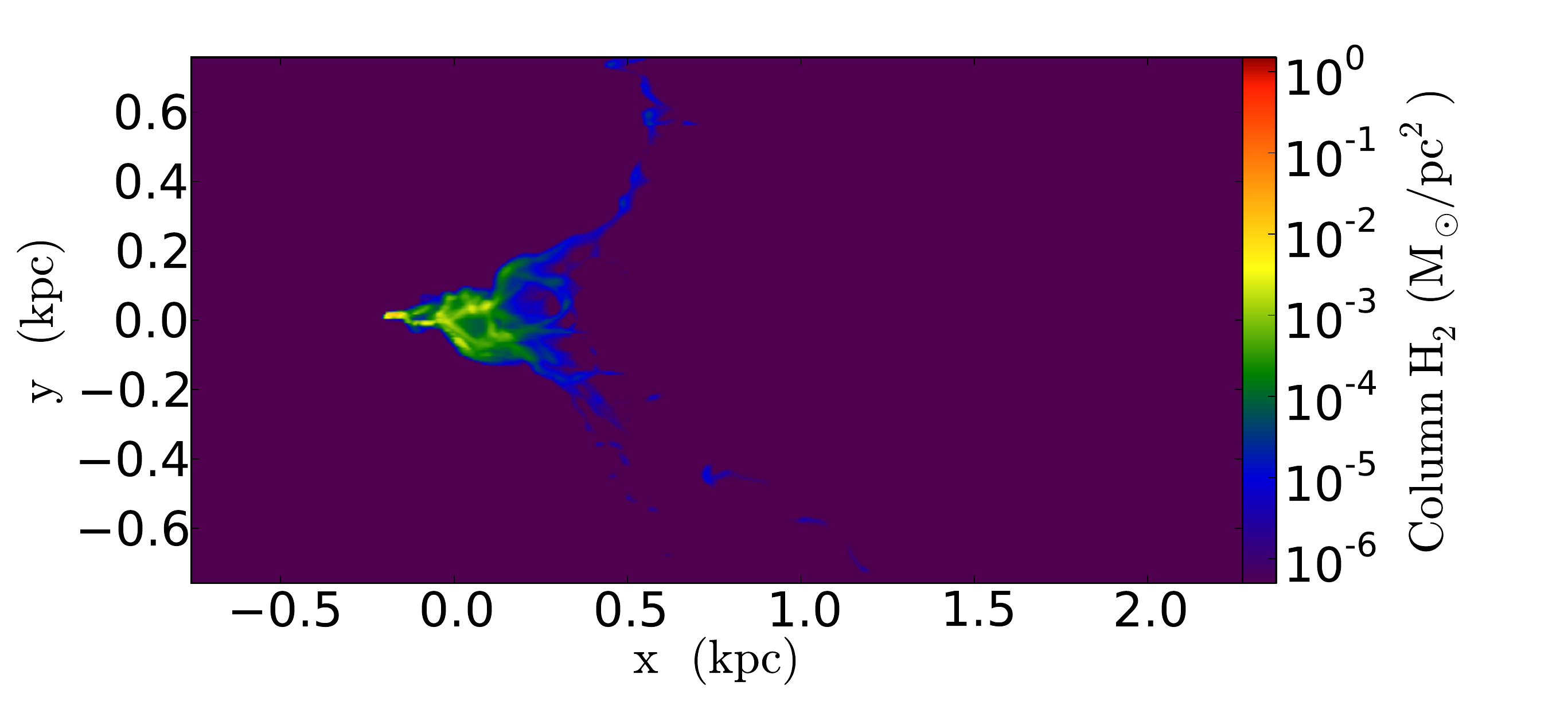}
\includegraphics*[scale=0.23, trim=95       0 162.5 21]{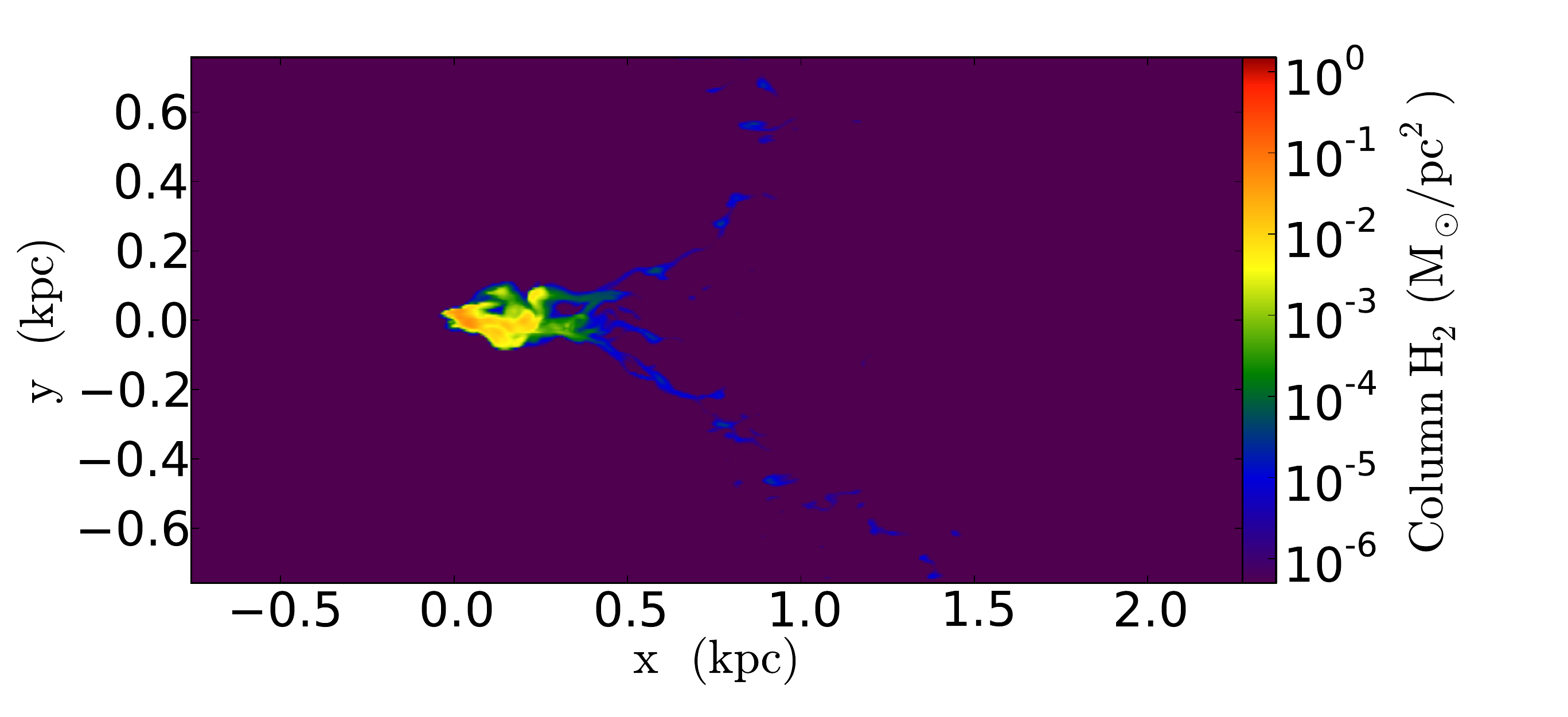}
\includegraphics*[scale=0.23, trim=95       0         0 21]{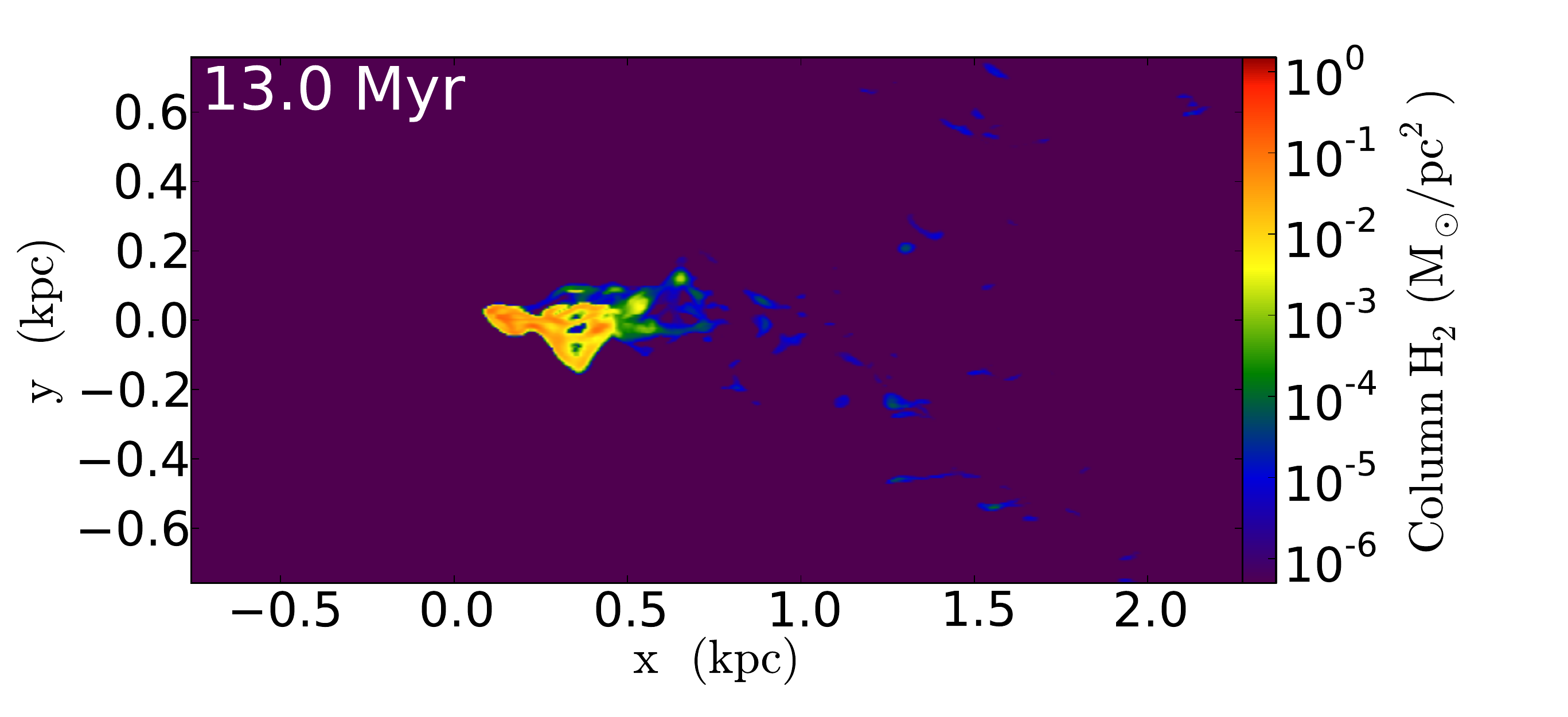}
\caption{\footnotesize{Evolution of the fiducial simulation, showing column density (first and fifth row), projection of density-weighted temperature (second and sixth row),
the column density of  metals  (third and seventh row), and  the column density of  molecular hydrogen (fourth and eighth row) at 0, 2.3, 5.0, 6.7, 9.0, and 13 Myr from left to right.}}
\label{fig_fid}
\end{figure*}

We determined if a fluid segment was bound  by comparing the $y$- and $z$-component of the velocity of the fluid with the $y$- and $z$-component of the 
minihalo's dark matter gravitational potential.  Here we assumed a 
NFW profile \citep{Navarro97}:
\begin{equation}\label{nfw}
\Phi(r) = -\frac{GM_{\rm vir}}{rF(c)}\log\left(1+c\frac{r}{r_{\rm vir}}\right),
\end{equation}
where $G$ is Newton's gravitational constant, $M_{\rm vir}$ is the virial radius, assumed to be the same as the minihalo mass, 
$r_{\rm vir}$ is the virial radius, set by the virial mass and redshift, $F(x) = \log(1+x) - x/(1+x),$ and $c$ is the concentration parameter, assumed to be 4, slightly less than 
the fiducial value in GS11B since we find the halos in our Gadget simulations are typically less concentrated (e.g., Richardson \etal\ 2013).  For each segment  determined to be bound to the $x$-axis, we added its gravitational potential into radial bins. We then determined the best-fit NFW profile for the radial gravitational potential, fitting for mass and clumping factor, using a Gauss-Newton algorithm. We 
then rechecked which gas segments were bound to the $x$-axis using the fitted NFW profile. We iterated this process until the profile fit
was self-consistent.

Once we were satisfied with our identification of the bound material, we tracked the variance
of the abundance of the particles to determine its uniformity. The cloud particles 
were then evolved ballistically for 200 Myr, including their mutual gravitational attraction and 
the fitted gravitational potential of the minihalo.  Finally, when one particle overtook a second one, we merged them into 
a single new particle, conserving mass and momentum and averaging their abundance while combining their 
variance. 

\section{Results}\label{Results}

\subsection{Fiducial Behavior}\label{fid}
\fig{fig_fid} shows the evolution of our fiducial run (see FID in \tabl{tab_sims}) at 6 characteristic times.  This run had a minihalo mass of $M_{\rm 6} = 2.72$ and  a virial radius of 505 pc. Our fiducial outflow had a shock speed of $v_{\rm s} = 226$ km s$^{-1}$ and surface momentum of $\mu_{\rm s} = 60.4 \Msun$ pc$^{-1}$ Myr$^{-1}$, consistent with a starburst-minihalo separation of $R_{\rm s} = 3.6$ kpc, shock surface density of $\sigma_5 = 2.62$ and energy of $E_{55} = 10$. We used a redshift of $z=8$, and a shock enrichment of 0.12 $\Zsun$.
\begin{figure*}[t!]
\centering
\includegraphics[scale=0.5]{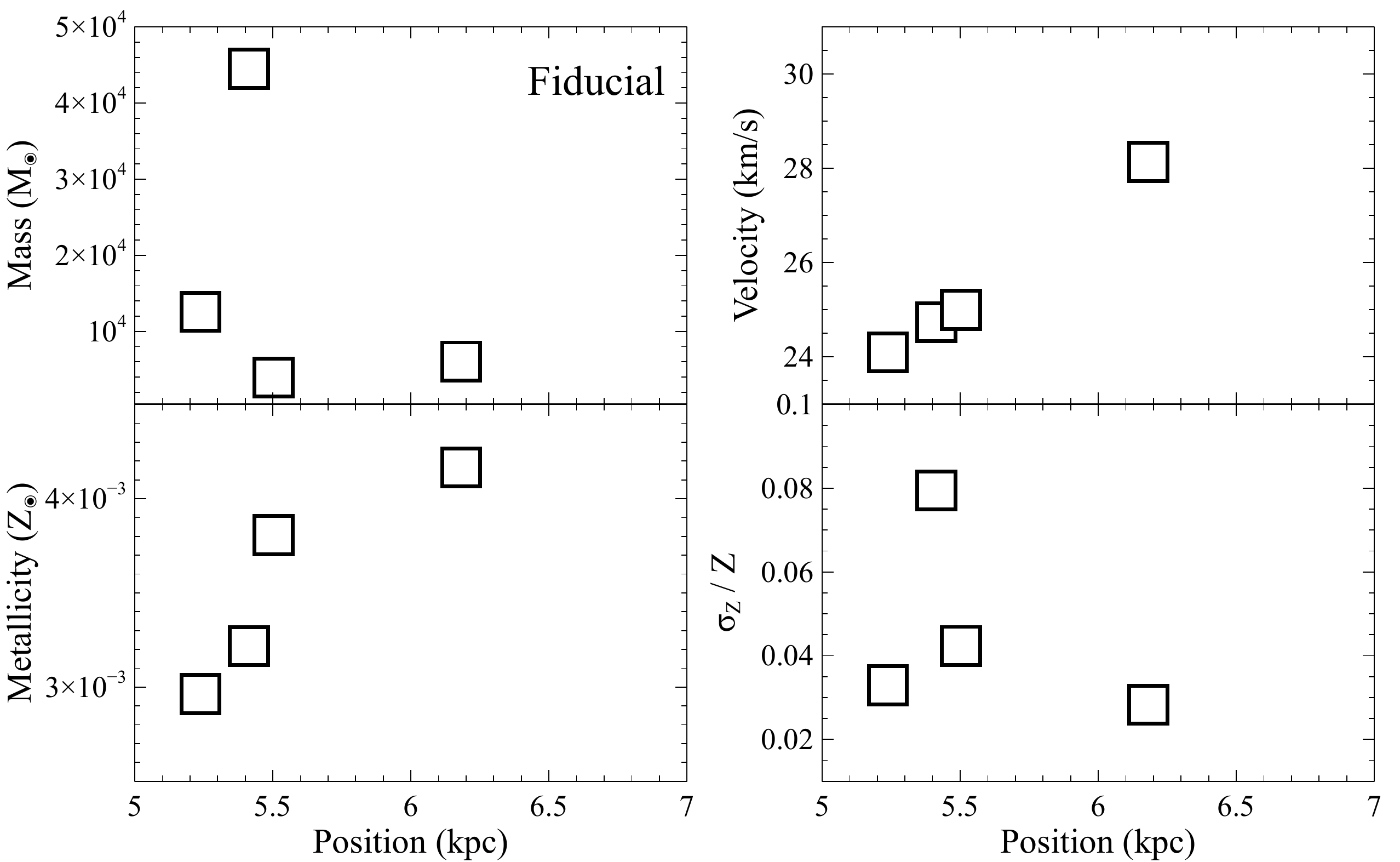}
\caption{\footnotesize{Distribution of cloud particle masses (top left), velocity (top right), metallicity (bottom left)
and relative metallicity dispersion (bottom right) vs particles positions after 200 Myr for the fiducial run.}}
\label{fig_part_fid}
\end{figure*}

As the shock enters the volume from the left boundary, it travels freely through the diffuse medium, but stalls slightly within the inner 100 pc as it moves along the denser filament.
At 2.3 Myr, the shock makes contact with the dense minihalo, and molecular hydrogen begins to form in the swept-up accretion lane. 
By 6.7 Myr, the shock has overtaken the minihalo in the diffuse medium, 
and has propagated roughly 60\% around the minihalo itself. The shock front has mixed in some metals and molecular hydrogen has formed along this front,
most notably along the periphery of the minihalo and along accretion lanes. The minihalo  begins to cool more efficiently due to molecular cooling.

At  9 Myr, the shock has propagated around the now collapsed minihalo. What remains of the minihalo is enriched to roughly 2\% of the value of the incoming material, or 
$2.5\times10^{-3} \Zsun$, and it contains a significant amount of molecular hydrogen. The minihalo begins to cool below $10^3$ K. Finally, by about 13 Myr, the
shock has passed through the box, leaving a stream of material that is not entirely bound to the $x$-axis. This is because anisotropies in the minihalo lead 
to mismatched times at which the shock reaches the antipodal point, resulting in some material being pushed away from the axis. This is different from what is seen
in GS10, GS11A and GS11B. This material is almost uniformly at a few 100 K, enriched to 2\% of the shock's metallicity, and has a mass fraction of $H_{\rm 2}$
of about half a percent. During this time, we find that the dark matter of the minihalo does not respond to the outflow, although it does move in towards the halo from the outer edges of the simulation volume. 
This is mostly due to our choice of isolated boundary conditions, and results in a small increase in the gravitational attraction on the gas, but this  
movement is small and well beyond the virial radius of the halo.

After determining the gas that is transferred to the ballistic particle scheme, we find 
only 24\% of the baryonic mass of the original minihalo is contained in these particles, while 76\% is blown away from the minihalo, out of the simulation volume.
This material is not bound to the $x$-axis, and does not coalesce into large clumps. For our fiducial case, we consistently get 24\% of the mass in particles, regardless of the number of ballistic bins, the density cutoff, or the timescale over which the gas must be bound to the $x$-axis.

\fig{fig_part_fid} shows the final particles 
 whose masses are above 1\% of the original minihalo's baryons, a limit we adopt in similar figures below.
 The momentum from the
outflow pushes this mass out of the dark matter halo, while small variations in the velocity of the original particles, along with
their self-gravity, allow the particles to merge, leading to a small population of high-mass clusters. 
For the fiducial case, most final particles are about $1$-$4\times 10^4 \Msun$, and have a velocity profile nearly following a free expansion
law, with $v \simeq 4.6(x/\mbox{kpc})\mbox{ km s}^{-1} \simeq x/200$ Myr, as expected. We also found a slight trend of increasing metallicity
with position, due to increased mixing on the backside of the minihalo.   

\subsection{Convergence}\label{converge}

\begin{figure*}[h!]
\centering
\includegraphics*[scale=0.245, trim=0 64.1 162.5 21]{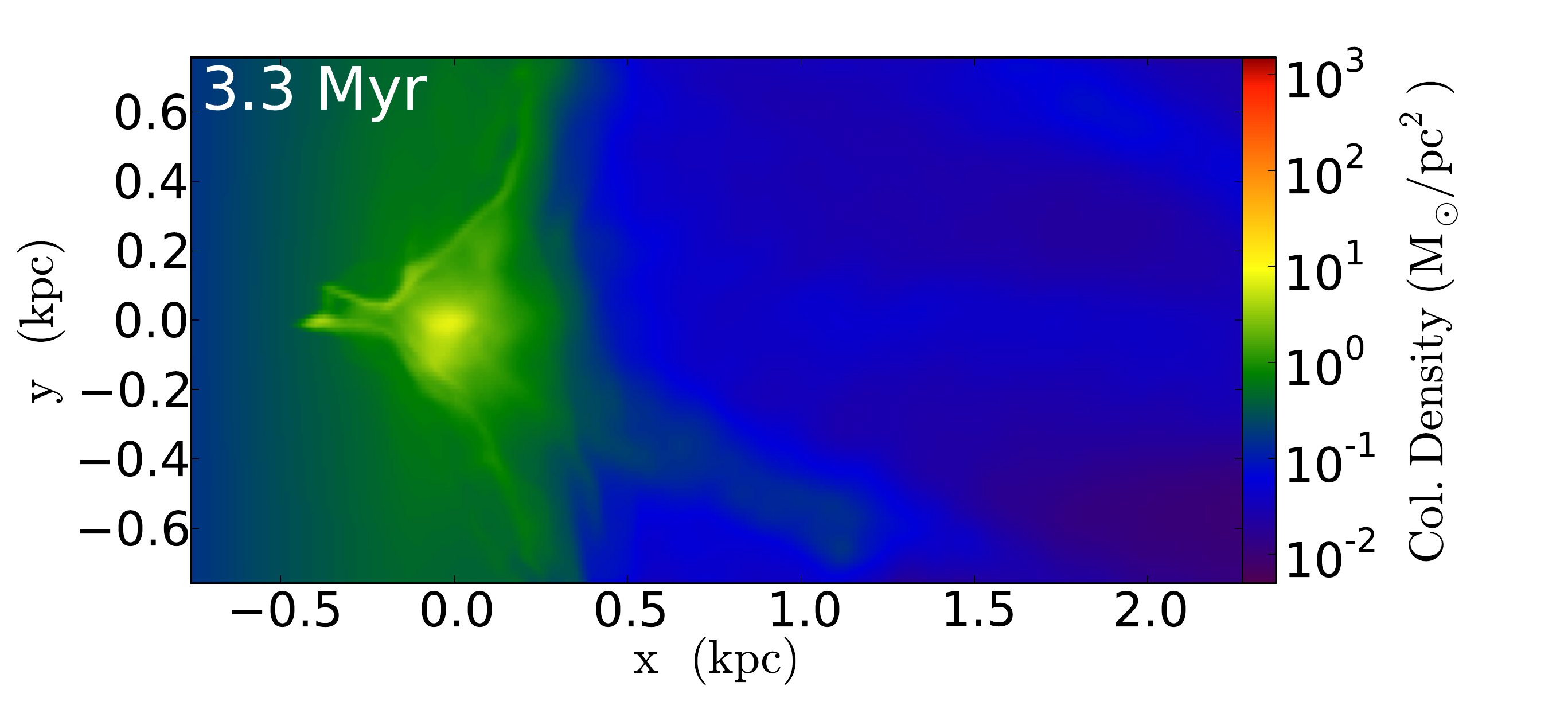}
\includegraphics*[scale=0.245, trim=95 64.1 162.5 21]{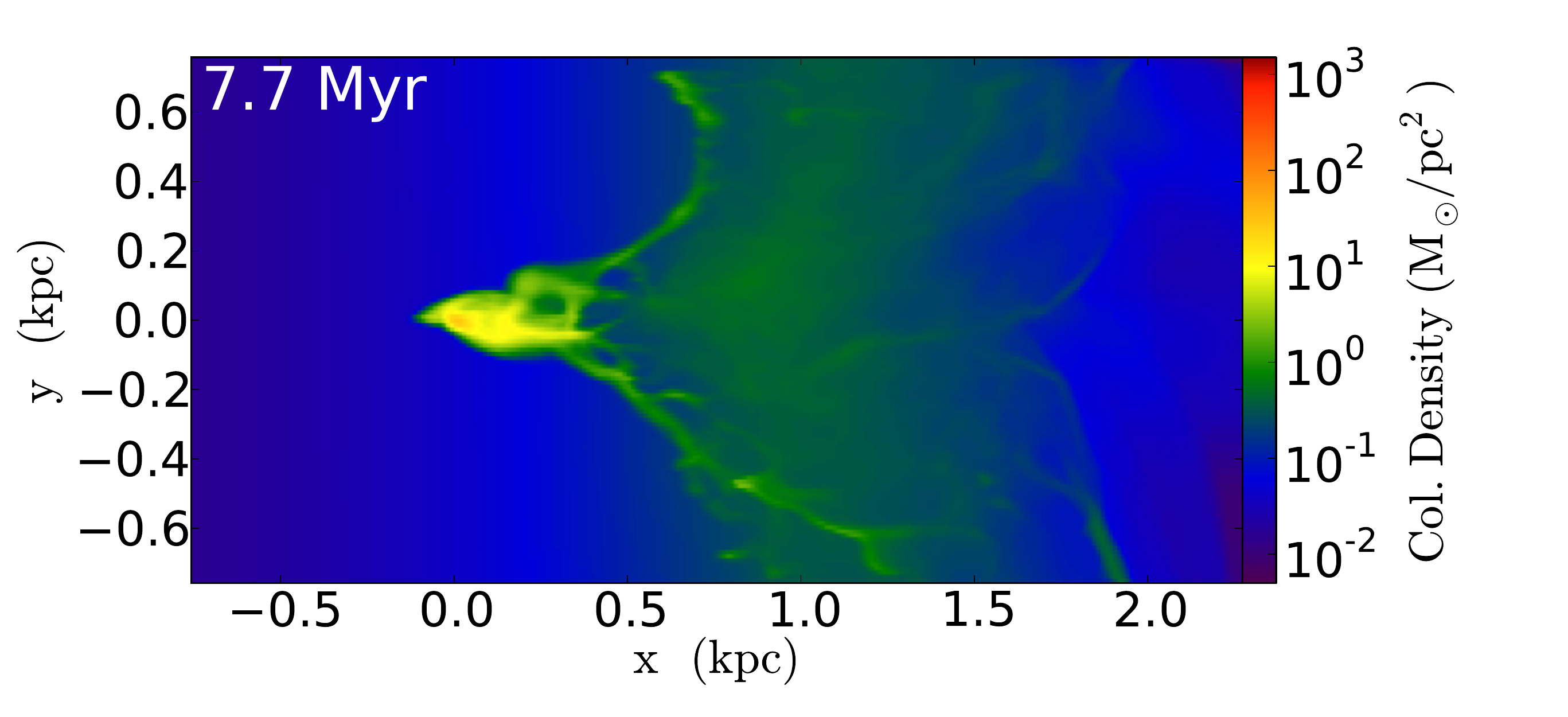}
\includegraphics*[scale=0.245, trim=95 64.1 0 21]{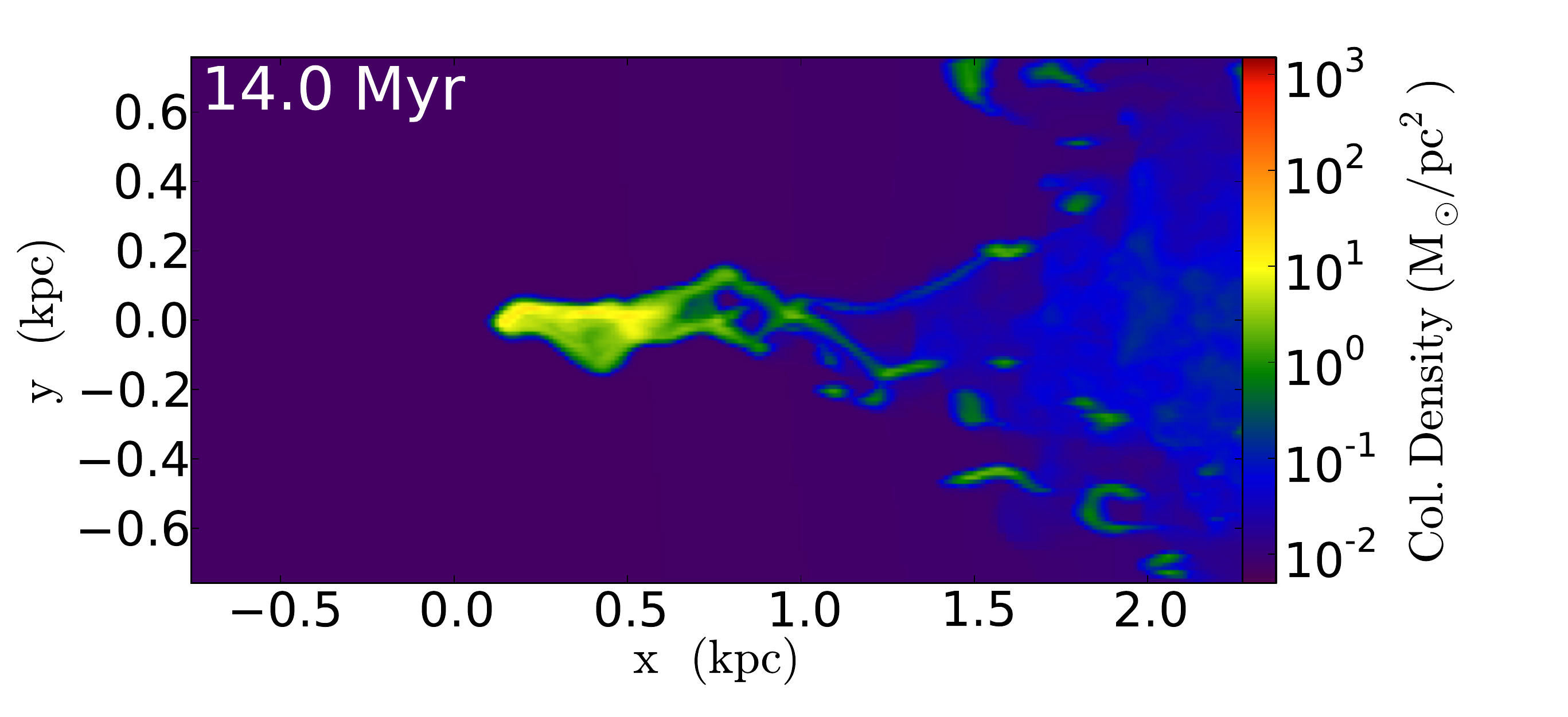}
\includegraphics*[scale=0.245, trim=0 64.1 162.5 21]{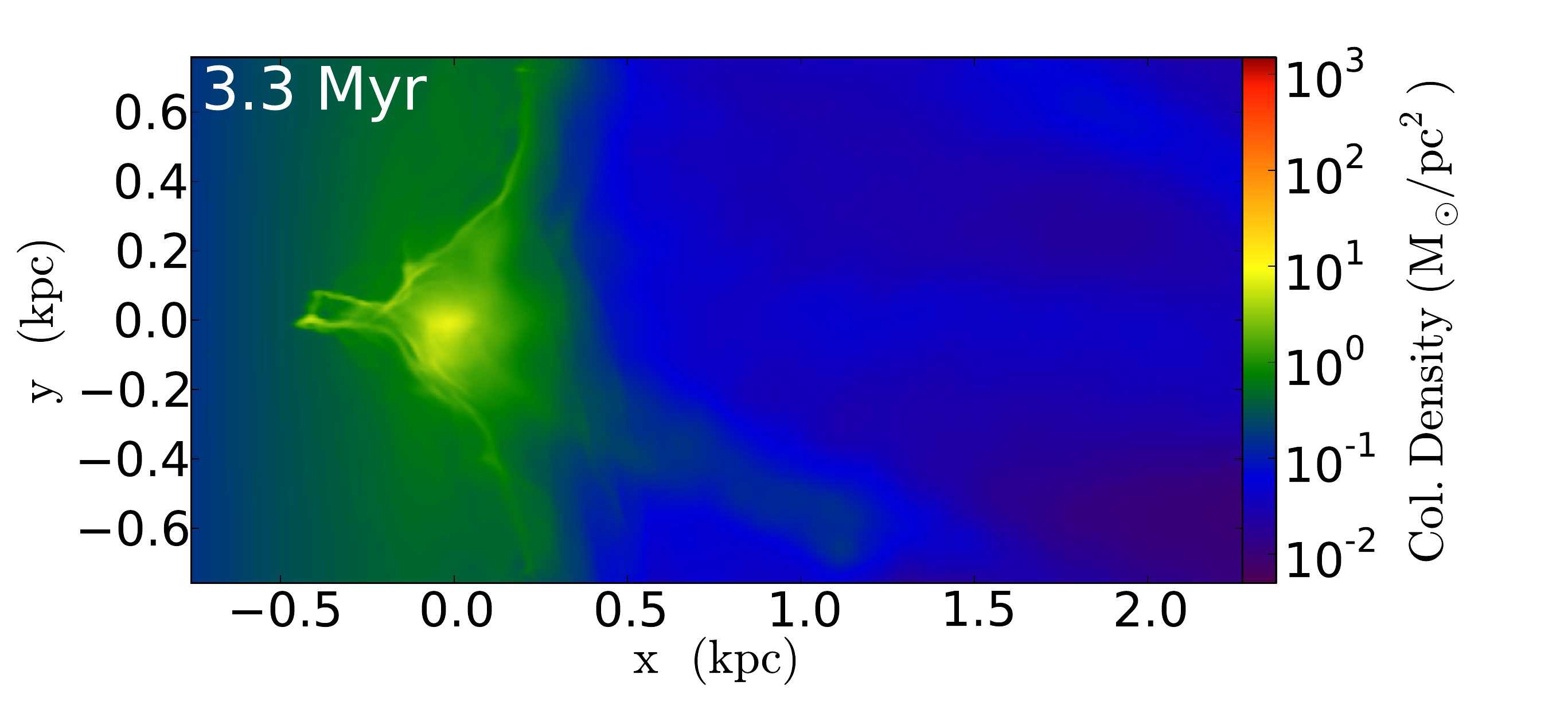}
\includegraphics*[scale=0.245, trim=95 64.1 162.5 21]{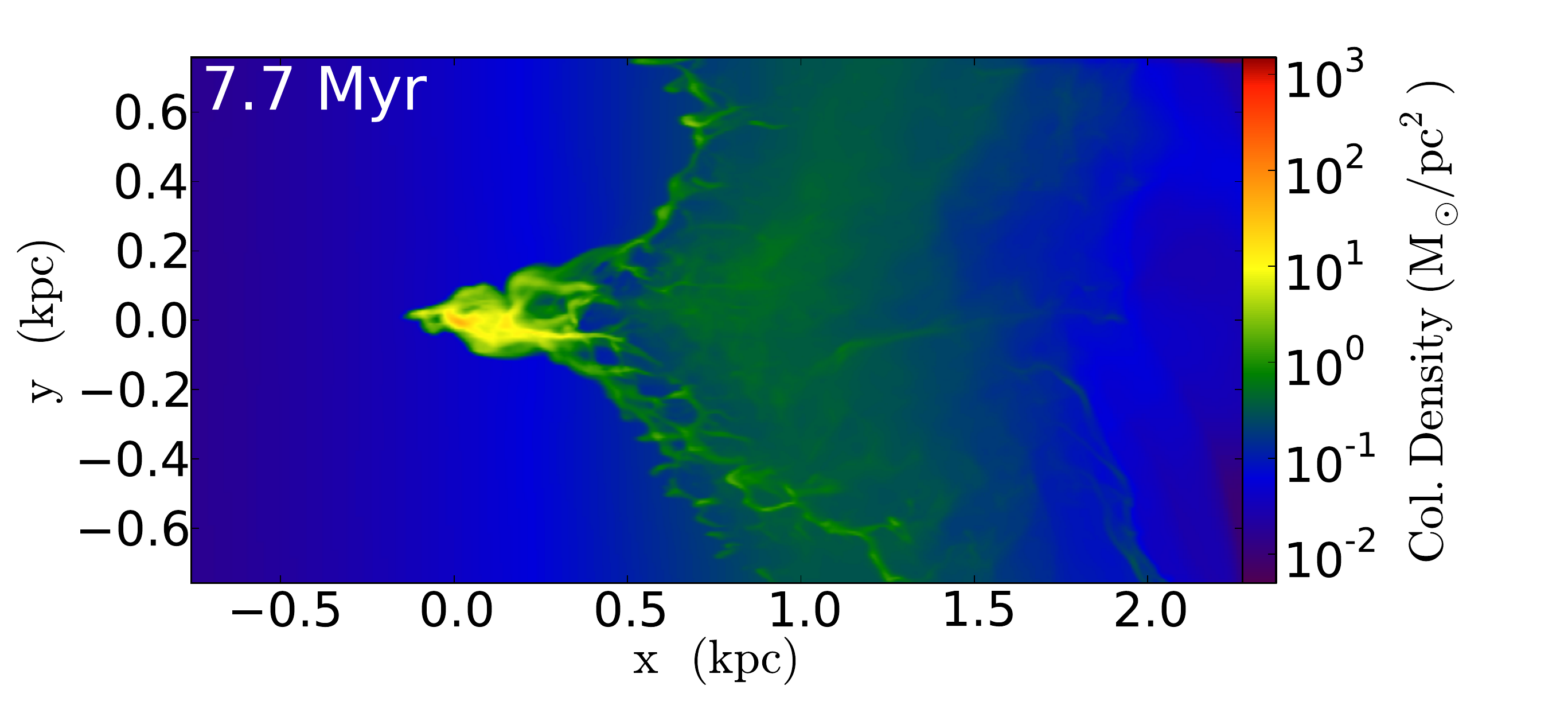}
\includegraphics*[scale=0.245, trim=95 64.1 0 21]{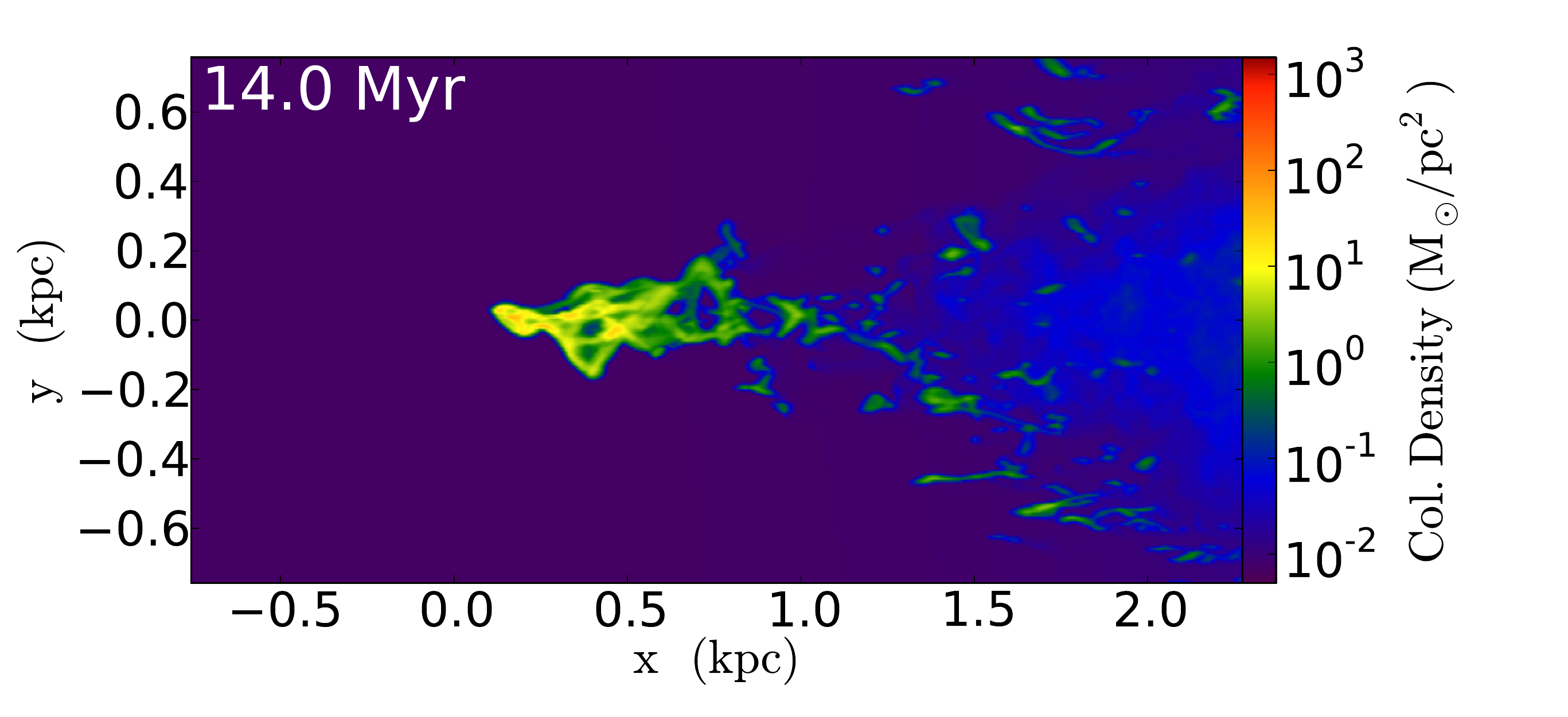}
\includegraphics*[scale=0.245, trim=0 0 162.5 21]{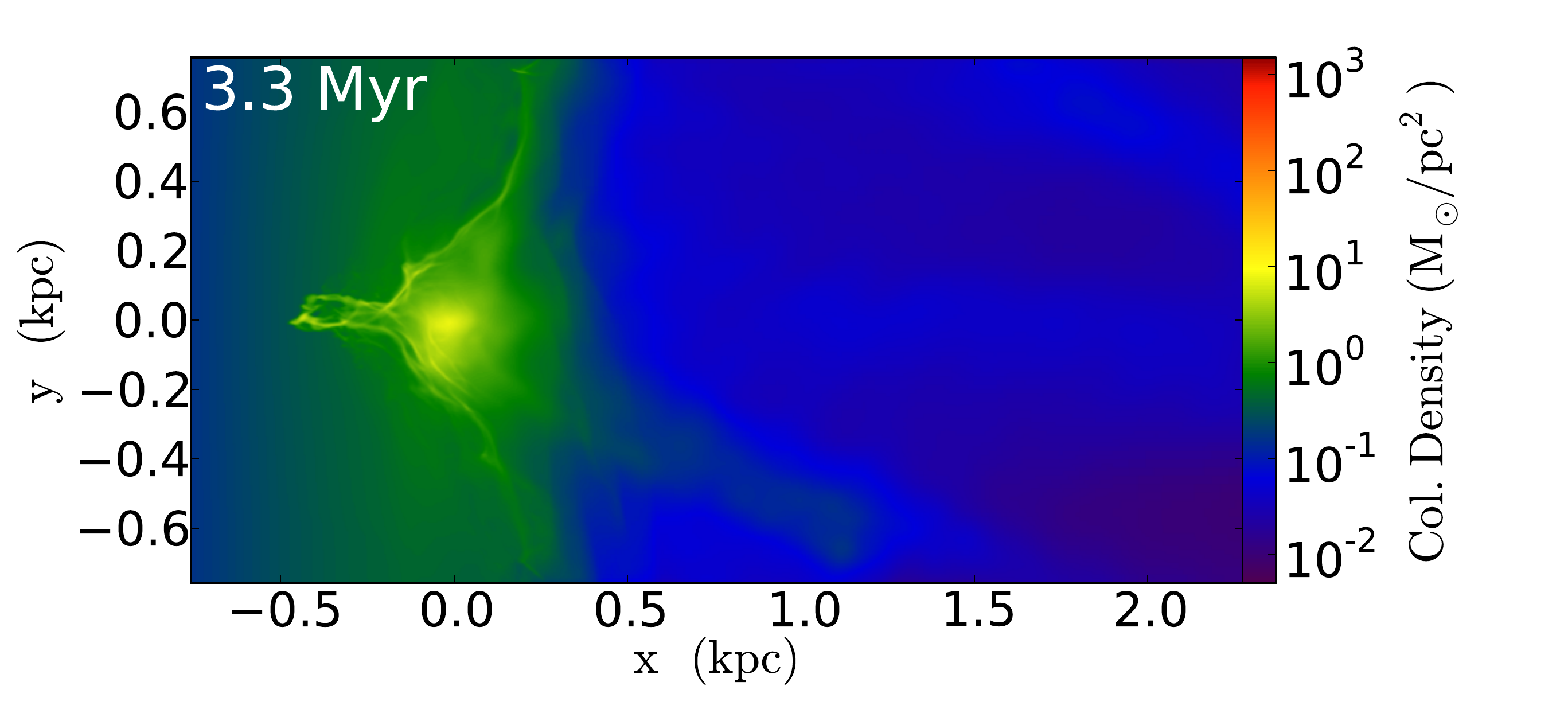}
\includegraphics*[scale=0.245, trim=95 0 162.5 21]{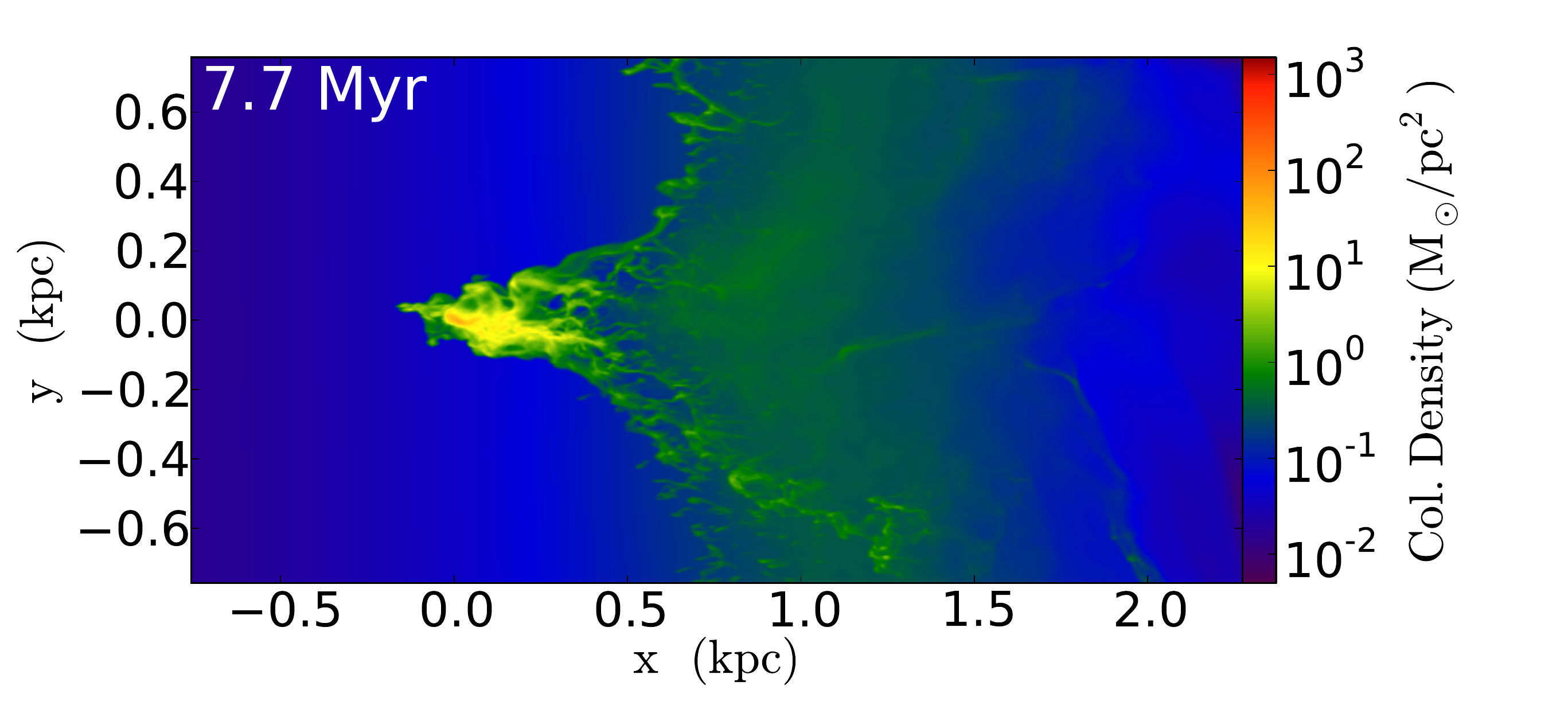}
\includegraphics*[scale=0.245, trim=95 0 0 21]{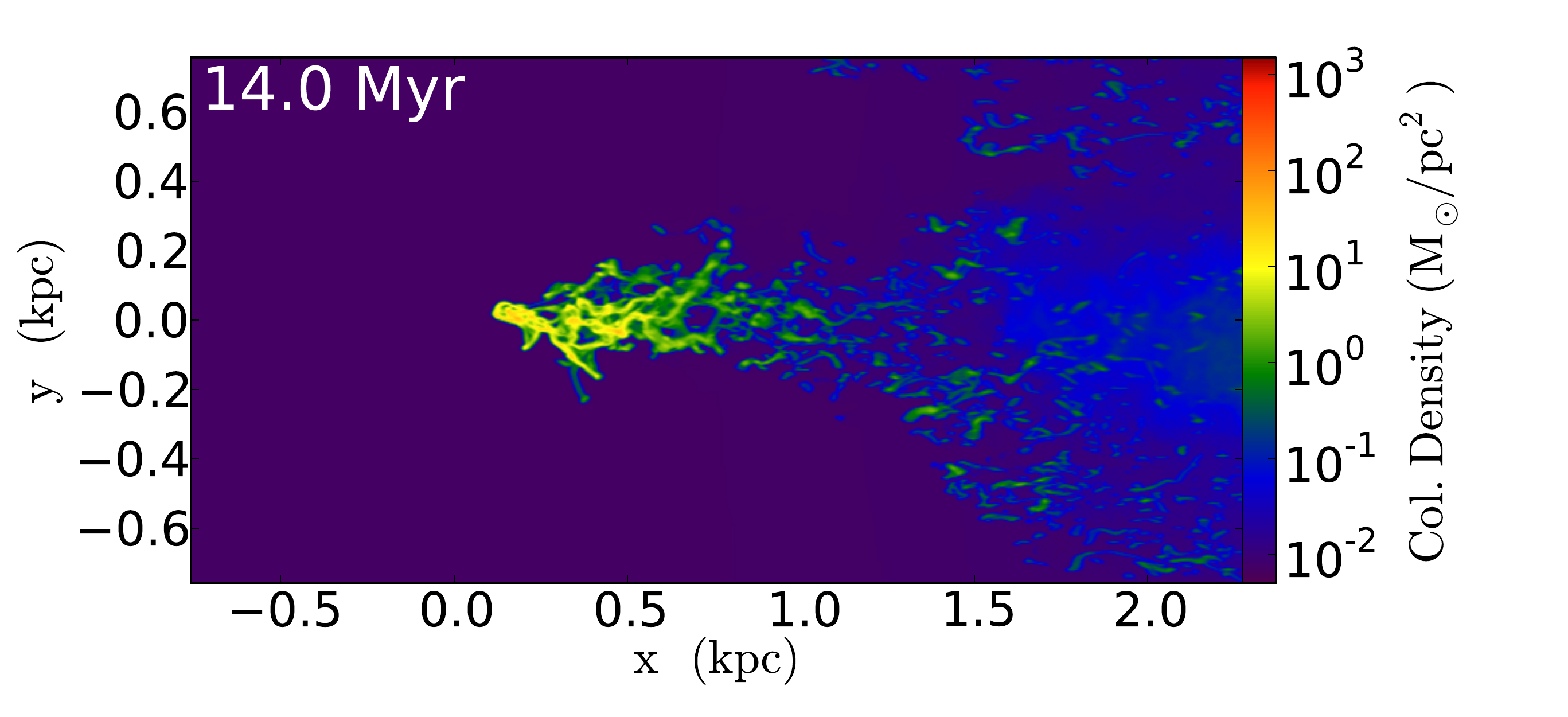}
\caption{\footnotesize{Demonstration of convergence of the fiducial simulation (FID; center) with the low-resolution (LR; top) and the high-resolution (HR; bottom) column density
results at 3.3, 7.7, and 14 Myr from left to right.}}
\label{fig_converge}
\end{figure*}

Simulations LR and HR were run with the maximum resolution in FLASH3.2 at half and double that of run FID (11.8pc and 2.46 pc, respectively). In the LR
there are regions that do not meet the Truelove criterion \citep{Truelove97}, as the Jeans length is not resolved by at least four fluid elements.
However, in both the FID and HR runs, this criterion is always met. \fig{fig_converge} shows the evolution of the column density for the LR, FID, and HR 
runs. The increased resolution is able to better resolve the fragmentation of the cloud caused by turbulence along the shock front.
This increased fragmentation leads to increased metal enrichment and general mixing of the material. 
Note that unlike GS11A we do not include a subgrid turbulence model in our calculations, meaning 
that the enrichment we compute is resolution dependent and should be considered a lower limit.

\begin{figure*}[h!]
\centering
\includegraphics[scale=0.5]{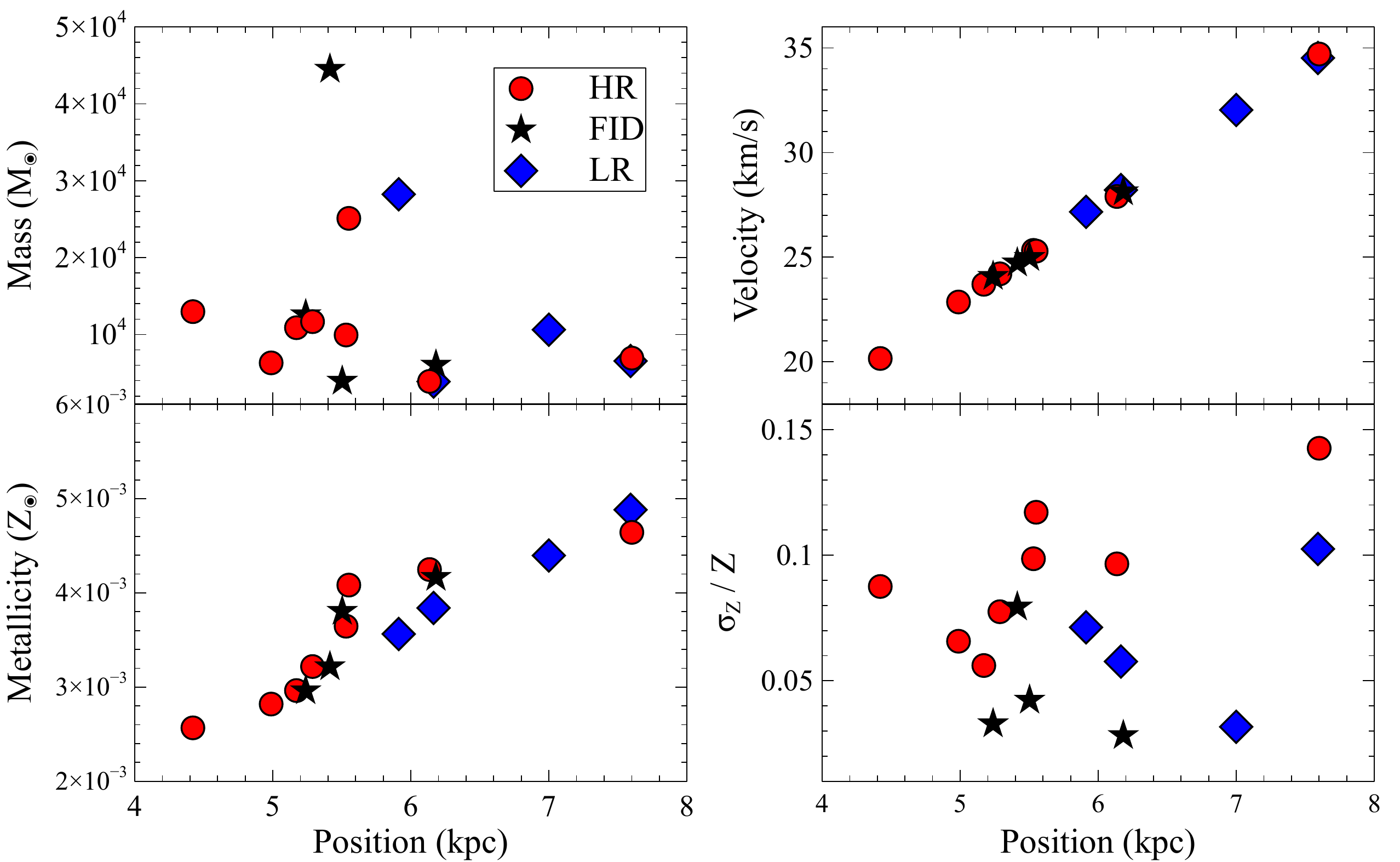}
\caption{\footnotesize{A comparison between the LR (blue diamonds), FID (black squares), and HR (red circles) particle masses (top left), velocity (top right), metallicity (bottom left)
and metallicity dispersion (bottom right) vs particles positions after 200 Myr.}}
\label{fig_converge2}
\end{figure*}

We also performed the same ballistics evolution for each of the simulations. We compare the resulting particles in \fig{fig_converge2}. The amount of the minihalo
baryons captured in these particles is 19\%, 24\%, and 26\% for the LR, FID, and HR runs, respectively. The increased resolution has lead to more fragmentation, leading 
to slightly increased enrichment of the final particles with a slightly larger dispersion, and whose mass is in a few more less massive clumps. However, the mass dispersion is roughly 0.3 dex
for all three simulations, while the average mass varies only by about 0.06 dex. Thus our statistics are still robust, even if the individual clumps are not identical with their higher
resolution counterparts. Likewise, we find little dependence on the abundance of the final clumps, in their evolution, so we do not expect the slightly-increased enrichment at higher resolution to influence our results. See \sect{metl} and \sect{glob} for more discussion on this. We are thus satisfied that
the FID resolution is sufficient to model this interaction for other parameters sets.

\subsection{Parameter Study}\label{param}
The simulations run in this parameter suite are detailed in \tabl{tab_sims}, with names corresponding to the parameter
that was altered from the fiducial value. PO1 has the outflow propagating through the IGM before striking the minihalo, while in all other simulations the
outflow propagates along a filament. In the following subsections we discuss both the results from the FLASH3.2 simulations for each
parameter as well as the evolution of the ballistic particles. \vspace{10mm}

\begin{figure*}[h!]
\centering
\includegraphics*[scale=0.245, trim=  0 43.7 162.5 21]{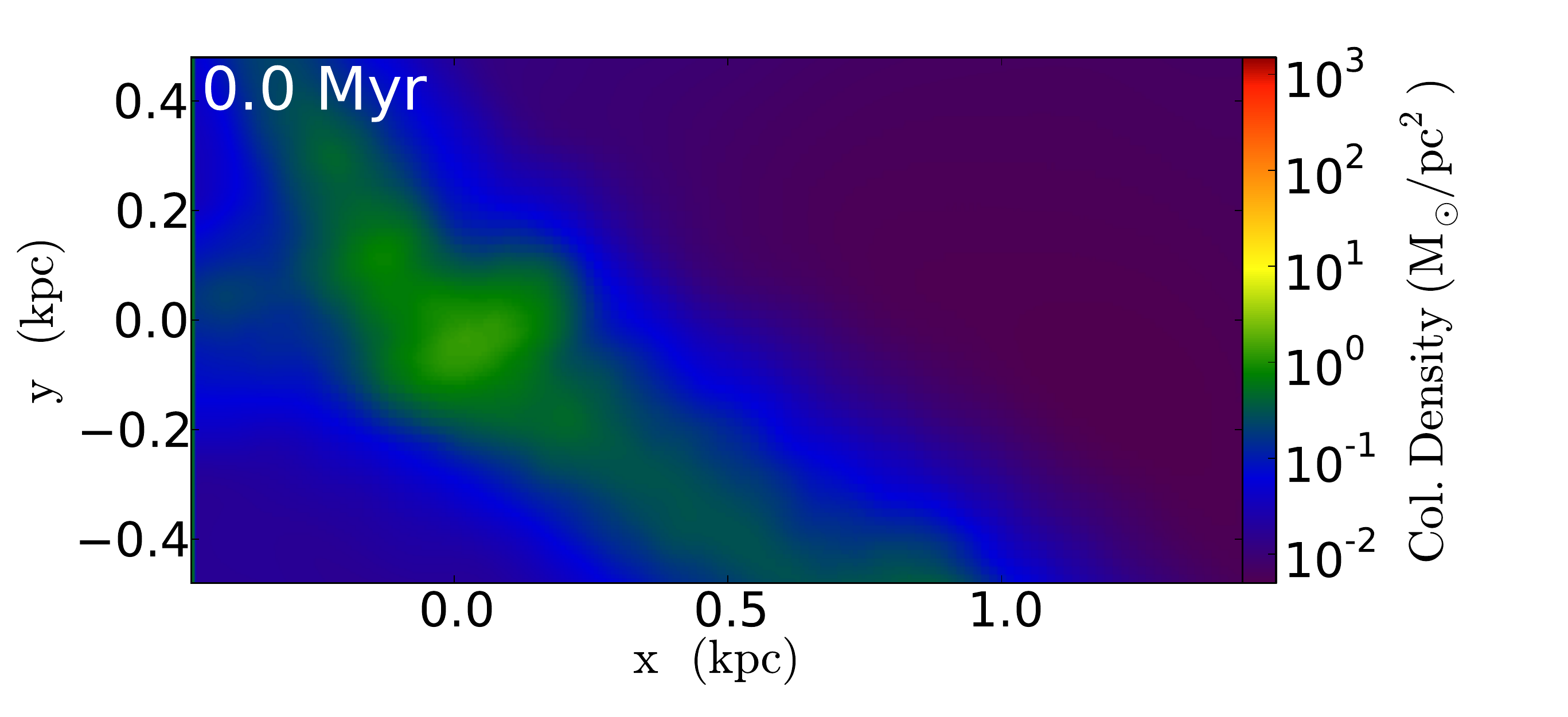}
\includegraphics*[scale=0.245, trim=95 43.7 162.5 14]{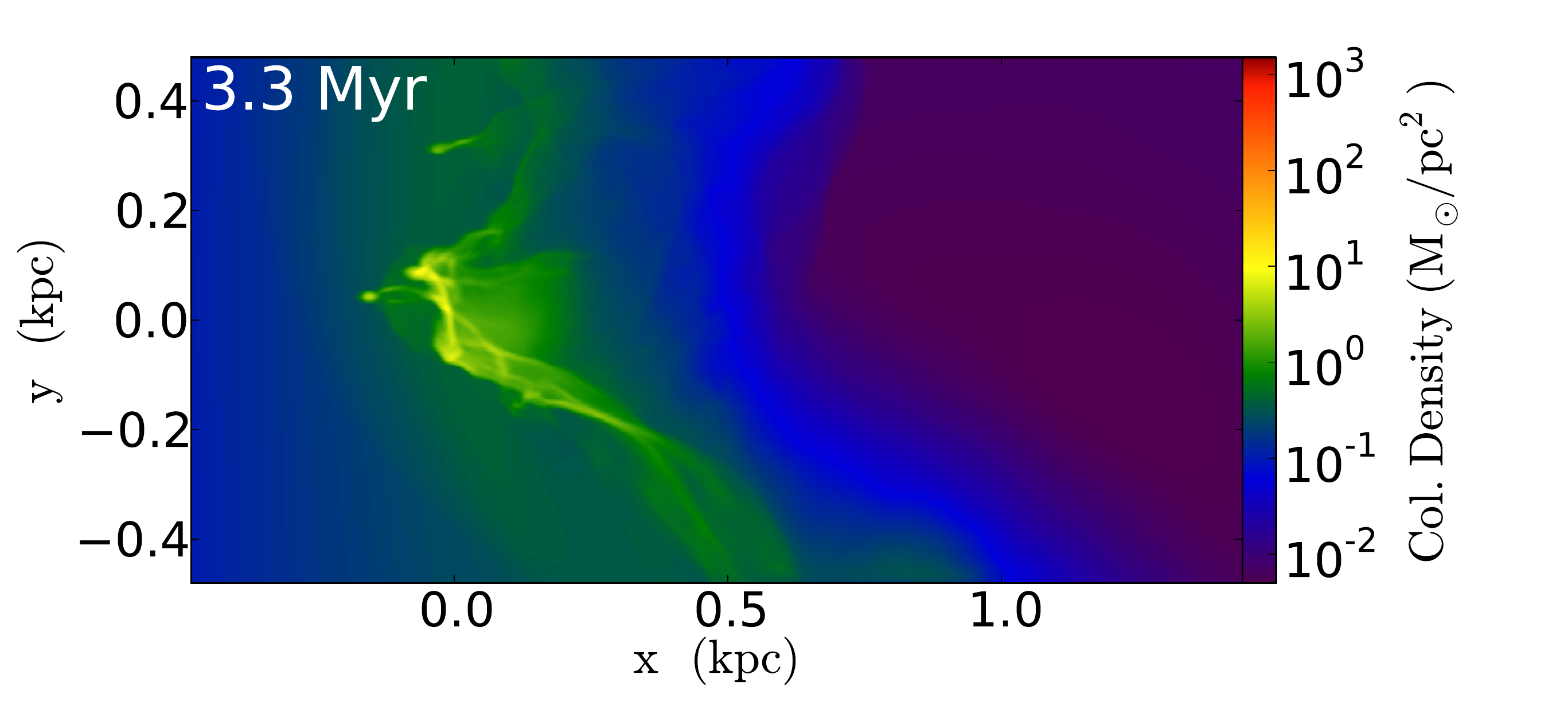}
\includegraphics*[scale=0.245, trim=95 43.7         0 14]{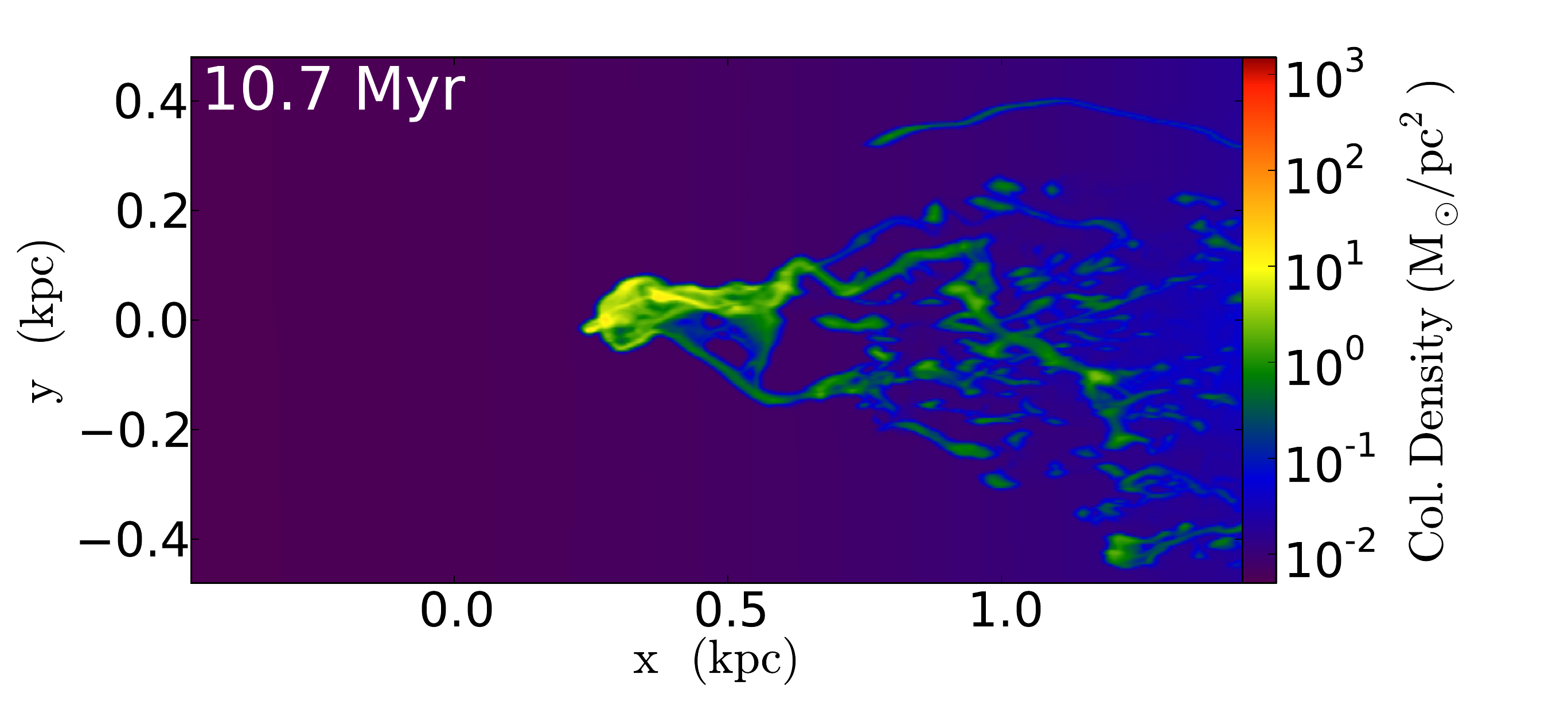}
\includegraphics*[scale=0.245, trim=  0 43.7 162.5 14]{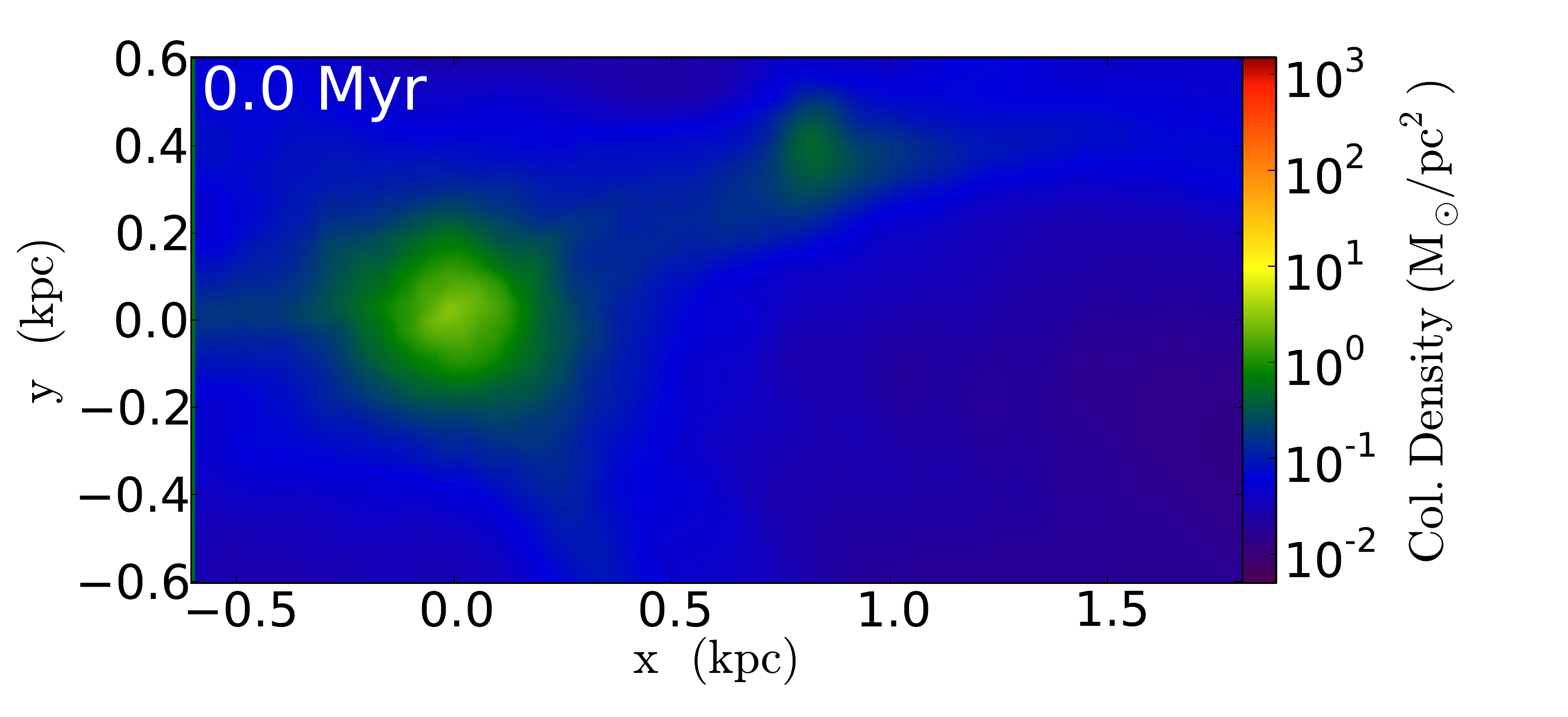}
\includegraphics*[scale=0.245, trim=95 43.7 162.5 14]{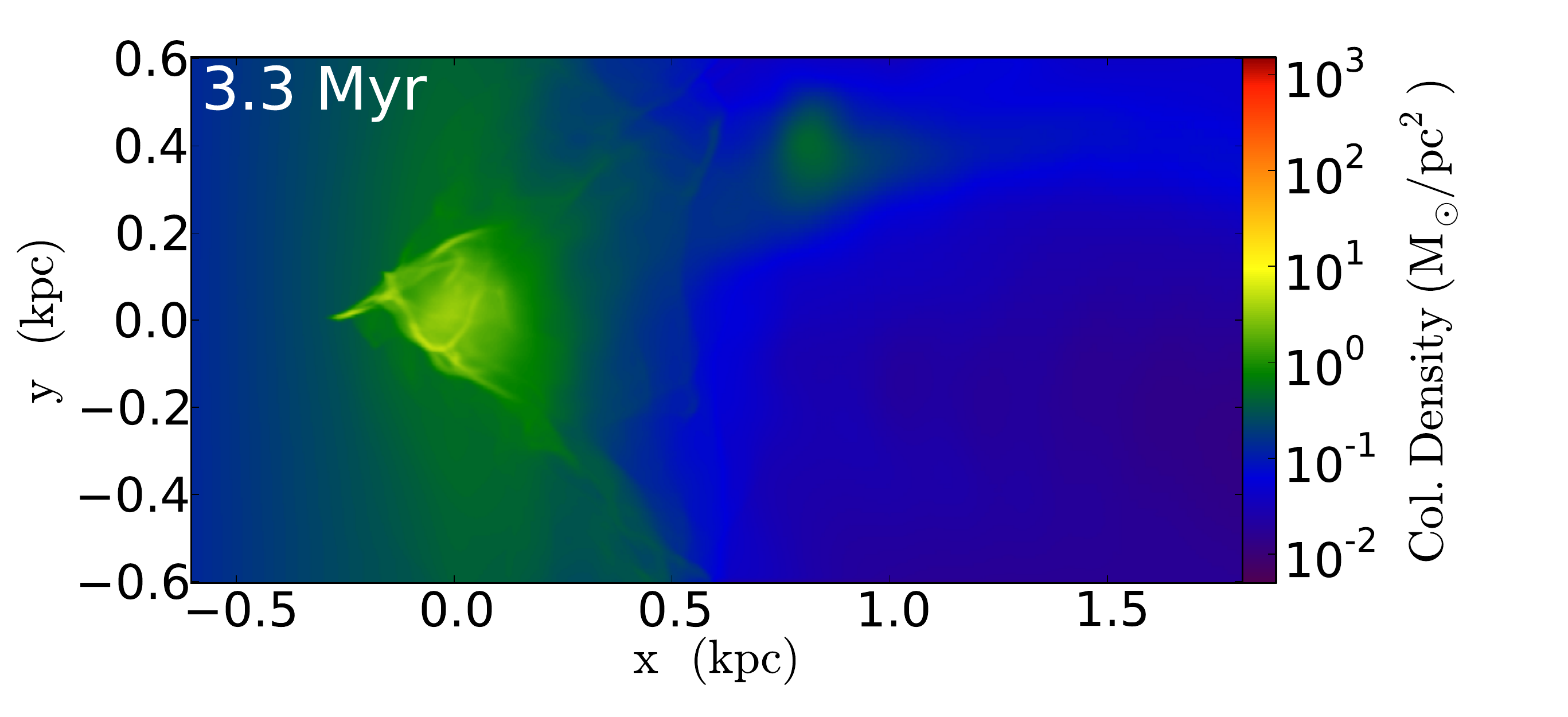}
\includegraphics*[scale=0.245, trim=95 43.7         0 14]{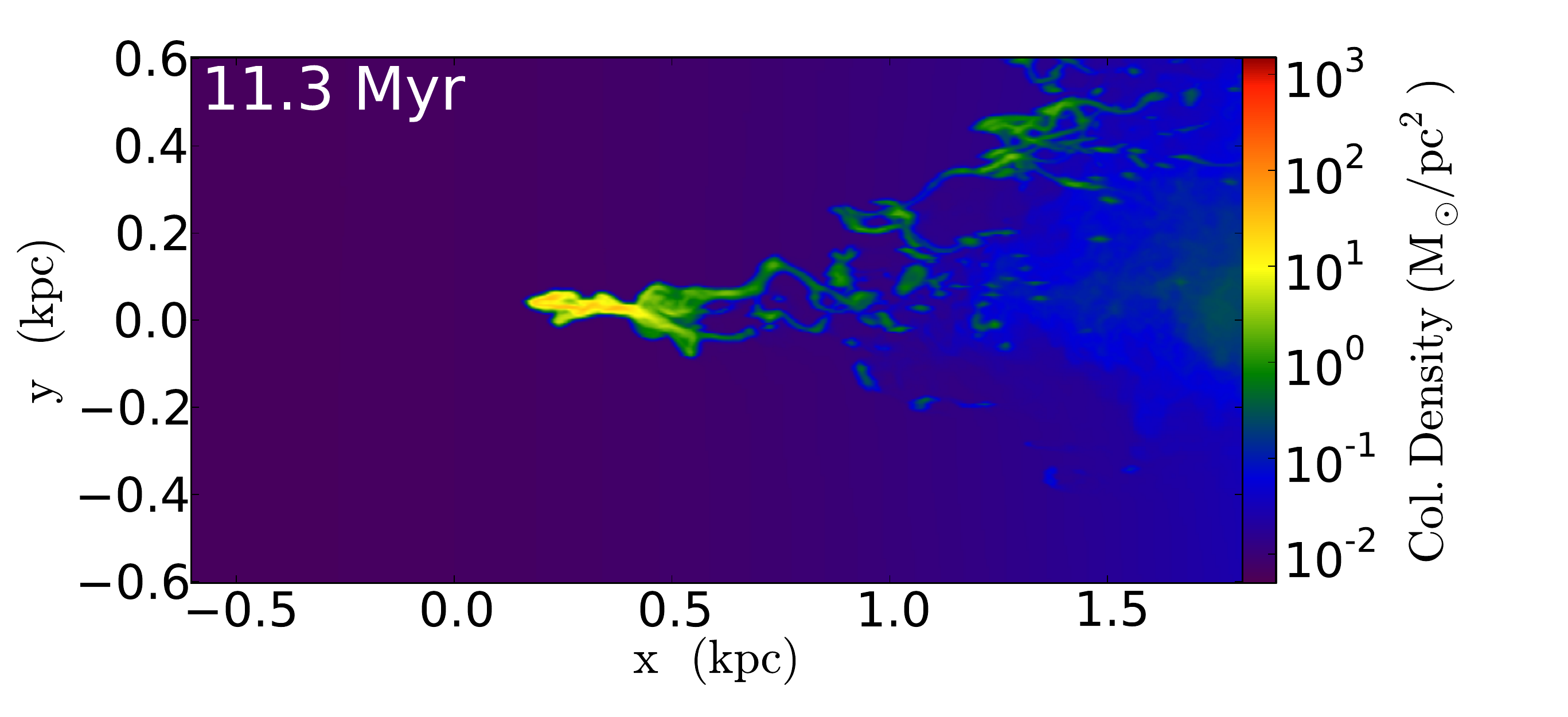}
\includegraphics*[scale=0.245, trim=  0 43.7 162.5 14]{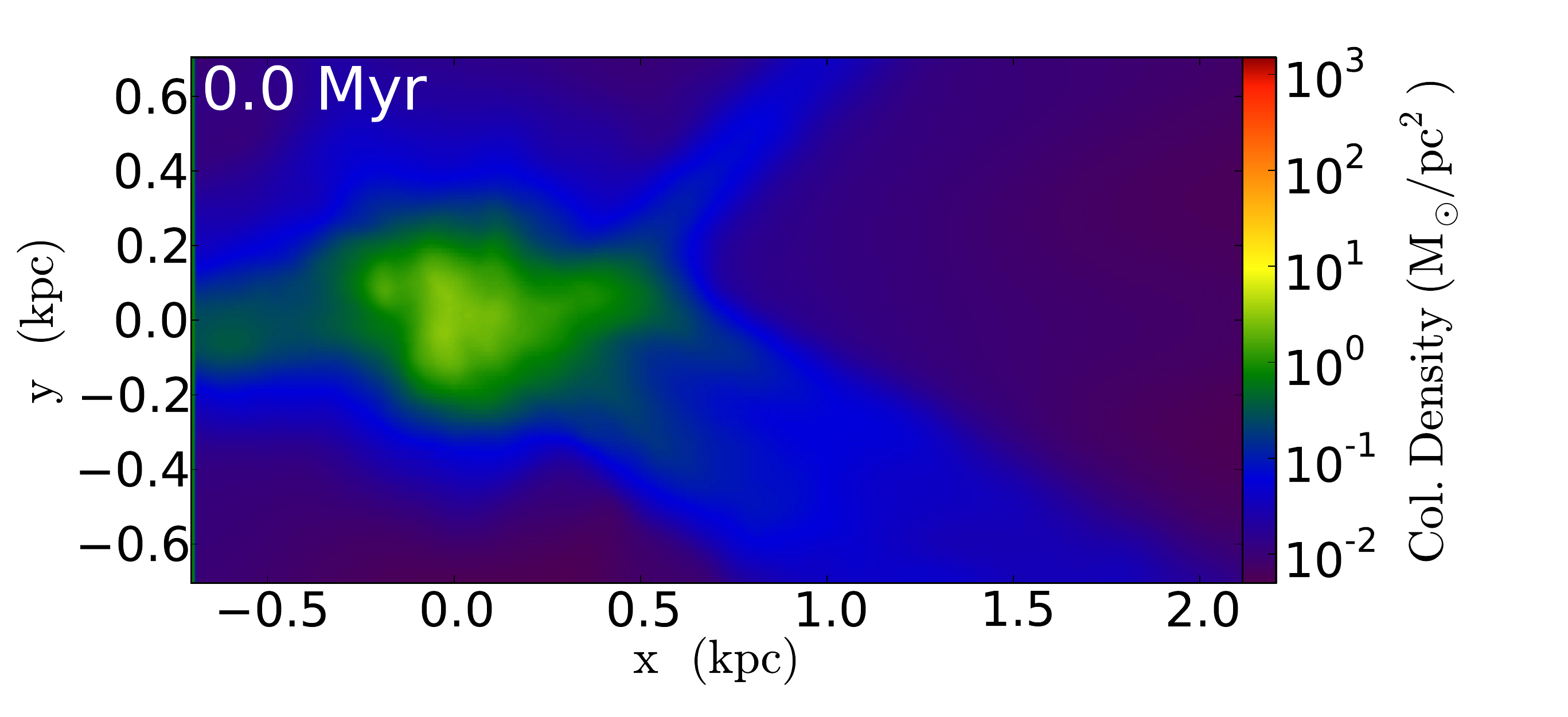}
\includegraphics*[scale=0.245, trim=95 43.7 162.5 14]{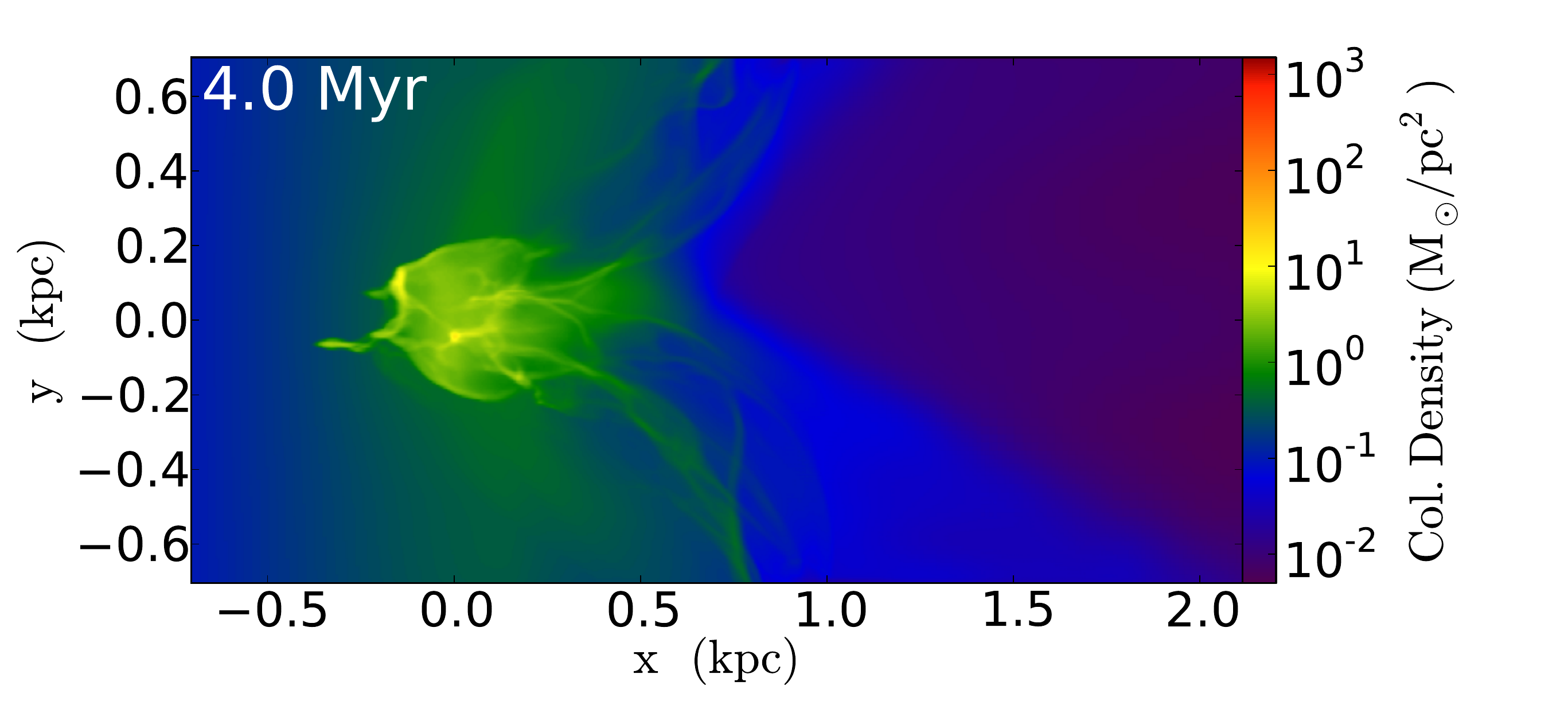}
\includegraphics*[scale=0.245, trim=95 43.7         0 14]{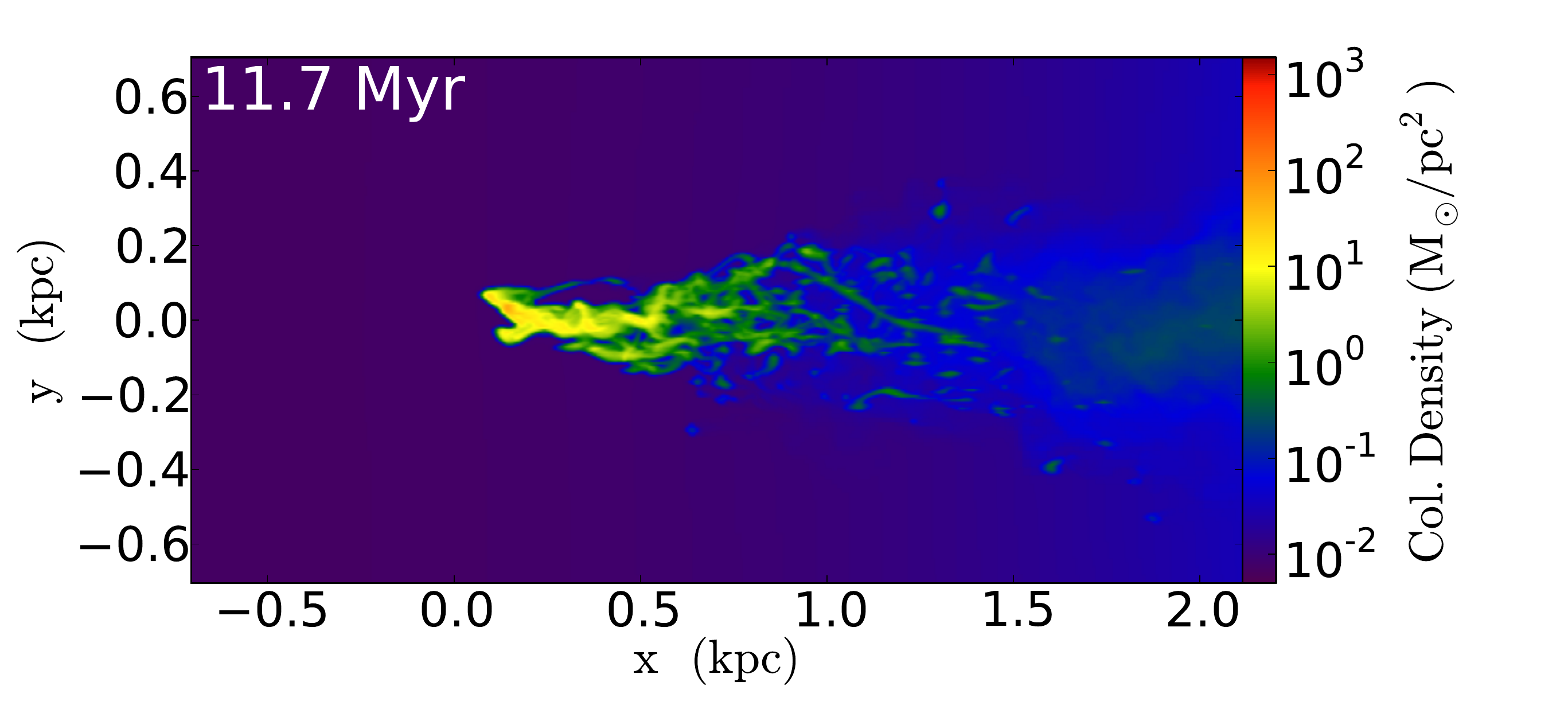}
\includegraphics*[scale=0.245, trim=  0 43.7 162.5 14]{Fid_Col_Dens_0_Proj.pdf}
\includegraphics*[scale=0.245, trim=95 43.7 162.5 14]{Fid_Col_Dens_15_Proj.pdf}
\includegraphics*[scale=0.245, trim=95 43.7         0 14]{Fid_Col_Dens_39_Proj.pdf}
\includegraphics*[scale=0.245, trim=  0 43.7 162.5 14]{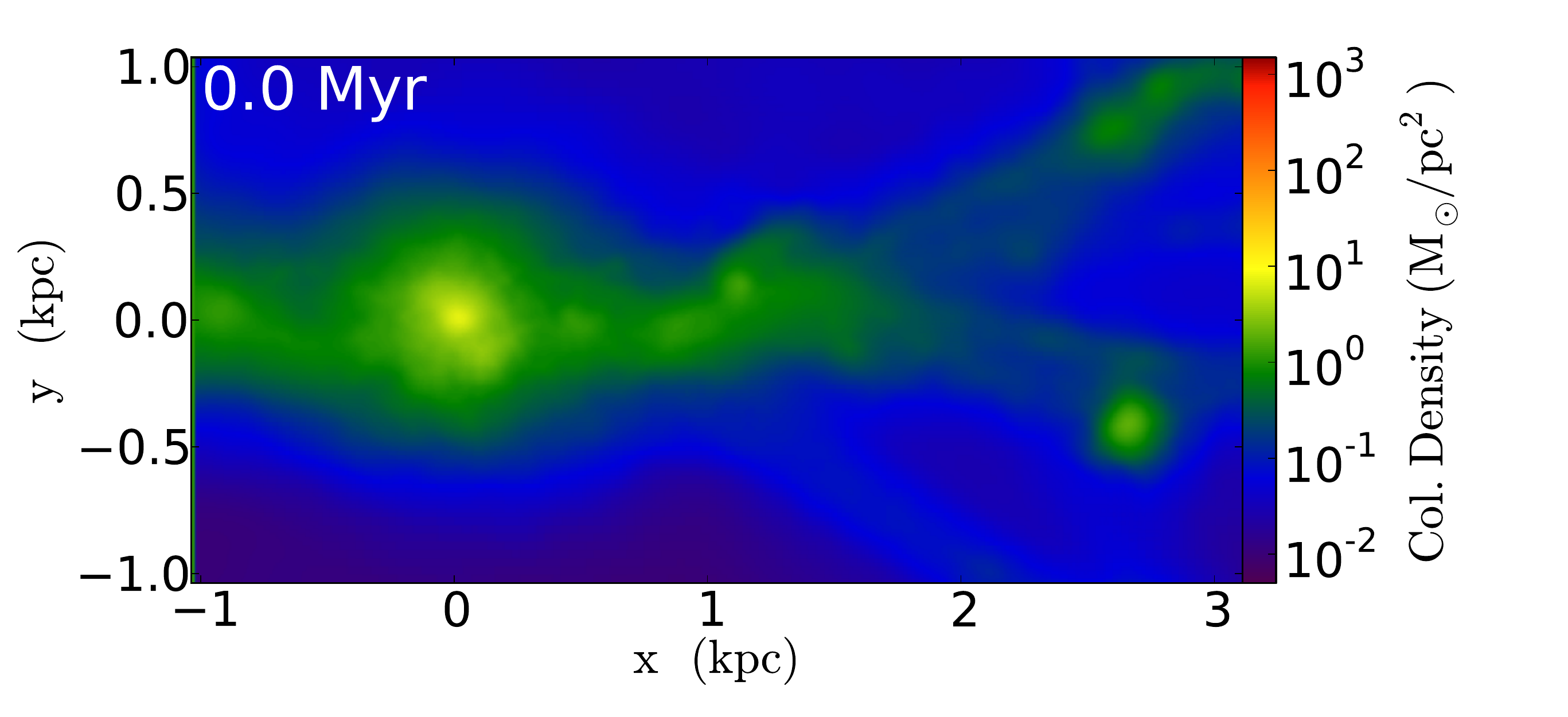}
\includegraphics*[scale=0.245, trim=95 43.7 162.5 14]{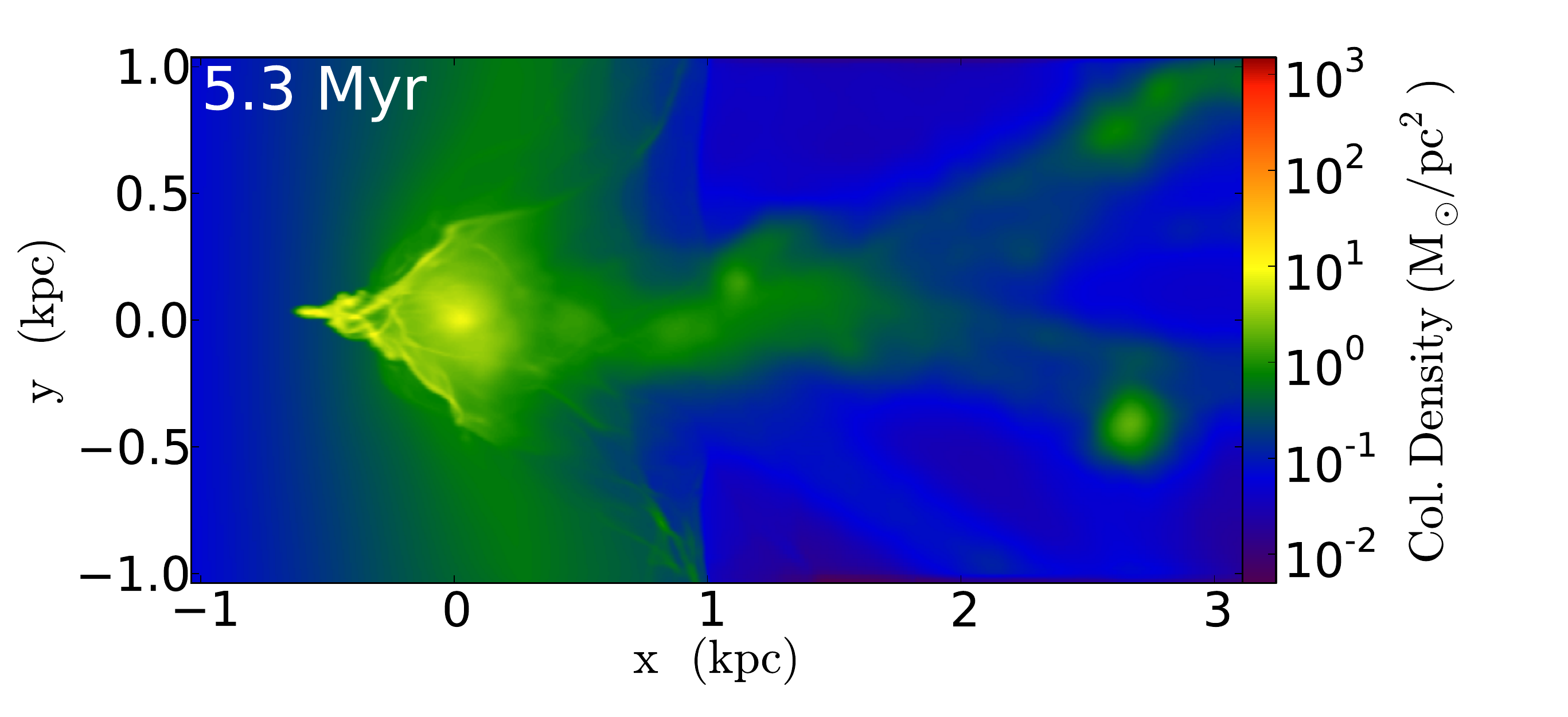}
\includegraphics*[scale=0.245, trim=95 43.7         0 14]{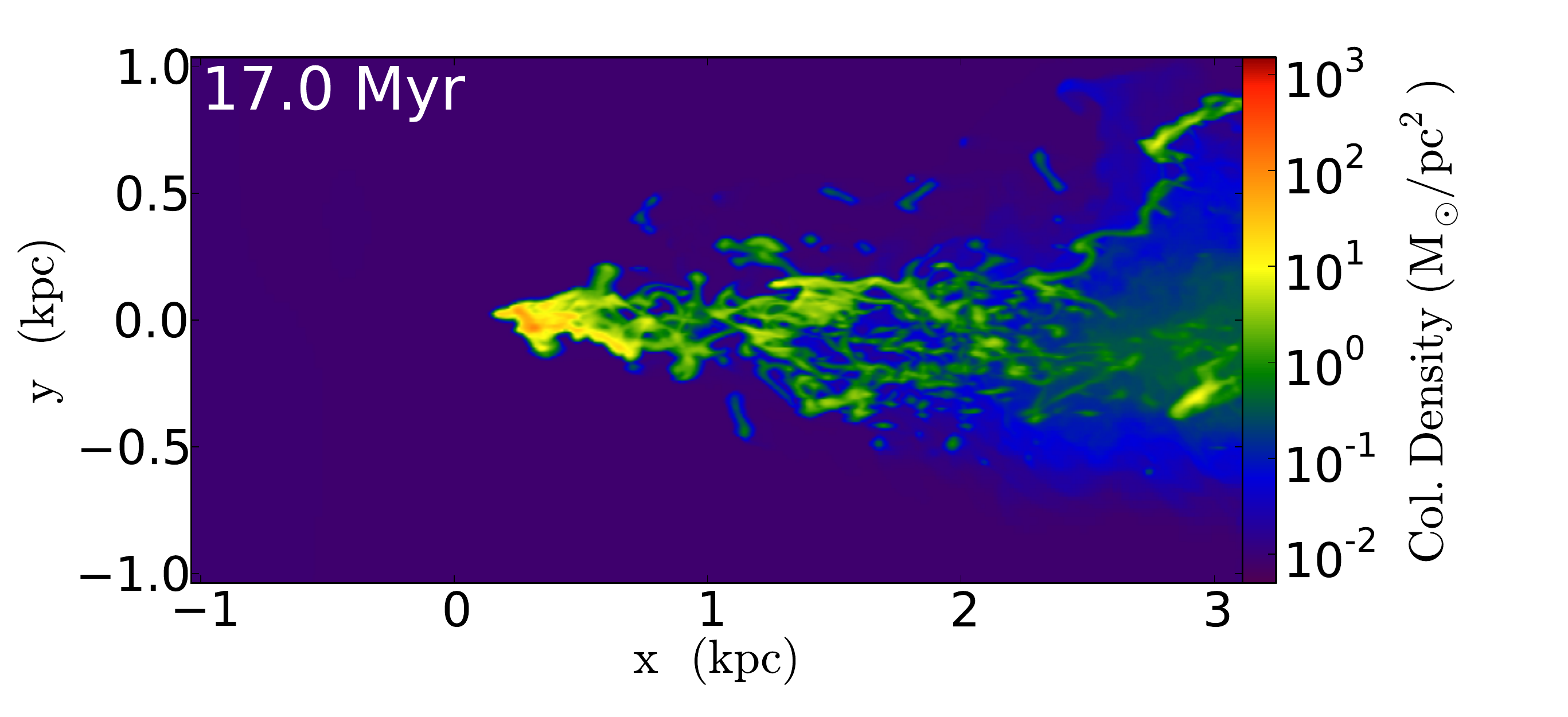}
\includegraphics*[scale=0.245, trim=  0       0 162.5 14]{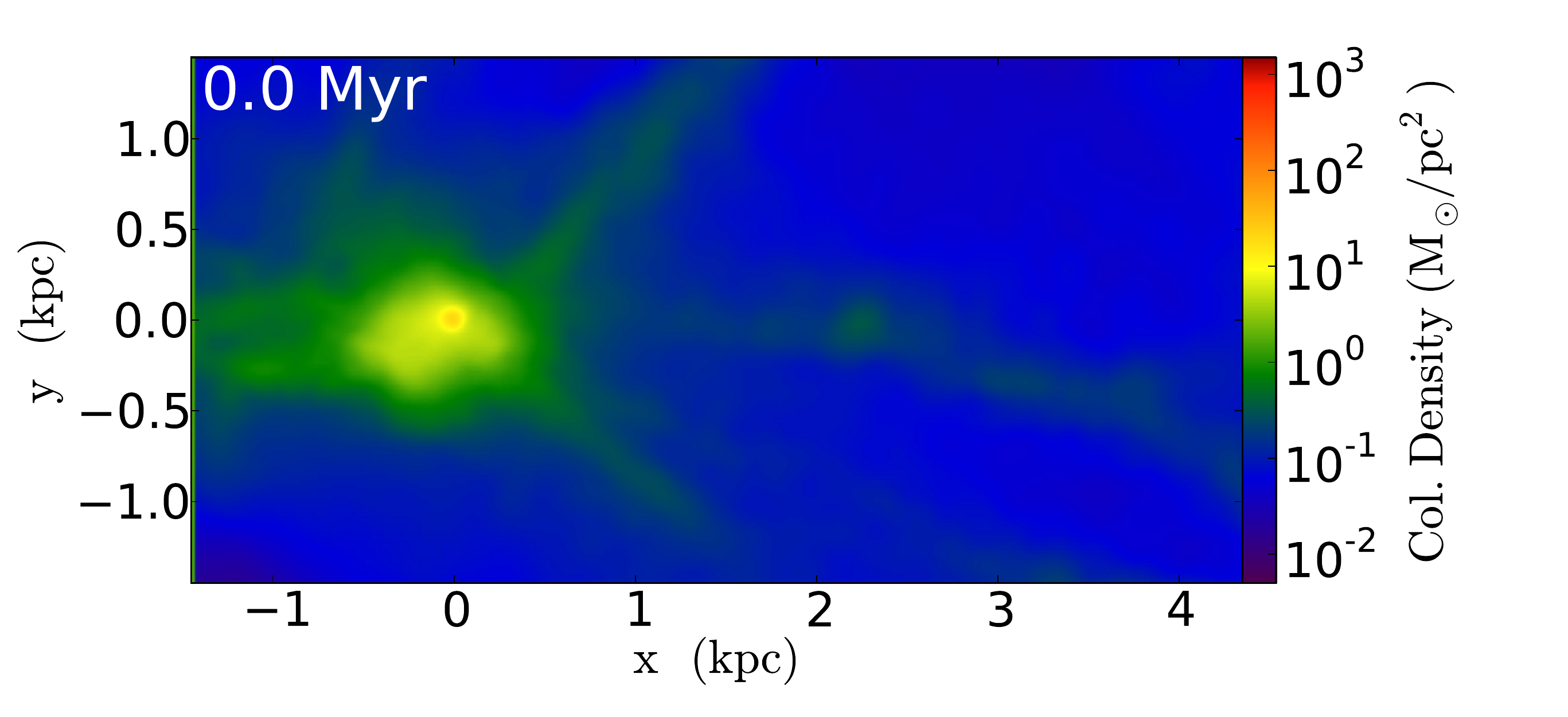}
\includegraphics*[scale=0.245, trim=95       0 162.5 14]{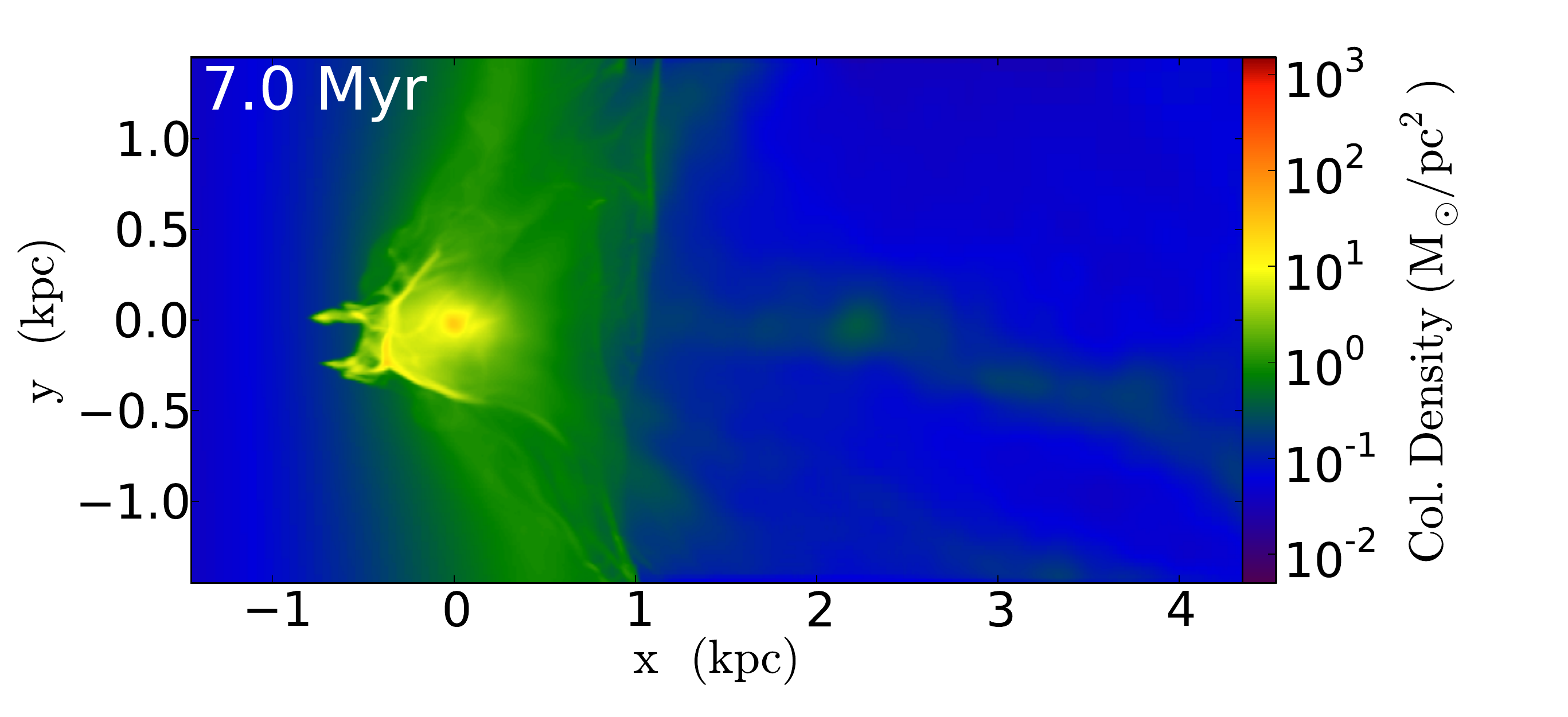}
\includegraphics*[scale=0.245, trim=95       0         0 14]{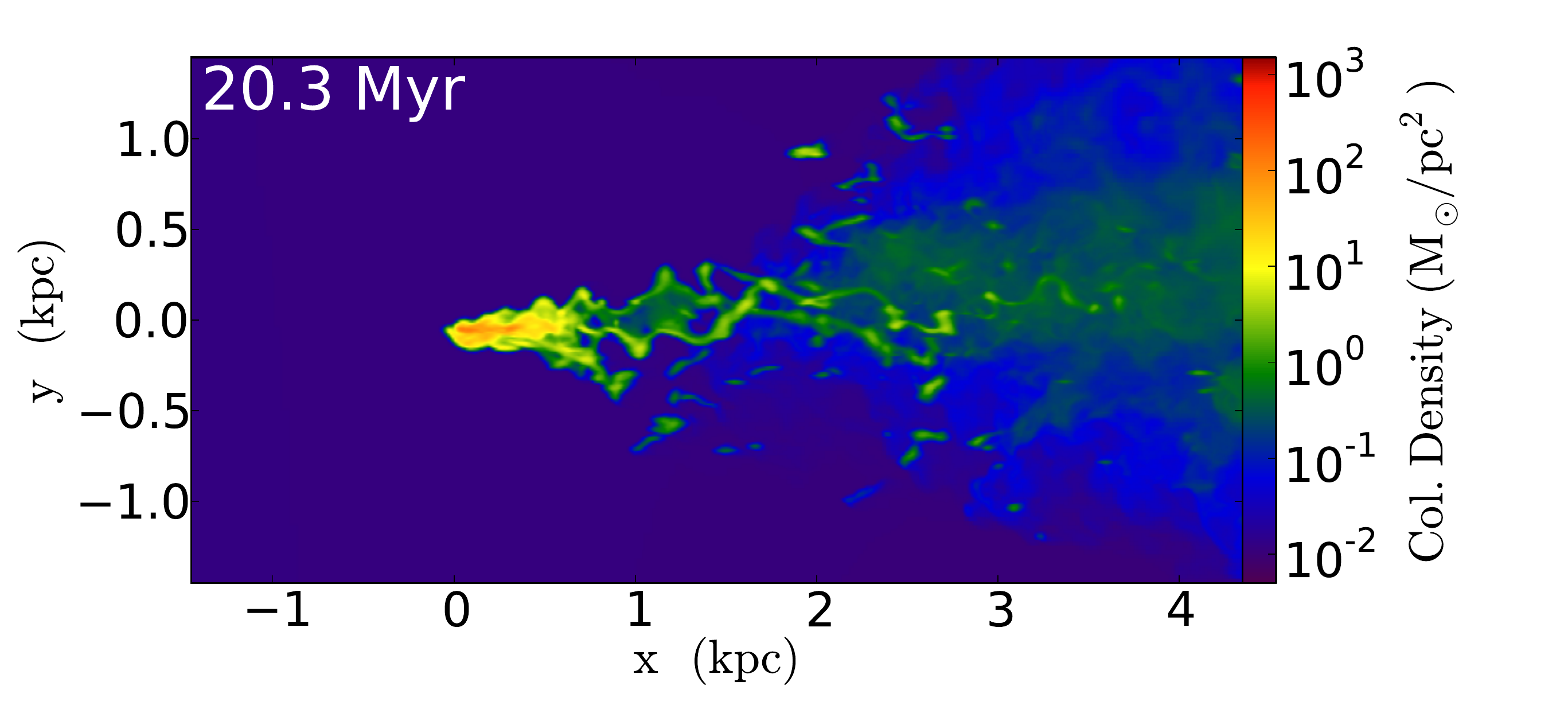}
\caption{\footnotesize{Column density of the simulations varying the minihalo mass, from top to
bottom PM07, PM1, PM23, FID, PM7, and PM19, respectively, with evolution increasing from the 
beginning on the left, to when the outflow is just passing the minihalo in the middle, to when the 
outflow reaches the end of the box on the right. All plots have $x$ ranging from $-1.5R_{\rm v}$ to $4.5R_{\rm v}$,
while $y$ ranges from $-1.5R_{\rm v}$ to $1.5R_{\rm v}$. Thus the physical scale varies from one
row to the next, but the characteristic halo scale matches. Similarly, the elapsed time varies as we increase mass
since it requires more time for the shock to traverse the minihalo.}}
\label{fig_mass}
\end{figure*}

\begin{figure*}[t!]
\centering
\includegraphics[scale=0.42]{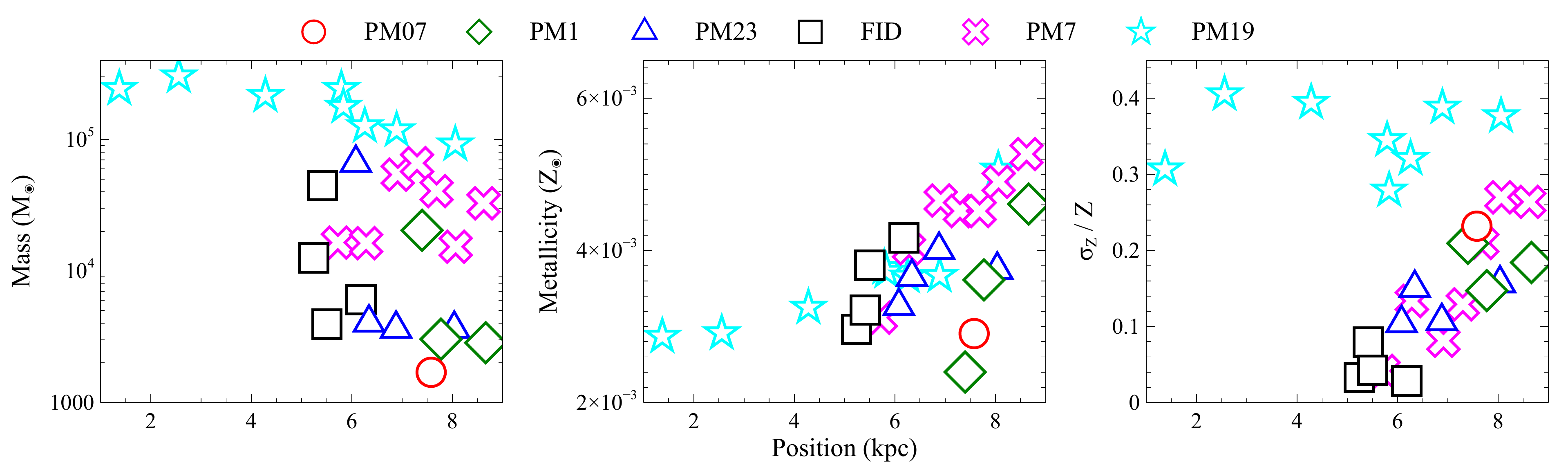}
\caption{\footnotesize{Comparison between the PM07 (red circles), PM1 (green diamonds), PM23 
(blue triangles), FID (black squares), PM7 (magenta crosses) and PM19 (cyan stars) simulations 
illustrating the dependence on minihalo mass. The particle masses (left), 
metallicity (middle) and relative metallicity dispersion (right) vs particles 
positions after 200 Myr are shown.}}
\label{fig_mass2}
\end{figure*}

\subsubsection{Minihalo Mass}\label{mass}

The minihalo mass is one of the most important parameters, as it sets not only the mass of baryons present, but
the scale of the interaction, the virial temperature of the minihalo, and the depth of the gravitational potential. 
How this parameter affects the evolution of the interaction is shown in \fig{fig_mass}.
Each minihalo is unique, chosen for its mass. Nevertheless, the interactions of the outflows with these 
minihalos are consistent. We see that the more massive minihalos create denser shock fronts as 
the outflow is stalled by the denser material, have denser ribbons of material after the shock 
overtakes the minihalo, and they have larger wakes of swept up material. The nature of the 
filament along which the outflow propagates appears to influence how the shock interacts with the 
minihalo itself, and in \sect{orientation} we explore  whether this can affect the produced particle 
clouds.

\fig{fig_mass2} shows the distribution of cloud particles  200 Myr after these simulations.
As the minihalo mass increases, the cloud particles are found closer to the dark 
matter halo, there are more final particles, and they have larger masses. The more massive 
minihalos have a deeper potential well, and the innermost, most massive cloud particles are 
unable to escape for the very largest minihalo. One exception to this appears to be PM7. We 
suspect in this case  the proximity of two smaller minihalos, as well as the orientation of a  second filament 
directly behind the minihalo allows the gas to better escape the minihalo, while entraining the 
secondary filament material. We find the percent of the original minihalo's 
baryons in  bound cloud particles is much larger for larger minihalo mass, with 
only 3.7\% of baryons contained in the collapsed gas for PM07, while 61\% of baryons are 
contained in PM19. 

The metallicity again increases with increasing position of the final cloud particles. 
The smaller mass minihalo results in less metals penetrating into the cloud, as the shock 
travels more quickly around the more tenuous material. The relative spread in metallicity, 
$\sigma_{\rm Z}/Z$, is mostly below 0.1 dex, 
except for PM19, whose increased gravity
allows for more material of different enrichment to be constrained in the final ribbon of material.

\begin{figure*}[t!]
\centering
\includegraphics*[scale=0.245, trim=0 64.1 162.5 21]{Fid_Col_Dens_0_Proj.pdf}
\includegraphics*[scale=0.245, trim=95 64.1 162.5 21]{Fid_Col_Dens_15_Proj.pdf}
\includegraphics*[scale=0.245, trim=95 64.1 0 21]{Fid_Col_Dens_39_Proj.pdf}
\includegraphics*[scale=0.245, trim=0 0 162.5 21]{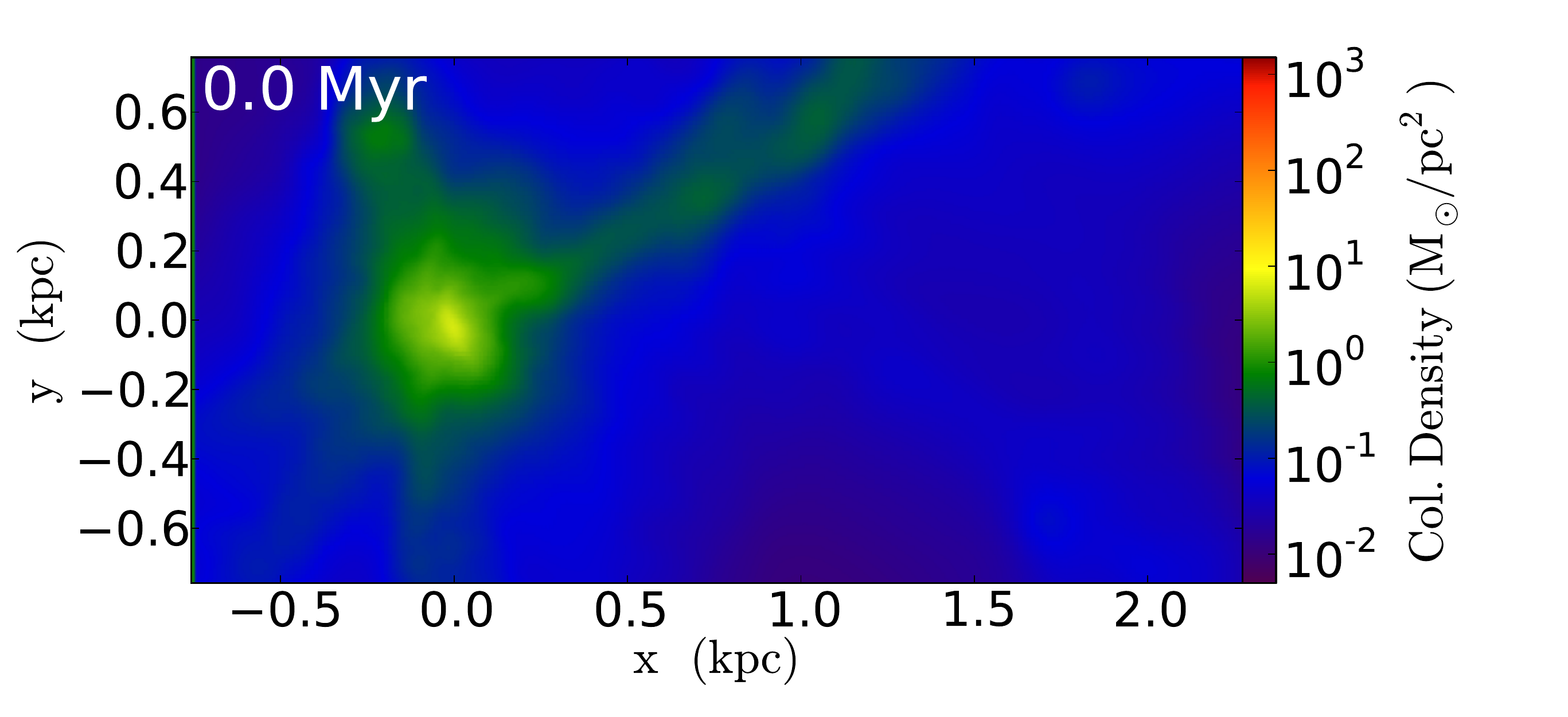}
\includegraphics*[scale=0.245, trim=95 0 162.5 21]{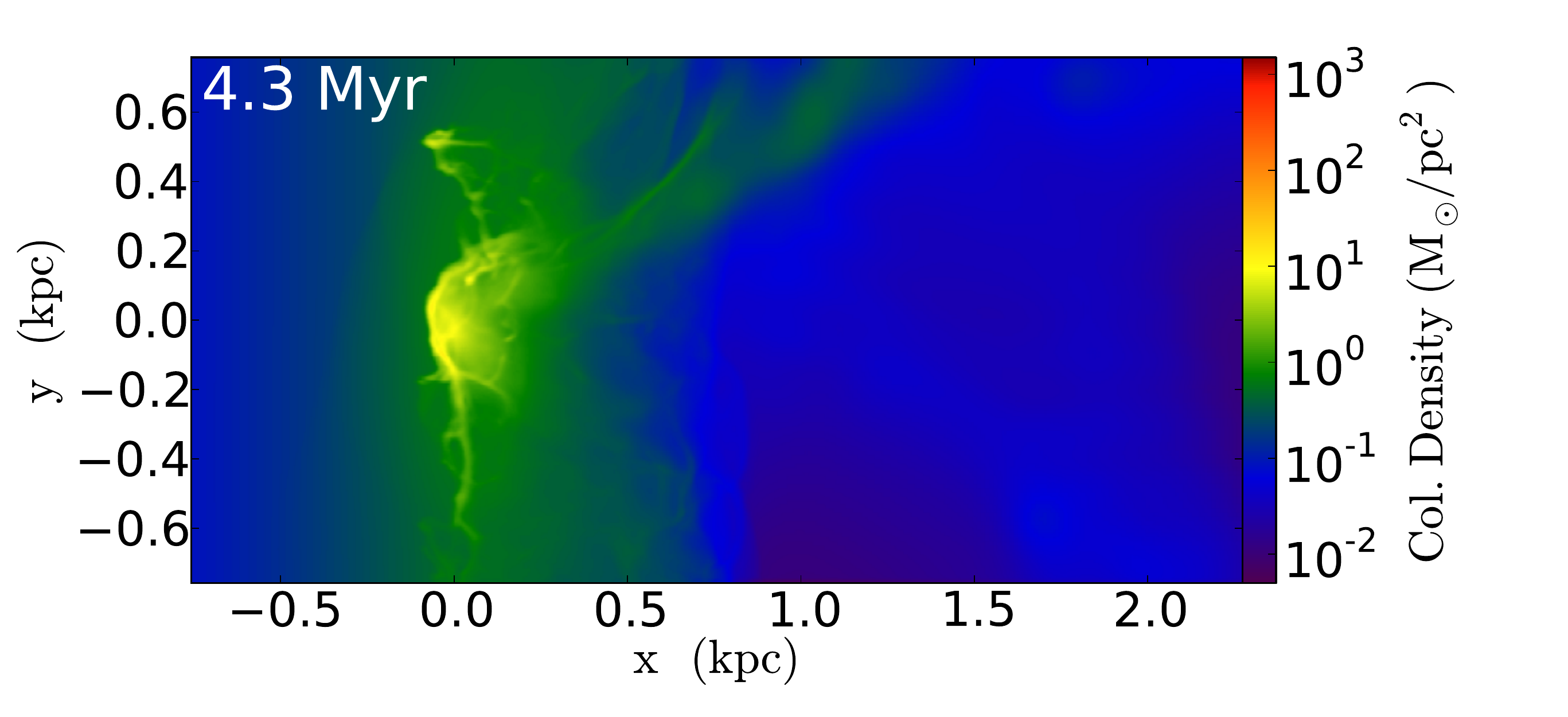}
\includegraphics*[scale=0.245, trim=95 0 0 21]{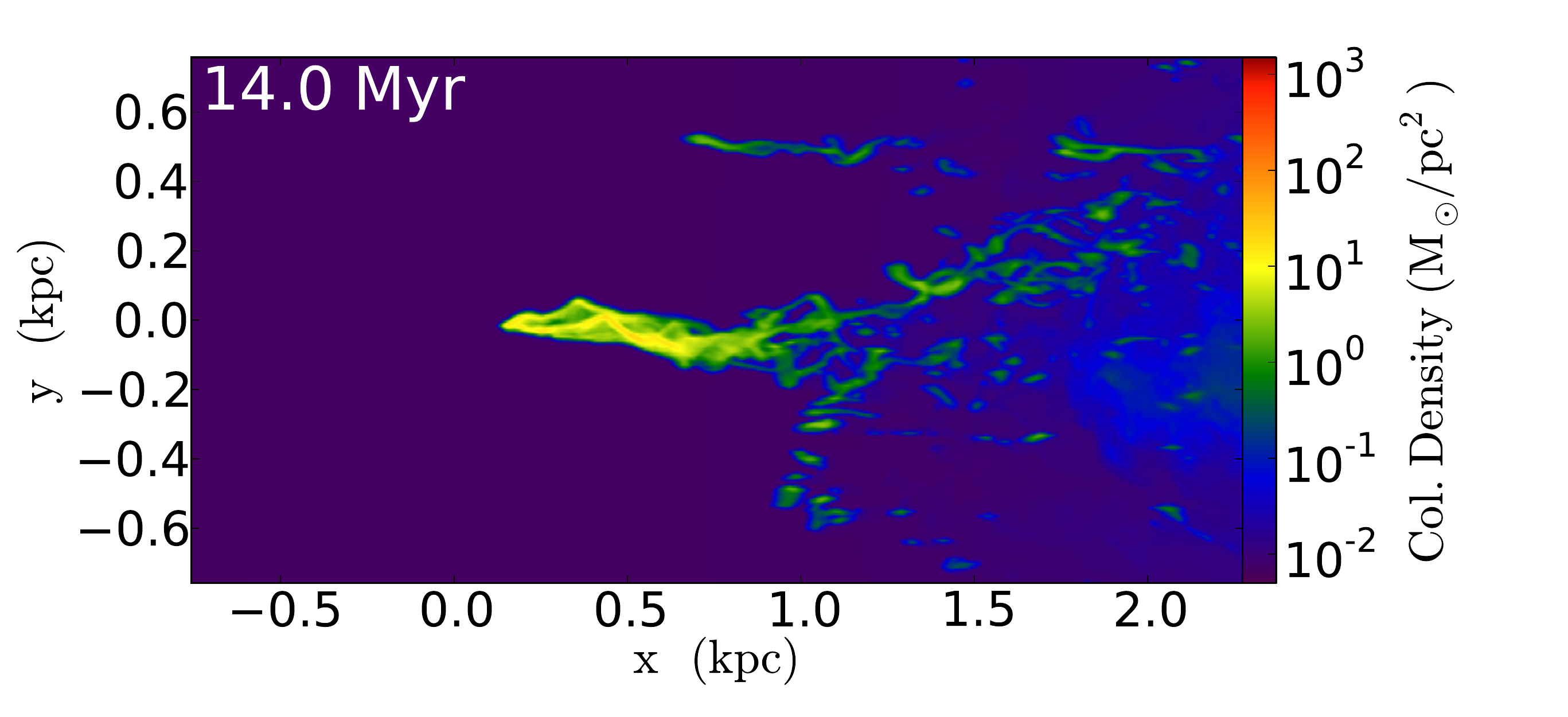}
\caption{\footnotesize{Column density of the simulations varying the minihalo-shock orientation, with FID on top, and PO1 on bottom, and evolution increasing from the 
beginning on the left, to when the outflow is just passing the minihalo in the middle, to when the 
outflow reaches the end of the box on the right.}}
\label{fig_orient}
\end{figure*}

\subsubsection{Orientation}\label{orientation}

A key component to anisotropic minihalos in this interaction is how the outflow is oriented with
respect to the minihalo and its accretion lanes. Simulation PO1 looks at the effect of orientation on the shock-minihalo interaction. \fig{fig_orient}
compares the evolution of the fiducial run with that of PO1, which has the outflow propagating through the 
low-density IGM, instead of along a filament.
In FID, the shock material in the filament is stalled, first collapsing the filament gas, before
striking the minihalo. In PO1, the shock material along the $x$-axis is first to hit the minihalo, 
without a reduction in kinetic energy. The specifics of this interaction will be fairly stochastic, 
dependent on the orientation of other filaments, and the geometry of the minihalo. In PO1, we find
the outflow triggers some molecular hydrogen formation in surrounding filaments, that cool and collapse, while being driven down towards the $x$-axis. Also, the minihalo baryons are in a  
more extended and bound ribbon of material at the end of the interaction. 

\begin{figure*}[t!]
\centering
\includegraphics[scale=0.42]{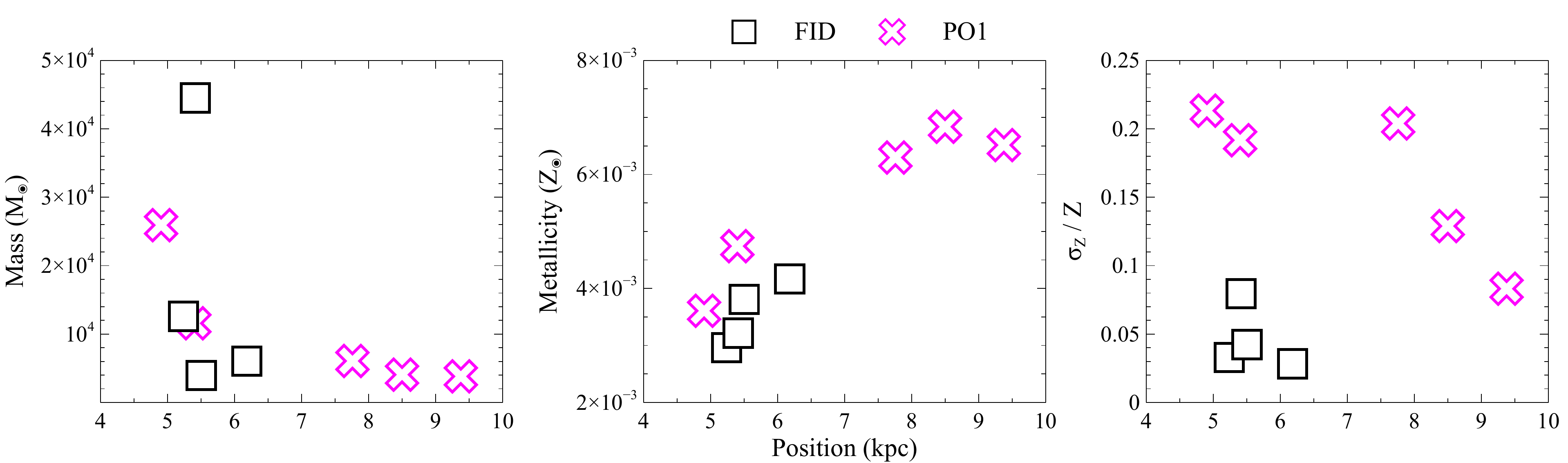}
\caption{\footnotesize{Comparison between the PO1 (magenta crosses) and FID (black squares) simulations illustrating the dependence on minihalo-shock orientation. The particle masses (left), 
metallicity (middle) and relative metallicity dispersion (right) vs particles 
positions after 200 Myr are shown.}}
\label{fig_orient2}
\end{figure*}

In \fig{fig_orient2} we show the ballistic particles after  200 Myr of evolution for the two different orientations. PO1 has only 14\% of the minihalo's baryons in these particles, compared with FID's 24\%. 
PO1 has its final particles much further out than FID, due to the undiminished kinetic energy of the
outflow. PO1 also has more final particles of lower mass, suggesting that they are less efficient
at merging. Regardless of orientation, metallicity increases with position outside the dark matter 
potential. 

\subsubsection{Shock Velocity}\label{vel}
\begin{figure*}[t!]
\centering
\includegraphics*[scale=0.245, trim=0 64.1 162.5 21]{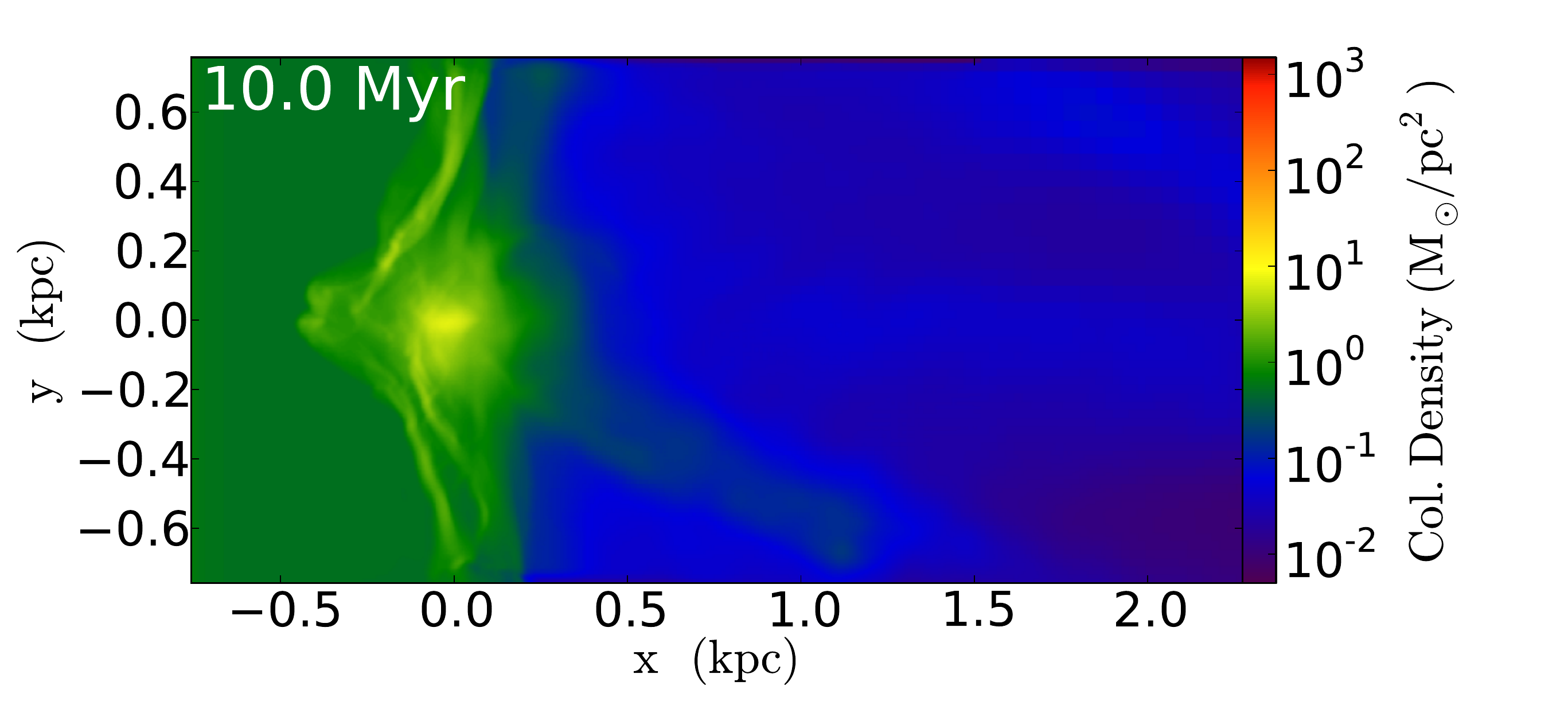}
\includegraphics*[scale=0.245, trim=95 64.1 162.5 21]{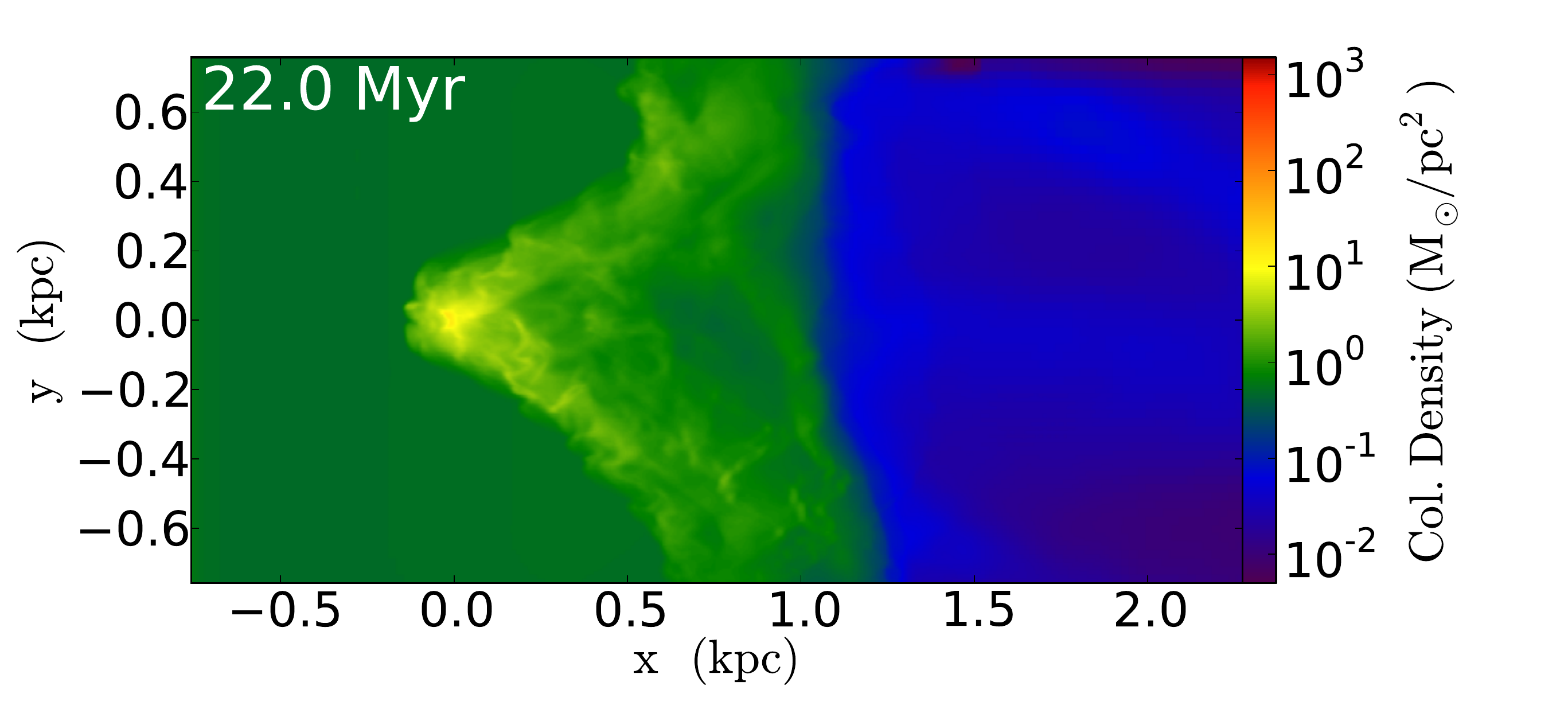}
\includegraphics*[scale=0.245, trim=95 64.1 0 21]{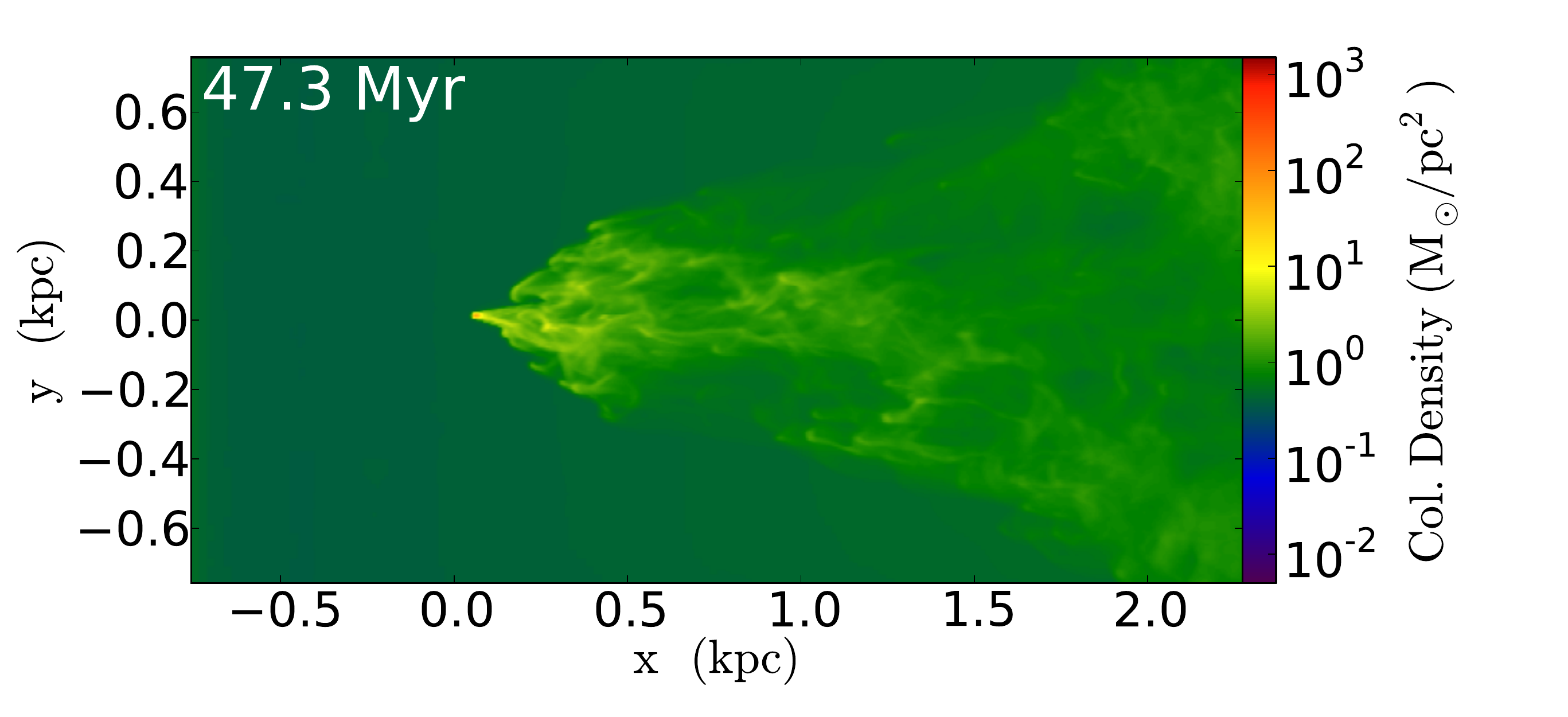}
\includegraphics*[scale=0.245, trim=0 64.1 162.5 21]{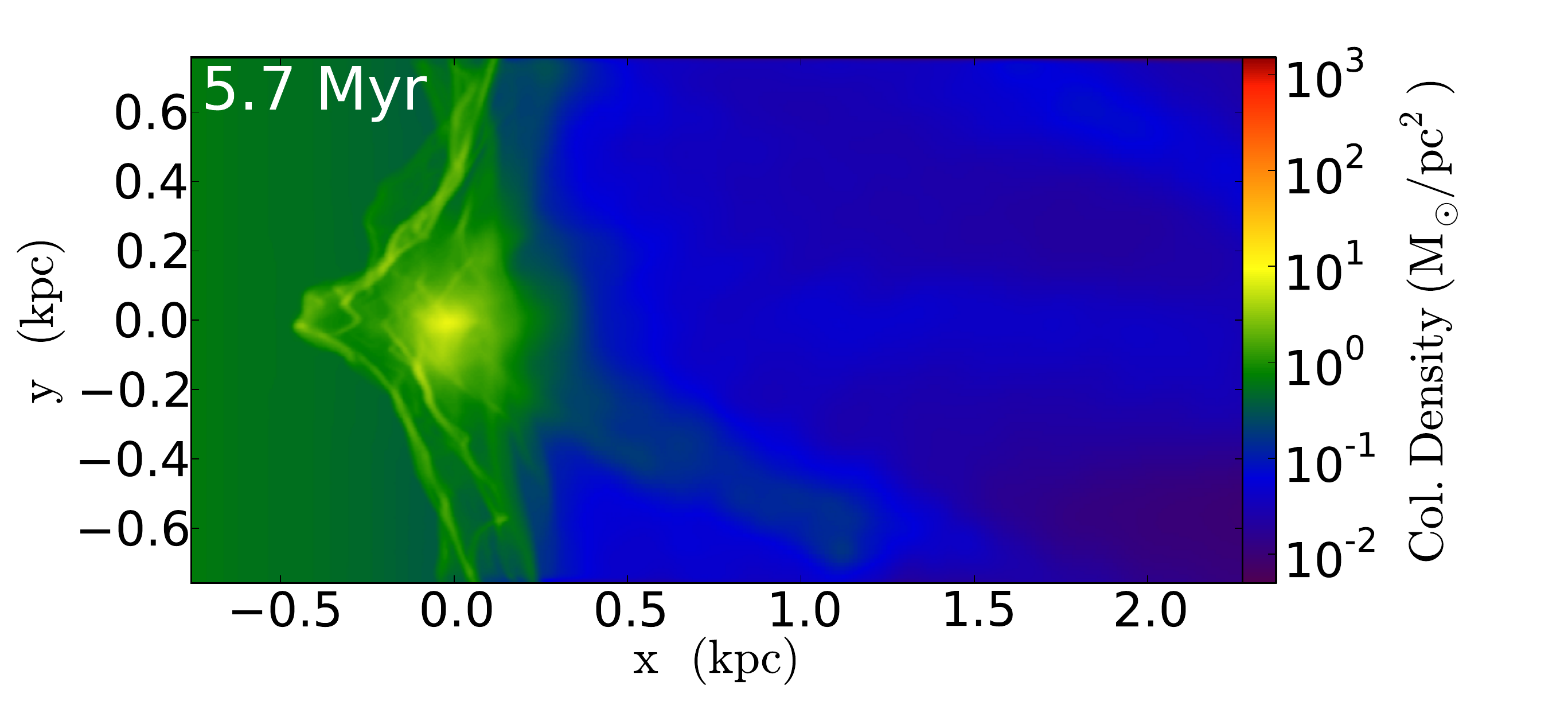}
\includegraphics*[scale=0.245, trim=95 64.1 162.5 21]{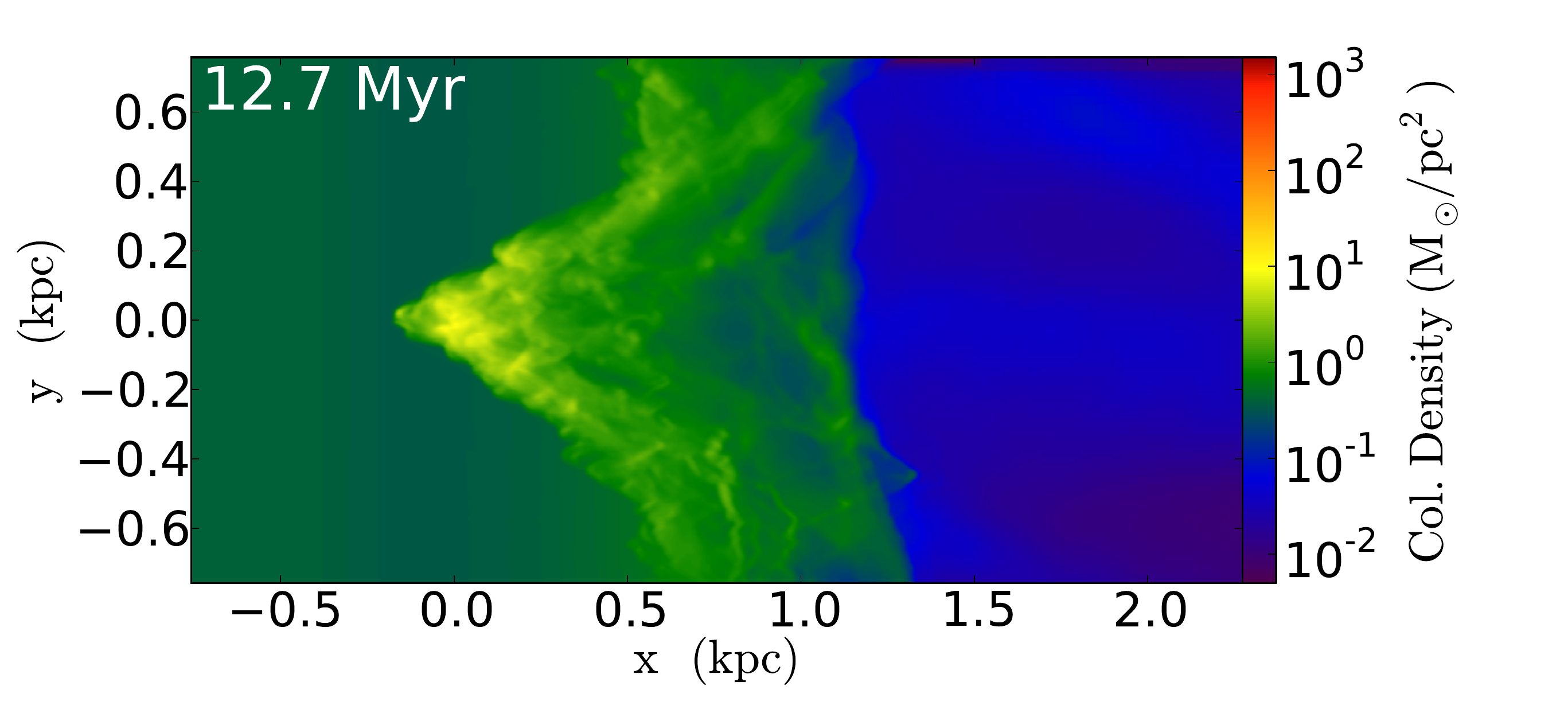}
\includegraphics*[scale=0.245, trim=95 64.1 0 21]{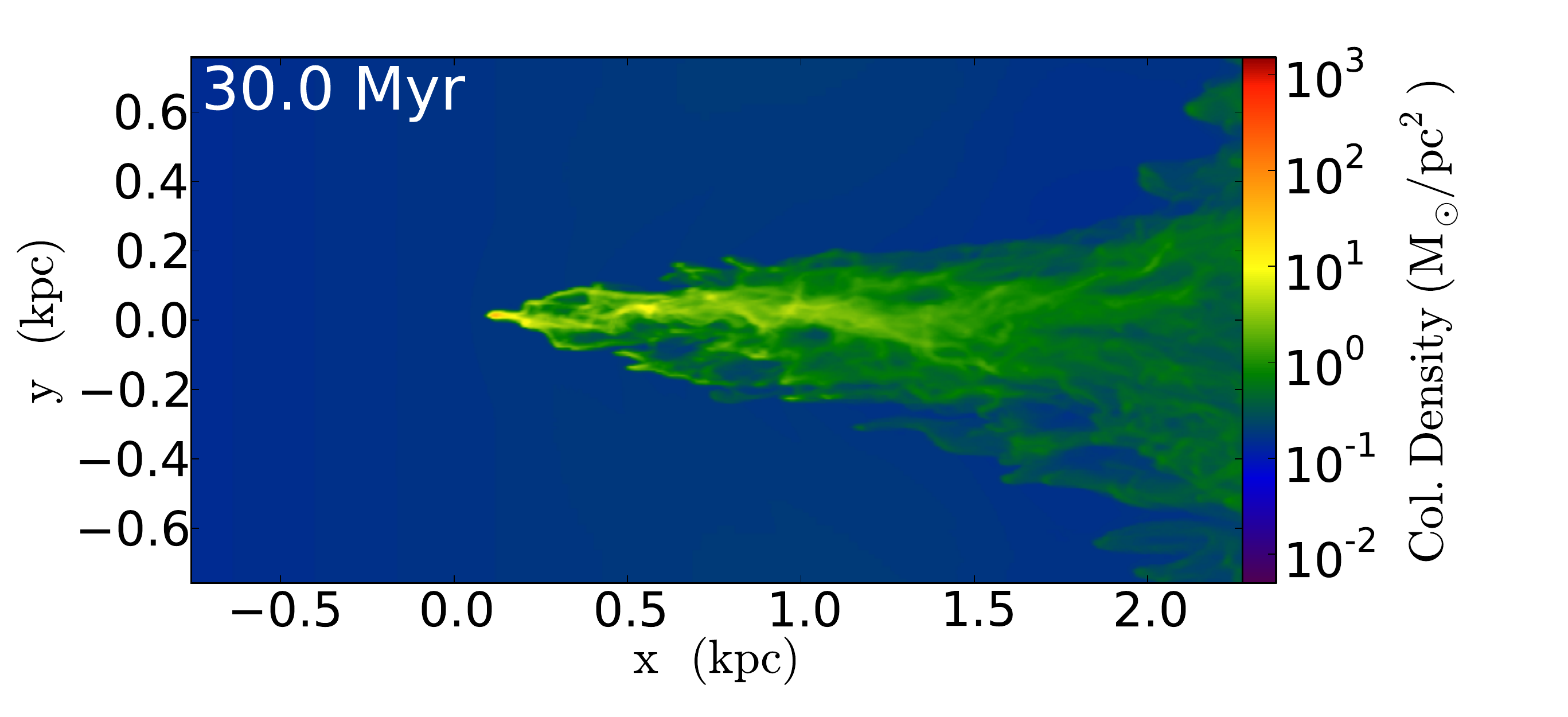}
\includegraphics*[scale=0.245, trim=0 64.1 162.5 21]{Fid_Col_Dens_10_Proj.pdf}
\includegraphics*[scale=0.245, trim=95 64.1 162.5 21]{Fid_Col_Dens_23_Proj.pdf}
\includegraphics*[scale=0.245, trim=95 64.1 0 21]{Fid_Col_Dens_39_Proj.pdf}
\includegraphics*[scale=0.245, trim=0 64.1 162.5 21]{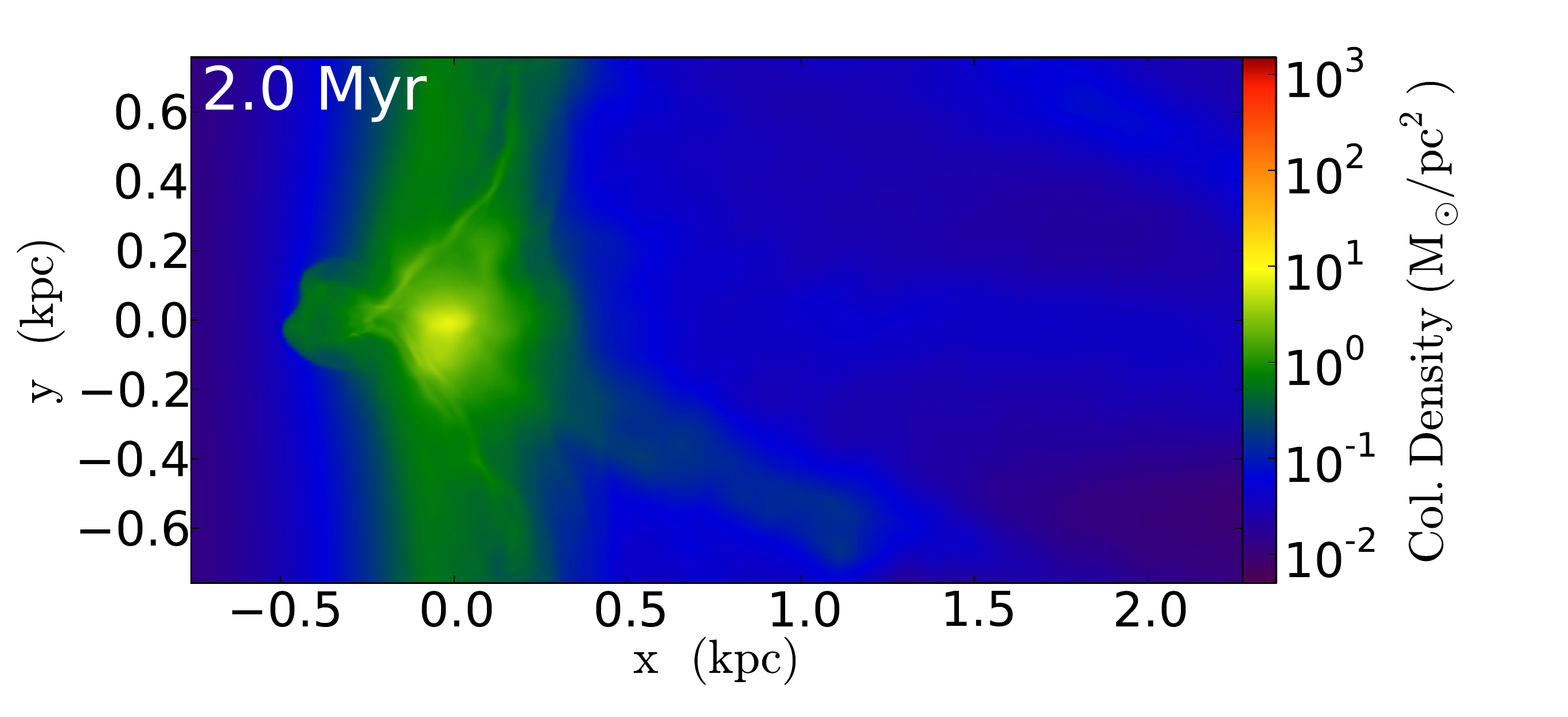}
\includegraphics*[scale=0.245, trim=95 64.1 162.5 21]{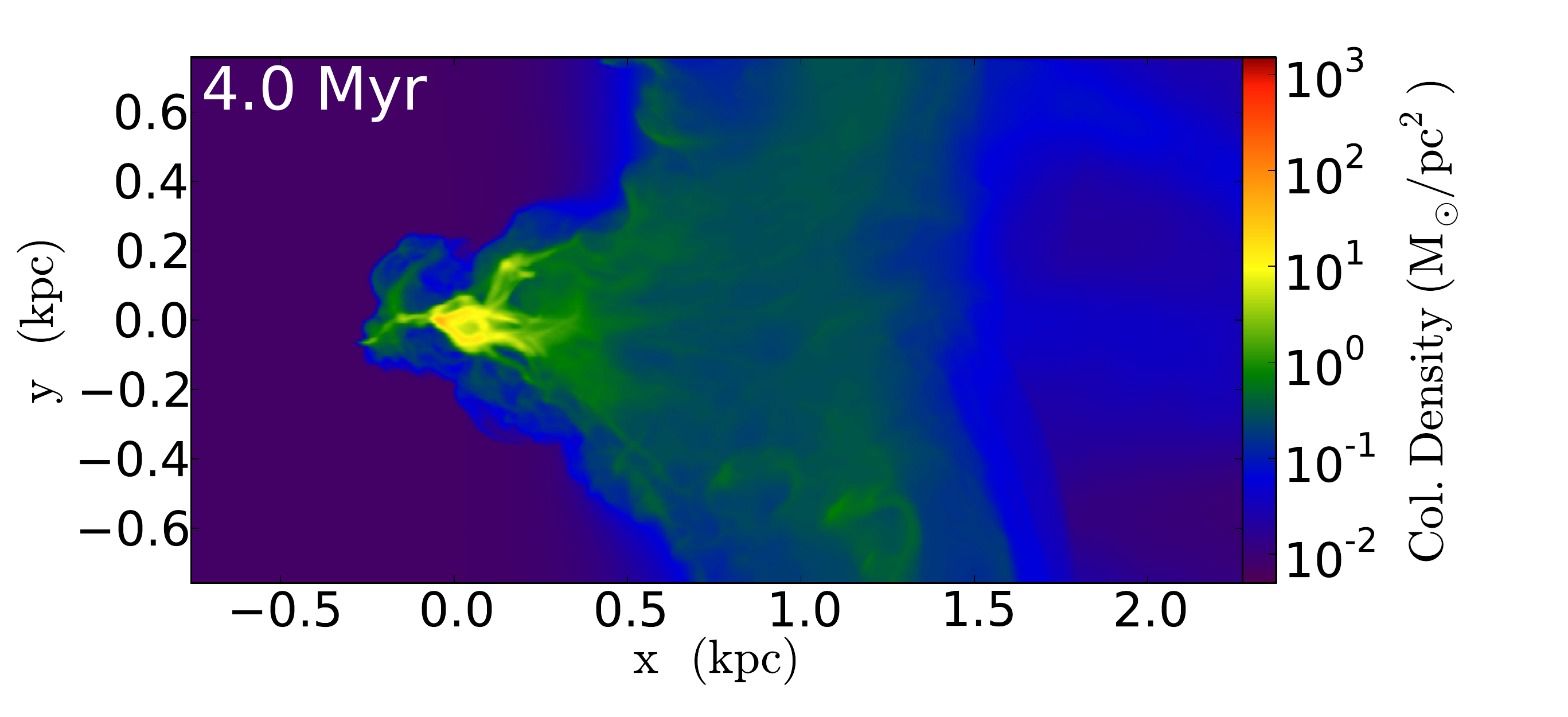}
\includegraphics*[scale=0.245, trim=95 64.1 0 21]{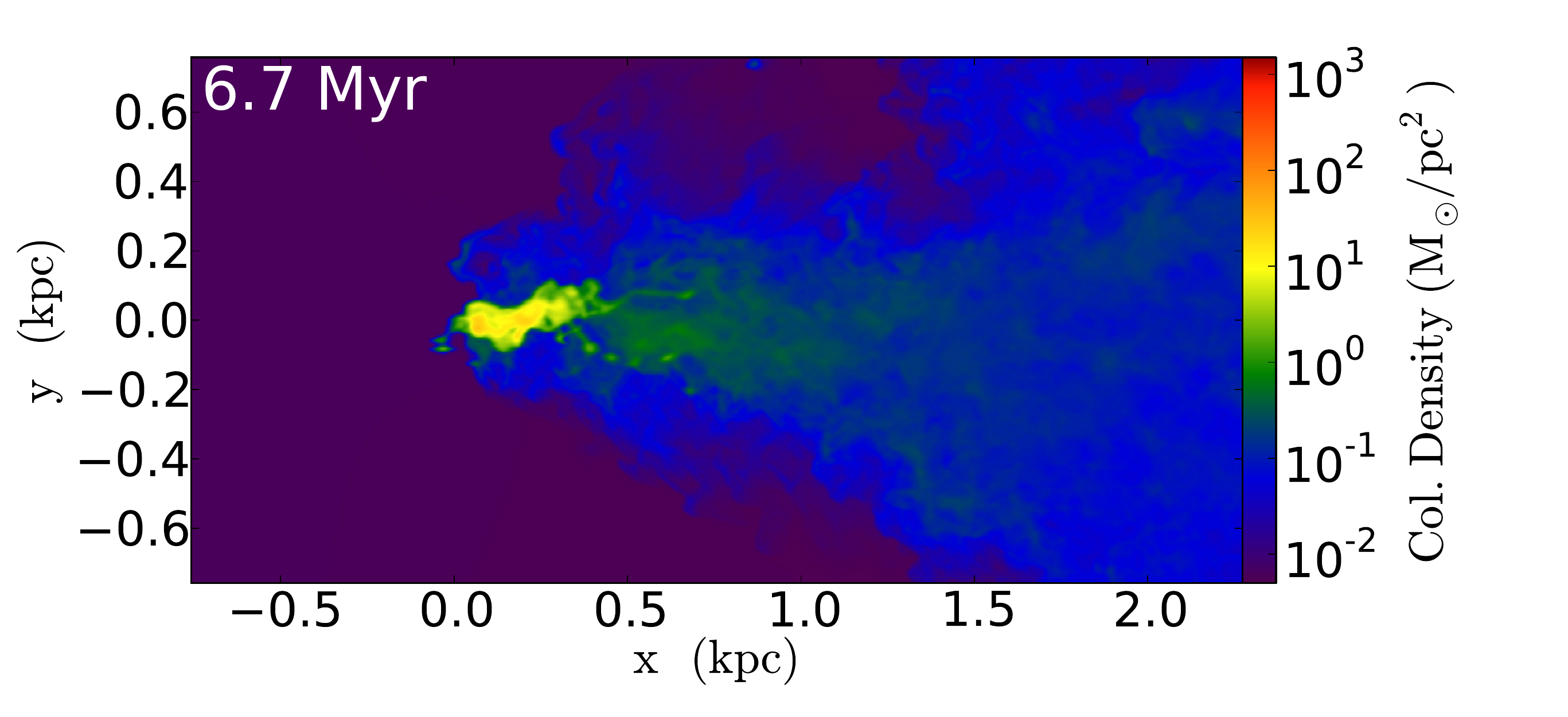}
\includegraphics*[scale=0.245, trim=0 0 162.5 21]{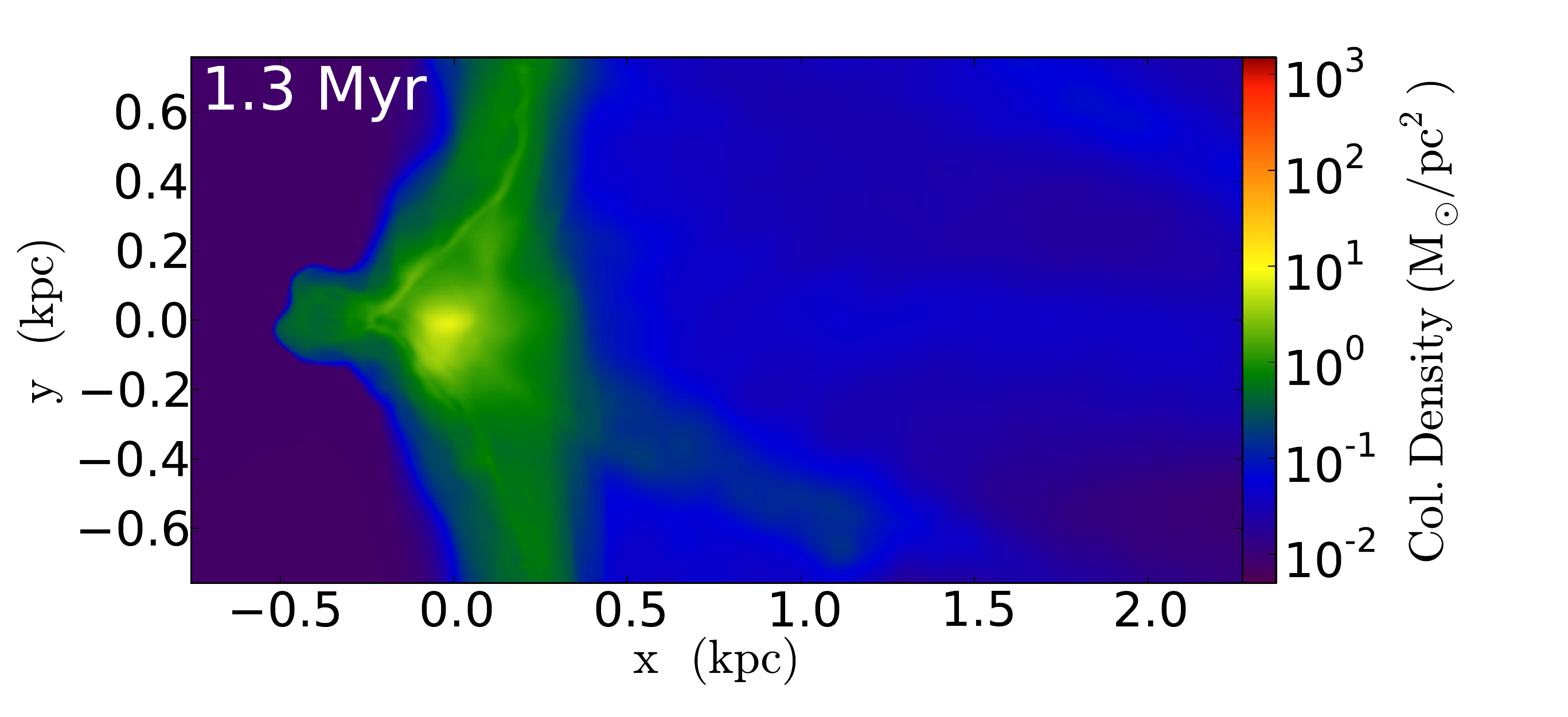}
\includegraphics*[scale=0.245, trim=95 0 162.5 21]{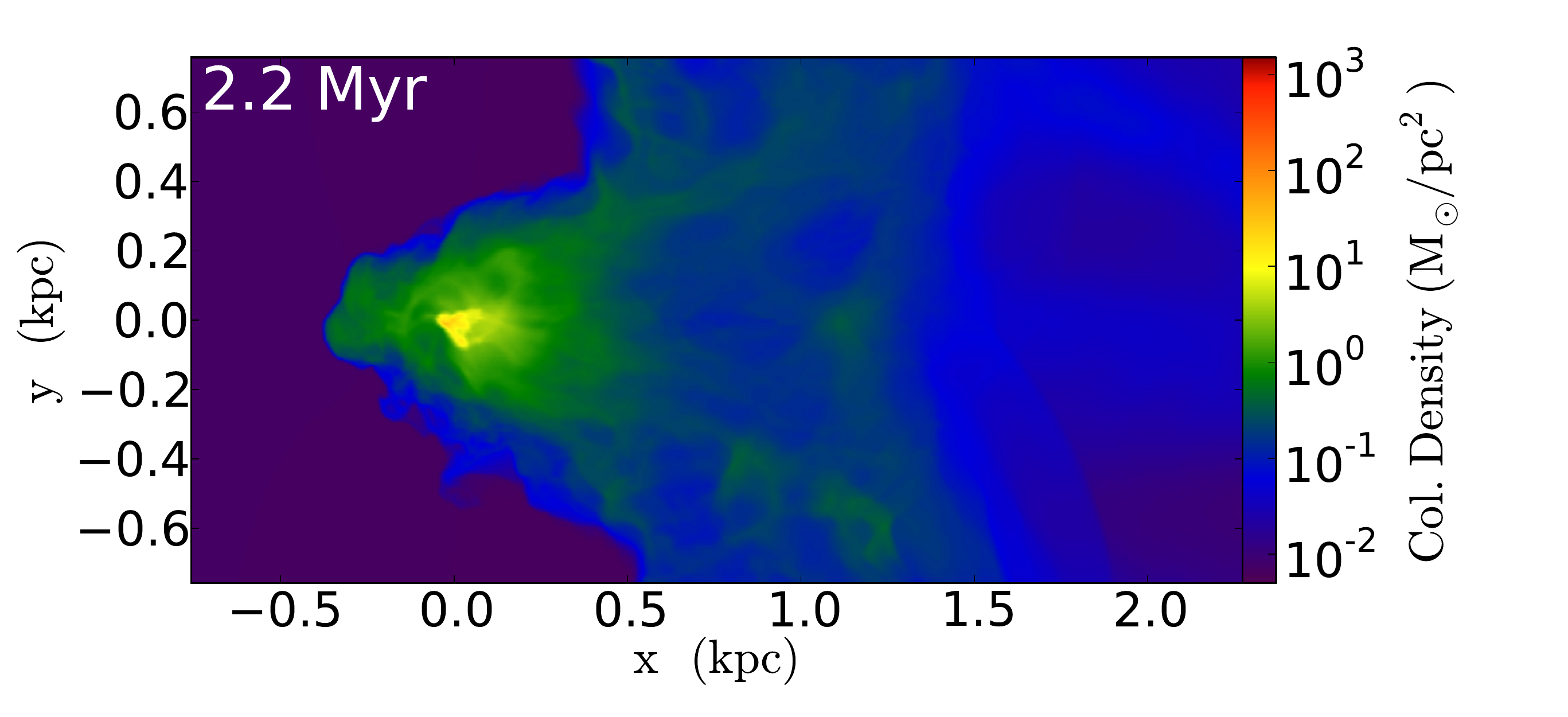}
\includegraphics*[scale=0.245, trim=95 0 0 21]{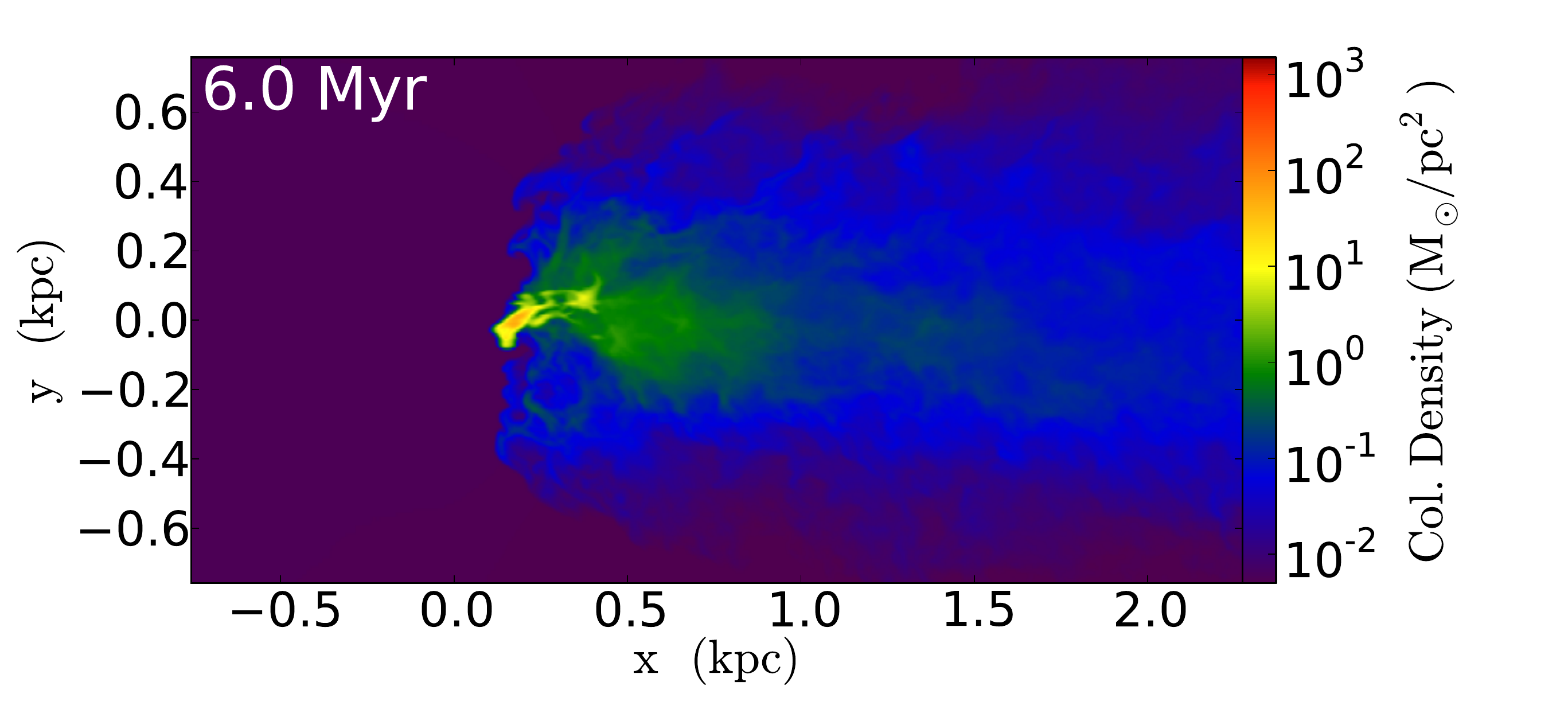}
\caption{\footnotesize{Column density of the simulations varying the shock velocity, $v_{\rm s}$, with Pv75, Pv125, FID, Pv340, and Pv510 from top to bottom, and evolution increasing from the 
first interaction with the minihalo on the left, to when the outflow is just passing the minihalo in the middle, to when the 
outflow reaches the end of the box on the right.}}
\label{fig_vel}
\end{figure*}

The shock velocity, $v_{\rm s}$, is one of the most significant parameters affecting this interaction. The 
outflow acts to remove the baryons from their dark matter potential well, sets the post-shock temperature, 
and catalyzes H$_2$ and HD formation by ionizing the gas. \fig{fig_vel} compares the evolution of the runs with 
different $v_{\rm s}$ values.
As the velocity of the shock increases, the post shock density decays quicker to maintain a constant surface momentum. At slower
speeds, the minihalo baryons are slowly pushed back. The material does not get constricted to the $x$-axis, and instead forms a
diffuse bow shock, with very little able to collapse or escape the dark matter potential. At intermediate velocities,
the material collapses into extended ribbons on the $x$-axis, with the length of this ribbon decreasing with increasing velocity. 
At the largest velocity, the most material is collapsed into a single dense clump, with a diffuse envelope of gas.

\begin{figure*}[t!]
\centering
\includegraphics[scale=0.42]{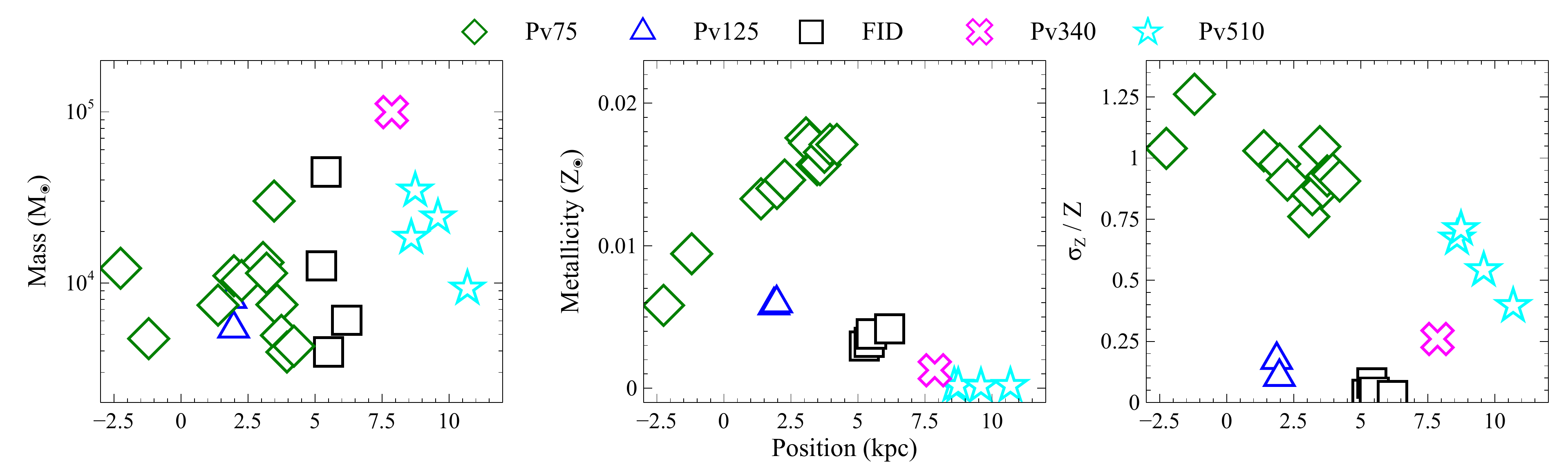}
\caption{\footnotesize{Comparison between the Pv75 (green diamonds), Pv125 (blue triangles), FID
(black squares), Pv340 (magenta crosses), and PE30 (cyan stars) simulations illustrating the dependence on shock velocity. 
The particle masses (left), 
metallicity (middle) and relative metallicity dispersion (right) vs particles 
positions after 200 Myr are shown.}}
\label{fig_vel2}
\end{figure*}

In \fig{fig_vel2} we show the ballistic particles after 200 Myr for the simulations that vary the shock speed. 
The slower outflows 
are unable to push the gas far from the dark matter potential, imparting little kinetic energy into 
the gas. As a result, these clumps are also less able to merge, resulting in more numerous cloud 
particles.
The slower the outflow, the longer the timescale of the interaction, which leads to an increase in the enrichment of the baryonic material in the slow velocity cases. 
The most energetic outflows, on the other hand, are unable
to appreciably enrich the baryons efficiently, which both  lowers the average metallicity of the clumps and increases the metallicity dispersion
of individual cloud particles. Finally, in Pv75 roughly 36\% of the minihalo baryons are contained in the final cloud particles. In Pv125,
only about 8\% of the baryons are in these particles, and then for even larger speeds we asymptote to about 25\%. At the lowest shock speed, the material
is not efficiently removed from the halo. At somewhat larger velocities the material is stripped from the minihalo potential, while not collapsed on to the $x$-axis. 
At even higher shock speeds, the material is collapsed on to the $x$-axis before it can be quickly removed from the potential, leading to a higher baryonic component.

\subsubsection{Outflow Surface Momentum}\label{mu}
\begin{figure*}[t]
\centering
\includegraphics*[scale=0.245, trim=0 64.1 162.5 21]{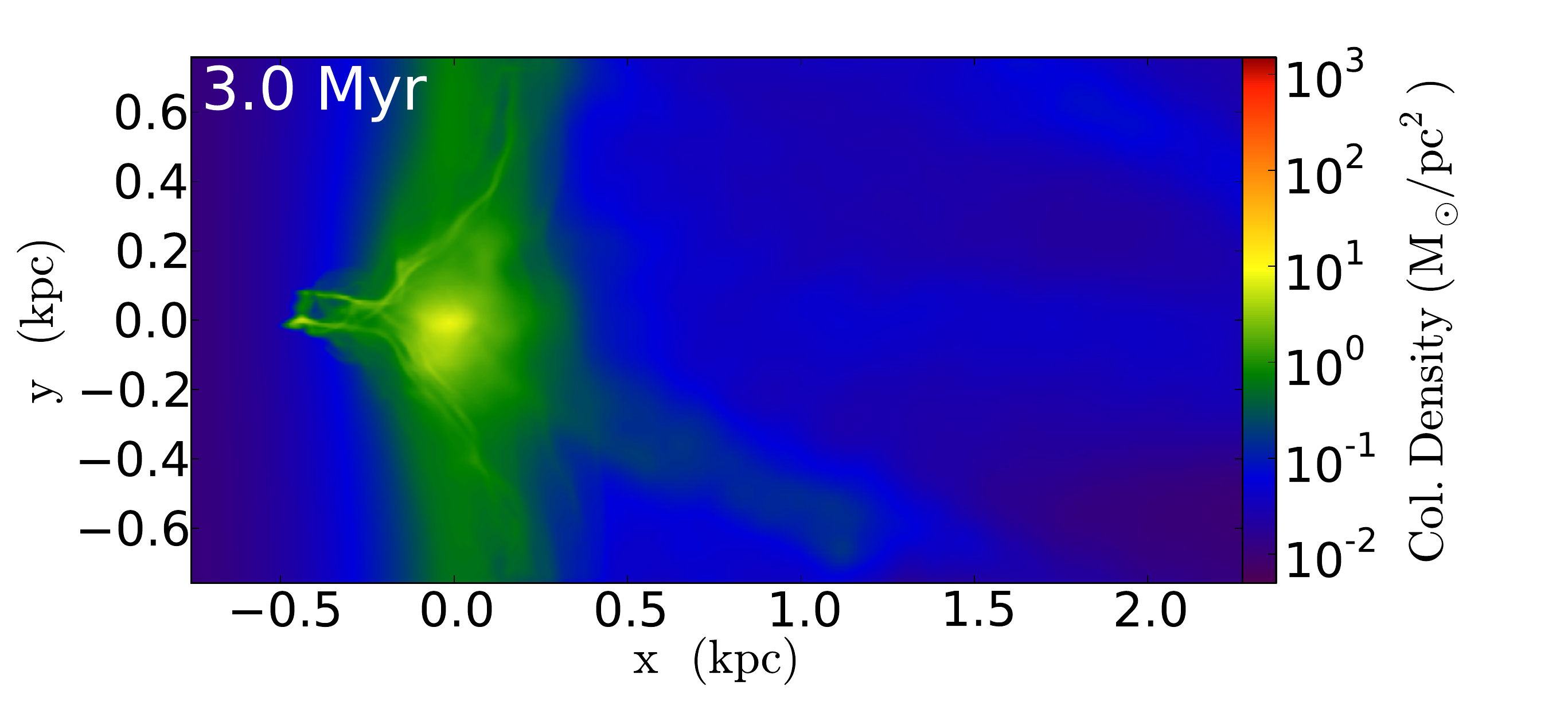}
\includegraphics*[scale=0.245, trim=95 64.1 162.5 21]{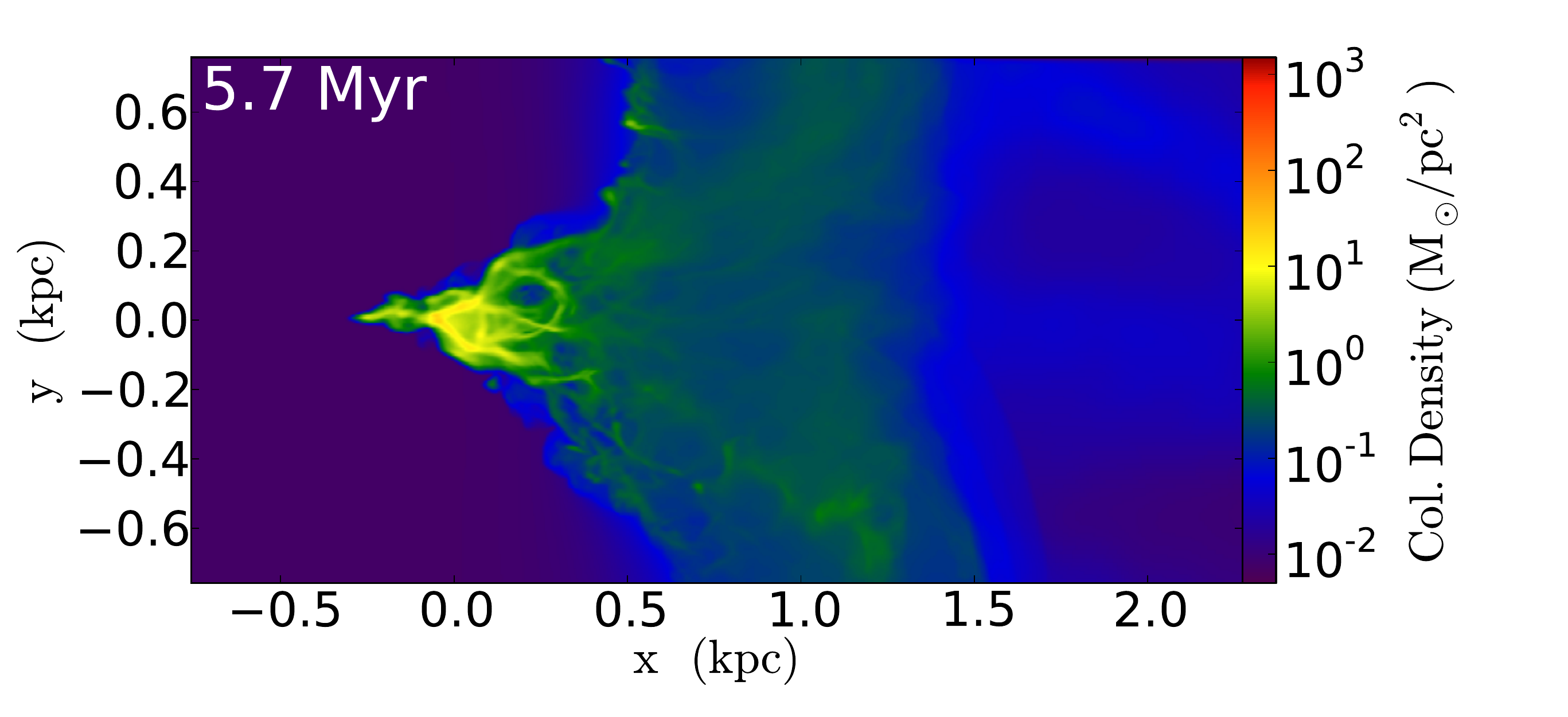}
\includegraphics*[scale=0.245, trim=95 64.1 0 21]{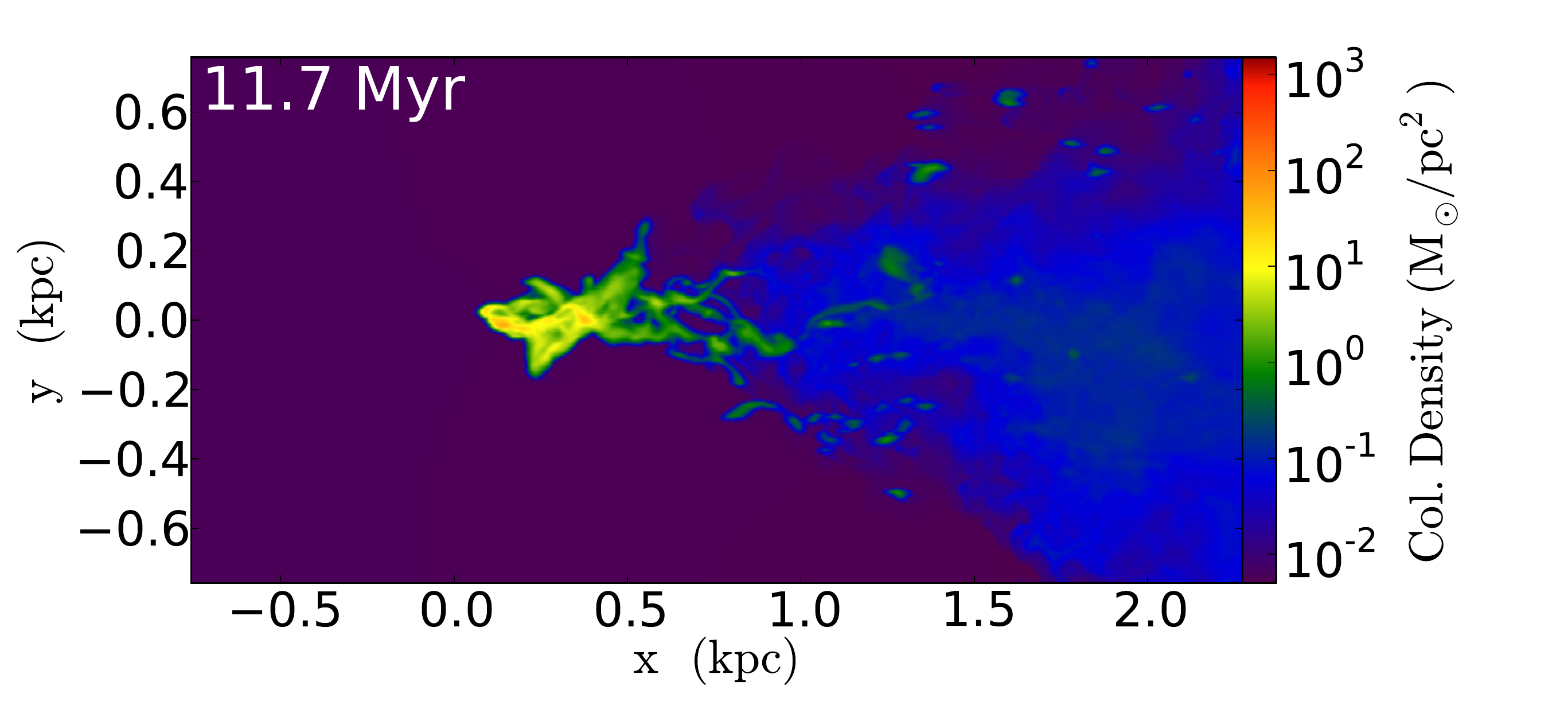}
\includegraphics*[scale=0.245, trim=0 64.1 162.5 21]{Fid_Col_Dens_10_Proj.pdf}
\includegraphics*[scale=0.245, trim=95 64.1 162.5 21]{Fid_Col_Dens_23_Proj.pdf}
\includegraphics*[scale=0.245, trim=95 64.1 0 21]{Fid_Col_Dens_39_Proj.pdf}
\includegraphics*[scale=0.245, trim=0 64.1 162.5 21]{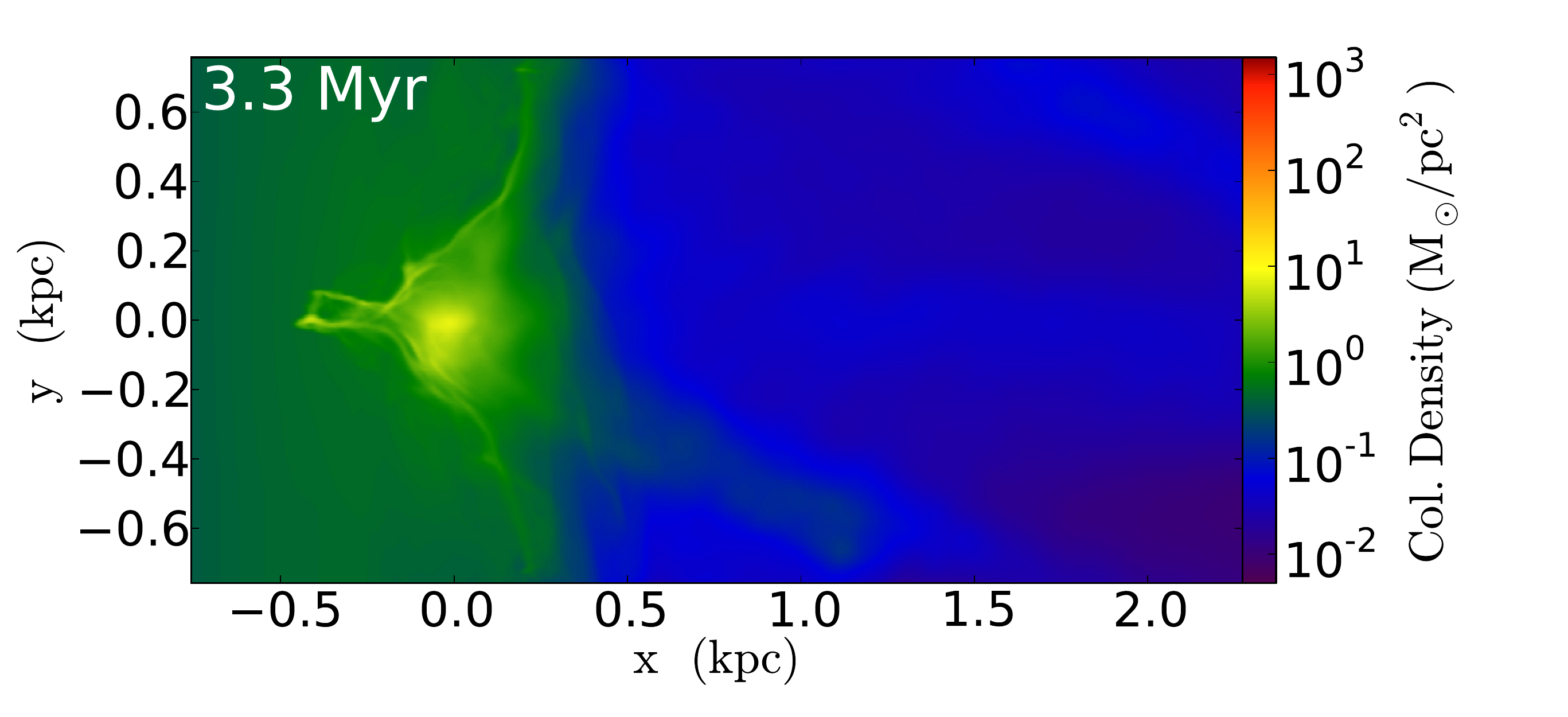}
\includegraphics*[scale=0.245, trim=95 64.1 162.5 21]{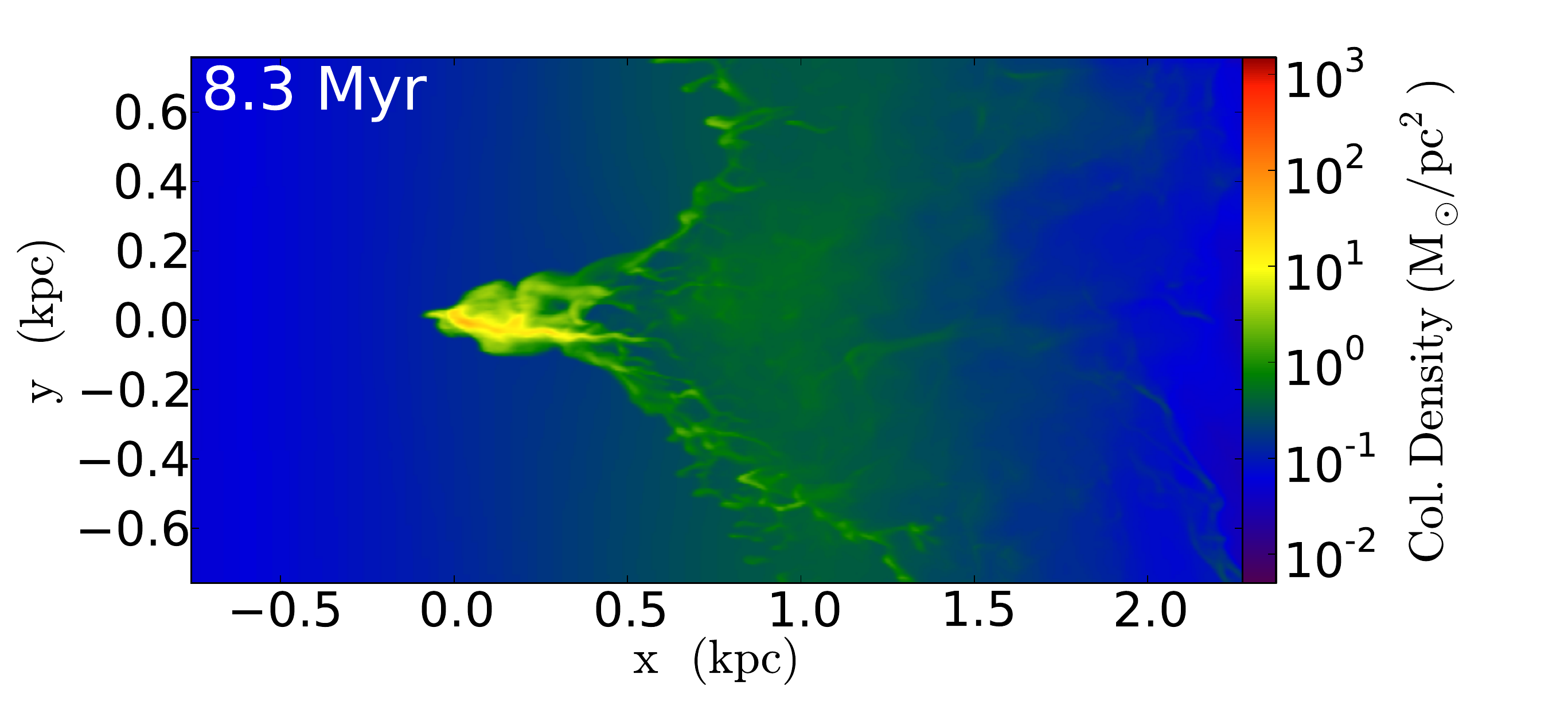}
\includegraphics*[scale=0.245, trim=95 64.1 0 21]{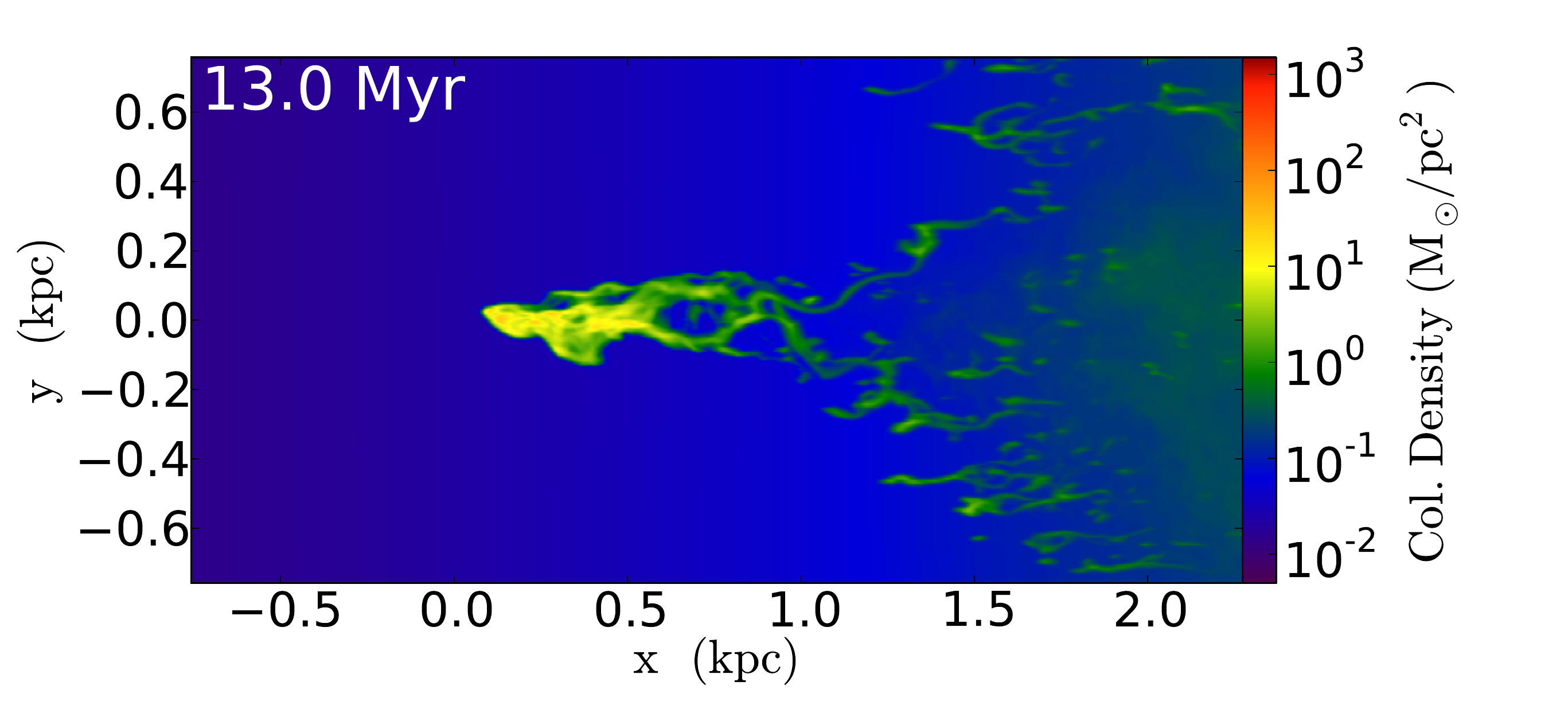}
\includegraphics*[scale=0.245, trim=0 0 162.5 21]{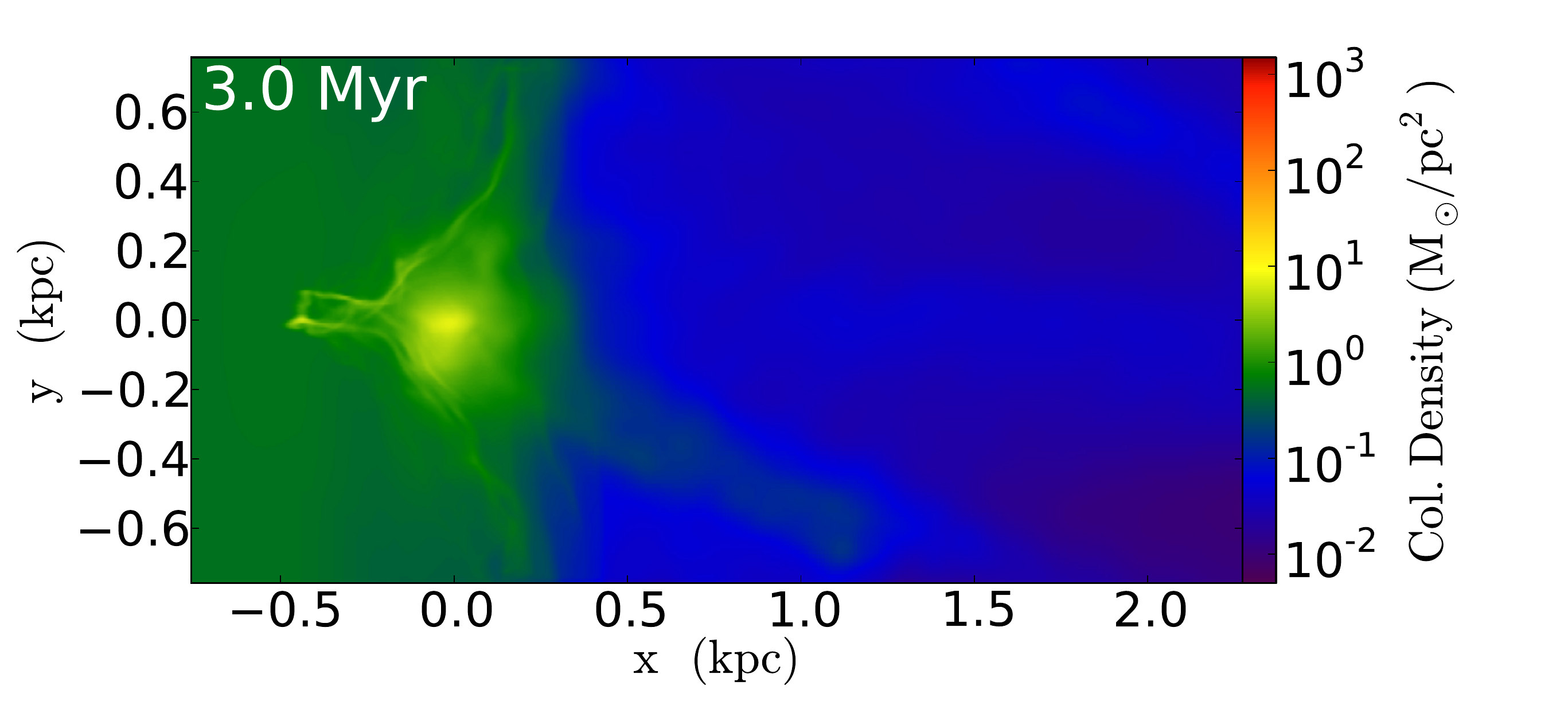}
\includegraphics*[scale=0.245, trim=95 0 162.5 21]{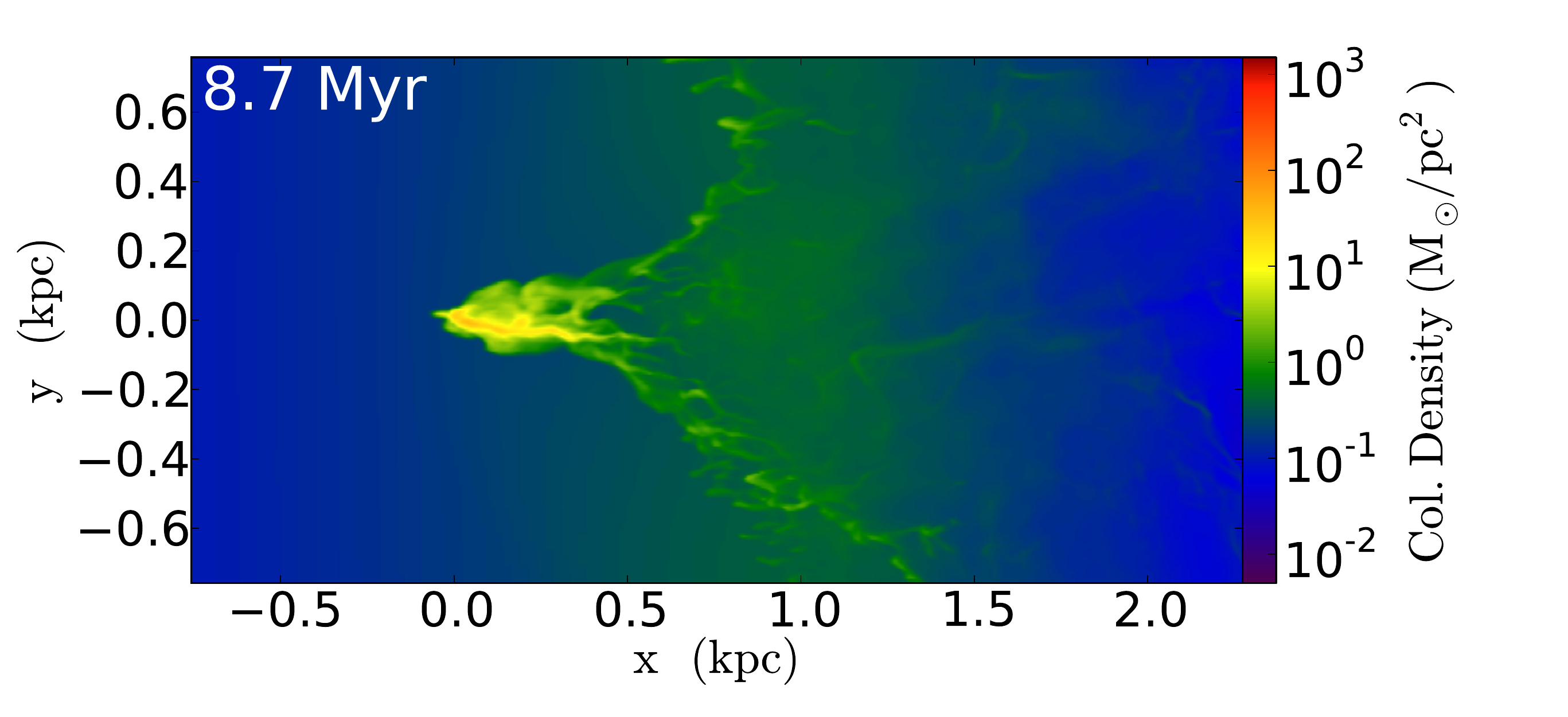}
\includegraphics*[scale=0.245, trim=95 0 0 21]{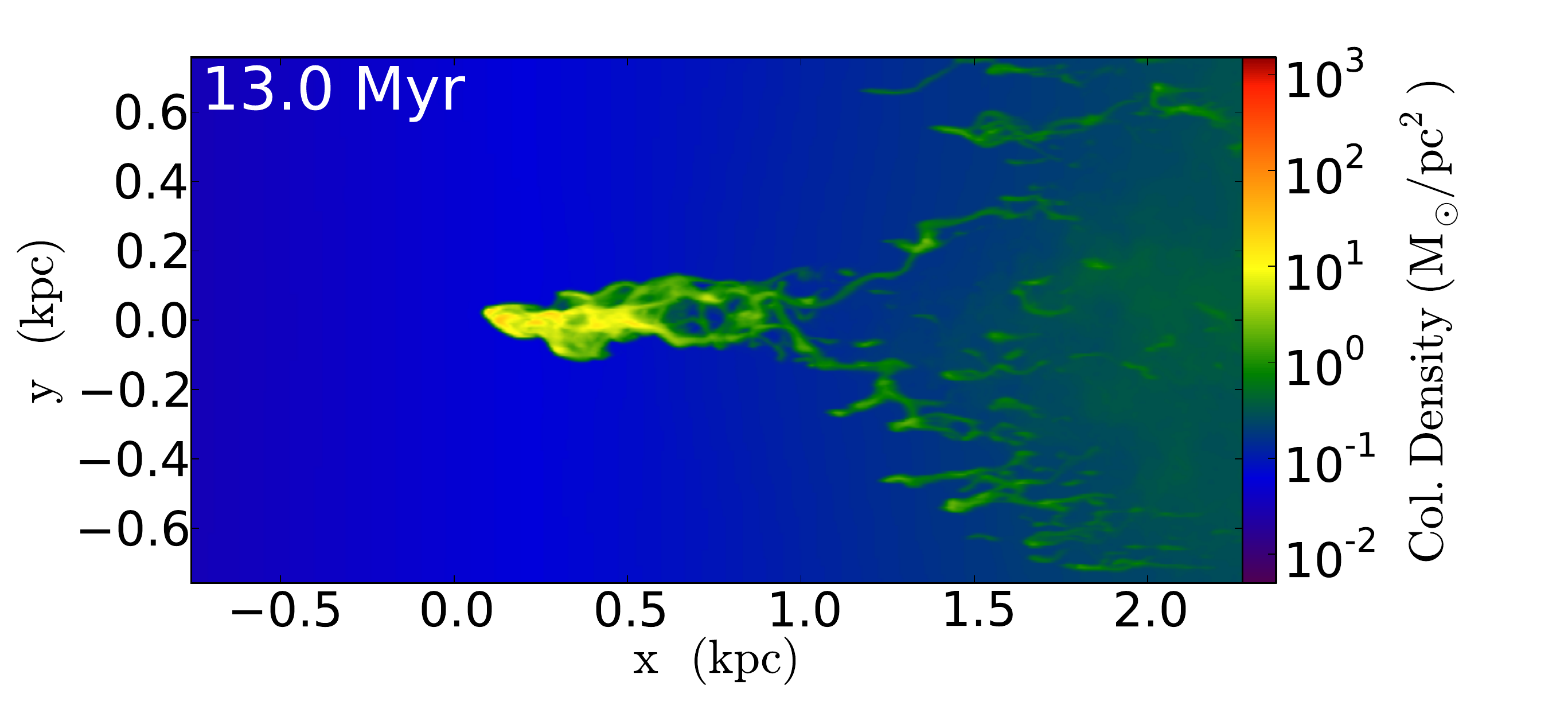}
\caption{\footnotesize{Column density of the simulations varying the shock surface momentum, 
$\mu_{\rm s}$, with P$\mu$3, FID, P$\mu$8, and P$\mu$9 on from top to bottom, and evolution increasing from the 
first interaction with the minihalo on the left, to when the outflow is just passing the minihalo in the middle, to when the 
outflow reaches the end of the box on the right.}}
\label{fig_surf}
\end{figure*}

The shock surface momentum is a significant parameter that along with the shock velocity, sets the initial energy driving the outflow, 
while simultaneously setting the surface density of the post-shock material. \fig{fig_surf} compares the evolution of the runs with 
different $\mu_{\rm s}$ values.
As the surface momentum of the shock increases, so does its surface density, leading to increased momentum transfer
to the minihalo gas. This leads to increased fragmentation of the shock front, and confinement onto the $x$-axis. At the lowest 
surface momentum, the final ribbon of material is more extended perpendicular to the $x$-axis, and less extended along the $x$-axis.
At the largest surface momentum, the final ribbon of material is the opposite, more constrained to the $x$-axis, and more extended along
the $x$-axis.

\begin{figure*}[t!]
\centering
\includegraphics[scale=0.42]{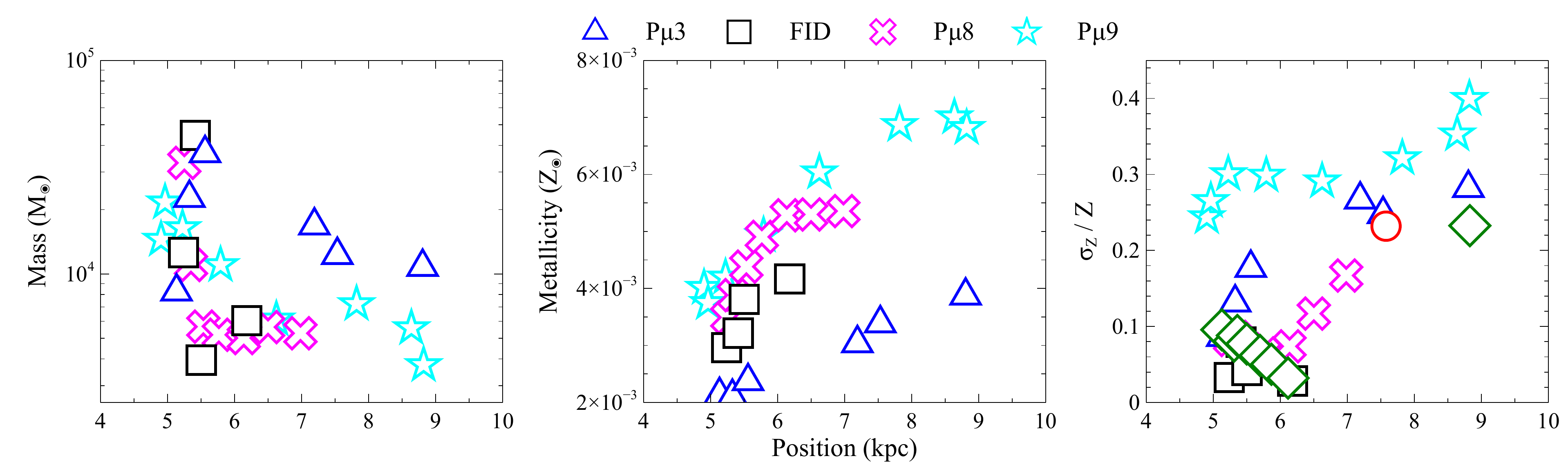}
\caption{\footnotesize{Comparison between the P$\mu$3 (blue triangles), FID (black squares), 
P$\mu$8 (magenta crosses), and P$\mu$9 (cyan stars) simulations illustrating the dependence on shock surface momentum.  The particle masses (left), 
metallicity (middle) and relative metallicity dispersion (right) vs particles 
positions after 200 Myr are shown.}}
\label{fig_surf2}
\end{figure*}

In \fig{fig_surf2}, we show the ballistic particles after 200 Myr of evolution for the simulations that vary the surface momentum.
The position of most of the final cloud particles is roughly the same for all surface momenta, indicating a greater dependence on the
shock speed. Nevertheless, the smallest and largest momentum cases have final particles further removed from the dark matter potential. The
primary cause of this in the low momentum regime is that more gas is bound to the $x$-axis at further distances, 
leading to larger clumps further away from the dark matter potential. Note that P$\mu$3 has 28\% of the minihalo baryons in the final cloud particles. 
With larger momentum, less gas is ultimately bound to the $x$-axis at further distance, despite the ribbon of material being initially more 
extended. P$\mu$9 has about 24\% of the minihalo baryons in the final cloud particles. Since we only show the final
particles with more than 1\% of the baryon component, there are many more extended particles for intermediate momentum that are not shown.

The mass of the final cloud particles is roughly independent of momentum. Again, the total mass is highest at lowest momentum. The enrichment
of these clouds again scales with position, for a given simulation, and increases with momentum, as does the dispersion. This is due to the increased
fragmentation and confinement of the baryons along the $x$-axis at larger momentum, leading to a higher penetration of enriched material.

\subsubsection{Outflow Abundance}\label{metl}

\begin{figure*}[t]
\centering
\includegraphics*[scale=0.245, trim=0 64.1 162.5 21]{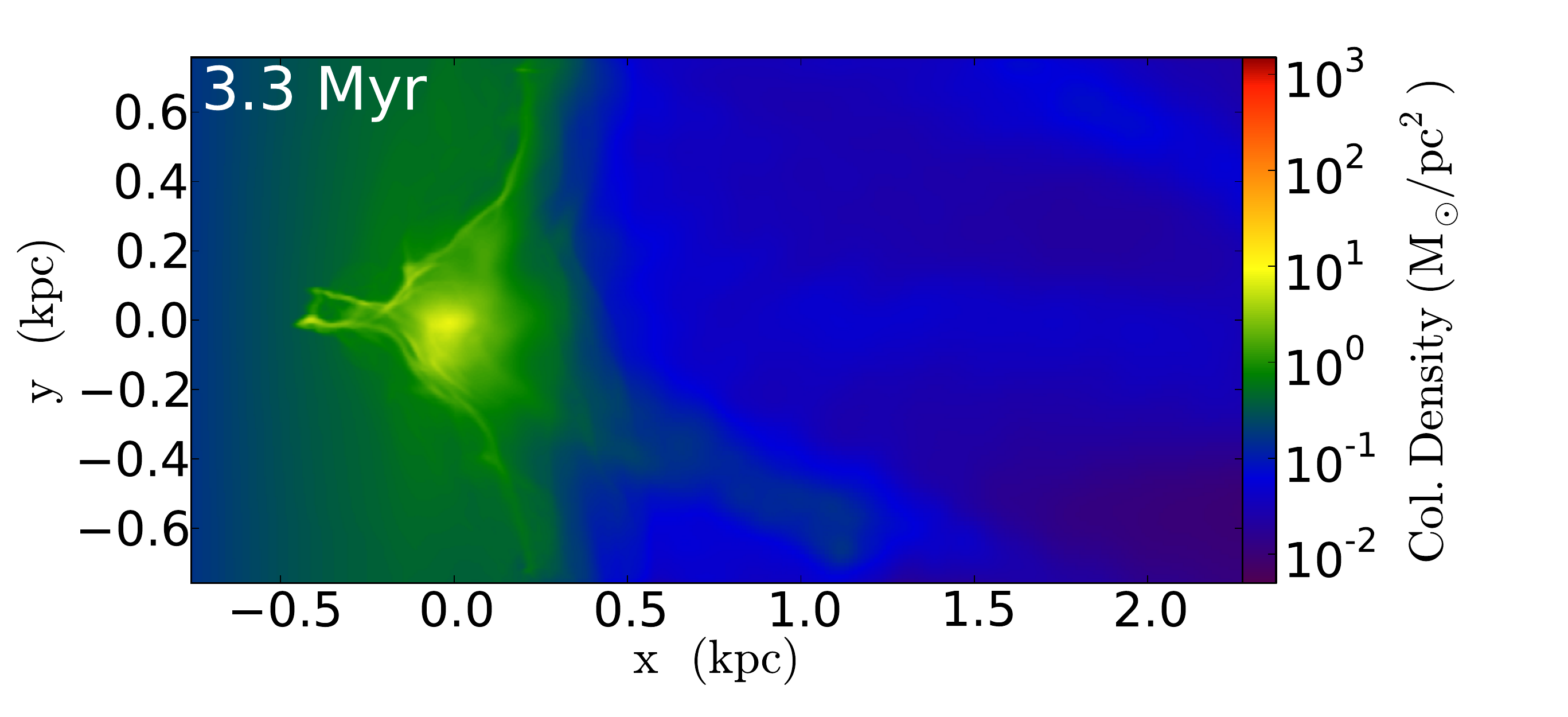}
\includegraphics*[scale=0.245, trim=95 64.1 162.5 21]{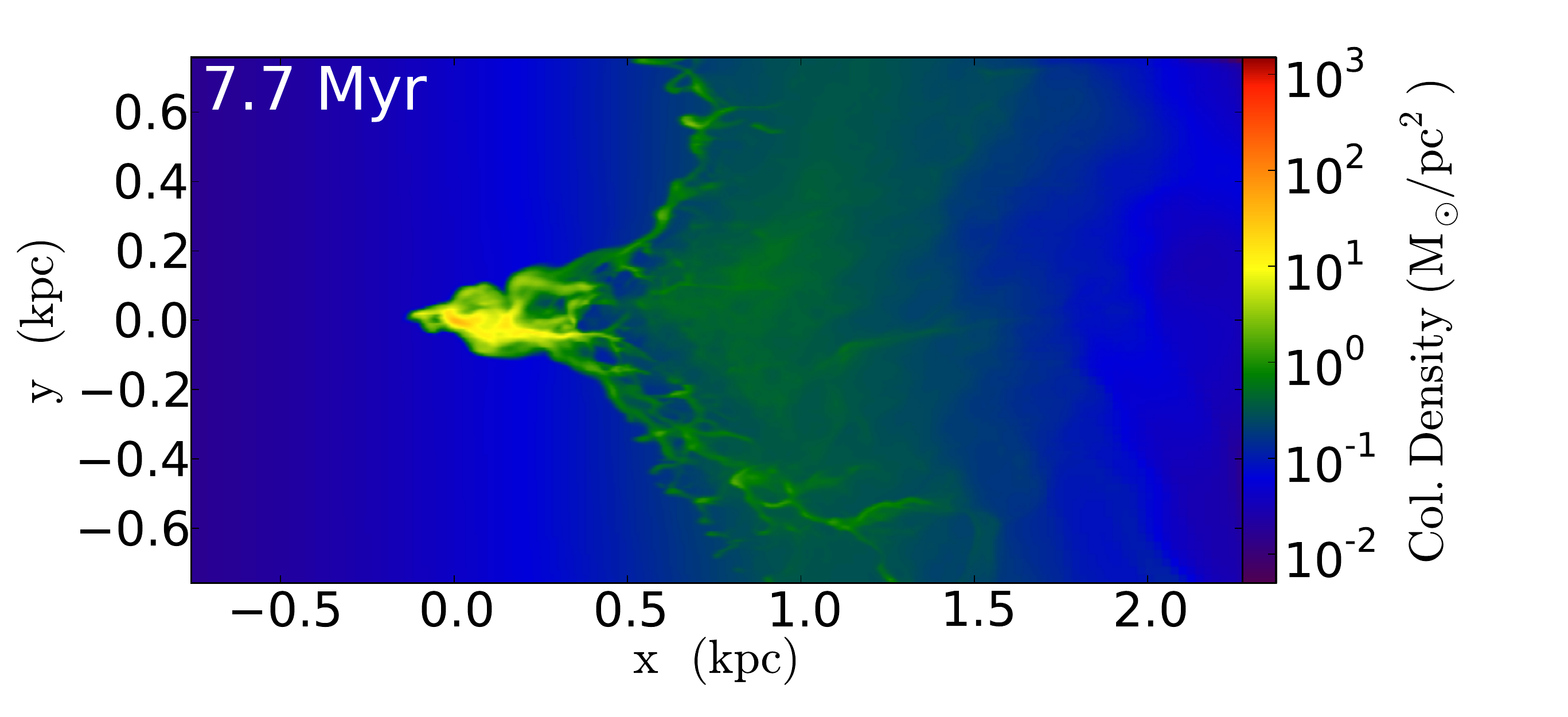}
\includegraphics*[scale=0.245, trim=95 64.1 0 21]{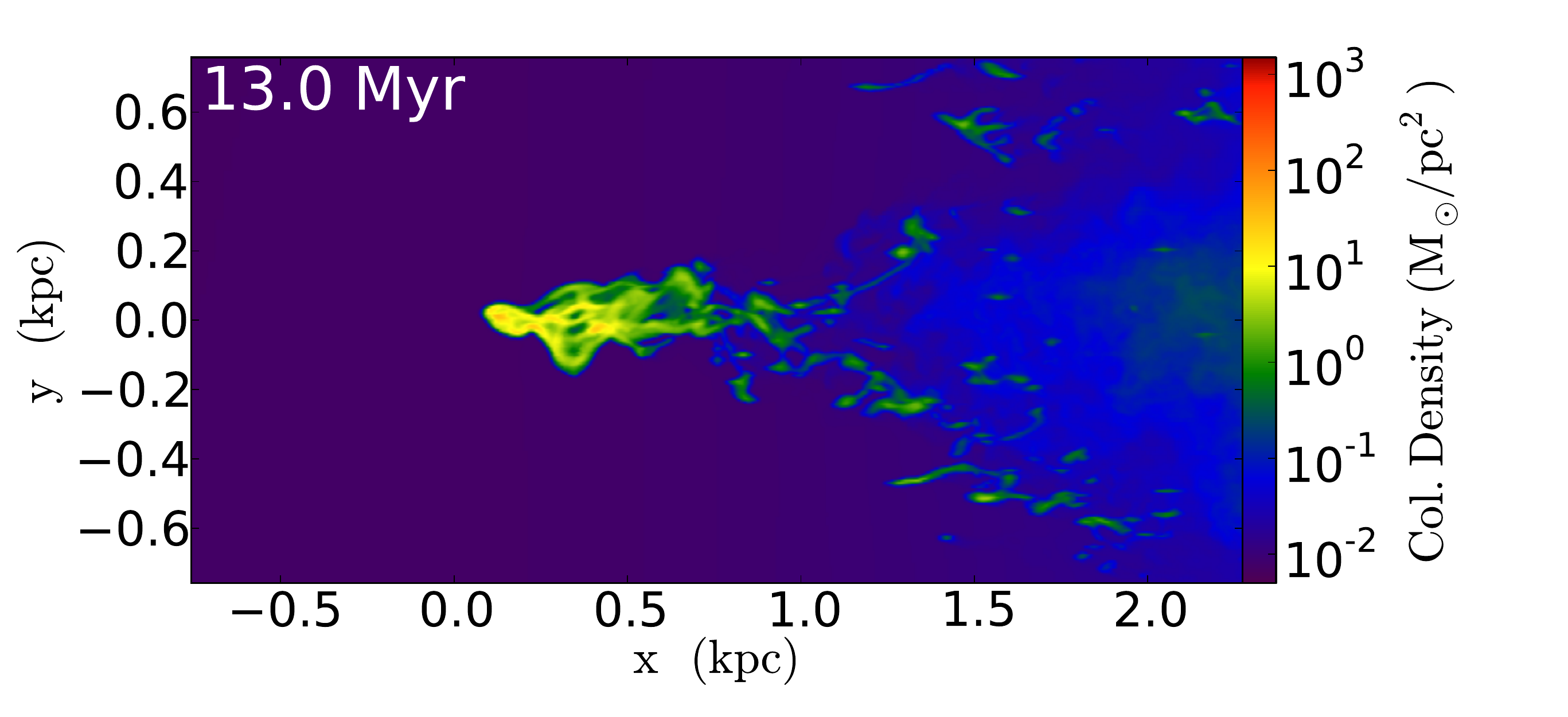}
\includegraphics*[scale=0.245, trim=0 64.1 162.5 21]{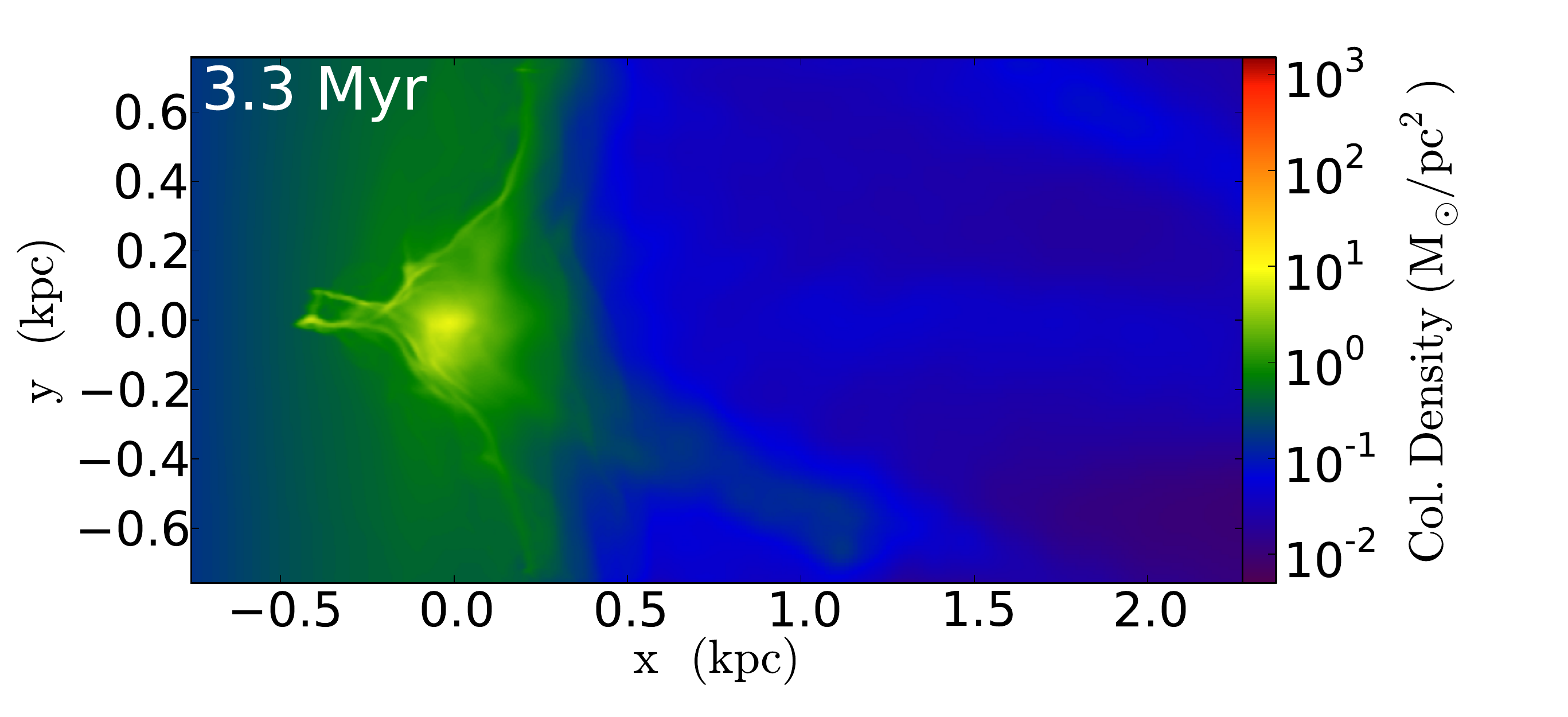}
\includegraphics*[scale=0.245, trim=95 64.1 162.5 21]{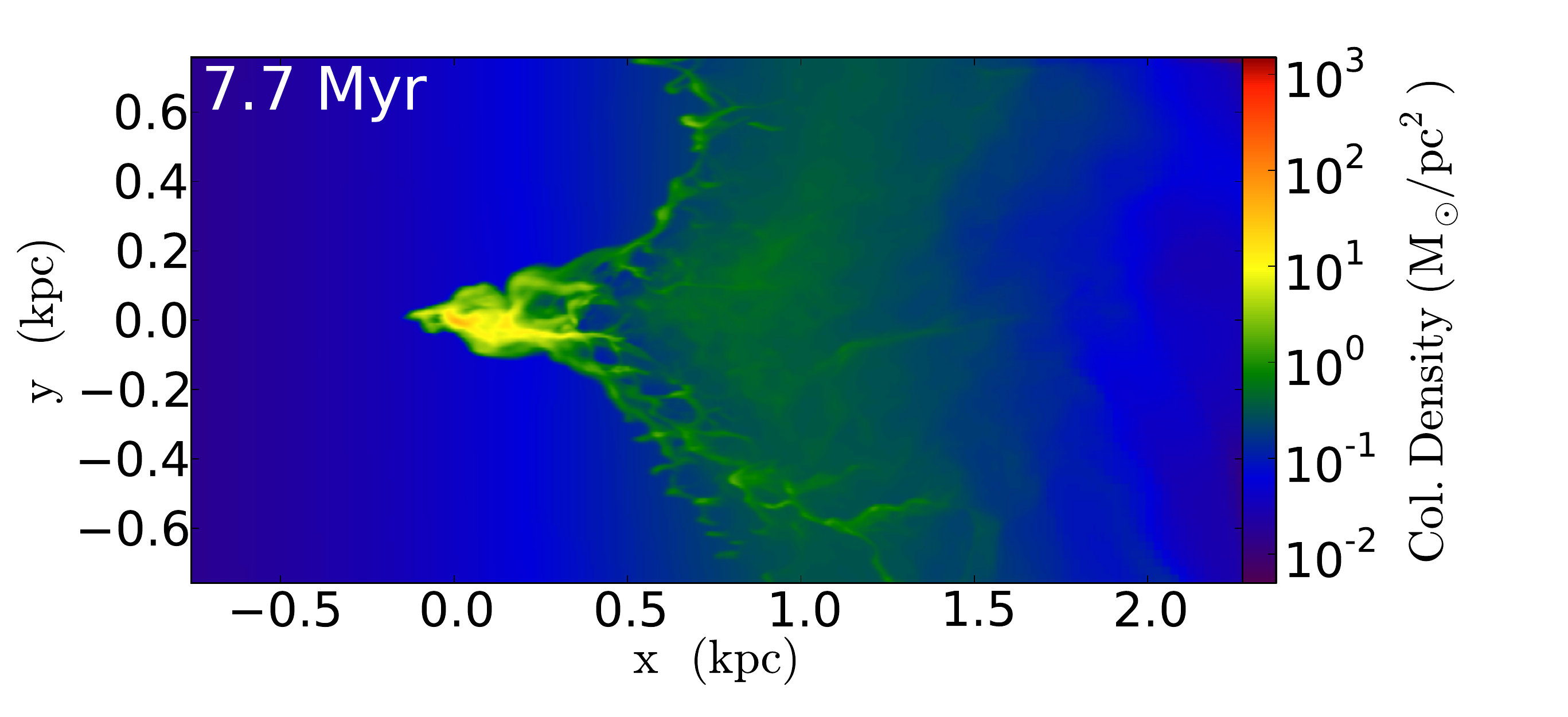}
\includegraphics*[scale=0.245, trim=95 64.1 0 21]{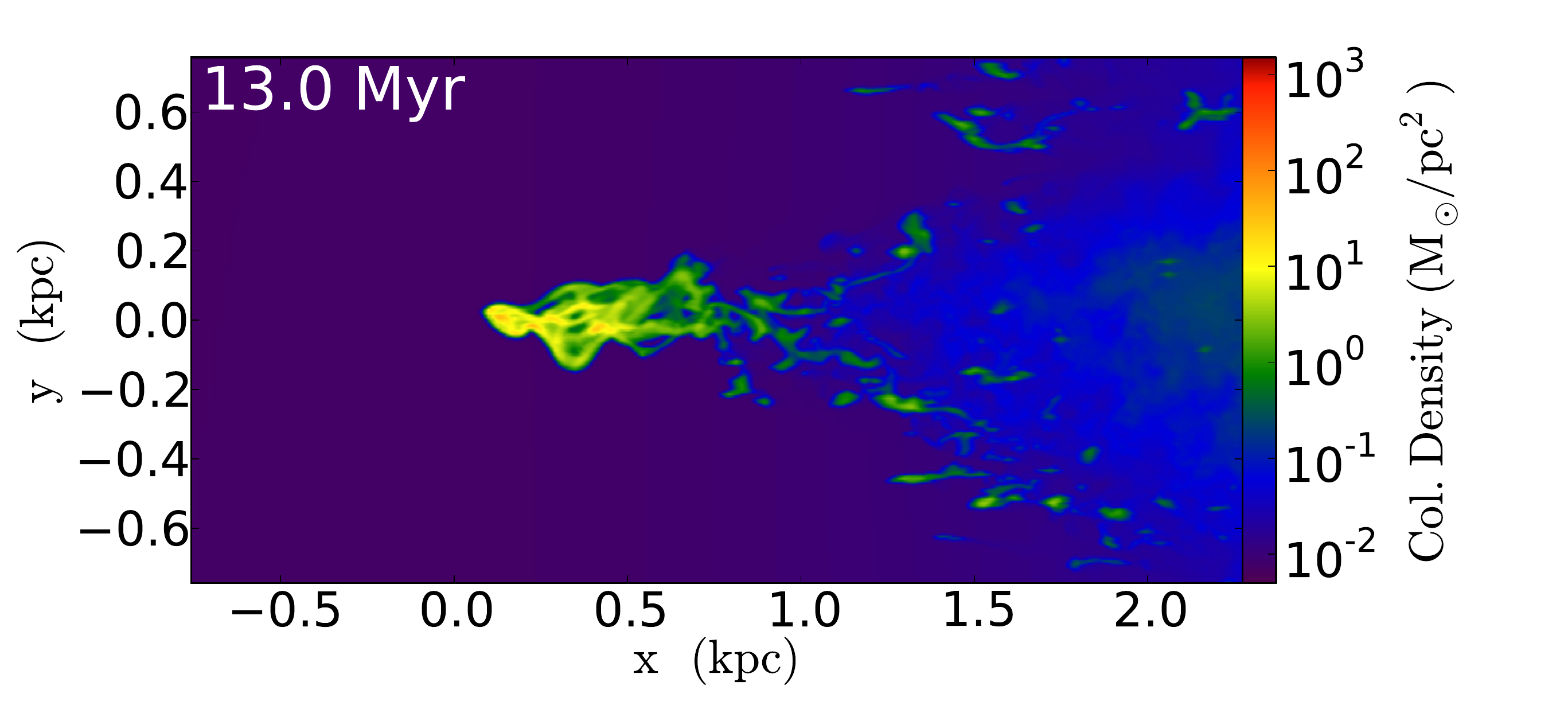}
\includegraphics*[scale=0.245, trim=0 64.1 162.5 21]{Fid_Col_Dens_10_Proj.pdf}
\includegraphics*[scale=0.245, trim=95 64.1 162.5 21]{Fid_Col_Dens_23_Proj.pdf}
\includegraphics*[scale=0.245, trim=95 64.1 0 21]{Fid_Col_Dens_39_Proj.pdf}
\includegraphics*[scale=0.245, trim=0 0 162.5 21]{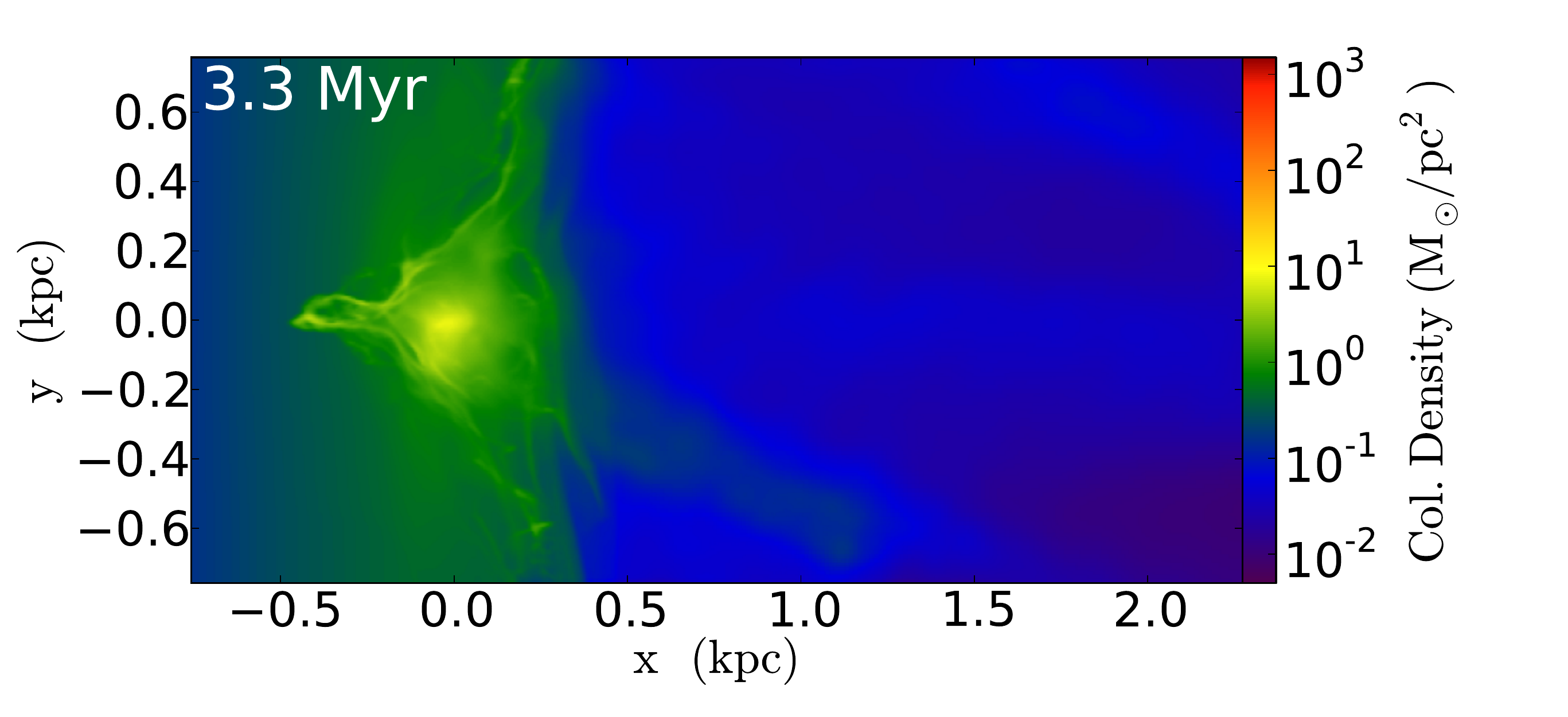}
\includegraphics*[scale=0.245, trim=95 0 162.5 21]{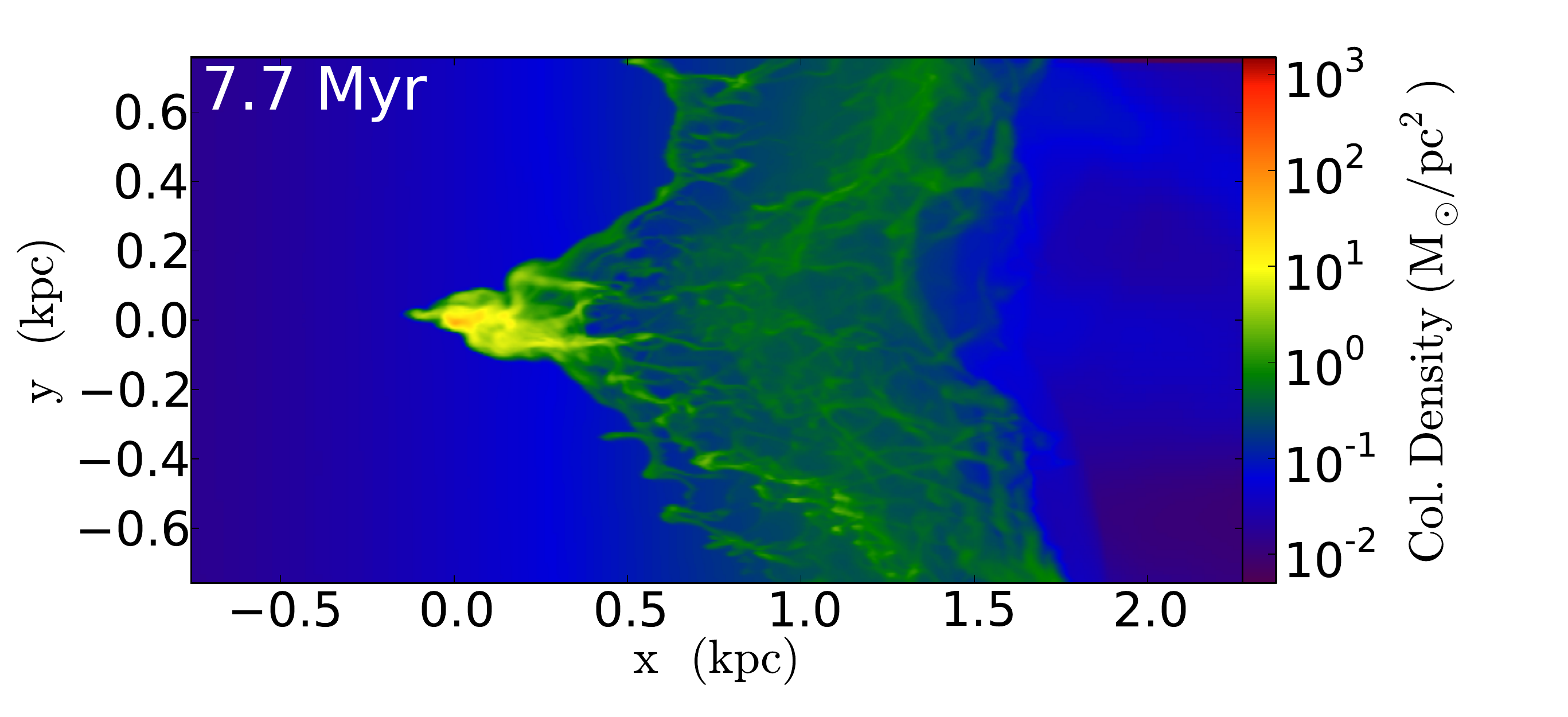}
\includegraphics*[scale=0.245, trim=95 0 0 21]{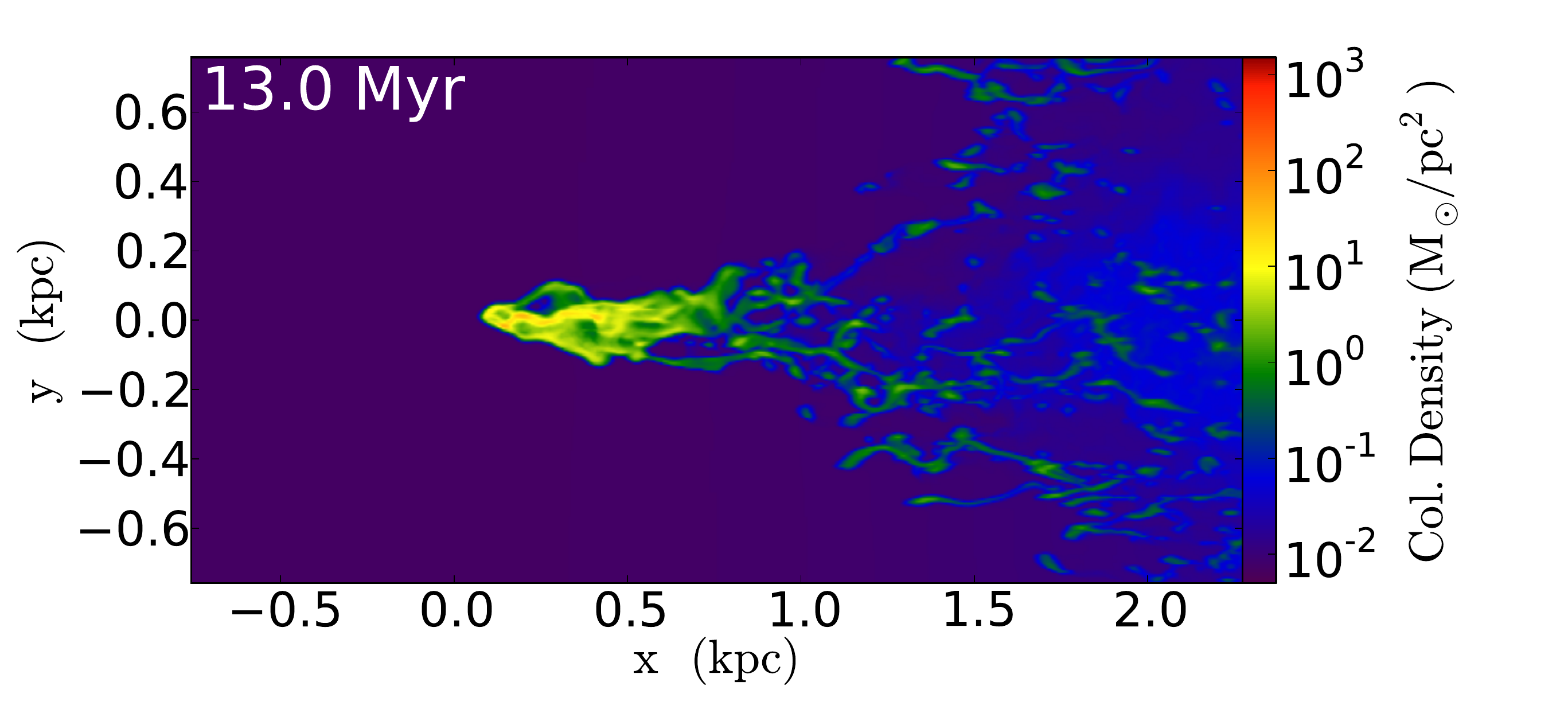}
\caption{\footnotesize{Column density of the simulations varying the shock enrichment, with PZ005, PZ05, FID, and PZ5 from top to bottom, and evolution increasing from the 
first interaction with the minihalo on the left, to when the outflow is just passing the minihalo in the middle, to when the 
outflow reaches the end of the box on the right.}}
\label{fig_enrich}
\end{figure*}

Throughout the simulations described above,
the level of enrichment of the final minihalo material has been consistently a few percent of the outflow abundance. Thus by changing 
the outflow abundance, we expect the enrichment of the minihalo baryons to vary, which may affect its cooling efficiency as well as the
metallicity of the final clusters where stars will be formed. Note that since the majority of the cooling is caused by the produced molecules, 
the increased metallicity may have little effect on the evolution. We explore this evolution in \fig{fig_enrich}.
We see no discernible difference between the different models, with the exception of PZ5. The most enriched outflow, once a significant amount
of metals has mixed into the minihalo, is able to contribute a non-negligent amount of cooling from metals, leading to a more collapsed structure starting at
7.7 Myr, and much more noticeable by 13 Myr. Also, the post-shock ambient medium is able to cool more, and fragments earlier than the metal-poor runs.

\begin{figure*}[t]
\centering
\includegraphics[scale=0.42]{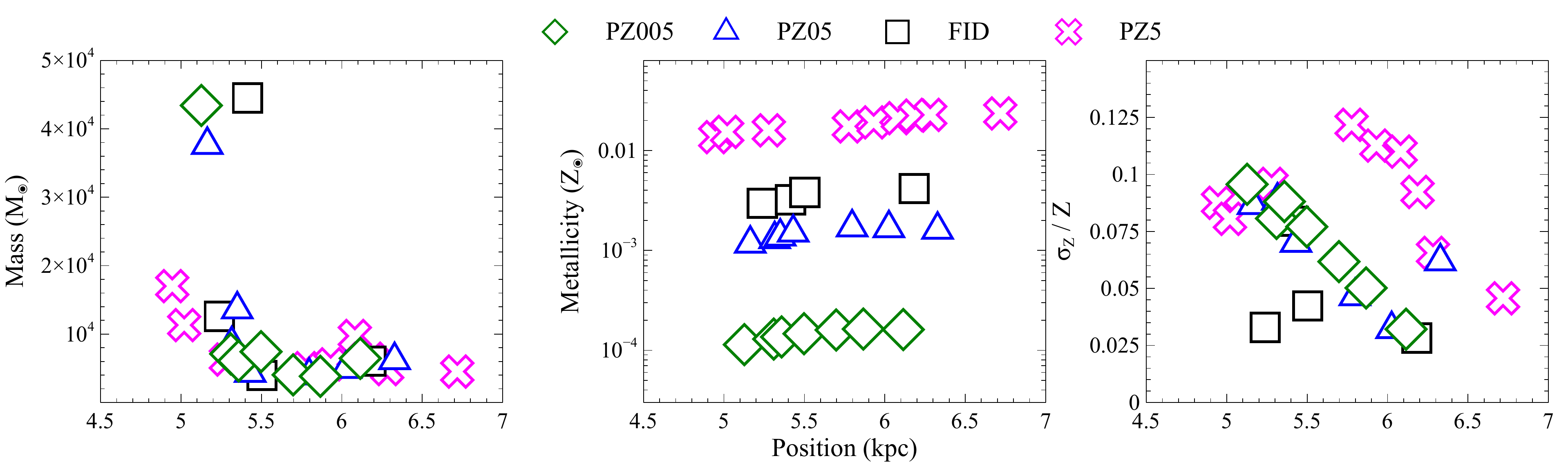}
\caption{\footnotesize{Comparison between the PZ005 (green diamonds), PZ05 (blue triangles),
FID (black squares), and PZ5 (magenta crosses) simulations illustrating the dependence on shock metallicity.  The particle masses (left), 
metallicity (middle) and relative metallicity dispersion (right) vs particles 
positions after 200 Myr are shown.}}
\label{fig_enrich2}
\end{figure*}

In \fig{fig_enrich2}, we show the ballistic particles after 200 Myr of evolution for the simulations that vary the outflow abundance.
The mass distribution of final particles is almost independent of enrichment. Only in the most enriched outflow is the additional cooling sufficient to cause increased fragmentation in the 
pre-ballistic gas, such that these clumps do not merge. In all simulations roughly 25\% of the minihalo baryons are bound in these cloud particles. The enrichment appears
to only affect the abundance of the clumps, which is always roughly 2\% of the outflow abundance, with a slight dependence on position of the final particles. The variance in abundance
is fairly uniform, although the most metal-rich outflow has a slight peak in its spread. This simulation's increased fragmentation leads to a larger variance of enriched proto-clumps,
which, through merging, results in clumps with higher variance.

\subsubsection{Redshift}\label{red}
\begin{figure*}[h!]
\centering
\includegraphics*[scale=0.245, trim=0 43.7 162.5 21]{Fid_Col_Dens_10_Proj.pdf}
\includegraphics*[scale=0.245, trim=95 43.7 162.5 21]{Fid_Col_Dens_23_Proj.pdf}
\includegraphics*[scale=0.245, trim=95 43.7 0 21]{Fid_Col_Dens_39_Proj.pdf}
\includegraphics*[scale=0.245, trim=0 43.7 162.5 21]{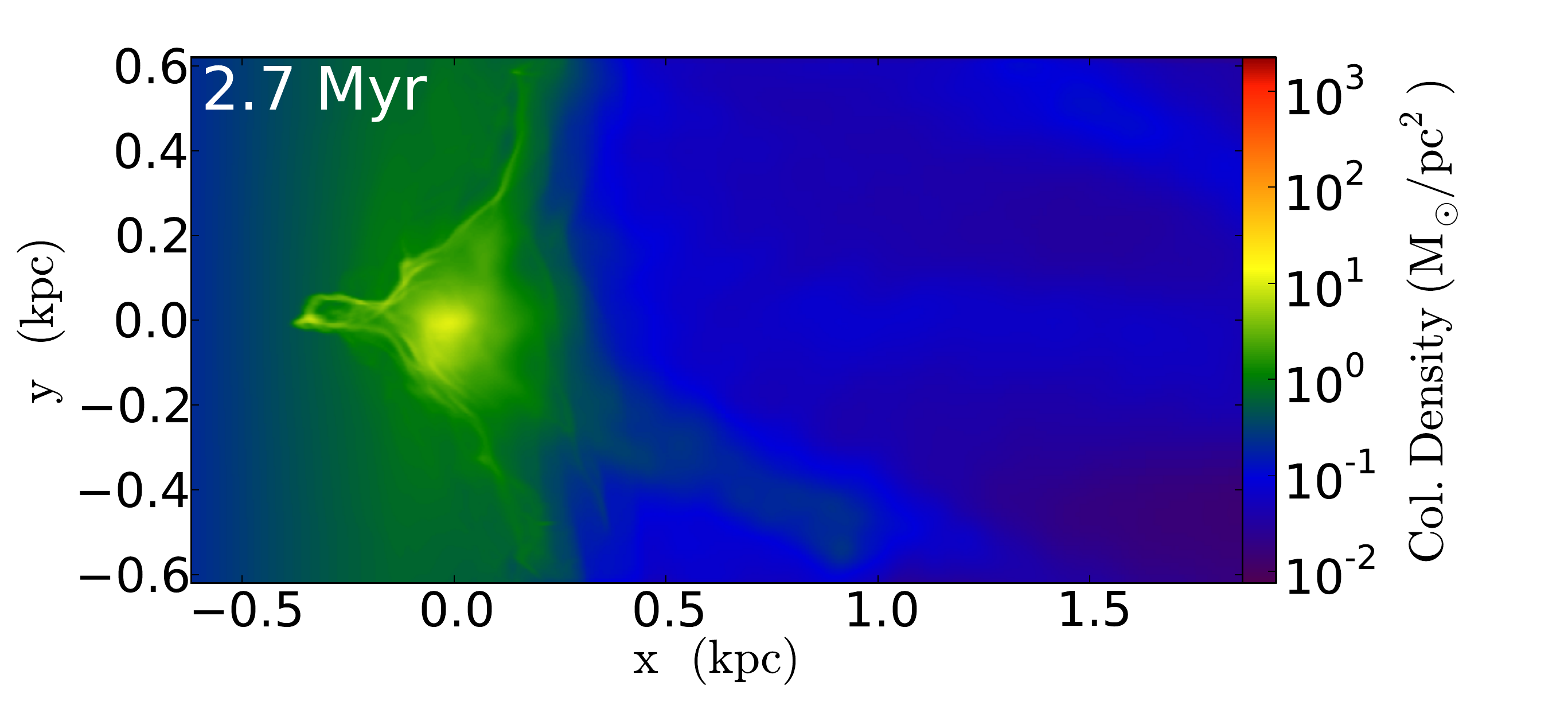}
\includegraphics*[scale=0.245, trim=95 43.7 162.5 21]{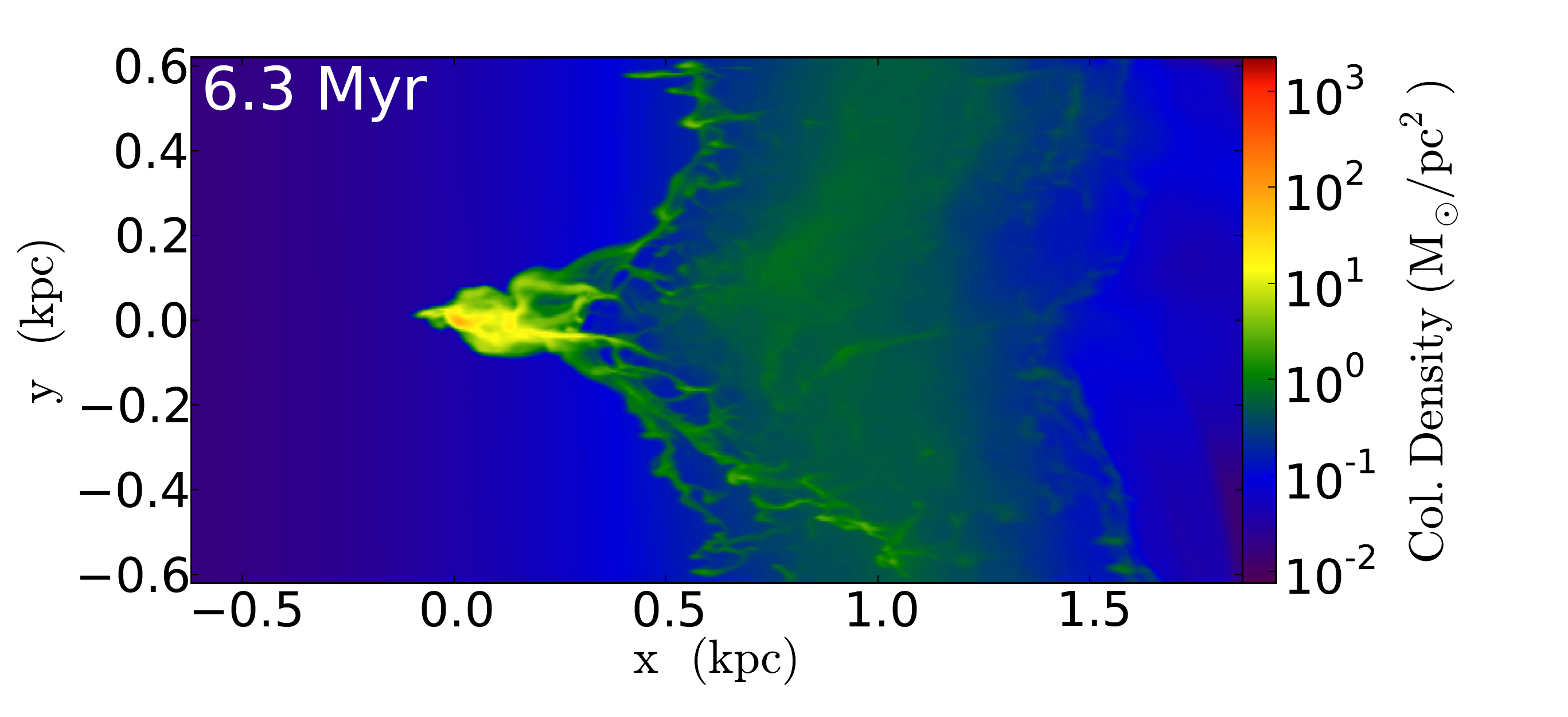}
\includegraphics*[scale=0.245, trim=95 43.7 0 21]{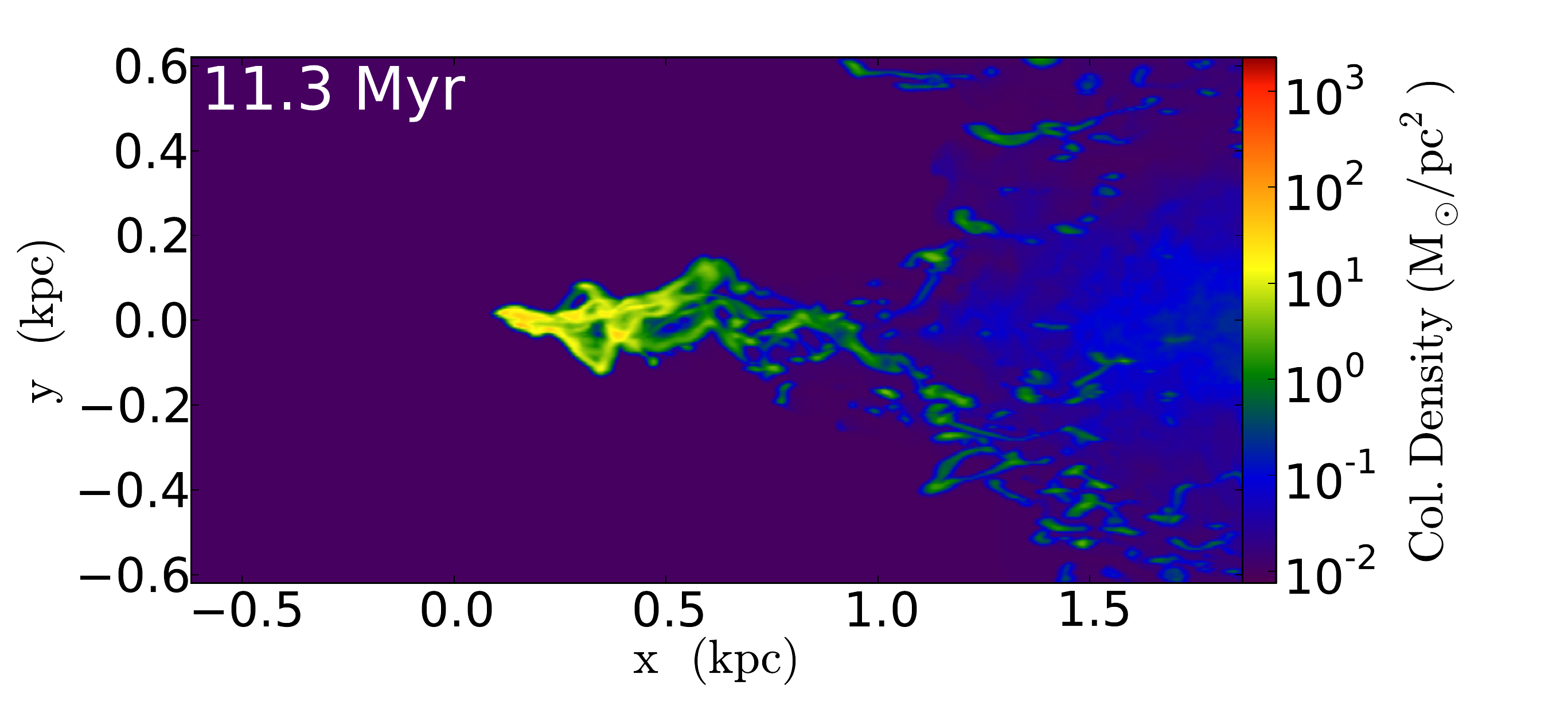}
\includegraphics*[scale=0.245, trim=0 43.7 162.5 21]{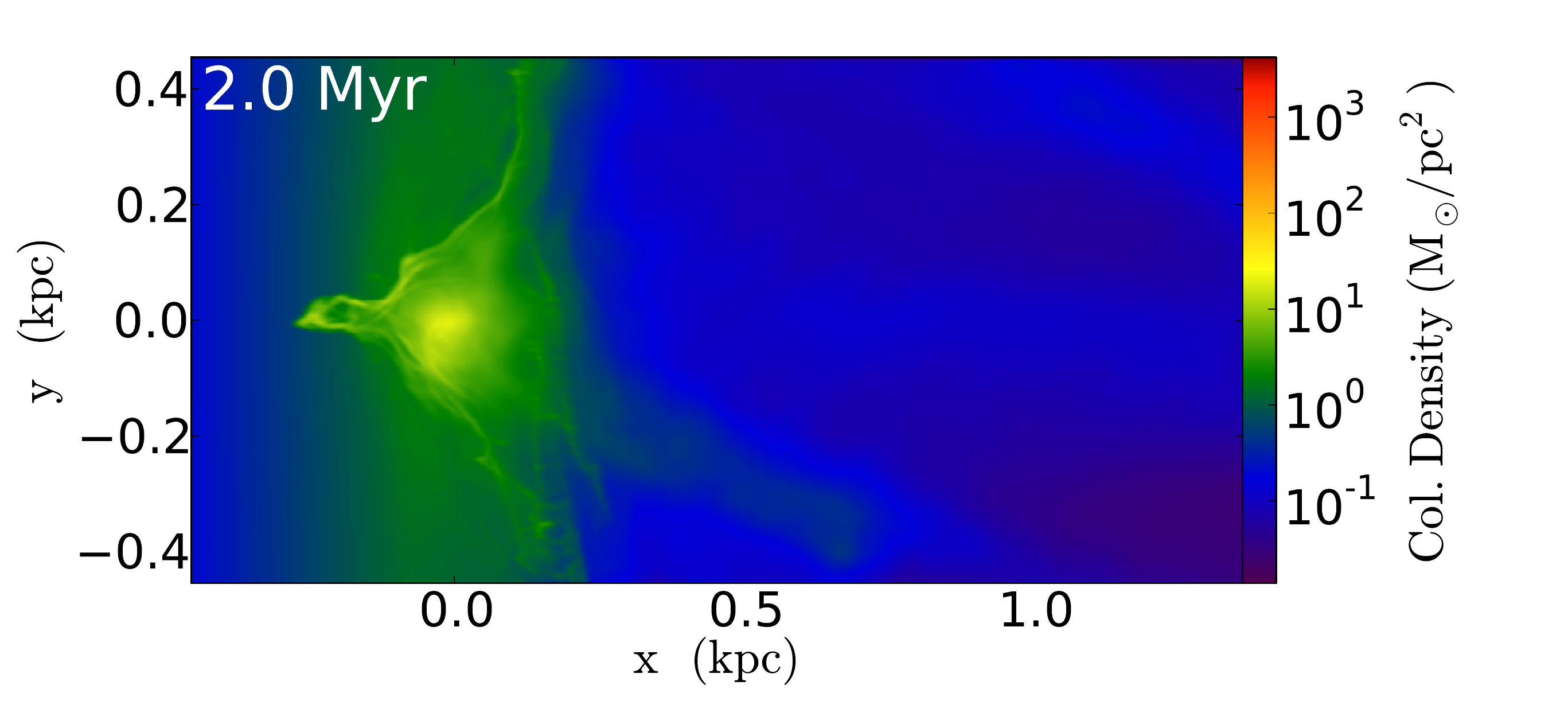}
\includegraphics*[scale=0.245, trim=95 43.7 162.5 21]{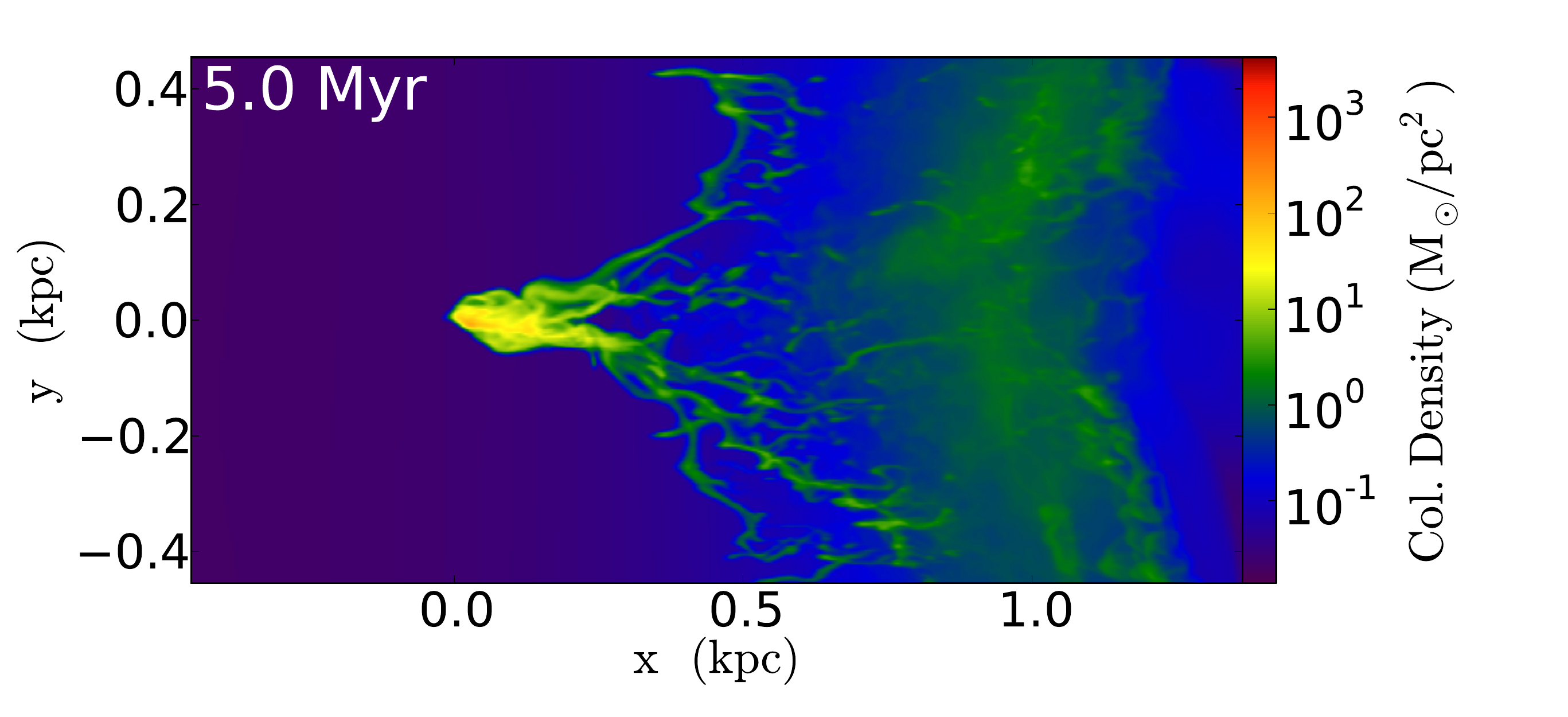}
\includegraphics*[scale=0.245, trim=95 43.7 0 21]{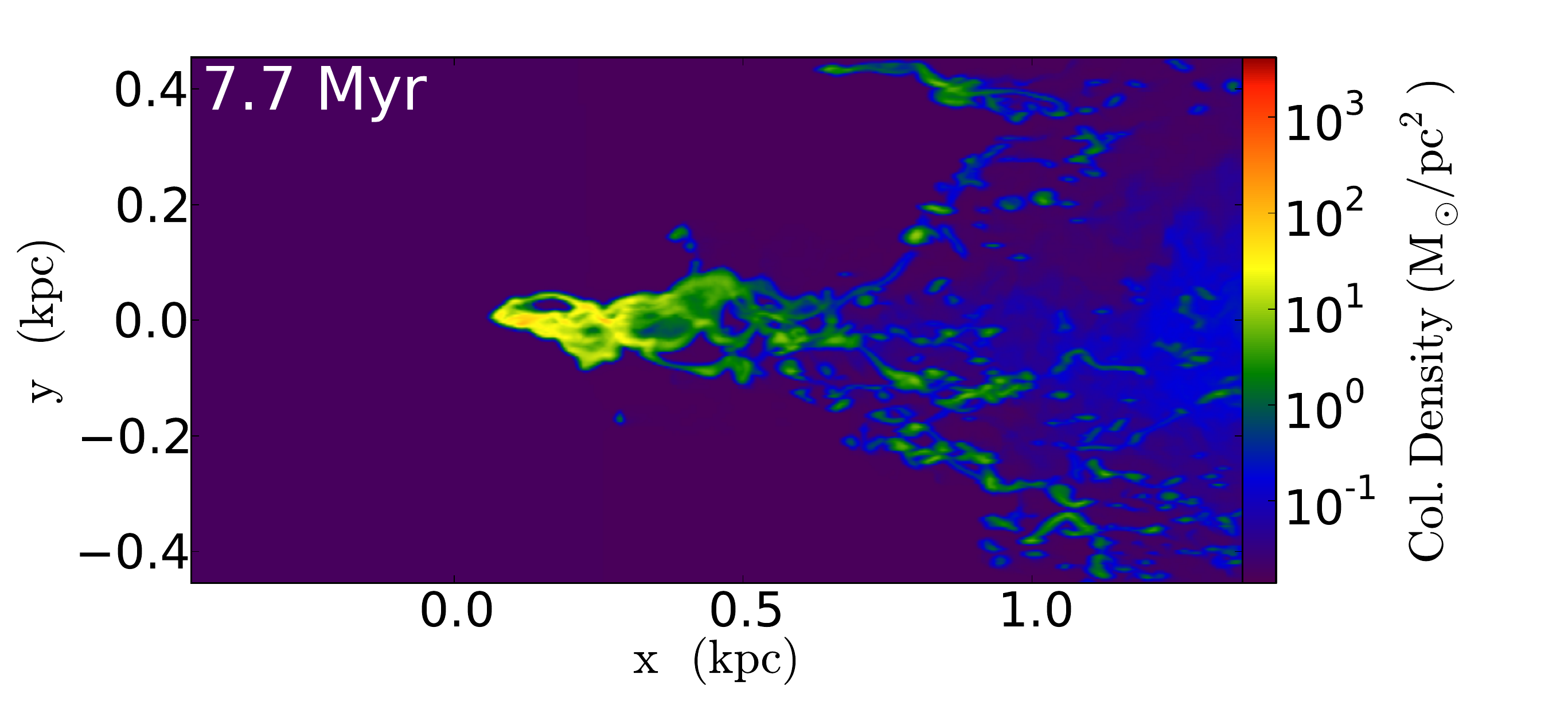}
\includegraphics*[scale=0.245, trim=0 43.7 162.5 21]{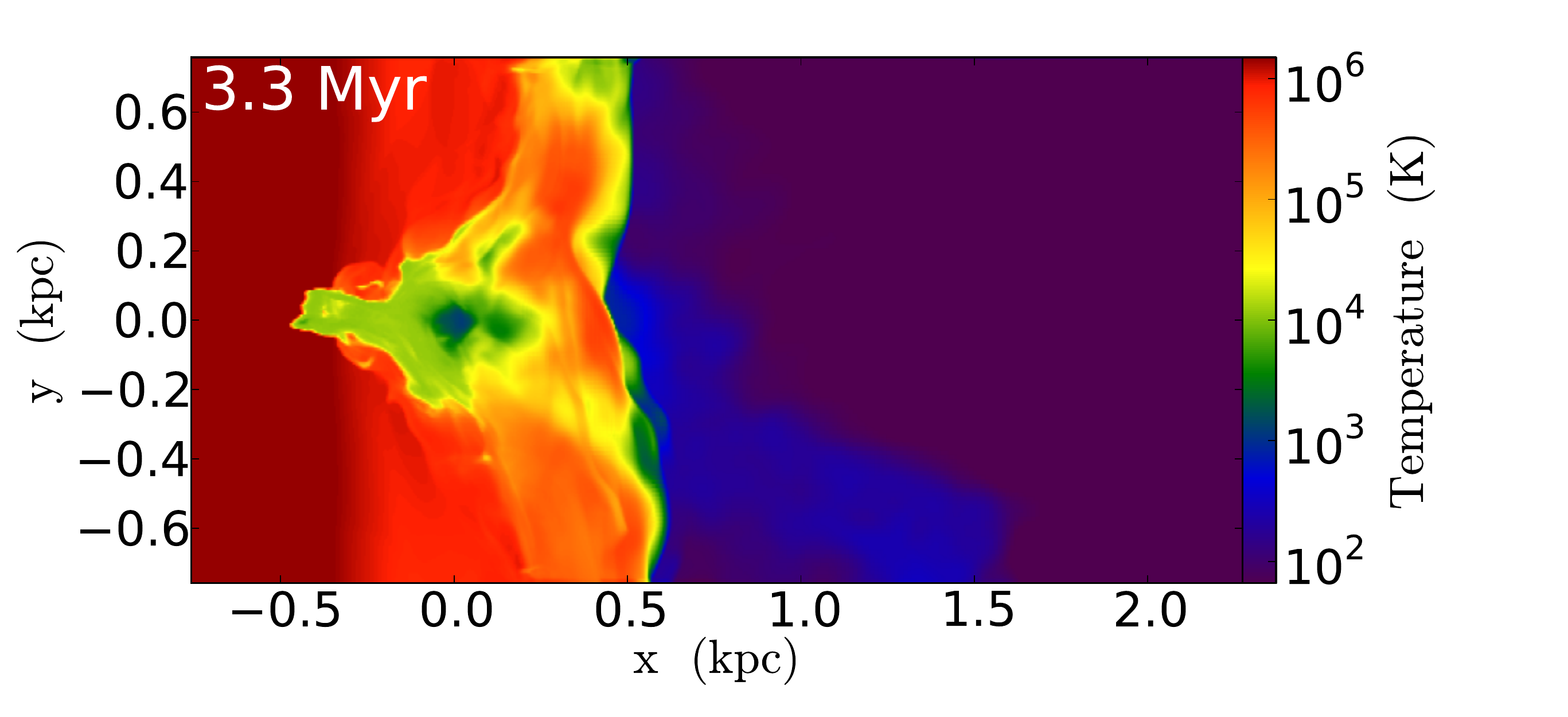}
\includegraphics*[scale=0.245, trim=95 43.7 162.5 21]{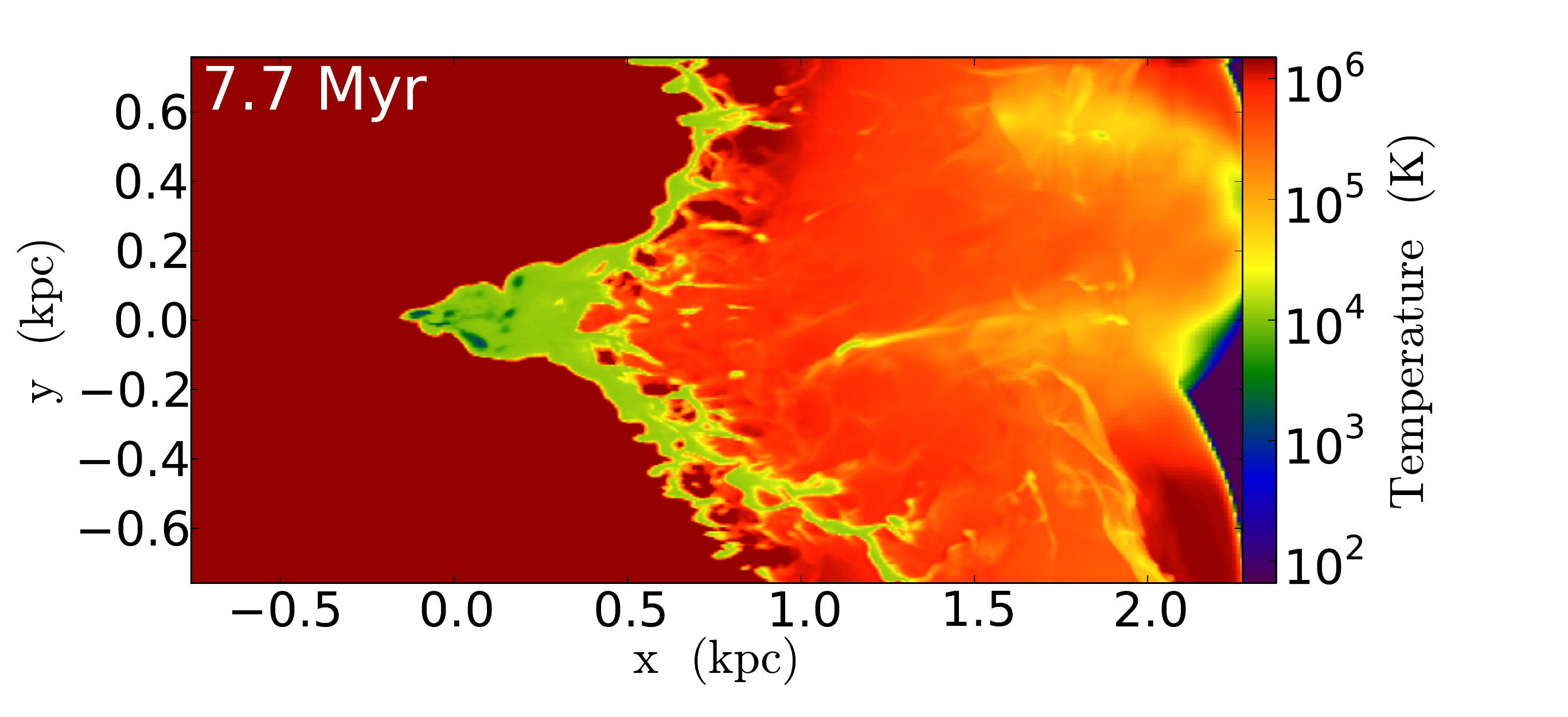}
\includegraphics*[scale=0.245, trim=95 43.7 0 21]{Fid_Temp_D3_39_Proj.pdf}
\includegraphics*[scale=0.245, trim=0 43.7 162.5 21]{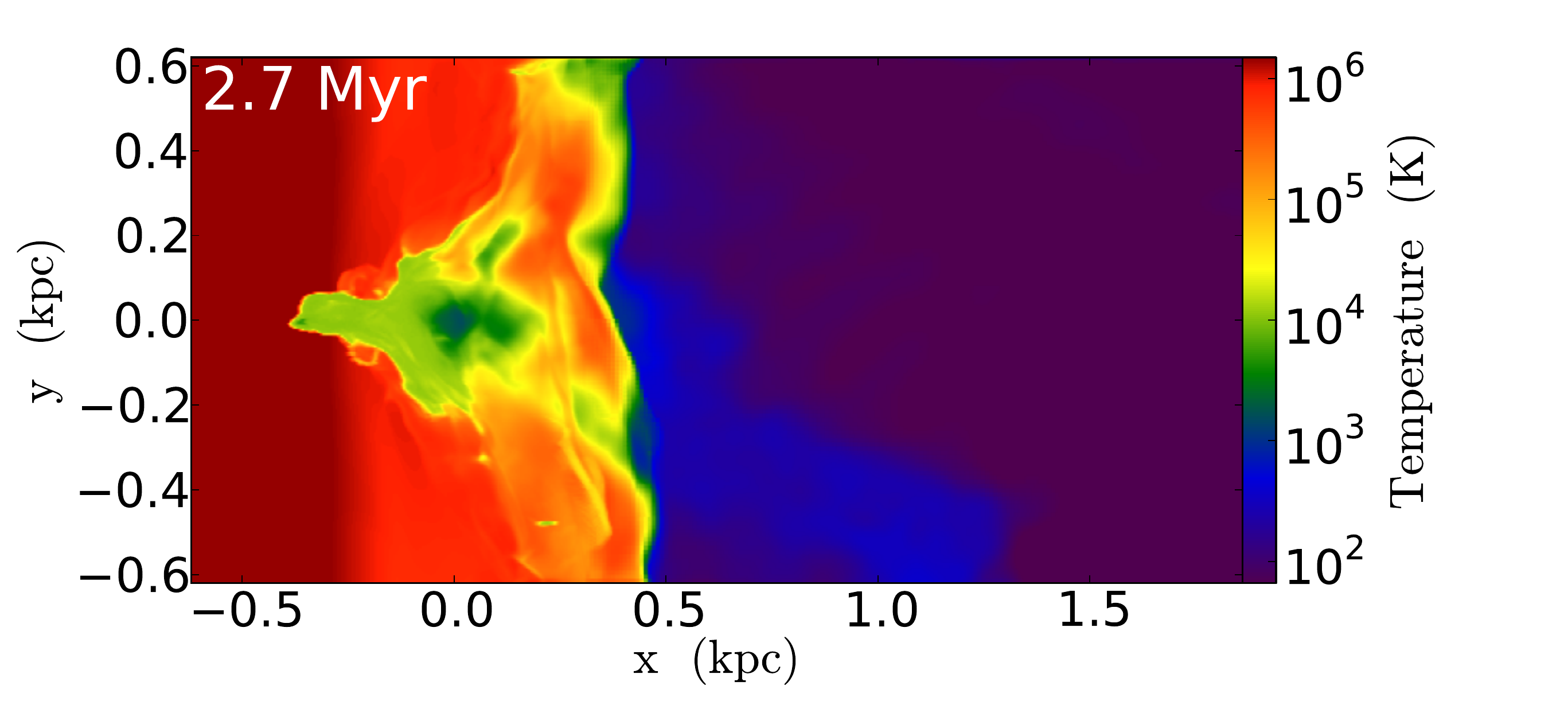}
\includegraphics*[scale=0.245, trim=95 43.7 162.5 21]{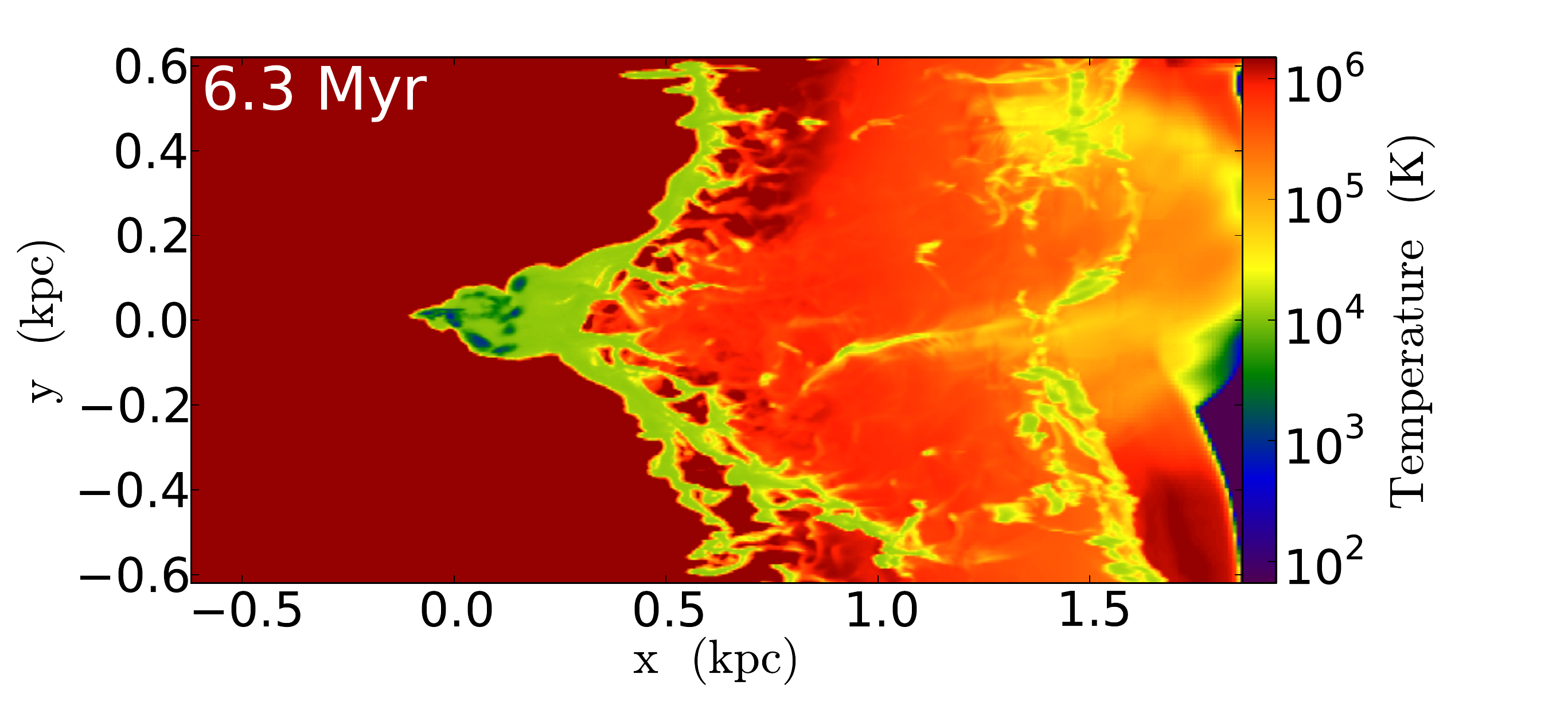}
\includegraphics*[scale=0.245, trim=95 43.7 0 21]{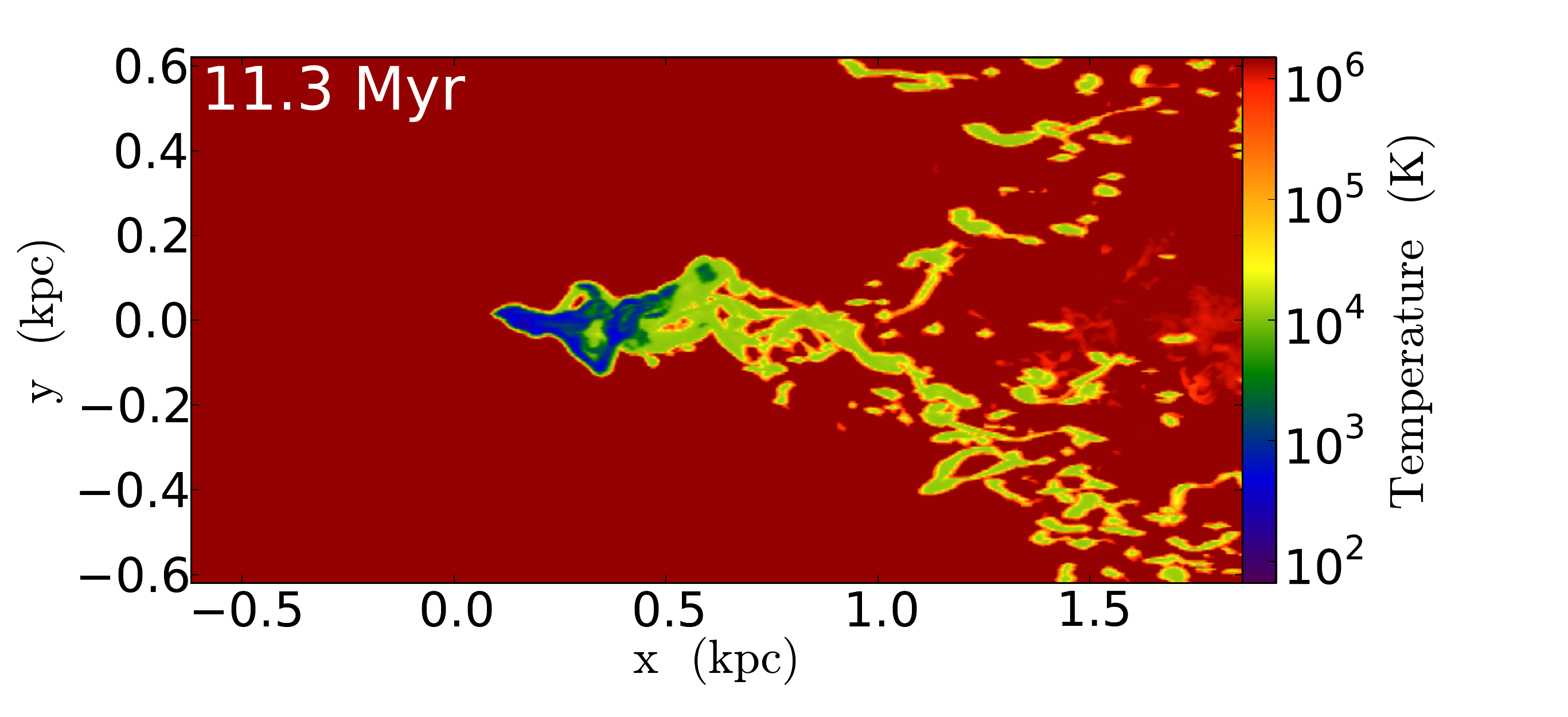}
\includegraphics*[scale=0.245, trim=0 0 162.5 21]{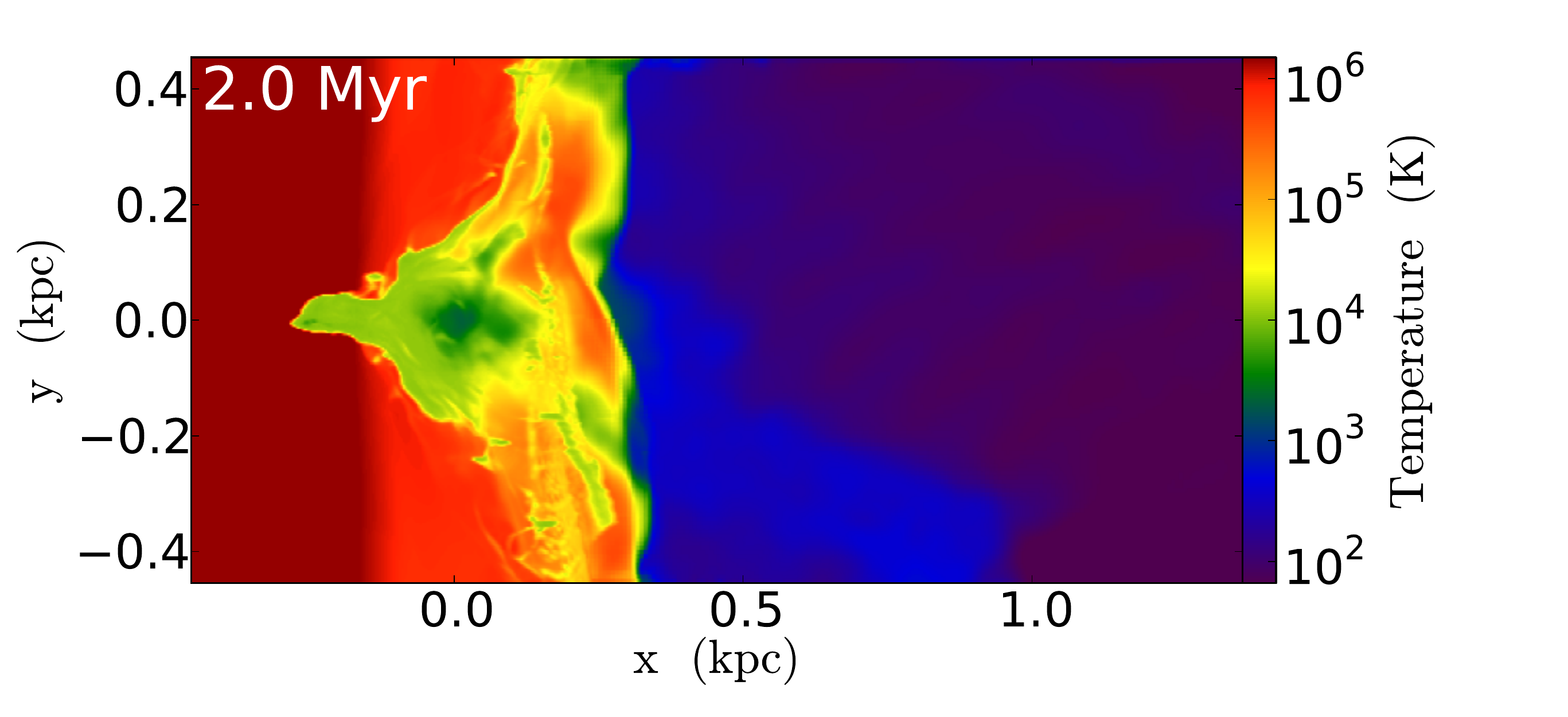}
\includegraphics*[scale=0.245, trim=95 0 162.5 21]{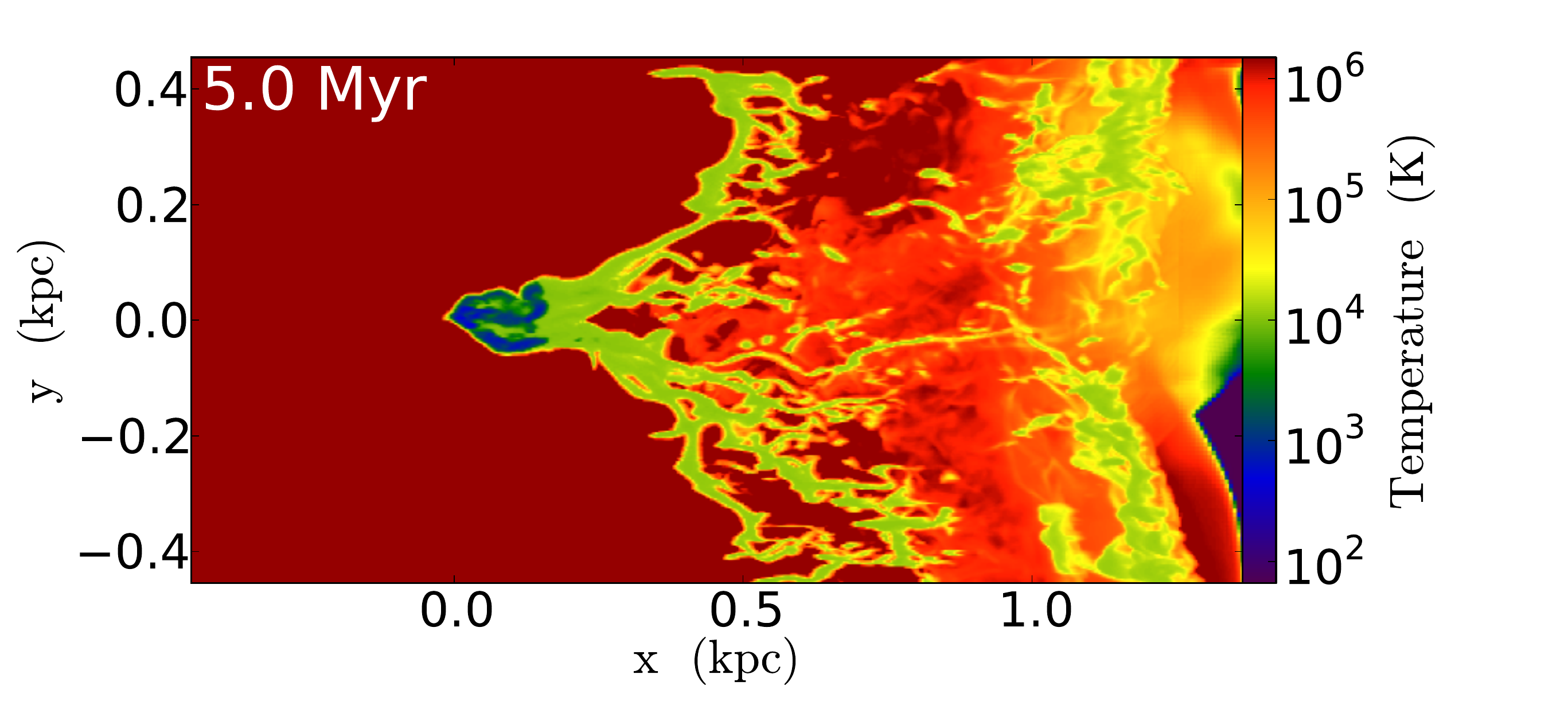}
\includegraphics*[scale=0.245, trim=95 0 0 21]{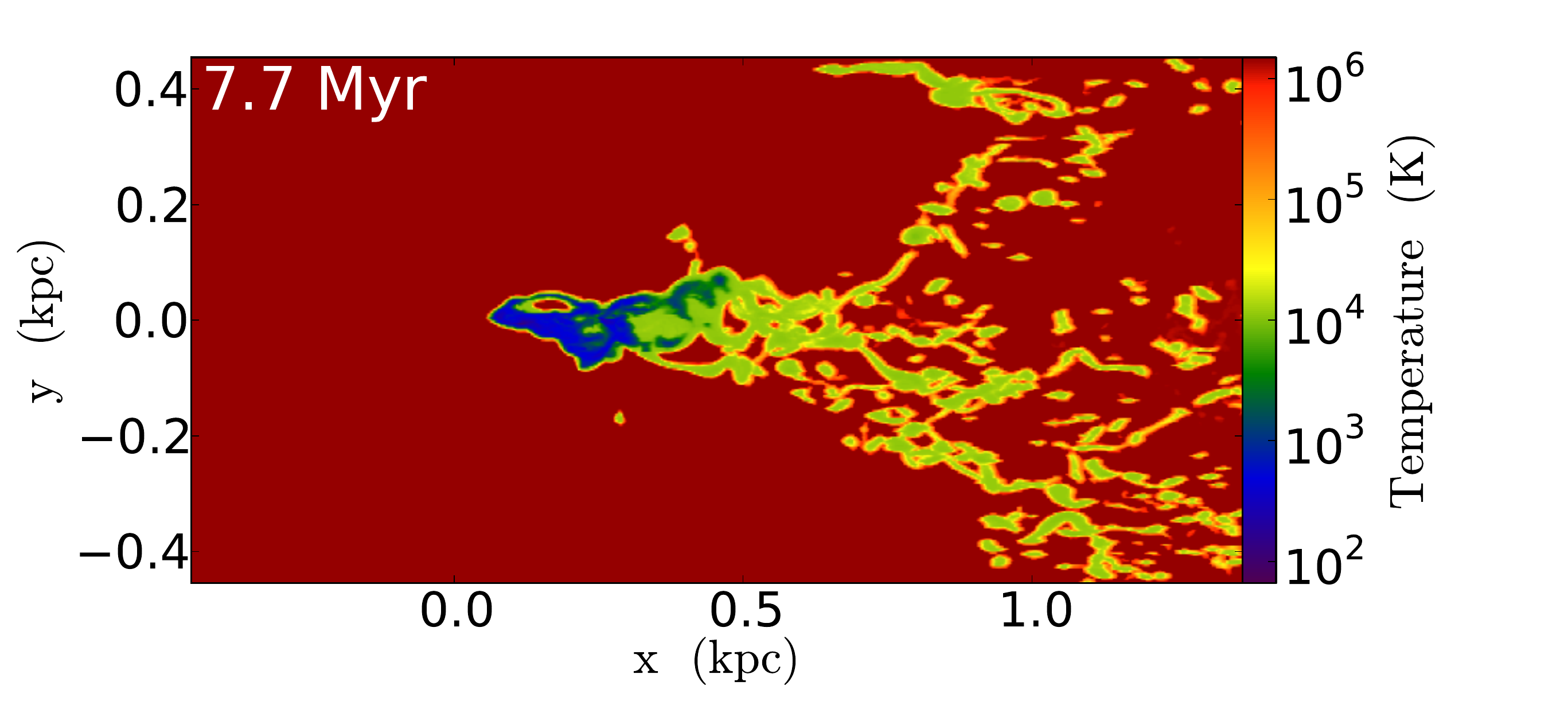}
\caption{\footnotesize{Column density (rows 1-3) and projected temperature (rows 4-6) of the simulations 
varying the redshift of the interaction, along with the surface momentum, with FID shown in rows 1 and 4, 
Pz10 shown in rows 2 and 5, and Pz14 shown in rows 3 and 6. Evolution increases from the 
first interaction with the minihalo on the left, to when the outflow is just passing the minihalo in the middle, to when the 
outflow reaches the end of the box on the right.}}
\label{fig_redshift}
\end{figure*}

The redshift of the interaction is a significant parameter that sets the minihalo density and temperature profiles, as well as its environment. The redshift also sets the post-shock
density as well as the surface momentum. We assume the surface density and momentum scales with $(1+z)^2$, while the shock velocity is invariant with redshift, as it scales with the supernovae
input energy, which we assume is independent of redshift. In \fig{fig_redshift} we illustrate the evolution of this interaction for various redshifts, while scaling the density color scheme with $(1+z)^2$.
We see little variation between the three simulations. The timescale of the interaction scales with $(1+z)^{-1}$ since $v_{\rm s}$ is the same, while
the physical scale of the minihalo is smaller by $(1+z)$. The cooling is more efficient at higher redshifts with larger densities, leading to 
slightly cooler, denser clumps, however this effect is minor.

\begin{figure*}[h!]
\centering
\includegraphics[scale=0.42]{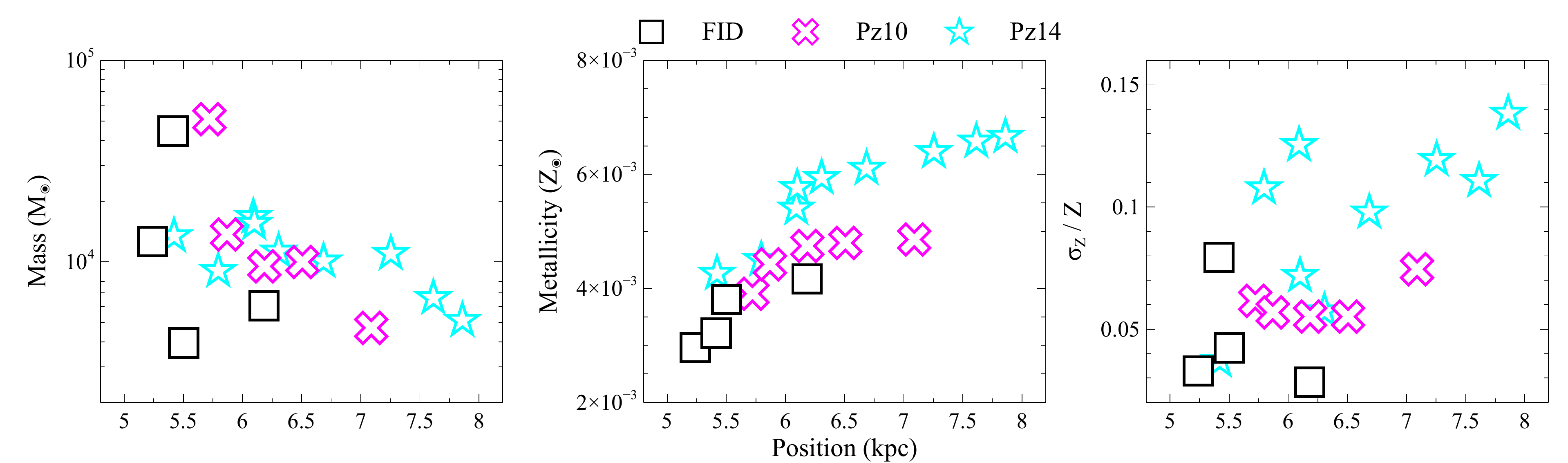}
\caption{\footnotesize{Comparison between the FID (black squares), Pz10$\sigma$ (dark blue 
circles), Pz10C (blue triangles), and Pz14C 
(magenta crosses) simulations illustrating the dependence on redshift. 
The particle masses (left), 
metallicity (middle) and relative metallicity dispersion (right) vs particles 
positions after 200 Myr are shown.}}
\label{fig_redshift2}
\end{figure*}

In \fig{fig_redshift2}  we show the ballistic particles after 200 Myr of evolution for these simulations.
Similar to the metallicity parameter study, we see subtle effects. The increased cooling at the largest redshift leads to 
an increase in fragmentation of the clumps before the ballistic evolution, and this increased fragmentation leads to less merging
between the final cloud particles and an increase in enrichment and enrichment variance.

\subsubsection{UV Background}
\begin{figure*}[t!]
\centering
\includegraphics*[scale=0.245, trim=0 64.1 162.5 21]{Fid_Col_Dens_10_Proj.pdf}
\includegraphics*[scale=0.245, trim=95 64.1 162.5 21]{Fid_Col_Dens_23_Proj.pdf}
\includegraphics*[scale=0.245, trim=95 64.1 0 21]{Fid_Col_Dens_39_Proj.pdf}
\includegraphics*[scale=0.245, trim=0 64.1 162.5 21]{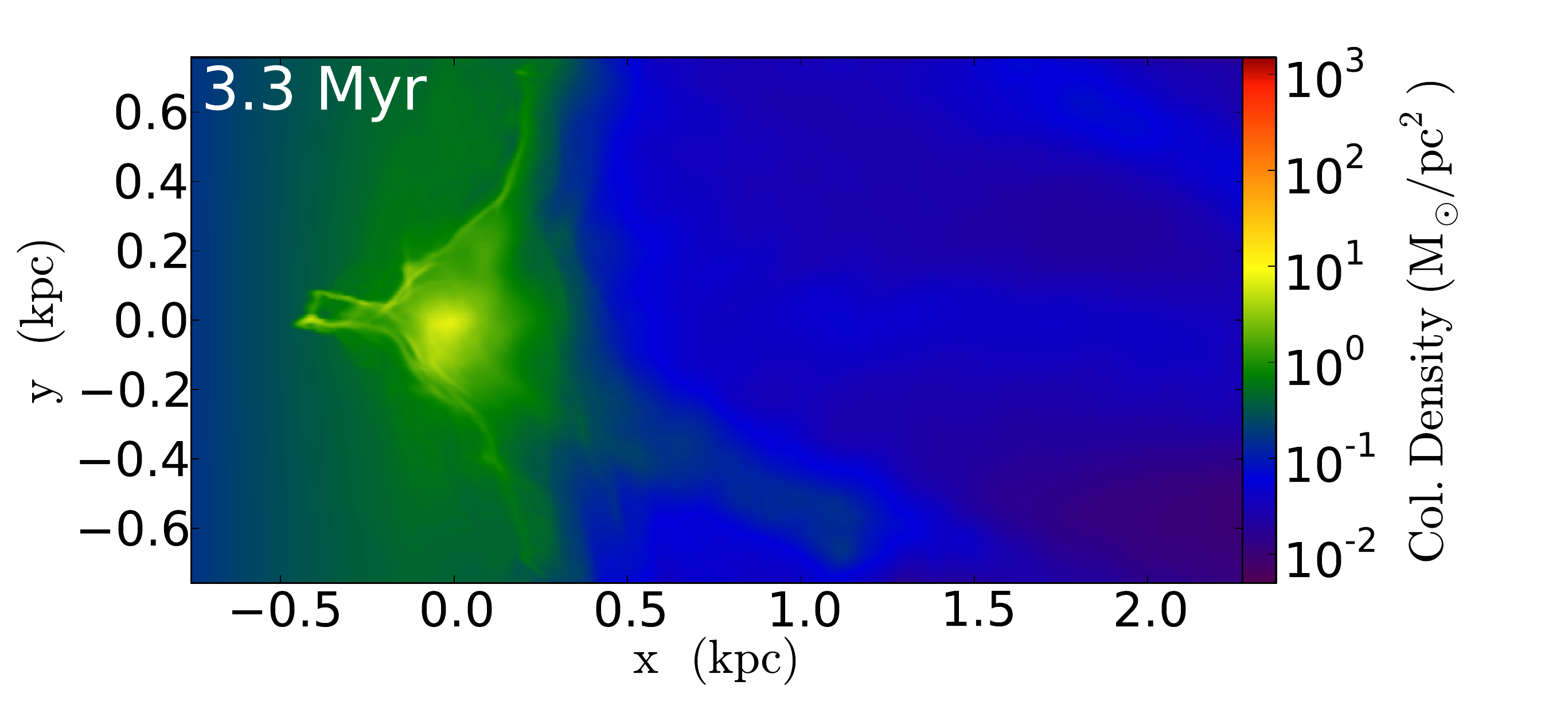}
\includegraphics*[scale=0.245, trim=95 64.1 162.5 21]{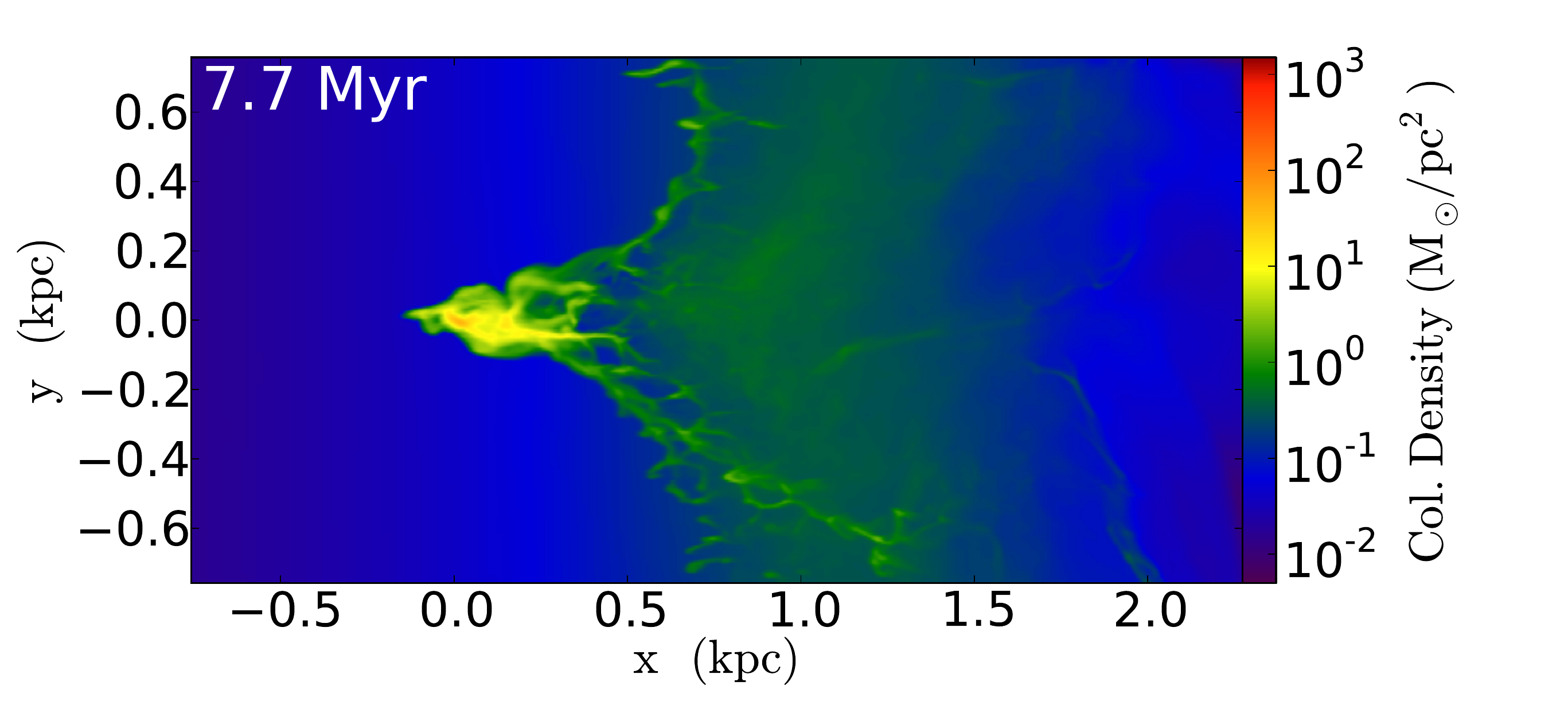}
\includegraphics*[scale=0.245, trim=95 64.1 0 21]{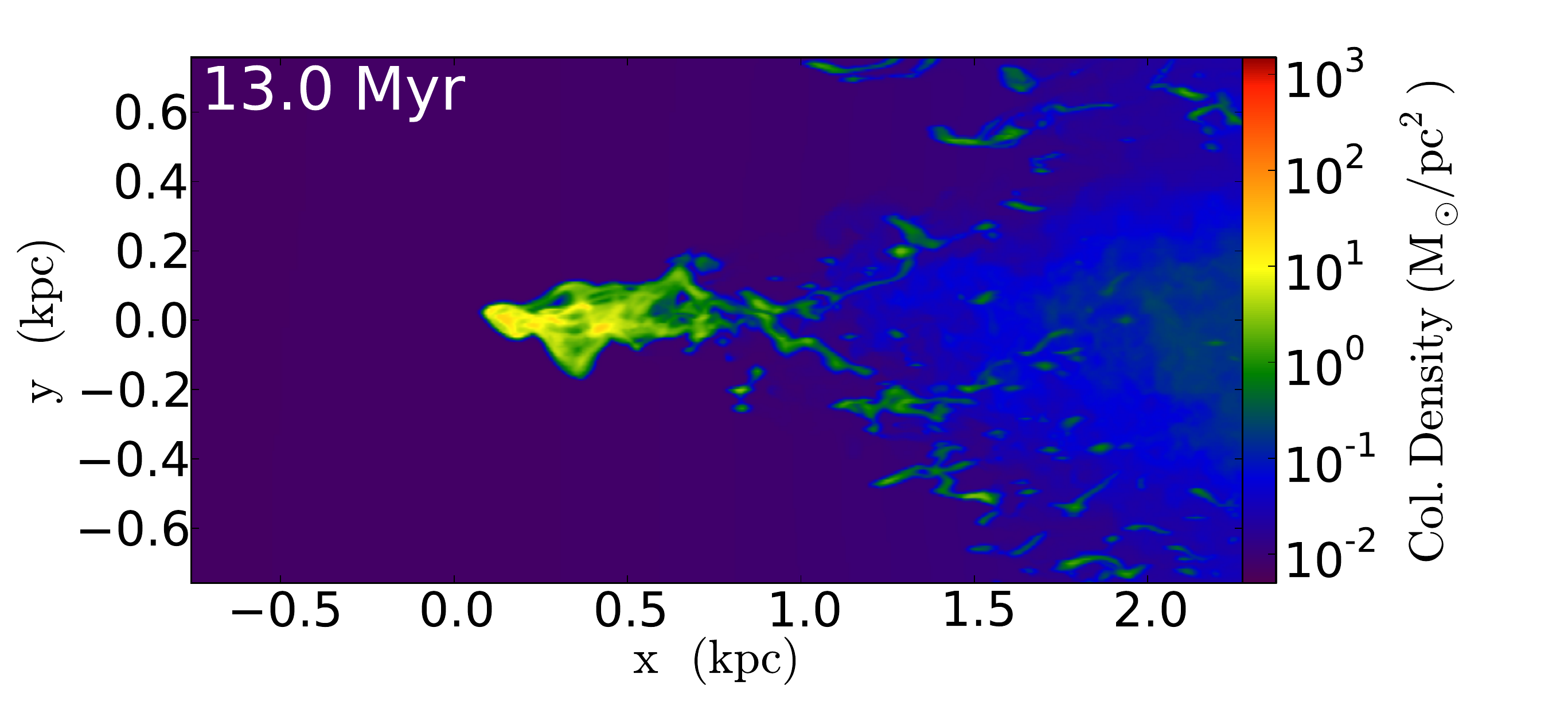}
\includegraphics*[scale=0.245, trim=0 64.1 162.5 21]{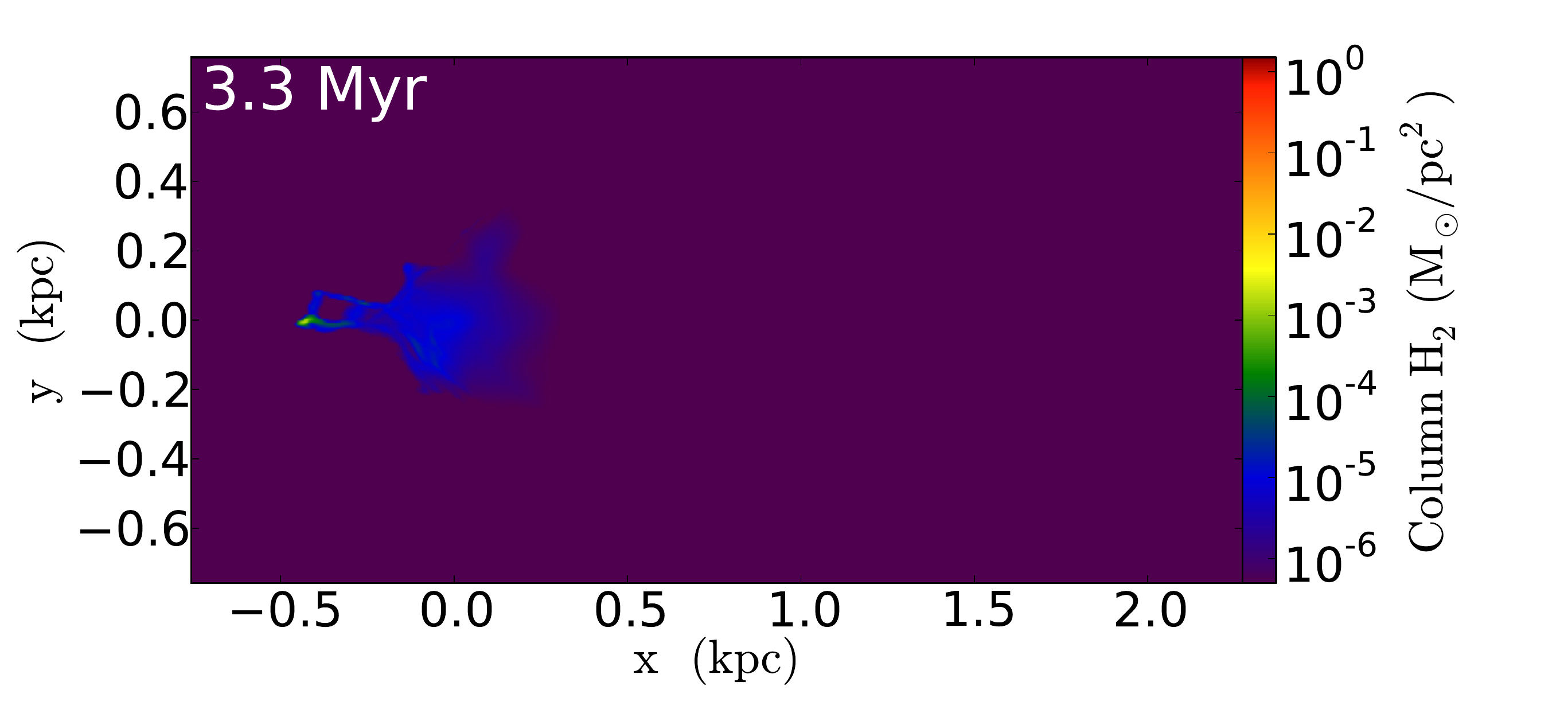}
\includegraphics*[scale=0.245, trim=95 64.1 162.5 21]{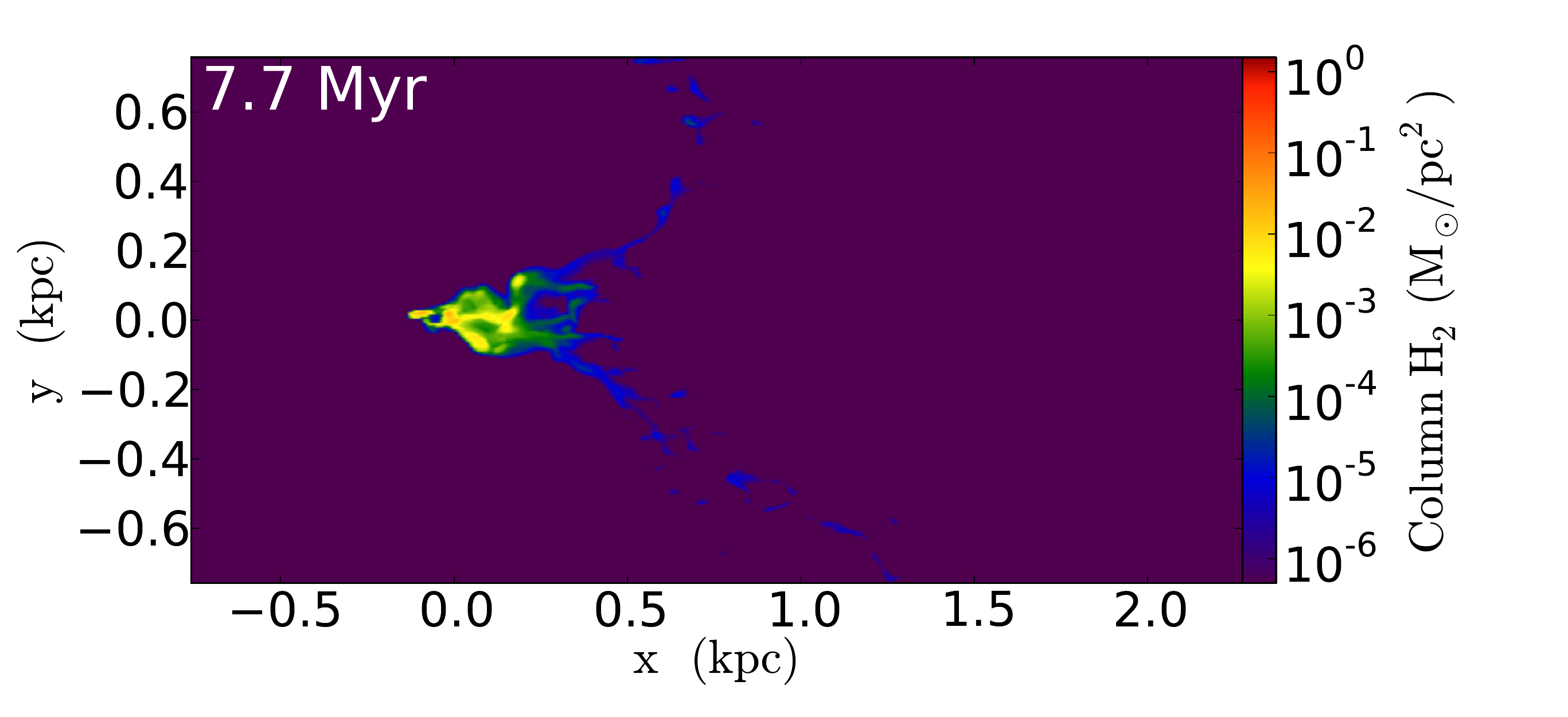}
\includegraphics*[scale=0.245, trim=95 64.1 0 21]{Fid_Col_Molh_39_Proj.pdf}
\includegraphics*[scale=0.245, trim=0 64.1 162.5 21]{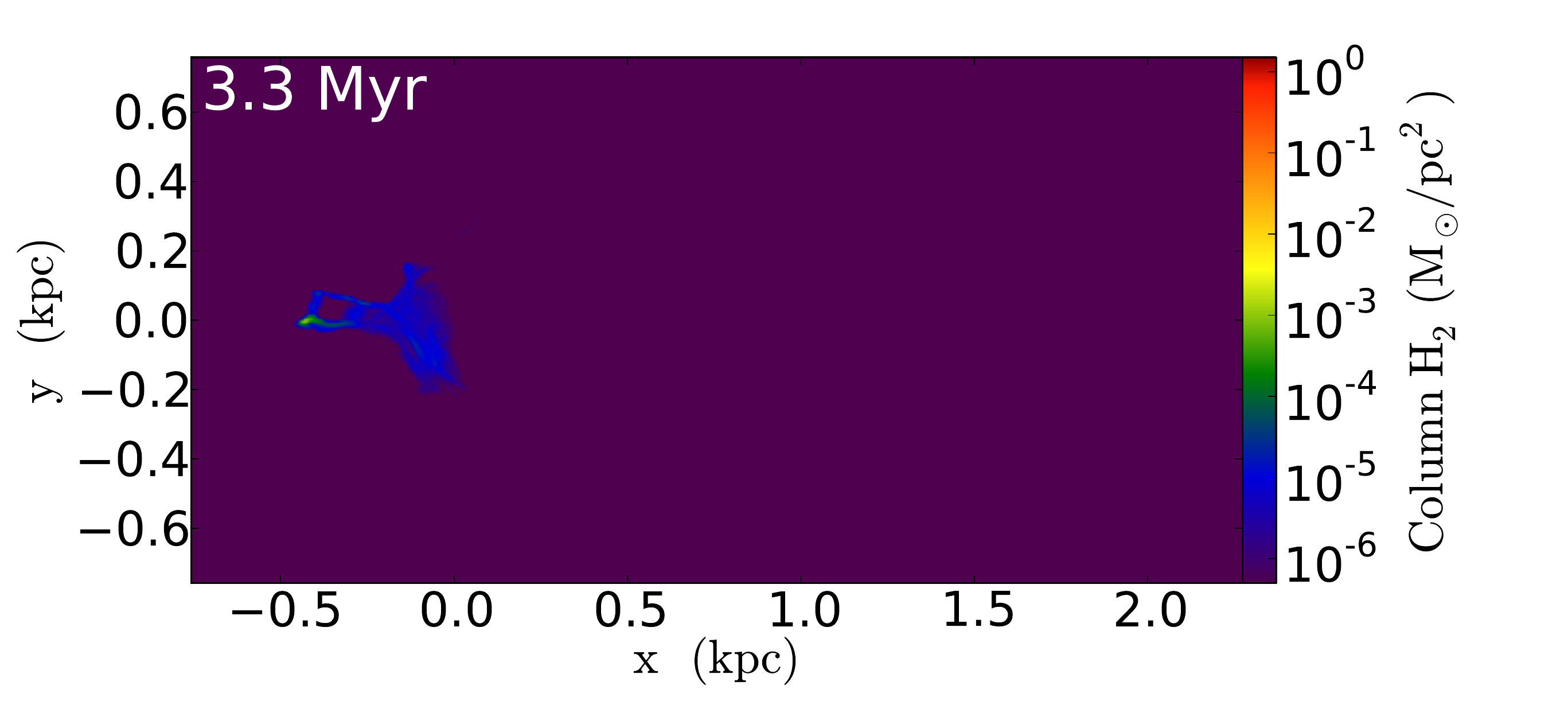}
\includegraphics*[scale=0.245, trim=95 64.1 162.5 21]{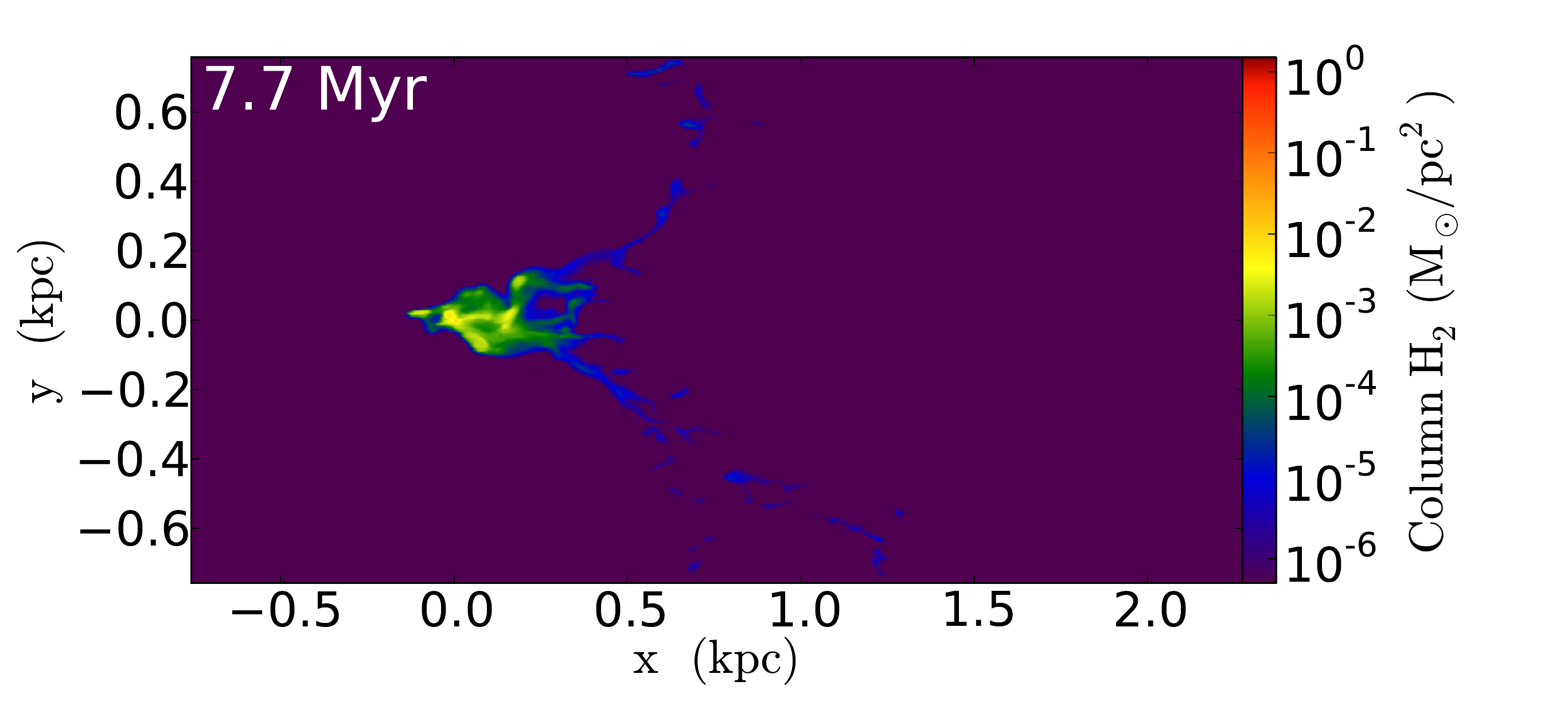}
\includegraphics*[scale=0.245, trim=95 64.1 0 21]{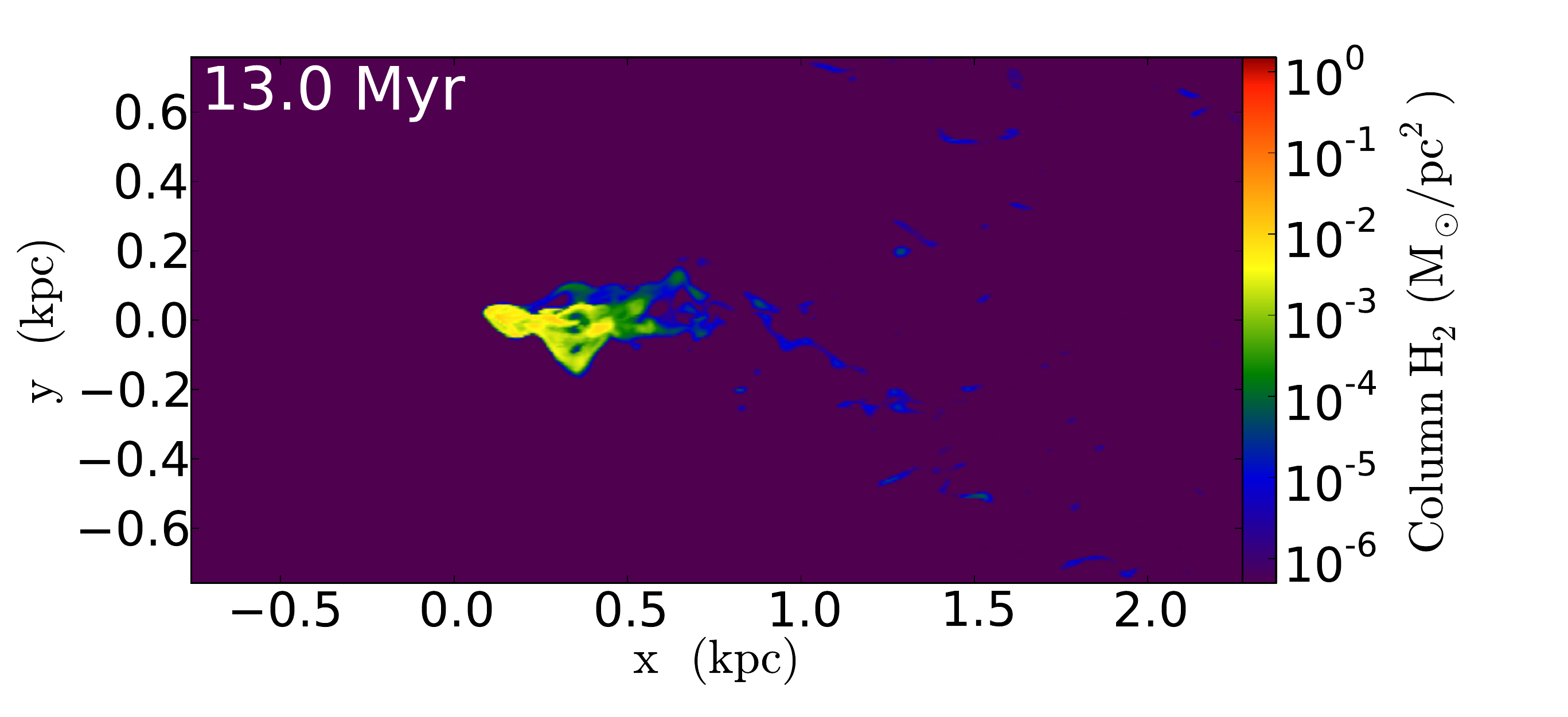}
\includegraphics*[scale=0.245, trim=0 64.1 162.5 21]{Fid_Temp_D3_10_Proj.pdf}
\includegraphics*[scale=0.245, trim=95 64.1 162.5 21]{Fid_Temp_D3_23_Proj.pdf}
\includegraphics*[scale=0.245, trim=95 64.1 0 21]{Fid_Temp_D3_39_Proj.pdf}
\includegraphics*[scale=0.245, trim=0 0 162.5 21]{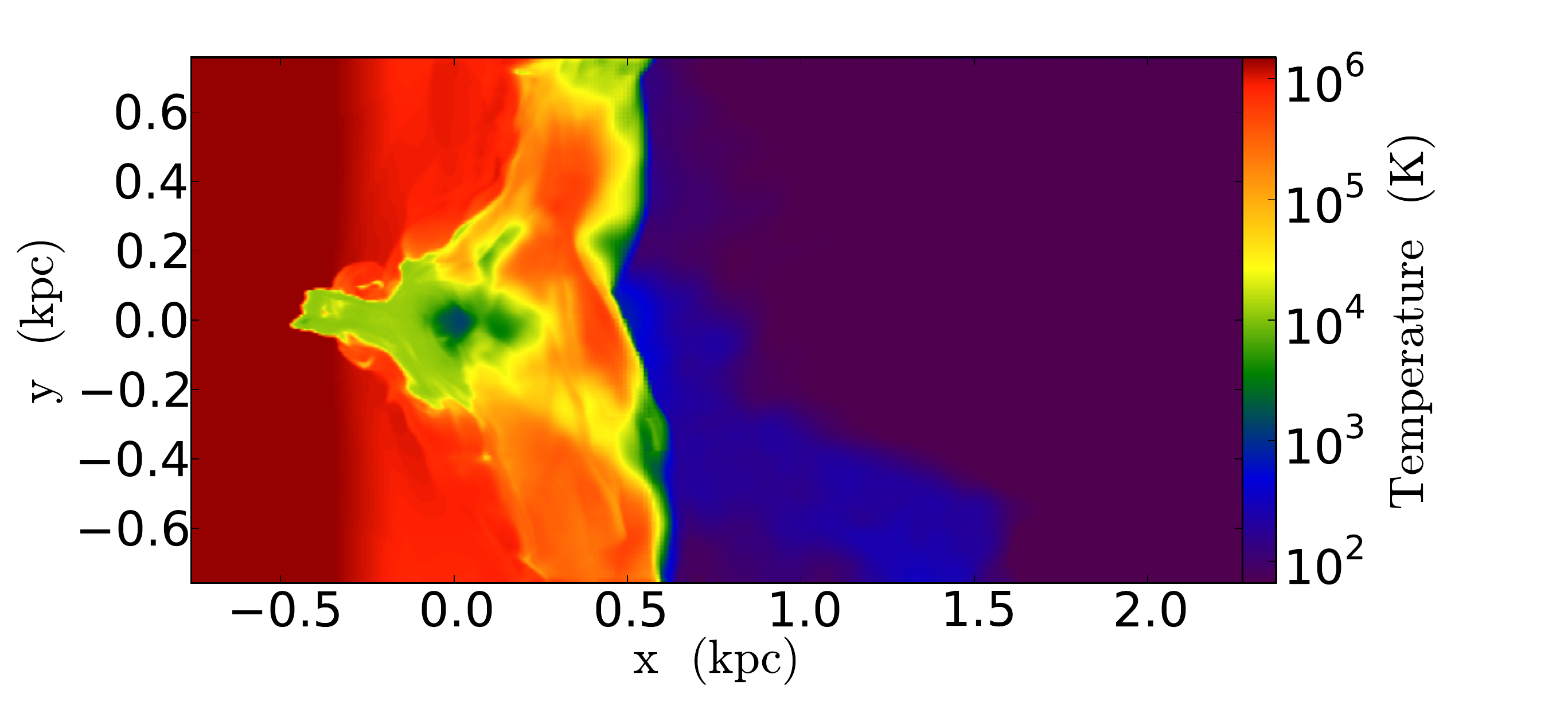}
\includegraphics*[scale=0.245, trim=95 0 162.5 21]{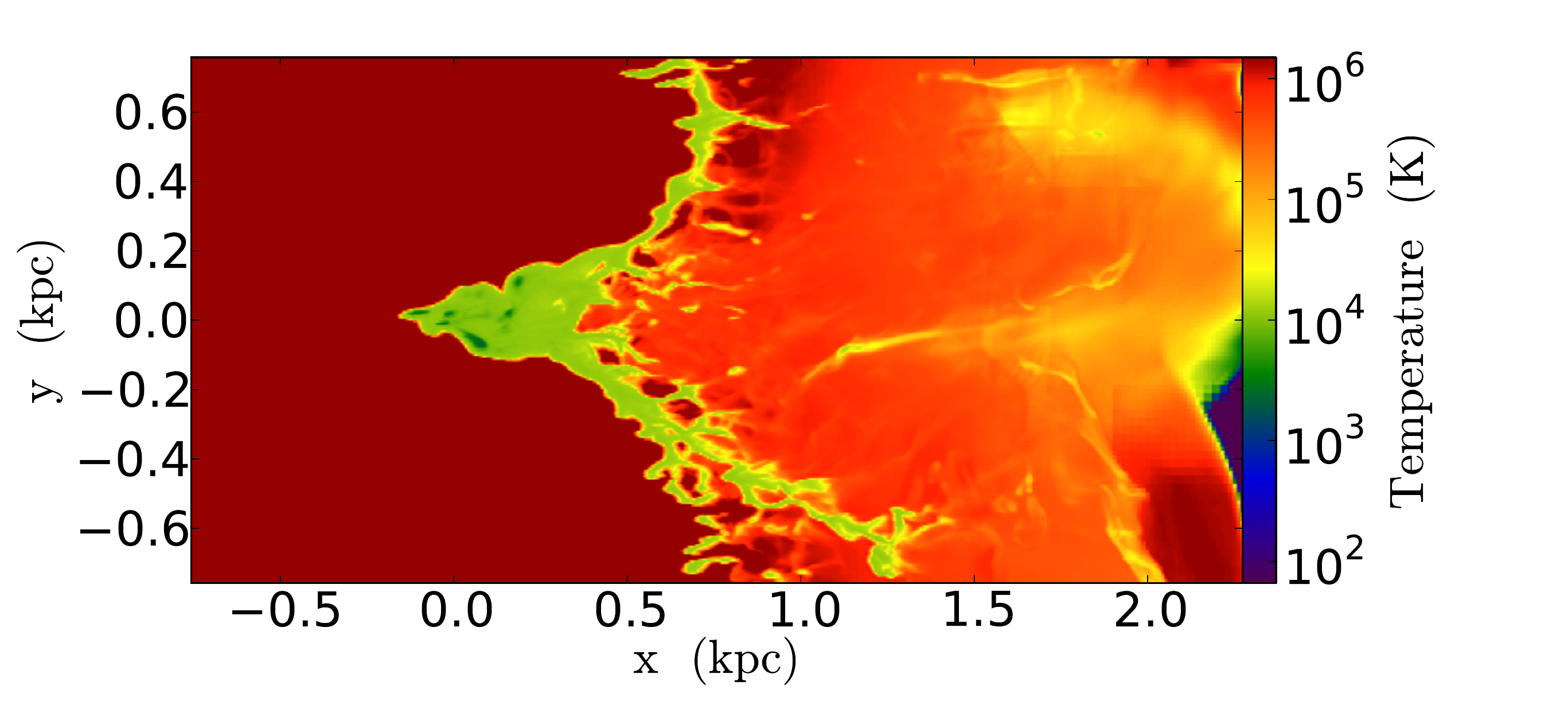}
\includegraphics*[scale=0.245, trim=95 0 0 21]{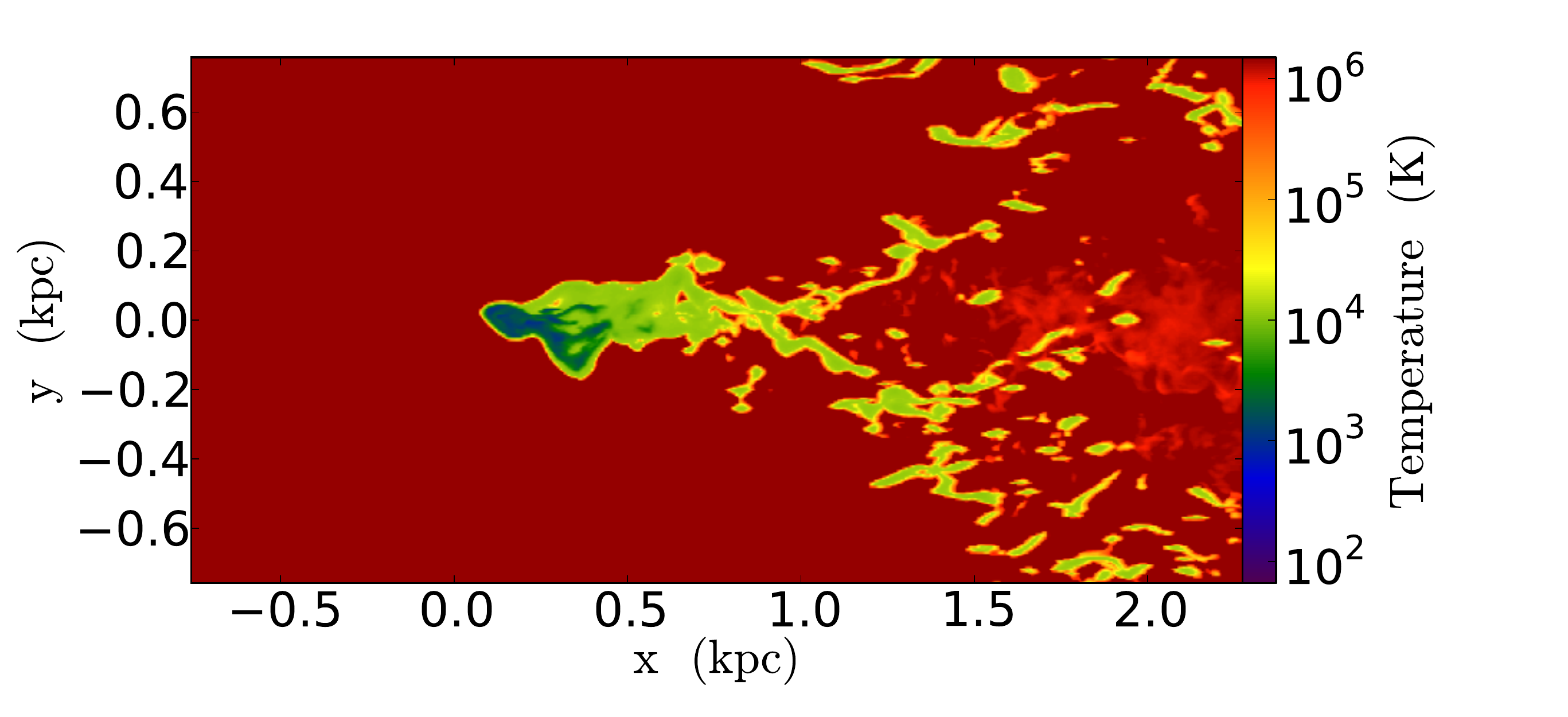}
\caption{\footnotesize{Column density (rows 1 and 2), column H$_2$ (rows 3 and 4) and projected temperature (rows 5 and 6) of the simulations varying the UV background, with FID 
shown in rows 1, 3, and 5, and PJ01 shown in rows 2, 4, and 6. Evolution increases from the 
beginning on the left, to when the outflow is just passing the minihalo in the middle, to when the 
outflow reaches the end of the box on the right.}}
\label{fig_j21}
\end{figure*}
We perform one model where a UV background is present. We allow this background to affect the chemistry rates in all cells, 
consistent with a optically thin medium everywhere. Thus, this is an upper limit to the effect of such a background, as the densest
clumps with an appreciable amount of H$_2$ and HD would self-shield, further delaying the effect of Lyman-Werner photons. We use 
a large UV flux of $J_{21} = 0.1$, as discussed in \sect{amr_out}. This value is taken from the fiducial value in GS10, where the effect of 
a UV background of this intensity coupled with the gas density in these models produces a disassociation timescale of about 1 Myr. This value
also is consistent with \citet{Ciardi05}, who find at such values you should just begin suppressing structures of mass $10^{6-7} \Msun$ from 
collapsing. In \fig{fig_j21} we compare the column density, H$_2$, and temperature 
of FID and PJ01. We see little variation between these two simulations. The PJ01 run has less H$_2$, as expected, resulting in a slightly
warmer collapsed gas. However, the molecular hydrogen forms sufficiently quickly so that the amount, even after accounting for loss to
disassociation, is large enough to cool the dense clumps faster than the dynamical time. We see little other differences between the two models.

\begin{figure*}[t!]
\centering
\includegraphics[scale=0.42]{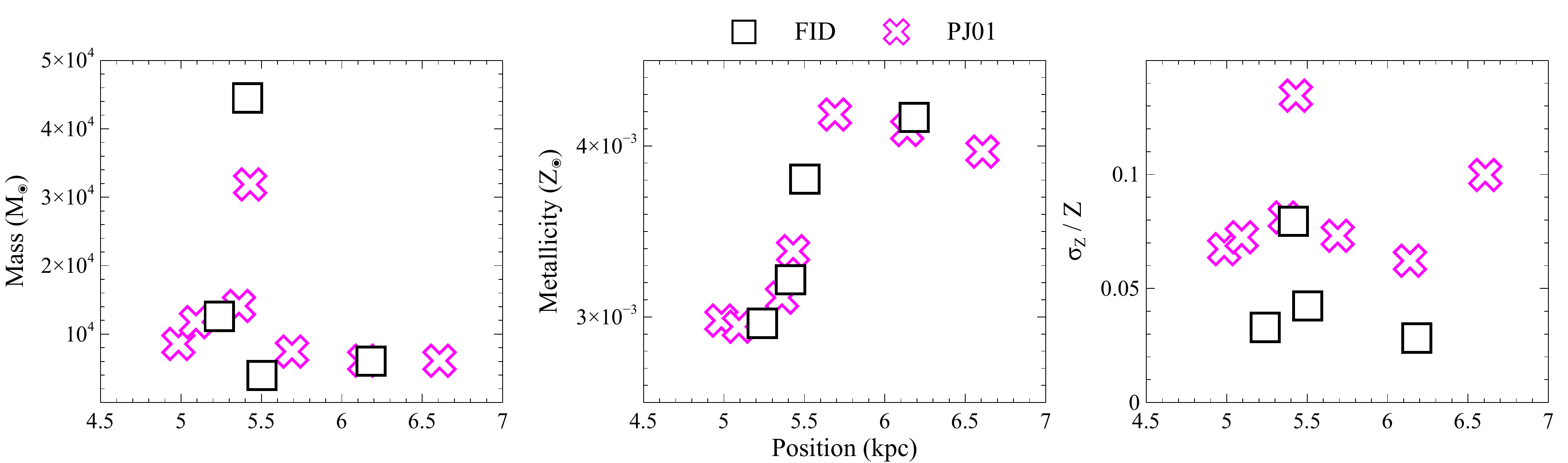}
\caption{\footnotesize{Comparison between the PJ01 (magenta crosses) and FID (black squares) simulations illustrating the dependence on the UV background. The particle masses (left), 
metallicity (middle) and relative metallicity dispersion (right) vs particles 
positions after 200 Myr are shown.}}
\label{fig_j212}
\end{figure*}
In \fig{fig_j212} we show the ballistic particles after 200 Myr of evolution for these simulations. We find very little differences between the two 
runs. PJ01 has 25.4\% of the minihalo's baryons collapsed into the cloud particles, slightly above the fiducial value. The UV background has slightly delayed the cooling
by molecular hydrogen, which delays the leading shock-minihalo interface from fragmenting, allowing for slightly more material to be compressed
along the main axis. This material is also slightly more enriched since it is the leading interface with the enriched outflow, thus the metals are
slightly increased in PJ01 compared with FID, and they have a larger dispersion. \vspace{10mm}

\section{Observational Signatures}\label{images}

Our various simulations of starburst-driven outflows interacting with minihalos consistently 
produce dense massive objects that we expect to form stars since molecular cooling should be sufficient to cool
these objects beyond the Jeans limit. While the resulting high-redshift cluster of stars are not observable with modern 
telescopes, their epoch of star formation should generate sufficient UV flux and Lyman-$\alpha$ photons that they may be
visible with upcoming observatories. Also, their compact nature should be sufficient to survive to modern day, such that 
we may be able to identify presently existing objects to compare with these clusters. Here we discuss both of these possibilities.

\subsection{Direct Observations}

To determine how bright the star-formation episode in our simulations would appear, we produced mock observations of these interactions. As we did not implement a star formation prescription 
in the simulations themselves, we implemented the post-processing technique of GS11B. We broke the forming ribbon of collapsed material in 175 stellar mass bins. Inside each bin we added the mass from
cells with density above $([1+z]/ 9)^3 \times10^{-23}$ g cm$^{-3}$, consistent with GS11B and the typical peak density of the ribbon. This assumes a star-formation
efficiency of 100\% for gas above this threshold, thus our estimates are upper limits. We interpolated this stellar
mass from one output to the next, yielding an effective star-formation history over the ribbon. We used the stellar population synthesis code bc03 \citep{Bruzual03}  
to estimate fluxes from a starburst population, and convolved these outputs with our ribbon's star-formation history as a function of frequency and age. We were also
able to make estimates on the Lyman-$\alpha$, H-$\alpha$ and H-$\beta$ lines by assuming a production of Ly$\alpha$ photons proportional to the star formation
rate, while the Balmer lines were estimate from case B \citep{Osterbrock89}. We do not consider extinction by dust, which combined with resonant scattering of
Ly$\alpha$ may result in a large optical depth. Again we stress that our final results are only upper limits. For more details the reader is referred to GS11B.

\begin{figure*}[h!]
\centering
\includegraphics[scale=0.6]{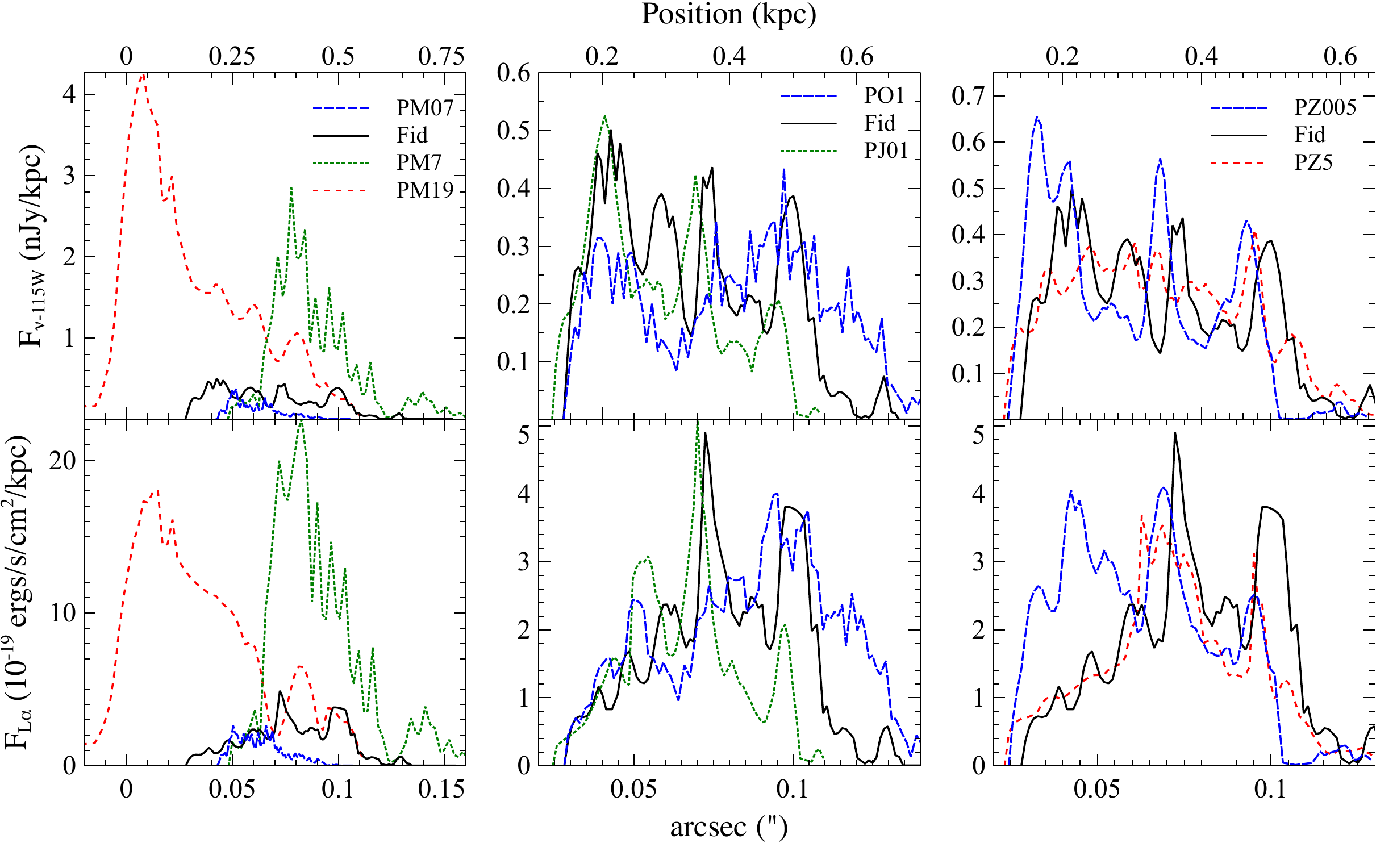} \ \ 
\includegraphics[scale=0.6]{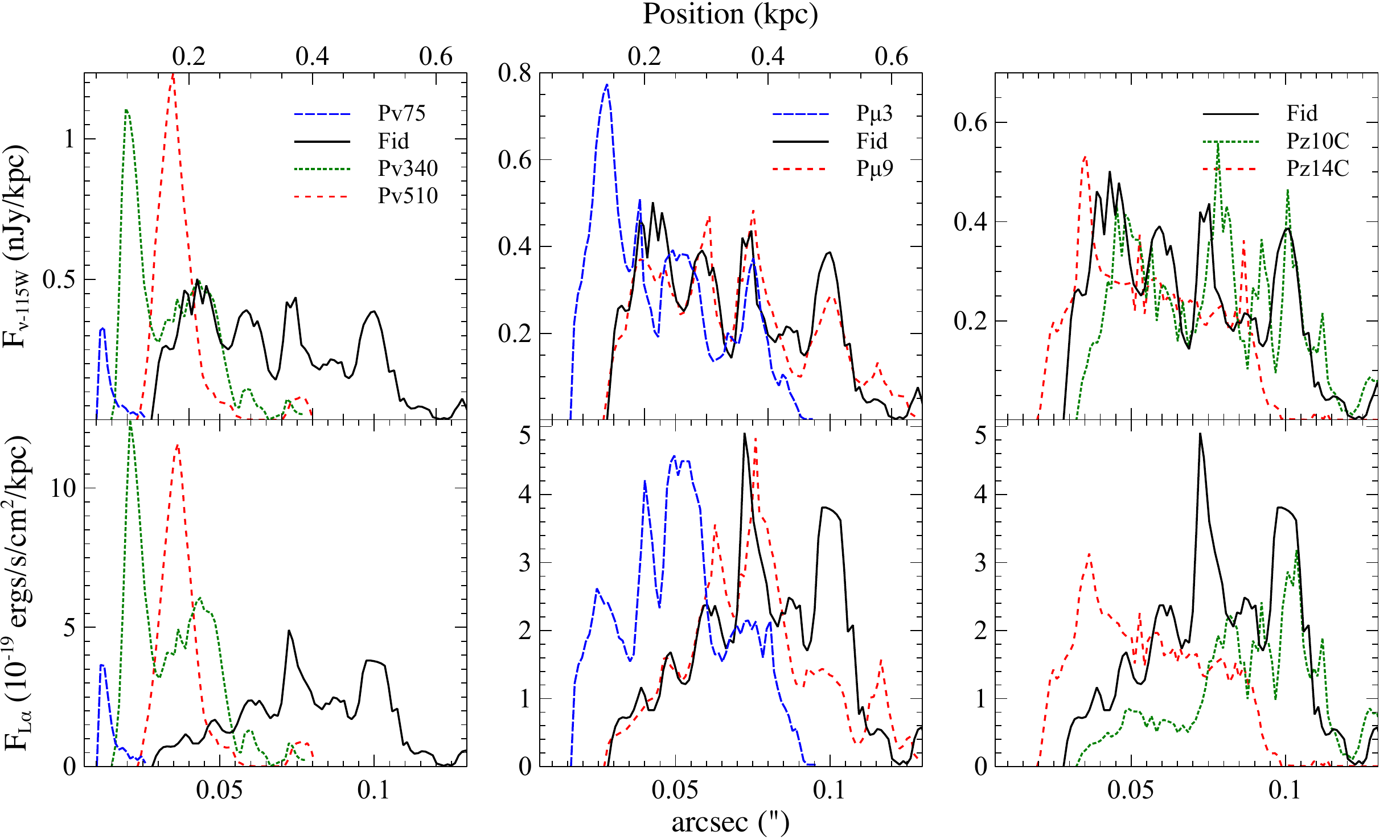}
\caption{\footnotesize{Comparison between the mock observations of several of our simulations 
as seen in the F115W wideband filter of \textit{JWST} (first and third row), and observed Ly$\alpha$ (second and fourth row). 
We illustrate the effect of different parameters in columns, with minihalo mass, shock orientation and UV background, shock enrichment, shock
velocity, shock surface momentum, and redshift, running from top left to bottom right, respectively.}}
\label{fig_observe}
\end{figure*}

We determine the
fluxes expected in each of the \textit{James Webb Space Telescope} (\textit{JWST}) bands, as well as from bright line emission that may be detected with future 
ground-based observing facilities, such as the \textit{Giant Magellan Telescope} (\textit{GMT}), the \textit{Thirty Meter Telescope} (\textit{TMT}), and the 
\textit{European Extremely Large Telescope} (\textit{E-ELT}). 
The expected F115W wideband filter detections and expected 
observed Ly$\alpha$ flux are presented in units of per ribbon length in  \fig{fig_observe} for several parameters. 
The physical and angular scales both assume an edge-on viewing angle.  Note that for the redshift parameter study, there is not a constant
mapping between position and observed angle, thus  we have set the $x$-axis to reflect the angular scale. 
We find that the final
ribbon of material is typically a fraction of a kpc long, slightly smaller than those found in GS11B. This is not surprising, as the
unidealized environment, including filaments, makes it more difficult for an efficient transfer of kinetic energy from outflow
to minihalo gas. 
We tabulate the integrated flux from all bands for each simulation in \tabl{tab_flux}. 
Note that this is the net flux from the star formation episode, and not the expected flux from the final globular cluster-like clouds.

We find that the expected flux in the \textit{JWST} bands is typically around 1 nJy/kpc, or just under 1 nJy when
integrated over the entire ribbon, with this flux greatest in the F115W band. These values are slightly below those of GS11B, and thus are not very optimistic for
detection by JWST, as typical observations with \textit{JWST} will have sensitivity down to only 10-20 nJy for a 10$\sigma$ detection
after a 10,000s integration \citep{Stiavelli08}. 

Fortunately, the expected fluxes from Ly$\alpha$, although a factor of a few less than those predicted in GS11B, are still well
above the expected detection limit of upcoming ground-based observing facilities. For example, the proposed
\textit{Near Infrared Multi-object Spectrograph} on
the \textit{GMT} is expected to find similar sources down to a flux limit of $10^{-20}$ erg cm$^{-2}$ s$^{-1}$
given 25 hr of integration (McCarthy 2008; GMT Science Case). Also, the
\textit{Infrared Imaging Spectrometer} on the \textit{TMT} will detect
Ly$\alpha$ sources at $z = 7.7$ with fluxes of $10^{-18}$ erg  cm$^{-2}$ s$^{-1}$
with a signal-to-noise ratio of 15 in only 1 hr of integration (Wright \& Barton 2009; TMT Instrumentation and Performance
Handbook). Finally, the planned \textit{Optical-Near-
Infrared Multi-object Spectrograph} for the \textit{E-ELT} will detect
sources with fluxes of $10^{-19}$ erg cm$^{-2}$ s$^{-1}$ with a signal-to-noise ratio of 8 in 40 hr
of integration \citep{Hammer10}. All three of these detectors would be sufficient
to observe the brightest of our objects.

These future ground-based observatories will employ the use of adaptive optics, hoping to get angular resolution in the range
of 0.1-0.3 arcsec (McCarthy 2008; Wright \& Barton 2009; Hammer \etal\ 2010), which means our objects will likely
be unresolved. However, using \eqn{rs} many should be within roughly 5 kpc of the starbursting galaxy driving the outflow, with masses $\gtrsim 10^8 \Msun$,
which will be easily detectable with \textit{JWST} broadband data. Thus we expect that future observations of starbursts with ground-based 
narrowband imaging may see barely resolved objects around the periphery of starburst galaxies 
extended away from the central  starburst galaxy.

\begin{deluxetable}{lcccc}[b!]
\tabletypesize{\scriptsize}
\tablewidth{0pc}
\tablecaption{Observed Flux
\label{tab_flux}}
\startdata
Simulation         &F115W$^a$  & Ly$\alpha$$^b$  &  Extent$^c$& Extent$^d$ \\ \hline \hline
FID                      &            11.9                        &            8.83                        &            0.766            &   0.155 \\ \hline
PO1                    &            11.3                        &            10.9                        &            0.871            &  0.176 \\ \hline
PM07                 &            2.90                        &            2.80                        &            0.460            &   0.0928 \\
PM1                   &            7.71                        &            7.52                        &            0.529            &   0.107 \\
PM23                 &            9.33                        &            5.38                        &            0.710            &   0.143 \\
PM7                   &            37.7                        &            39.2                        &             1.44             &  0.292 \\
PM19                 &            95.7                        &            52.4                        &            0.747            &   0.151 \\ \hline
Pv75	                   &            0.866                     &            1.02                        &            0.169            &  0.0341 \\
Pv125                &            2.87                        &            2.55                        &            0.962            &   0.194  \\
Pv340                &            9.47                        &            10.6                        &            0.362            &   0.0731 \\
Pv510                &            7.26                        &            7.01                        &            0.393            &   0.0793 \\ \hline
P$\mu$3            &            11.4                        &            8.62                        &            0.492            &   0.0993 \\
P$\mu$8            &            12.1                        &            8.05                        &            0.610            &   0.123 \\
P$\mu$9            &            11.3                        &            7.74                        &            0.616            &   0.124 \\ \hline
Pz10	                   &            9.51                        &            4.73                        &            0.652            &   0.152 \\
Pz14	                   &            6.16                        &            4.20                        &            0.472            &   0.140 \\ \hline
PZ005                &            12.3                        &            9.66                        &           0.625             &   0.126 \\
PZ05                  &            12.1                        &            9.00                        &            0.613            &   0.124 \\
PZ5                    &            12.2                        &            7.46                        &            0.689            &   0.139 \\ \hline
PJ01                  &            8.99                       &             6.38                        &           0.542             &   0.109
\enddata
\tablenotetext{}{$^a$ $10^{-2}$ nJy \ \ \  $^b$ $10^{-20}$ erg cm$^{-2}$ s$^{-1}$ \ \ \  $^c$ kpc \ \ \  $^d$ arcsec}
\
\end{deluxetable}

Finally, the observability of these objects are not very dependent on our parameters. The notable exceptions are minihalo
mass and shock speed. For large minihalo masses, there are significantly more photons emitted for larger mass. This is because
the high-mass runs  result both in significantly more collapsed baryons and higher densities, leading to
increased cooling and collapse. Similarly, there are significantly more photons emitted when
the shock is faster. This is because the increased shock speed results in more compressed material along the $x$-axis,
leading to denser material, and thus more stars.

\begin{figure*}[t]
\centering
\includegraphics[scale=0.5]{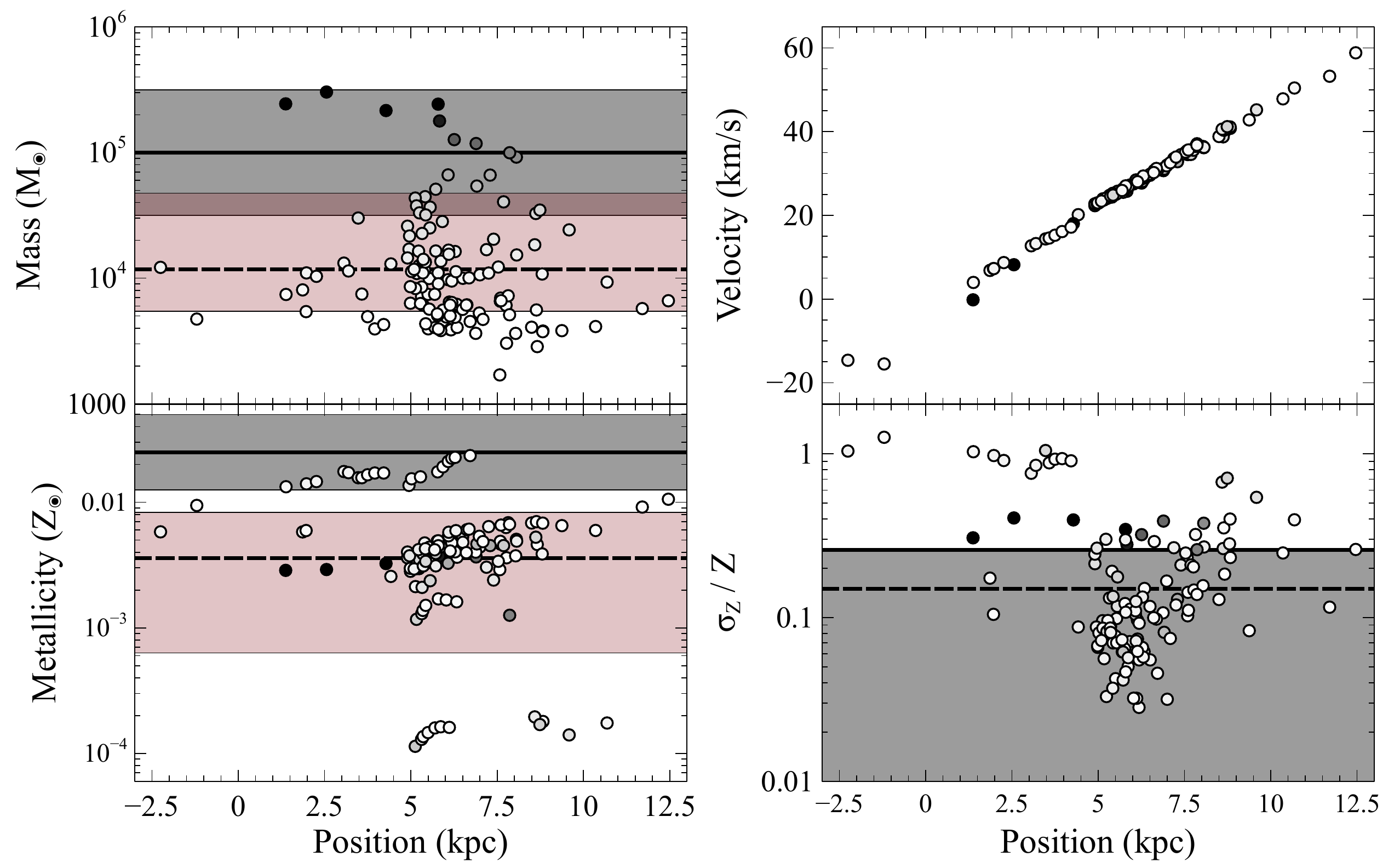}
\caption{\footnotesize{The distribution of final cloud particles for all of our simulations. 
The particle masses (top left), velocity (top right), metallicity (bottom left) and relative 
metallicity dispersion (bottom right) vs particles positions after 200 Myr are shown. The circles' color is indicative of the 
particle mass, darker for more massive particles. The solid lines show the average values for observed halo globular clusters and include a shaded
region corresponding to the 1-$\sigma$ spread in the observations. The dotted line and pink shaded regions mark
the asymmetric log-normal distribution of our results.}}
\label{fig_summary}
\end{figure*}

\subsection{Modern-day Analogs}\label{glob}

\fig{fig_summary} shows the 
distribution of final cloud particles whose masses are at least 1\% of their minihalo's baryon mass for all
of the simulations performed in this work. 

We find that the mass of our final cloud particles range from $\approx 10^{3.5}$ to  $10^{5.5} M_\odot,$ while the
metallicity of the final cloud particles are consistent with an asymmetric log-normal distribution of $\log Z = -2.44^{+0.36}_{-0.76}$. Our dispersion in metallicity
is typically about 0.06 dex. Our final cloud particles can be as far removed from their dark matter halo as 12 kpc, with the typical particles found removed by about 6 kpc.

The statistical sample of the final cloud-particles from all of our simulations shares many traits with modern
day globular clusters. First, the majority of our clusters are unbound from their dark matter halo, consistent with observations
of globular clusters which limit the amount of dark matter at less than twice the amount of gas (e.g., Moore 1996; Conroy et al. 2011; Ibata et al. 2013).
Also, their enrichment is significantly homogenous, with less than 0.1 dex for most particles. However, there is a tension between the metallicities of our final cloud 
particles and observations around the Milky Way and Andromeda, which find a consistent value of log$Z = -1.6 \pm 0.3$ (Zinn 1985; Ashman \& Bird 1993).
If most starburst outflows at these high redshifts were more metal-rich than we assume here (cf., Scannapieco et al.\ 2004), with values of $Z \simeq 0.3\Zsun$, 
then our results would be more consistent with these observations. Additionally, we have purposefully neglected to include subgrid turbulent mixing, as we were concerned 
with overestimating this process. This may also explain why we only found an enrichment of about 2\% of the outflow's metallicity, which may be
much lower than actually occurs in this interaction.

Finally, our typical mass of $\sim 10^4 \Msun$ has a large scatter, and although it is below the average mass of halo globular clusters in the Milky Way, with an observed distribution 
much closer to log$M = 5.0 \pm 0.5$ \citep{Armandroff89}, this typical mass is not unreasonable given our fiducial mass. Our process of making globular cluster-like clumps seems most efficient at
higher mass, and our largest mass, PM19, produced multiple final particles with masses of a few $\times 10^5 \Msun$. It would be interesting to study this parameter space further around such high-mass
minihalos, keeping in mind that provided $M_6 < 52 ([1+z]/10)^{-3/2}$, then the virial temperature of the halo will be less than $10^4$ K, and will be unable to cool on its own. That our objects are typically
less massive and less enriched than modern-day globular clusters may be consistent with some observations that show that their metallicity may scale slightly with mass (e.g., Ashman \& Bird 1993; 
Harris \etal\ 2006, 2009; Mieske \etal\ 2006; Strader \etal\ 2006; Peng \etal\ 2009). 
Additionally, since structure formation is hierarchical, we would expect a large abundance
of smaller mass objects. Destructive processes then act to destroy these smaller clusters. First, the minimum radius as a function of mass
is bound by mechanical evaporation (e.g., Spitzer \& Thuan 1972). Second, the maximum radius as a function of mass is bound
by ram-pressure stripping as the clusters move through the plane of the Galaxy (e.g., Ostriker et al.\ 1972). Thus, only
the largest of these objects should survive to today. Since the destructive processes set the minimum mass of the globular cluster mass distribution for a given age of the host galaxy,
we would expect that at higher redshift the globular cluster mass distribution should approach the distribution found in our very high redshift models.

\section{Summary and Conclusions}\label{Conc}

The early Universe hosted a large population of small dark matter `minihalos' that were too small to cool and form stars on their own.
These existed as static objects around larger galaxies until acted upon by some outside influence. Outflows, which have been 
observed around a variety of galaxies, can provide this influence in such a way as to collapse, rather than disperse the
minihalo gas. 

Here we performed SPH cosmological simulations using the GADGET code to produce realistic minihalos
and their environment  at $z\simeq 10$. We then implemented a derefinement technique for dark matter particles in the
AMR code FLASH. With this we mapped the SPH minihalos into AMR datasets and conducted a parameter 
suite studying the effect of energetic outflows, similar to those originating from high-redshift starbursting galaxies,
impacting inert primordial minihalos. This was a continuation of the idealized work of GS10, GS11A, and GS11B.

We endeavored to determine what effect the minihalo mass, outflow-minihalo environment,  outflow speed,  
outflow surface momentum, outflow metallicity, redshift, and UV background had on this interaction. We found that the general
interaction proceeded by first shock-heating the front of the minihalo, catalyzing the production of molecular hydrogen.
As the shock traversed the rest of the minihalo and molecular hydrogen continued to form, increasing the efficiency of cooling, while 
simultaneously compressing the minihalo towards the $x$-axis. The compressed minihalo gas continued to cool courtesy of the molecular hydrogen, 
while it was pushed out of the dark-matter halo. The shock then dissipated, removing roughly 75\% of the baryons from the system,
and collapsing the remaining 25\% into a cool, relatively homogenous ribbon that was no longer bound to the dark matter
halo. To compare, the idealized interaction studied in GS11B found upwards of 100\% of the baryons condensed into this ribbon. 
Typically the resulting ribbon of gas was enriched to $\simeq$ 2\% the metallicity of the outflow, and this material was then treated ballistically,
allowing for merging between separate clumps.

The most influential parameters are the minihalo mass, orientation, and shock velocity. The minihalo mass is essential in two respects.
First, changing this parameter requires changing the particular minihalo and environment mapped from the cosmological simulation. This produced
stochastic effects whereby individual minihalo asymmetry and environment  influenced the behavior of the interaction as well as its
final cluster distribution. This was made clear by comparing models FID and PM2, which varied subtly in mass, but had significantly different shapes
and environments. Second, the minihalo mass sets the abundance of gas available to the interaction, as well as the original minihalo temperature and
density profile. As the minihalo mass increased in out simulations, more mass collapsed along the $x$-axis, but less was able to escape the dark matter potential, 
a trend similar to that found in GS11B.

The orientation is also important as it sets the medium the shock interacts with before striking the minihalo. If an outflow was propagating along a filament,
in our simulations, it acted to delay the initial impact, leading to less momentum transfer. Also, this delay allowed the surrounding shock to strike the 
periphery of the minihalo first, causing it to preferentially collapse before being pushed along out of the halo. 
Thus, we find that when the shock is oriented along a filament, we have a larger fraction of baryons collapsed into the final cloud particles.

Finally, the shock velocity plays a dramatic role in the interaction. At very low velocities the shocks were incapable of collapsing much of the baryons
into a coherent ribbon on the $x$-axis, leading to a large number of low-mass, enriched clumps that were still bound to the dark matter potential. 
As the shock speed increased, baryons were condensed more efficiently, leading to more massive final particles that were more removed from the dark 
matter potential.

Our models have only considered a primordial chemistry network, and our UV background has assumed an optically thin medium everywhere. 
It would be interesting for future work to include additional chemical networks and a full treatment of different abundances, instead of a
simple tracer for metallicity. Although the amount of enriched material mixed in to the minihalo material is roughly two orders of magnitude below the amount of molecular
hydrogen formed during the shock interaction, the subsequent chemistry and cooling could have a non-negligible effect. This would be even more important
when a UV background is considered, which reduces the overall impact from molecular hydrogen. Simultaneously, a full treatment of the metal chemistry
could yield insights into atypical abundances in oxygen, carbon, and nitrogen, $\alpha$-elements, sodium and aluminum,  and heavier elements, 
some of which are thought to occur from enrichment from type II SNe, proton capture in the cores of massive stars, and
multiple stellar populations (e.g., Pilachowski et al. 1980; Fran\c{c}ois 1991; Colucci et al. 2013; Kacharov et al. 2013). We also encourage 
future work to explore a wider range of UV background intensities. These would need to be coupled with delays of incidence to insure the
minihalo is not photo-evaporated before the interaction begins, and should attempt to include self-shielding, so as to better constrain the ionizing
photons' effect in the densest regions.

We also produced simulated observations of this interaction using the post-processing technique from  GS11B. We find that the final
ribbon of material is typically a fraction of a kpc long, slightly smaller than those found in GS11B, and by using \eqn{rs} can be found around 5 kpc
from the starburst galaxy driving the outflow.
We find that the expected flux in the JWST bands is typically around 1 nJy/kpc, or just under 1 nJy when
integrated over the entire ribbon, well below the expected sensitivity of the telescope. However, we find expected Ly$\alpha$ fluxes of
 around $10^{-19}$ erg cm$^{-2}$ s$^{-1}$,
well above the expected detection limit of upcoming ground-based observing facilities. Although likely unresolvable, these objects should be visible within 
$2"$ of the starburst driving the interaction. Thus future observations of dense bright clumps around the periphery of starburst galaxies 
will be a clear demonstration of outflows driving star-formation in surrounding minihalos.

Finally, the statistical sample of the final cloud-particles from all of our simulations share many traits with modern
day halo globular clusters. They are often unbound from their host dark matter halos by about 6 kpc, but by as much as 12 kpc, 
and are chemically homogenous. Our mass distribution has a typical value of $10^4 \Msun$ with a large scatter, while the most massive cloud particles
have masses of a few $\times 10^5 \Msun$, consistent with modern day halo globular clusters. This indicates that most present-day globular clusters are analogues of the objects created from
minihalos with $M_6 \sim 20-60$, where we show the production of larger final clumps to be more efficient. Also, the enrichment of our objects is lower than seen in globular clusters. This suggests that a
way of modeling subgrid mixing must be included, or that starburst galaxies may drive winds as enriched as $\sim 0.3 - 0.5 \Zsun$.

We conclude that the interaction of starburst outflows with primordial minihalos is an energetic event that leads to several dense, uniformly-enriched
clumps of gas that would be expected to undergo star formation. Such interactions may be visible with the next generation of ground-based telescopes,
and may produce clusters that are the progenitors of modern-day halo globular clusters. Future observational work searching for globular clusters around younger galaxies where we expect 
their typical mass to be less than that found around the Milky Way, and theoretical work exploring
more of the high-mass range of the parameter space, coupled with a more complete treatment of chemistry, will be crucial 
for further demonstrating the connection between halo globular clusters and minihalos.

\

M. L. A. R. was supported by NSF grant AST11-03608 and the National Science and Engineering Research Council of Canada. 
E. S. was also supported by the National Science Foundation under grant AST11-03608 and NASA theory grant NNX09AD106. 
The authors would like to acknowledge the Advanced Computing Center at Arizona State University (URL: http://a2c2.asu.edu/), 
the Pittsburg Supercomputer Center (PSC) (URL: http://www.psc.edu/) and the Texas Advanced Computing Center (TACC) at 
The University of Texas at Austin (URL: http://www.tacc.utexas.edu) for providing HPC resources that have contributed to the 
research results reported within this paper. The authors would like to thank the Extreme Science and Engineering Discovery 
Environment for allocation time on TACC and PSC resources. We would like to thank Robert Thacker and Paul Ricker for discussions that greatly improved this study.
This work performed under the auspices of the U.S. Department of Energy by Lawrence Livermore National Laboratory under Contract DE-AC52-07NA27344

\end{document}